# WSO-UV FCU

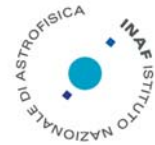
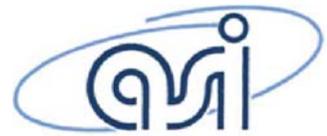

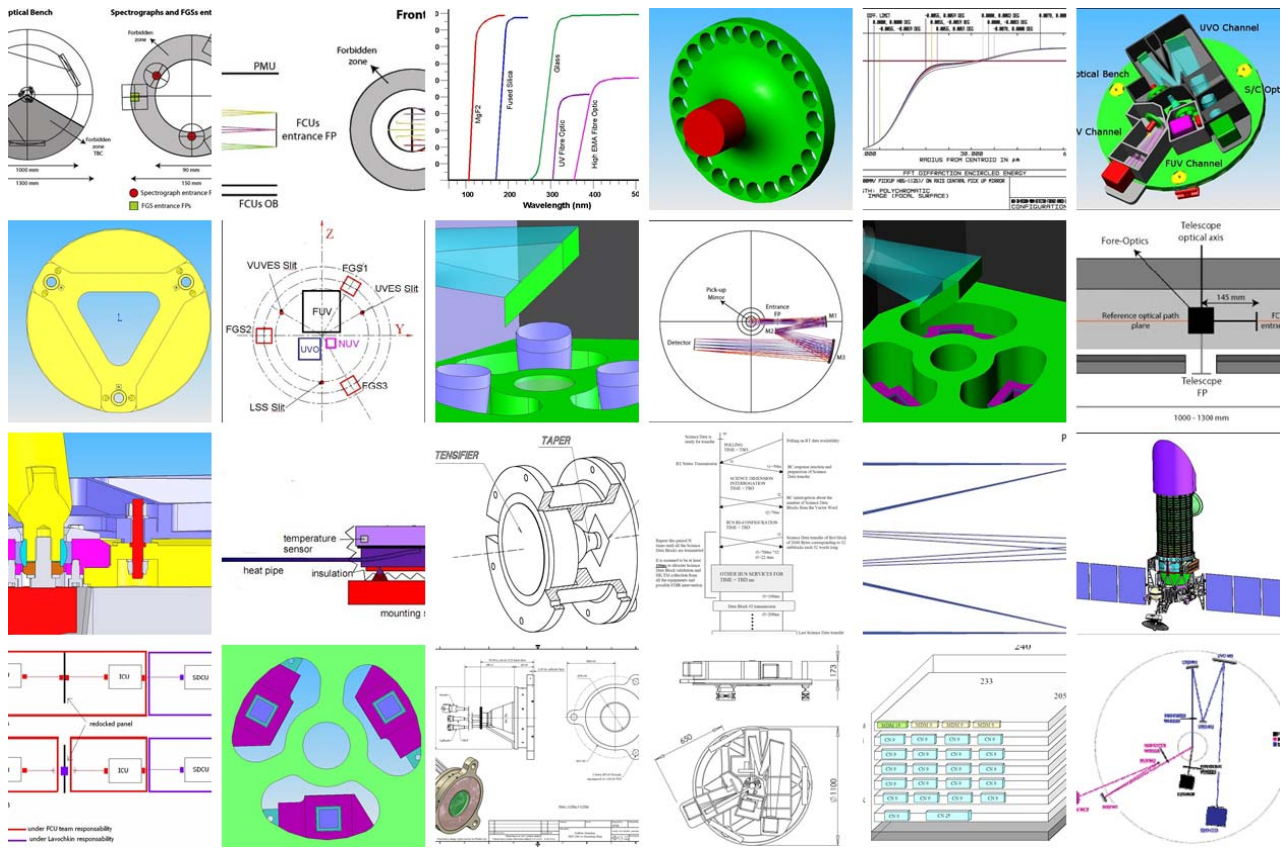

# Field Camera Unit
## Phase A Study Report

edited by
I. Pagano, R. Claudi, G. Piotto,
S. Scuderi, and M. Trifoglio

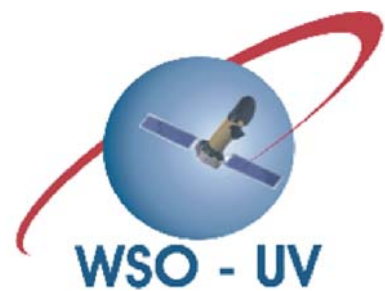

WSO - UV

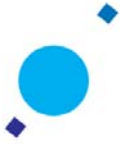

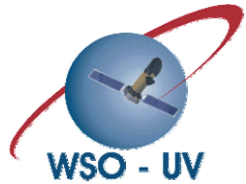

# Field Camera Unit

## Phase A Study Report

I. Pagano[1], F. Bacciotti[2], L. Bedin[3], F. Bracciaferri[4], E. Brocato[5], A. Bulgarelli[6], L. Buson[7], C. Cacciari[8], A. Capetti[9], A. Cassatella[10], E. Cavazzuti[4], R. Claudi[7], D. De Martino[11], G. De Paris[12], F. Ferraro[13], M. Fiorini[14], L. Gambicorti[15,16], A. Gherardi[15], F. Gianotti[6], D. Magrin[7], S. Marchi[17], G. Mulas[18], M. Munari[1], M. Nonino[19], E. Pace[15], M. Pancrazzi[15], E. Pian[19], G. Piotto[17], C. Pompei[20], C. Pontoni[1], G. Preti[20], S. Scuderi[1], S. Shore[21], M. Trifoglio[6], M. Turatto[7], M. Uslenghi[14]

1. INAF – Osservatorio Astrofisico di Catania
2. INAF – Osservatorio Astrofisico di Arcetri, Firenze
3. STScI, Baltimora, USA
4. ASI, Roma
5. INAF – Osservatorio Astronomico di Teramo
6. INAF – Istituto di Astrofisica Spaziale e Fisica Cosmica, Bologna
7. INAF – Osservatorio Astronomico di Padova
8. INAF – Osservatorio Astronomico di Bologna
9. INAF – Osservatorio Astronomico di Torino
10. INAF – Istituto di Astrofisica Spaziale e Fisica Cosmica, Roma
11. INAF – Osservatorio Astronomico di Capodimonte, Napoli
12. INAF, Sede Centrale, Roma
13. Dipartimento di Astronomia, Università di Bologna
14. INAF- Istituto di Astrofisica Spaziale e Fisica Cosmica, Milano
15. Dipartimento di Astronomia e Scienze dello Spazio, Università di Firenze
16. Consiglio Nazionale delle Ricerche, Istituto Nazionale di Ottica Applicata, Firenze
17. Dipartimento di Astronomia, Università di Padova
18. INAF – Osservatorio Astronomico di Cagliari
19. INAF – Osservatorio Astronomico di Trieste
20. Galileo Avionica, Firenze
21. Dipartimento di Fisica "Enrico Fermi", Università di Pisa

The *Phase A/B1 Study for the Italian Participation to WSO-UV* has been funded by the Italian Space Agency.

The activities have been carried out in the framework of the ASI-INAF Contract No. I/085/06/0.

**Participating Institutes and Industries**

INAF - Osservatorio Astrofisico di Catania
INAF - Osservatorio Astrofisico di Arcetri, Firenze
INAF - Osservatorio Astronomico di Bologna
INAF - Osservatorio Astronomico di Cagliari
INAF - Osservatorio Astronomico di Capodimonte, Napoli
INAF - Osservatorio Astronomico di Padova
INAF - Osservatorio Astronomico di Teramo
INAF - Osservatorio Astronomico di Torino
INAF - Osservatorio Astronomico di Trieste
INAF - Istituto di Astrofisica Spaziale e Fisica Cosmica, Bologna
INAF - Istituto di Astrofisica Spaziale e Fisica Cosmica, Milano
INAF - Istituto di Astrofisica Spaziale e Fisica Cosmica, Roma
INAF - Sede Centrale, Roma
Università di Bologna, Dipartimento di Astronomia
Università di Firenze, Dipartimento di Astronomia e Scienze dello Spazio
Università di Padova, Dipartimento di Astronomia
Università di Pisa, Dipartimento di Fisica "Enrico Fermi"
Galileo Avionica
Thales-Alenia Space Italia, Milano

# Table of contents

# List of Figures

# List of Tables

# PREFACE

The World Space Observatory UltraViolet (WSO-UV) project is an international space observatory designed for observations in the ultraviolet domain where many important astrophysical processes can be efficiently studied with unprecedented sensitivity.

WSO-UV is a multipurpose observatory, made by a 170 cm aperture telescope, capable of UV high-resolution spectroscopy, long slit low-resolution spectroscopy, and deep UV and optical imaging. With a nominal mission life time of 5 years, and a planned extension to 10 years, from a geosynchronous orbit with an inclination of 51.8 degrees, WSO-UV will provide observations of exceptional importance for the study of many unsolved astrophysical problems.

WSO-UV is implemented in the framework of a collaboration between Russia (chair), China, Germany, Italy, Spain, and Ukraine.

This book illustrates the results of the feasibility study for the *Field Camera Unit (FCU)*, a multi-spectral radial instrument on the focal plane of WSO–UV.

The book provides an overview of the key science topics that are drivers to the participation of the Italian astronomical community in the WSO-UV project. The science drivers here illustrated have been used to define the technical requirements for the conceptual and architectural design of the Field Camera Unit (FCU) focal plane instrument.

In Chapter I we show that WSO-UV will give a significant contribution to solve the key astronomical problems individuated by the ASTRONET consortium, and which are driving the European Space Agency "Cosmic Vision" program. Chapter II elucidates the scientific requirements for WSO-UV FCU instrument, discussed in Chapter I, which are translated in a list of verifiable top level requirements usable to make the conceptual design of the FCU instrument.

Chapter III is dedicated to the Field Camera Unit opto-mechanical design, its detectors and electronics subsystems. Finally, Chapter IV outlines the AIV and GSE plans and activities for the FCU instrument.

# EXECUTIVE SUMMARY

Access to the UV range is of fundamental importance in astrophysics, since most of the emission of hot (T>10,000 K) thermal processes occuring in a wide variety of astrophysical enviroments peak in this range. Moreover, UV spectroscopic and imaging capabilities are a fundamental tool to study plasmas at temperatures in the 3,000-300,000 K range. Also, the electronic transitions of the most abundant molecules in the Universe ($H_2$, CO, OH, CS, $CO_2^+$, $CO_2$) are in the UV range. The UV radiation field is also a powerful astrochemical and photoionizing agent.

Optical observations provide most of the standards for the study of stellar populations, and, coupled with UV observations, the rest frame fundamental information for the population synthesis models, and therefore the study of stellar population of high-z objects.

The Italian astronomical community has always been involved in scientific projects based on space-based UV-optical observations, and has an internationally recognized and highly competitive expertise in this branch of modern astronomy. The same community is therefore strongly interested to the development of a project aimed at building a space based observatory with spectroscopic and imaging capabilities in the UV-optical region, and strongly supports it.

The present document illustrates the scientific programmes that would strongly benefit from the use of WSO-UV and that the Italian community considers of primary importance. This document shall not be considered neither balanced nor fully representative of the scientific priorities either of the astronomical communities of the WSO-UV funding countries nor of the international community at large that will use WSO-UV. However, the scientific case here presented has driven to the precise definition of the science requirements for the imagers instruments on board of WSO-UV: the **Field Camera Unit (FCU)**.

Moreover, this document illustrates the science that can be explored thanks to the other focal plane instruments on board of WSO-UV:

**HIRDES**: The HIgh Resolution Double Echelle Spectrograph (PI, Prof. K. Werner, University of Tübingen, Germany – Industrial Contractor: Kaiser Threde – Funding Agency: DLR), that delivers spectra at R~55,000 in the102-180 nm range (VUVES), and in the 178-320 nm range UVES).

**LSS**: The Long Slit Spectrograph (PI: Prof. Gang Zhao, National Astronomical Observatories of China Academy of Science - Funding Agency: CNAS), that delivers spectra in the range 102- 320 nm, with spectral resolution R~1500–2500 with possible use of a long slit of 1x75 arcsec.

The scientific plans for WSO-UV are very ambitious, and span all of the astronomical research branches. WSO-UV will be operating in the second decade of this century, and it will be a fundamental tool for the development of astronomical knowledge, fully integrated with the many other space and ground-based observatories. In this document, we will show that WSO-UV will give a significant contribution to solve the key astronomical problems identified by the ASTRONET consortium, and which are driving the European Space Agency "Cosmic Vision" program and the NASA "Origins" program.

In particular, the data collected by WSO-UV will be used to answer to the following questions (as formulated by the ASTRONET consortium):

_Do we understand the extreme of the Universe?_ Important inputs for this problem will come from WSO-UV observations of: supernovae, gamma ray bursts, interacting binaries (millisecond pulsars, X-ray binaries, cataclysmic variables, blue stragglers, etc.), as well as observations of active galactic nuclei, their surrounding environment, and the accretion and outflow processes in their central black hole.

*How do galaxies form and evolve?* Important inputs for this problem will come from WSO-UV surveys in far-UV and near-UV of galaxy clusters at different redshifts, of the Virgo cluster, and from the extension to the far-UV and near UV of the GOODS ultra deep field survey. Surface brightness fluctuation technique will be used to investigate the unresolved stellar population in distant galaxies, while far-UV and near UV observations of galaxies in the local universe up to z>1 will be used to solve the still open problem of the UV-upturn in early-type galaxies and in bulges. Particular effort will be devoted to the origin and early evolution of our Galaxy, using as fundamental probe, among others, the Galactic globular cluster system.

*What is the origin and evolution of stars and planets?* Important inputs for this problem will come from WSO-UV observations of the stellar population in different environments (stellar associations, open and globular clusters, Local Group galaxies, including the dwarf satellites of our Galaxy). WSO-UV observations will provide fundamental inputs in the study of young stellar objects, including the processes of accretion and outflow. Additional information on stellar structure and evolution will come from asteroseismological studies, from the study of stellar magnetic activity, and of stellar variability. Efforts will also be devoted to the study of the interstellar medium. WSO-UV will also provide important data for the study of the extrasolar planet atmospheres (including a number of biomarkers). Finally, the study of Solar System bodies will benefit from observations with the different instruments onboard of WSO-UV.

The science team has concluded that the main scientific drivers of interest for the Italian astronomical community, and of the international astronomical community as well, can be carried out by a design of a Field Camera Unit (FCU) that employs three channels:

**FUV channel**: to satisfy the high sensitivity requirement below 200 nm, this channel is optimized in the range 115-190 nm, with 0.2 arcsec/pixel scale. Its design minimizes the numbers of optical elements in order to maximize the throughput. The FUV channel will be equipped by a 2kx2k pixel MCP detector and will have a FoV=6.0x6.0 arcmin[2]. This channel will be equipped with a filter wheel hosting broad and narrow band filters, neutral filters, and a R~100 light disperser, as described in RD4.

**NUV channel**: this channel provides close to diffraction limit images at 200 nm, with a 0.03 arcsec/pixel scale. It operates in the range 150-280 nm. The NUV channel will be equipped by a 2kx2k pixel MCP detector and will have a FoV=1.0x1.0 arcmin[2]. This channel will be equipped with two filter wheels hosting broad and narrow band filters, neutral filters, polarizes and a R≥100 grism, as described in RD4.

**UVO channel**: this channel is a near ultraviolet-visual diffraction limit imager, operating in the interval 200-700 nm, with a 0.07 arcsec/pixel scale, equipped with a 4kx4k pixel CCD, providing a field of view of 4.6x4.6 arcmin[2]. This channel will be equipped with a set of broad band and narrow band filters, a ramp filter, polarizers and a R~250 grism, as described in RD4.

In addition to the newly observed targets, with the synergy of the HST archive, WSO-UV will allow long term photometric and spectroscopic monitoring of a variety of astronomical objects. Moreover, the extended temporal coverage (up to 25-30 years using the HST archive) with high angular resolution UV-optical imaging will allow the measurement of relative and absolute proper motions with unprecedented accuracy. We estimate an accuracy of up to ~5 microarcsec for relative proper motions, and up to 10-20 microarcsec for absolute proper motions. These astrometric applications, just started with HST, de facto open a new, completely unexplored research branch. Astronomers have just begun to exploit this possibility. Equipped with cameras with diffraction limit imaging capabilities, WSO-UV will give the opportunity to fully develop this research activity, allowing proper motion measurements with the same accuracy of GAIA, but in a fully complementary way. WSO-UV+HST proper motions will be measured down to much fainter targets than GAIA, and in much more crowded environments. It is therefore a matter of fact that WSO-UV can be fully complementary, and in some cases competitive with GAIA, and this will happen long before (10 years at least) the full GAIA catalog will be available.

Because of the wavelength coverage, and because of the detector quality (for UVO) WSO-UV will have astrometric performances very similar to what is expected for JWST. WSO-UV will be complementary (in wavelength coverage) to JWST, and it will investigate a region of the electromagnetic spectrum more appropriate for the study of high energy phenomena, and of the local Universe stellar populations.

Because of the geosyncronous orbit, and of the operation mode, WSO-UV will be particularly suited for target of opportunity observations, allowing the study of a number of transients (including supernovae and gamma ray bursts) in a region of the electromagnetic spectrum almost completely unexplored.

The planned time-tagged capability in the far and near UV will allow high temporal resolution observations of various kinds of variable targets: from X-ray sources in the Local Group, to a number of exotica produced by the dynamical evolution in clusters and by the evolution of binaries, to YSOs and solar-like magnetic active stars which produce highly variable UV photoionizing and photexciting radiation fields that impact on the evolution of the protoplanetary systems and planetary atmospheres.

It must be also noted that WSO-UV will allow far-UV wide field (6.0x6.0 arcmin$^2$) surveys with unprecedented high angular resolution, very important in the study of star forming regions, and starburst galaxies. In this respect, the FCU will allow follow-up observations of the GALEX most interesting fields.

In conclusion, WSO-UV will provide a fundamental observational support for the development of our astronomical knowledge, in full synergy with the other groundbased and space observatories that shall be operating during the second decade of this century. Moreover, WSO-UV will allow to fully exploit the Italian expertise in UV-optical observations, it will allow to further develop it, and it will put Italian astronomers in a highly competitive track for solving a number of exciting, but challenging astrophysical open problems.

# Chapter I.

# Key Science Drivers to the Italian participation to WSO-UV

## 1. INTRODUCTION

The dramatic progress of astronomical discoveries of the past few decades is closely connected with the advances in technology and with the access to the entire electromagnetic spectrum thanks also to the development of space astronomy.

In this context, the impact of UV instruments in modern astronomy is of particular relevance, as testified by the undoubtful success of missions like IUE, HST, GALEX, FUSE, etc. In particular, the HST mission has shown the extraordinary importance of coupling UV-optical observations with high angular resolution imaging and spectroscopy. HST has provided a number of cornerstone results which allowed significant progress in all fields of astronomy (from planetary science to early epoch Universe). HST has perfectly filled the gap in the electromagnetic spectrum between the space gamma and X-ray missions and the near-IR IR space observatories, and this is another, often forgotten merit of that mission.

HST contribution to our astronomical knowledge is not only a commonly shared opinion among professional astronomers, but also by more general public. The attraction to astronomy and science in general by HST is of primary importance, beyond the scientific achievements of the project.

Unfortunately, even in case of full success of the SM4 servicing mission, HST operations will be terminated at the beginning of next decade and an UV-optical mission able to take the heritage of HST is badly needed and requested by the worldwide astronomical community.

The WSO-UV project will perfectly fill this gap. With a telescope with just half the collecting area of HST, but taking advantage of the modern technology for astronomical instrumentation, and of a high altitude, high observational efficiency orbit, WSO-UV will provide UV-optical astronomical data quantitatively and qualitatively comparable to the exceptional data base collected by HST.

It must be clear, anyway, that WSO-UV has not to be intended as a continuation of the HST project. Surely, the demand of the astronomical community of UV-optical high angular resolution space-based data and UV spectroscopy capabilities is still enormous, and it would justify a new analogous mission by itself. Even with half of its instruments working, and even if it can presently offer only technologically obsolete and worn out by the long permanence in space instruments, HST is heavily oversubscribed (by a factor >7 in 2007, Cycle 16).

In addition to the newly observed targets, with the synergy of the HST archive, WSO-UV will allow long term photometric and spectroscopic monitoring of a variety of astronomical objects. Moreover, the extended temporal coverage with high angular resolution UV-optical imaging will allow the measurement of relative and absolute proper motions with unprecedented accuracy, de facto opening a new, completely unexplored research branch. Astronomers have just started to exploit this possibility. If properly equipped with cameras with diffraction limit imaging capabilities, WSO-UV will give the opportunity to fully develop this research activity, allowing proper motion measurements with the same accuracy of GAIA, but in a fully complementary way. WSO-UV+HST proper motions will be measured down to much fainter targets than GAIA, and in much more crowded environments. And all this will be done many years before the final catalog of GAIA will be available.

In the following sections the **WSO-UV Italian Science Team** (see Appendix A) present the scientific case which has driven to the precise definition of the science requirements for the imagers instruments on board of WSO-UV: the **Field Camera Unit (FCU)**.

Moreover we will illustrate the science that can be explored thanks to the other focal plane instruments on board WSO-UV:

**HIRDES**: The HIgh Resolution Double Echelle Spectrograph (PI, Prof. K. Werner, University of Tübingen, Germany – Industrial Contractor: Kaiser Threde – Funding Agency: DLR), that delivers spectra at R~55,000 in the 102-180 nm range (VUVES), and in the 178-320 nm range UVES).

**LSS**: The Long Slit Spectrograph (PI: Prof. Gang Zhao, National Astronomical Observatories of China Academy of Science  - Funding Agency: CNAS), that delivers spectra in the range 102- 320 nm, with spectral resolution R~1500–2500 with possible use of a long slit of 1x75 arcsec.

# 2.  THE KEY SCIENTIFIC DRIVERS

Access to the UV range is fundamental for astrophysics since thermal phenomena at temperatures T>10,000 K occur in a wide range of astophysical events, with flux emission mostly in the UV. Moreover, UV spectroscopic and imaging capabilities are a fundamental tool to study plasmas at temperatures in the 3,000-300,000 K range. Also, the electronic transitions of the most abundant molecules in the Universe ($H_2$, CO, OH, CS, $CO_2^+$, $CO_2$) are in the UV range. The UV radiation field is also a powerful astrochemical and photoionizing agent.

Optical observations provide most of the standards for the study of stellar populations, and, coupled with UV observations, the rest frame fundamental information for the population synthesis models, and therefore the study of stellar population of high-z objects.

Optical groundbased observations are hampered in many ways by the atmospheric effects, and necessarily complemented by space observations. Adaptive Optics, and MultiConiugate adaptive optics can provide very high angular resolution observations on ground-based facilities, but in limited fields of view, with unstable PSF (which make photometric and astrometric measurements quite difficult), and only in near-IR spectral domain.

WSO shall provide high angular resolution UV-optical observations in stable conditions as only space observations allow.

The scientific plans for WSO-UV are very ambitious, and span all of the astronomical research branches. WSO-UV will be operating in the second decade of this century, and it will be a fundamental tool for the development of astronomical knowledge, fully integrated with the many other space and ground-based observatories (including the new generation ELT telescopes) operating in the same temporal interval. The space missions operating in the WSO-UV era will provide observational data both at shorter wavelengths (e.g. Symbol-X, possibly the extended XMM and Chandra missions, etc.) and longer wavelengths (e.g. GAIA, JWST, Herschel etc.) than those covered by an UV optimized mission like WSO-UV. WSO-UV UV and optical observations are a necessary, fundamental complement of the data set collected by presently operating and planned space and ground based observatories.

An effort to individuate the main astrophysical problems to be solved in the next decade has been done by the European ASTRONET group, and described in the "A Science Vision for the European Astronomy" document. The key astronomical questions individuated by the ASTRONET consortium, are the same ones adopted by the European Space Agency in its "Cosmic Vision" program, and by NASA for the "Origins" program. Our science team has adopted the ASTRONET document as reference.

The ASTRONET working group has proposed four fundamental questions to be addressed in the future astronomical research, and a number of related sub-questions. In the next sections, we will present the specific contribution of WSO-UV to the solution of these main problems. In this presentation, we will follow as much as possible the structure of the ASTRONET final recommendations.

The present document provides the main interests of the Italian astronomical community in the use of WSO-UV. Therefore, this document shall not be considered neither balanced nor fully representative of the scientific priorities either of the astronomical communities of the WSO-UV funding countries nor of the international community at large that will use WSO-UV.

## 2.1 Do we understand the extreme of the Universe

Astrophysics offers unique possibilities for probing fundamental physics in detail, beyond the level that can be explored in the laboratory. The obvious examples are gravity in the strong-field regime, and particle physics at energies above the TeV scale. From this point of view, astrophysics becomes a prime arena where the frontiers of physics can be advanced.

These extreme applications of astrophysics deal with grand and general themes. At the greatest extreme of scale, we find questions of cosmology: how the Universe came to exist in its current form, and the nature of its contents.

The origin of the Universe is arguably the greatest challenge in strong-field gravity, but it remains to verify classical strong gravity in the Universe today. Many candidate black holes exist, but so far there is no direct evidence for an event horizon. Equally, gravitational waves are inferred only indirectly from binary systems containing pulsars. One probe of the very centre of a black hole may very well come from the phenomena of jets and outflows; another may come from a better understanding of the most powerful celestial explosions: supernovae and gamma-ray bursts. These phenomena may be associated to a more general type of astrophysical objects, the interacting binaries, which may manifest themselves under many different forms.

With these ideas in mind, the ASTRONET working group has formulated the following fundamental questions.

1) How did the Universe begin?

2) What is dark matter and dark energy?

3) How do supernovae work?

4) How do gamma-ray bursts work?

5) How do black hole accretion, jets and outflows operate?

6) Can we observe strong gravity in action?

While it is certain that WSO-UV can provide fundamental inputs for the answer to all of these questions, in the following we will further exploit the main contribution of WSO-UV towards a solution of the 3), 4), 5), and 6) astronomical problems.

## 2.1.1 How do supernovae work?

Supernovae (SNe) are the final events in the evolution of massive stars (core-collapse SNe, observationally subdivided in types II, Ib, Ic and IIn) as well as of moderate mass stars in binary systems (thermonuclear SNe, observationally called of type Ia). Thanks to the very high luminosity and rather regular properties Type Ia SNe (SNIa) have been extensively used as cosmological distance indicators to trace the structure of the Universe at high redshift. SNe are also the primary contributors to the chemical evolution of the Universe and play a crucial role in triggering/quenching star formation.

In this framework UV observations are particularly important. They allow to probe the latest evolutionary stages of the evolution of stars of different masses, to determine the metallicity of the precursor stars, to detect the progenitor stars of core-collapse SNe on prediscovery images, to study the kinematics of the outer ejecta and its interaction with the circumstellar matter, to study the ISM and IGM in the direction of the parent galaxy.

A major problem to UV observations of SNe is due to the unpredictability of their appearance, which therefore requires ToO capabilities. For this reason, despite the wide interest, to date UV observations have been obtained only in a few special cases. In a recent review Panagia (2003, Supernovae and Gamma- ray Bursters, ed K. Weiler, Springer) reminds that only 36 SNe have been observed in the UV with IUE and HST and only a fraction got sufficient temporal coverage (>5 spectra) among which are 10 type Ia, 6 type Ib/c, 4 type II including SN1987A, and 3 type IIn. Considering that most of these observations have been collected with IUE satellite in the 80's, it is fair to say that the behaviour of SNe in UV is largely unknown. An attempt to perform a large, coordinated observational effort with HST by the worldwide SN community has been turned down by the failure of STIS.

In the last years a significant contribution to UV observations of SNe has been given by SWIFT satellite (Gehrels, N., et al. 2005, ApJ, 611, 1005), especially designed for quick response observations of GRBs but successfully used also for SNe. UVOT (the UltraViolet-Optical Telescope on-board of Swift) has collected interesting data on SNe following all the early stages with well sampled observations. Unfortunately its small aperture (30cm) and limited spatial (2 arcsec at 3500A) and spectral resolution (R~75) constitute a strongly limitation.

WSO-UV with its relatively large size (1.7m), high spatial and spectral resolution, and good ToO capabilities constitutes a unique possibility to explore the UV properties and make a significant step forward in understanding the physics of SNe. The main scientific targets for SN observations with WSO-UV are summarized as follows.

### 2.1.1.1 SN behavior in the UV (present vs. past)

#### Scientific background

The determination of the properties, rate, environments, and energy output of high redshift SNe is one of the main science drivers of HST legacy programs, as well as of future missions like the JWST. However, observations of high redshift SNe observed by HST in the optical, or by JWST in the near- IR, actually sample the rest-frame UV of these objects. Therefore, the characterization of the UV properties of nearby SNe in UV is of utmost importance to fruitfully exploit the potential of present and future high-z SN observations. In fact, knowledge of the SED, of the spectral features and their behaviour with time, are required in order to determine the type, estimate the luminosity class and the properties of the SNe.

Although sometimes it is argued that the properties of SNe in the UV can be obtained from optical observations at z~0.5-1, it is clear that these observations cannot account for the evolutionary effects that we are expecting to take place over such long time intervals.

**Key observations with WSO-UV**

Broad and intermediate band photometry (with the FCU) of nearby (D<20 Mpc) SNe is required to this aim. Magnitudes range between B=12-15 at maximum to 20 one year later and require therefore exposures from seconds to minutes for a 1.7m telescope. High angular resolution is required for disentangling the targets from the (typically) complex background of the parent galaxies and obtain accurate measurements. Low resolution spectroscopy (R~2000) over the entire spectral range will also be extremely useful. According to ETC of HIRDES exposure times should range between 0.5 and 3h to obtain S/N~20. ToO mode (i.e. flexible scheduling) is mandatory over time intervals of the order of two months.

### *2.1.1.2 The progenitors of SNe*

**Scientific background**

An embarrassing situation in SNe research is that we do not yet know much on SN progenitors. For a handful of normal SNII, whose progenitors are red supergiants, high resolution optical pre-explosion images have allowed to detect the exploding stars providing information on the mass and evolutionary status. However because of the short wavelength baseline, the errors on the mass determination are large.

UV images of nearby galaxies with FCU of WSO-UV will allow the a-posteriori identification of the precursors stars of other types of core-collapse SNe, like SNIb an Ic which are associated to long-GRBs. Indeed, the evolutionary tracks of very massive stars (with initial mass M>40 solar masses) foresee the explosions when the stars have lost their H and He envelopes and are in the WR regime. Their detection is therefore prohibitive in the optical and IR, but accessible in the UV.

For thermonuclear SNe we do not know how the WD reaches the Chandrasekhar limit (accretion form a main-sequence/red-giant companion or merge with another WD), whether it even reaches the Chandrasekhar limit, and how it explodes (deflagration, detonation or delayed-detonation). Establishing the SN Ia progenitor systems and explosion mechanisms will give greater confidence in the use of SNe Ia as cosmological probes, and will quantify potential evolutionary trends from progenitor age and initial composition.

Calculations of the explosion mechanism of SNIa have led to predictions that can be directly tested observationally. Specifically, state-of-the-art radiative transfer and spectral synthesis codes can be used to place constraints on acceptable progenitors and explosion mechanisms. For example, Mg II 2800 is an indicator of the boundary between explosive C and O burning. Elements produced by incomplete Si burning are used to identify the regions of complete and incomplete Si burning. These temperature transitions are critical parameters in models of SNe Ia.

**Key observations with WSO-UV**

The collection of a database of deep UV images of nearby galaxies will allow the identification of the blue progenitors of type Ib/c SNe. These images need to be of the highest spatial resolution and in several bands to allow reconstructing the SED of the precursor stars. As an example a star of initial mass Min=40 Msun explodes at an age of 5x10e6 yr, with absolute magnitude M(UV)= -5 to -3 depending on the band (F170W to F336W, Girardi private communication). According to the expected preliminary performances of the FUV, NUV and UVO arms of the FCU (ETC by S. Scuderi) it is possible to detect such stars for all galaxies closer than the Virgo cluster with exposure time of the order of 1-2h, thus verifying the theoretical evolutionary models.

Detailed UV spectroscopy of SNIa in the earliest epochs past explosion can be combined with ground-based spectroscopy and spectropolarimetry to study the evolution of the high- velocity spectral features, and therefore to provide constraints on the nature of the CSM around SNe Ia. High-res spectroscopy can probe the presence of CSM as in the case of SN 2006X (Patat et al. Science submitted).

High-quality UV spectra will permit measurements of explosion products, such as MgII 2800A, that will break degeneracies in current optical/near-IR observations, and will constrain models for the effects of temperature, density, and non-thermal ionization. Moreover early times UV observations can also help to determine the extent of C/O in the outer layers; this provides a discriminant between burning modes which leave different amounts of unburned C and O. In addition, early-time UV observations may show whether Ni and Co (best seen in the UV) from the combustion are present in the outer layers (high velocities) as predicted by some models.

Because of the large expansion velocities involved, low-res (R= 1500-2500) spectroscopy with LSS is ideal.

### 2.1.1.3 Interaction with circumstellar material

#### Scientific background

Ejecta-CSM interaction has been studied in several SNe allowing deriving the latest years of mass loss history of the exploding stars.  The interaction reveals itself in the photometric behaviour with broad light curves, whose shape depend on the CSM distribution, and in the spectra with high ionization lines, complex line profiles and blue continua.

The interaction with the CSM has been repeatedly studied around core collapse SNe, whose massive progenitors are expected to undergo strong stellar winds. The possible detection of similar phenomena in thermonuclear SNe might eventually shed light on their progenitor systems, but so far the search of similar phenomena have revealed unproductive (the peculiar case of SN2002ic being questioned by Benetti et al., 2006 ApJ,653,129). The only exception is  SN 2005ke, which has been tentatively detected in the X-ray by Swift (Immler et al. 2006ApJ, 648, 119I) from which it has been deduced an upper limit to the stellar wind from the progenitor's companion star with a mass-loss rate of Mdot=0.3 x 10e-6 Msun/yr. The same SN has been monitored also by Swift UVOT and showed a clear deviation from the normal monotonic trend of SNIa in the UVW1 and UVW2 bands. No such excess was observed in the optical band light curves. Possible effects of CSM interaction on UV emission are inverse Compton scattering of photospheric photons by hot electrons or a reduction of UV line blanketing due to heating and ionization by circumstellar radiation, but the phenomenon still deserve full comprehension. There is, therefore, indication that UV photometric observations of SNe might provide new insight on the ejecta-CSM interaction.

#### Key observations with WSO-UV

The overall UV monitoring of the UV light curves are essential to discern which targets show evidence of CSM interaction and to trigger more detailed spectroscopic observations. For the nearest (D<5-10 Mpc) SNe it is possible to study the interaction spectroscopically at high-res with the two HIRDES arms, UVES (178-320nm) & VUVES (103-180nm). Detailed analyses of UV spectra with IUE and HST have revealed that the lines can be produced in the CSM itself or within the ejecta depending on the epoch of observation, energy of the explosion, density and distribution of the material. Well sampled observations allow therefore a sort of tomography of the ejecta and the CSM, revealing the last few years of mass loss history of the exploding stars.

### 2.1.1.4 Determination of the reddening laws in external galaxies using SNe

#### Scientific background

The determination of the extinction in the Galaxy and in external galaxies is of paramount importance in several fields of astrophysics. Indeed only through a precise determination of the extinction it is possible to determine the energetics of celestial bodies. Moreover the determination of the extinction law provides clues on the nature of the Inter-Stellar Medium, in particular, on the dust particles.

A common measure of the slope of the extinction curve in the optical region is the dimensionless quantity Rv = Av / (Ab - Av) = Av / E(B-V). Determinations of the reddening law for many lines of sight in the local Milky Way indicate an average extinction law for diffuse clouds with Rv=3.1 (Cardelli, Clayton & Mathis, 1989, ApJ, 345, 245 - hereafter CCM). However, for a few directions, values ranging from Rv~2 to 5.5 have been found (Fitzpatrick, 2004 ASPC, 309, 33; Geminale & Podowski, 2005). CCM demonstrated also that a link exists between the UV and the optical/IR extinction in the sense that the extinction curves over the entire wavelength range in general constitute a one-parameter family, the parameter being Rv.

Little is known on the extinction law in other galaxies and our information derives basically from the Large and Small Magellanic Clouds, and M31. Supernovae are extremely bright objects reaching absolute magnitudes at maximum M~-19.5 and in principle can be used as lighthouses for the determination of the extinction properties between them and the observer. Several studies on the general properties of SNe have yielded in the past evidence for low values of Rv,. Although these results were suggestive of interesting properties of the intervening dust, the statistical nature of these works and the heterogeneity of the samples prevented the application to specific cases.

More recently new tools have been developed which make use of SNIa, the brightest and most homogeneous type of SNe. Indeed the comparative study of reddened and unreddened SNIa provides clues on the extinction law and the very nature of the dust inside the parent galaxies. A precise analysis has been performed on a few highly reddened objects by making simultaneous use of the spectral energy distribution (SED) both from spectra and photometry (Elias-Rosa, 2007 PhD thesis). This method so far applied to 3 highly reddened SNe [SN~2003cg, Elias-Rosa et al. 2006, MNRAS, 369, 1880; SN~2002cv Elias-Rosa et al. 2007, MNRAS, submitted; SN~2006X, Elias-Rosa (2007, PhD thesis) has revealed very low values of Rv (in the range 1.6 to 1.8) indicating the presence of dust grains of small size along the line-of-sights inside the host galaxies. This finding raises new questions. Is the peculiar extinction law derived in the direction to these SNIa common to all the highly-reddened objects? Do low-reddening SNIa follow similar extinction laws? What is the effect on the calibration of nearby SNIa? Are the extinction laws the same at low and high redshift? Is this effect "local", i.e. somehow related to the nature of the progenitor system, or due to the overall dust properties of the host galaxy?

**Key observations with WSO-UV**

To answer these questions one has to reduce the large uncertainties of the available SED and to extend it down to the UV where the attenuation is stronger. This can be easily achieved by the three arms of the FCU for all SNe discovered in the local Universe and, for the brightest targets, by the LSS.. It will be possible, therefore, to have more precise determination of Rv for the more reddened objects and to apply the method also to mildly reddened SNe.

Moreover, repeated observations along the entire evolution of the SNe will allow studying possible time evolution of the reddening due to local effects, like dust formation.

## 2.1.2 How do gamma-ray bursts work?

GRBs are brief (durations range from few milliseconds to hundreds of seconds) flashes of soft gamma-rays, produced in extragalactic explosions. They have initially bright and rapidly fading multiwavelength counterparts. GRBs have been found in a range of redshifts from locally (z = 0.0085) to the edge of the Universe (z = 6.29). In the optical, the flux can reach up to V~9 at some tens of seconds to few minutes after the explosion (optical flash), then it subsides following a temporal power-law and connects with the afterglow. However, the majority of counterparts are at the level of V~19-20 or fainter at this epoch, and then they fade further. Our knowledge of the GRB counterparts is much more limited in the UV than in the optical domain, due to the lack of flexible and sensitive UV facilities. Yet, the UV continuum emission is an efficient tracer of the synchrotron mechanisms - the most relevant active mechanism responsible for the spectral continuum - especially at early epochs; the UV absorption spectra carry much more critical information on the GRB environment and on the intergalactic medium than the optical spectra do, thanks to the high number of transition lines detectable at the UV wavelengths; finally the UV study of the GRB close environment is much more effective than in the optical, especially for the long-duration variety of GRBs, which are associated with massive stars, because these are expected to be located close to OB associations, and are therefore better investigated through their UV light.

**Key Observables**

The majority of GRBs have a redshift in the range 1.5-2. The optical/UV telescope UVOT onboard Swift detects about 25% of the accurately localized GRBs. About 10% of the GRBs localized by Swift have been also detected in at least one of the UV filters.

Two Swift GRBs are particularly important: they have been detected both in the UBV range and at NUV wavelengths (1930-2600 Å): GRB060218 (z = 0.03) is the second closest GRB (or better X-ray Flash, for the marked softness of its spectrum) ever localized. Like 3 previously detected low-redshift GRBs, it is accompanied by a Type Ic supernova. The bright flare detected by UVOT in all filters starting 150 seconds after the trigger time and lasting more than 1 day (before the SN rise) has been interpreted as SN shock breakout emission. While this interpretation is very much controversial, the data underline the potential of a rapid reacting and sensitive UV facility both in detecting the elusive initial signature of a core-collapse SN and the early physics of afterglows, such as the development of a cocoon, due to interaction of the relativistic blast with the circumstellar medium.

GRB060614 is also rather close, z = 0.125, and has been detected at thousands of seconds after the explosion by UVOT and NUV filters at a level of 50 $\mu$Jy in few minutes exposure and with 5 sigma significance (Holland 2006).

Prior to UVOT, a few UV studies of GRBs have been performed by HST.

While the turnaround time of HST is rather long and makes efficient and rapid follow-up very difficult, in a couple of cases the availability of the NUV band proved to be important or even crucial.

The first determination of the redshift GRB000301C has been made using a STIS-MAMA spectrum in the NUV region, 5 days after the GRB explosion, thanks to the identification of the Lyman break at the redshift of the source, z = 2.067 ± 0.025 (Smette et al. 2001). This has been later confirmed from the ground.

The host galaxy of GRB030329 (z = 0.168), observed with the HST ACS equipped with the High Resolution Camera and the NUV filter, appears morphologically very different than at optical wavelengths, due to its highly star forming character (Fruchter 2003).

The GRB targets of WSO-UV will consist in: **1) the low-redshift GRBs** (z < 0.5), which represent a minority, but will be studied in very good detail: UV imaging and spectroscopy of the point-like afterglow source, in both total and polarized light, will give details of the physics, including those related to a possible SN component, and of the chemistry of the close and intervening environment; UV imaging, and possibly low-resolution spectroscopy, of the host galaxy will allow an accurate study of the larger scale environment; **2) GRBs with 0.5 < z < 2.5**, for which WSO-UV will sample the rest-frame far-UV emission. These observations will be limited by the onset of the Lyman alpha forest and Lyman continuum and will be used to map the highly variable spectral energy distribution of the afterglow, especially at early epochs, and to determine the star-formation rate of the host, thus contributing to constraining this important parameter in high redshift galaxies.

WSO-UV, with a collecting area about 50 times larger than that of UVOT and with a significantly faster reaction time than HST can afford, represents a step forward with respect to those facilities for the future study of GRBs. In the following, we will describe in some detail the specific issues related to GRB science that WSO-UV imaging and spectroscopy can address. With the exception of the host galaxy study, all goals require time-critical observations.

### GRB Rates

The approximate estimates of the GRB detection and localization rates during WSO-UV lifetime are the following:

*GLAST* (starting in 2007) will localize ~200 GRB per year with error boxes of 5 to 15 square degrees. Only the brightest ones will have arcminute radius localizations (arcminutes). Among GLAST GRBs, about 20 per year will be simultaneously detected and localized by Swift.

*AGILE* (2007 and on) will localize with arcmin precision about 12-24 GRB per year.

*EDGE*, starting in 2015, will localize about 100 GRBs per year with arcsecond precision.

*SVOM* (former ECLAIR) will fly in 2011 and will localize about 100 GRBs per year with 10-arcmin precision.

*EXIST* (foreseen to be in orbit in 2020, astroph/0606065+Washington) will have a 2-5 times better sensitivity for GRBs than Swift; it will localize ~600 GRB per year.

## Observations with WSO-UV Imagers

### Standard afterglow model

The early spectrum of a GRB afterglow is due to synchrotron radiation and the measurement of the characteristic frequencies (absorption, peak, and cooling frequency) allows the determination of the physical parameters. The UV band is usually not sampled, because of the limited sensitivity of the available small facilities and to the long reaction times of HST, beside the extinction, both Galactic, intervening, and intrinsic to the GRB host, which makes UV observations more difficult than optical/IR. The optical source is usually brighter than the V = 22 in the first 5 days, so that a 2m class space telescope equipped with UV cameras should be able to follow the evolution of the synchrotron spectrum with exposure times of minutes to 1 hour.

Among other things, an important measurement consists in the determination of the light curve break, i.e. a change of slope thought to be related to the physical aperture of the GRB jet. If the source is bright, longer monitoring is possible, to follow the late time evolution and understand the interaction of the blast with the interstellar medium.

Imaging polarimetry is almost completely pioneering, and would represent an invaluable progress here, because it is very poorly known in optical and totally unexplored in FUV-NUV.

NUV imaging polarimetry presents the advantage, over optical polarimetry, that, especially at early epochs, the afterglow synchrotron spectral peak is expected to occur at FUV wavelengths, rather than NUV-optical, and therefore, for z ~1.5-2, UV polarimetry is an excellent diagnostic of the non-thermal phenomena active in the early phases. Moreover, at early epochs the rest-frame NUV-optical emission is mixed with the reverse shock emission, peaking at rest-frame optical-NUV wavelengths, which makes the rest-frame FUV a much "cleaner" tracer than the highly relativistic shocks typical of the proper afterglow (the reverse shock is only mildly relativistic and represents the interaction of the blast wave with the ejecta, not with the circumstellar medium). For this observing mode, reaction times of no more than 10-12 hours are necessary.

### Host galaxies

Hosts galaxies of GRBs are usually less luminous than L*, they are small and compact and can only be investigated with space telescopes. UV observations are scanty. The best targets are the low redshift hosts. These have usually fluxes of about 50 $\mu$Jy and can be studied by a 2m class telescope in the UV with exposures of the order of 1-2 hours. It is particularly important to study in the UV the regions of the explosion, to explore their relationship with star forming regions and OB associations.

*GRB-SN connection*

Observing the early (within the first minutes to hours) counterparts of closeby long GRBs and XRFs may allow the detection of the extremely elusive UV flash caused by shock breakout in the core-collapse SN originating the GRB. Such an observation would be of unique importance, especially considering the debate about the nature of the UV flare accompanying the XRF060218 and detected by UVOT in all UV and optical filters.

**Observations with WSO-UV Spectrographs**

*Determination of redshifts*

The case of GRB000301C has demonstrated that the NUV band may be critical for the redshift measurement. In particular, for GRBs located at $1.3 < z < 2.5$, so that their redshift cannot be easily detected via optical emission line spectroscopy (because the host galaxy [O II] and Ly$\alpha$ emission lines are redward and blueward of it, respectively), the redshift can be measured via Lyman break detection in the observed NUV band. Low-resolution spectroscopy with the STIS MAMA in NUV range with HST of the optical counterpart of GRB000301C *5 days* after the explosion, when the target had R = 21.5, allowed the detection of a spectral discontinuity with an exposure time of 8000 s.

*Physics*

In general, more rapid turnaround times are necessary, because the counterparts are on average fainter than that of GBR000301C. Fast reaction is certainly an asset if spectropolarimetry is adopted. Since the counterpart is fading with time as a power-law if index 1-2, this observing mode requires a reaction time of no more than 5-6 hours. Spectropolarimetry allows us to obtain information on the initial orientation of the magnetic field and on the role of the beaming, which is highly relevant and dramatically variable at these early epochs.

*Intervening and host galaxy absorption systems*

Redshifts can be measured also via high S/N absorption line spectroscopy of the early bright optical counterpart. Provided the afterglow is sufficiently bright and the Galactic and intergalactic extinction not too heavy, numerous absorption lines can be detected in the observed UV. Beside redshift measurement, this also allows the study of the intervening medium, including the close environment (host galaxy and circumburst region) of the GRB. A timely acquisition of the target is necessary (within 1-2 days of the explosion). High dispersion spectroscopy requires a very bright target, therefore reaction times no longer than 1 day are necessary.

In general, UV information is fundamentally important at $z < 1.6$, since then the optical spectrum cannot cover the Ly$\alpha$ region, making metallicity measurements impossible. In these cases, a spectrograph with a medium-high resolution (R=10,000 or higher) on a 2m-class telescope, provided the slew time is no longer than about 12 hours.

## 2.1.3 How do black holes accretion, jets, and outflows operate?

Active galaxies are an important class of extragalactic objects for many reasons. They are extremely interesting in their own right, representing one of the most energetic astrophysical phenomena, involving the challenging task of studying physical phenomena in extreme conditions and affected by the strong relativistic effects. Moreover the study of active galaxies attained a new fundamental role after the recent developments in our understanding of the close connection between nuclear activity and the process of galactic evolution and formation. It is in fact becoming clear that most (if not all) massive galaxies host a super-massive black hole (SMBH) in their centres (e.g. Kormendy & Richstone 1995). The tight relationships between the SMBH mass and the stellar velocity dispersion (Ferrarese & Merritt 2000, Gebhardt et al. 2000) as well as with the mass of the spheroidal component of their host galaxies (e.g. Marconi & Hunt 2003) indicate that they follow a common evolutionary path. There is also increasing evidence that in such evolution the feedback from nuclear activity plays a major role. In fact, the vast amount of energy released by the formation and growth of SBMHs must have had a major impact on how gas cooled to form galaxies and galaxy clusters (Silk & Rees 1998). The intense nuclear emission and, when present, the collimated jets, can influence the star formation history, and the overall evolution of their host galaxies, and represent a significant fraction of the total energy input of the ISM and IGM. Indeed, it has been recently suggested that the whole process of galaxy formation is part of a self regulated process in which the active nucleus plays a major role (Di Matteo et al. 2005) accounting, for example, for the high-mass truncation of the early-type galaxy luminosity function (Best et al. 2006). Since the growth of a SMBH manifests itself in one of the many different forms known as Active Galactic Nuclei, AGN represent our best tool to investigate formation, evolution, and later growth of SMBH, and its influence on the build-up of galaxies structures. These exciting developments have led to a renewed interest in AGN as they encompass many of the major cosmological and astrophysical issues we confront today: the ubiquity and growth of supermassive black holes, their relationship to the host, and the physics of the central engine itself.

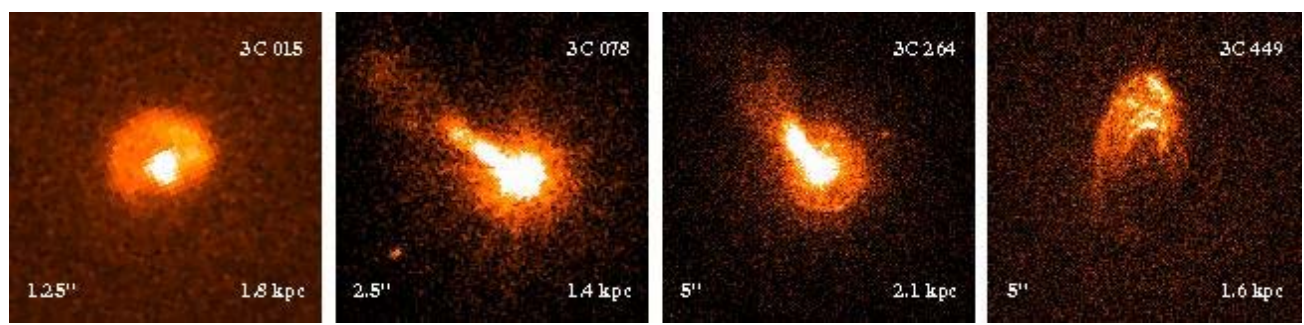

Figure 1: Four examples of the rich UV morphologies seen in HST images of nearby active galactic nuclei at sub-arcsec scale. Note the presence of low luminosity UV nuclei, of disk-like UV emission associated to a circum-nuclear starbust, and of UV synchrotron jets. The numbers at the bottom give both the angular and physical size of the images.

However, despite this breakthrough in our understanding of the SMBH/galaxy system, we still lack a clear answer to many key questions. First of all, why only a very small fraction of galaxies is active despite the fact that they all host a SMBH? Which is the triggering mechanism of activity and what sets its level? Which process maintains the link between the growth of galaxies (via star formation) and SMBH (via accretion) over cosmic time? The basic physical mechanisms of the accretion process are also still very unclear. For example, the AGN population is dominated by low luminosity objects, likely accreting at a very low Eddington rate, where the nature of the accretion flow might change to a radiatively inefficient regime (also known as RIAF, Narayan & Yi 1995). However, such flows are poorly understood theoretically and observational constraints do virtually not exist. Do RIAF actually exist? What sets the transition between RIAF and standard disks? And why only in about 10% of the AGN a relativistic jet is launched from the disk? Unveiling the origin of these dichotomies is of great importance not only for the study of AGN physics but also to understand the different manifestation of the coupling between nuclear activity and host galaxies.

## Key observables

### Observations with WSO-UV Imagers

Opening a window of high resolution UV observations represents a key for the study of AGN. In fact, the accretion flow radiates the bulk of its emission in this observing band, producing the so-called UV bump. Only observations in this crucial band (when combined in a broad-band view including optical imaging) can enable one to distinguish between different accretion modes and various models proposed. The high-resolution UV images will spatially separate the contribution of the AGN from their host galaxy also in low luminosity objects. This will extend the range of luminosity available for a detailed study from the high redshift objects, where the rest frame UV band is accessible from the ground, to these low luminosity AGNs. WSO-UV imagers should allow studying AGN down to a UV flux of $\sim 10^{-30}$ erg/s/cm$^2$/Hz with exposure times of $\sim 10^3$ s. This translates into a UV luminosity of L $\sim 10^{37}$ erg/s (at the distance of the Virgo cluster) several orders of magnitude fainter even than classical nearby Seyfert galaxies (e.g. L $\sim 10^{43}$ erg/s for NGC 4151, Bentz et al. 2006). But there is also the exciting possibility to unveil a totally unexplored population of even fainter AGN harboured in closer galaxies, effectively closing the gap between AGN and quiescent galaxies, when one considers that the bolometric luminosity of the nuclear source in our own Galaxy, Sgr A*, is L $\sim 10^{36}$ erg/s (Yuan et al. 2003).

Also the growth of AGN host galaxies, via star formation, produces its clearest signature in the UV band and WSO-UV observations shall then enable us to explore the synchronous growth of AGN and host galaxies. In particular, with high resolution WSO-UV imaging it shall be possible to directly see the presence of star forming regions associated to nuclear activity, and, via a broad-band spectral synthesis, to date and to quantify the star formation event. Surveys of star forming regions in different classes of AGN will enable us to compare quantitatively the star formation rate with the level (and possibly with the duty-cycles) of black-hole accretion.

Finally, there are several spatially extended components of AGN that can be optimally studied by a combination of high angular resolution UV and optical images, such as extended jets, AGN host galaxies and emission line regions. In particular line emission associated to AGN is often extended on a sub-arcseconds scale, out to quite large redshift. By studying their morphology with narrow-band imaging and by combining with long slit spectroscopy, one can quantitatively estimate the energy transfer from the active nucleus to the ISM via jets and winds (e.g. Capetti et al. 1999). In radio-loud AGN it will also be possible to study the properties of their UV and optical jets. In particular, imaging polarimetry provides us with detailed information on the structure of the magnetic field and on the processes of particle acceleration. This mode of observations represents also the only way to disentangle scattered (and thence polarized) light from emission produced in situ and it already proved to play a crucial role to test the unified models of AGN, based on orientation and scattering of otherwise hidden sources (Antonucci 1993). With this same technique it will be also possible to separate the contribution of genuine host galaxy emission in QSO from the nuclear light scattered into our line of sight.

### Observations with WSO-UV Spectrographs

Spectroscopic observations shall also enable to separate AGN emission from compact nuclear star forming regions looking for the absorption characteristic of the photospheres and winds of young stars, such a CIV λ1550 and SiIV λ1400 (Gonzales Delgado, et al., 1998). Ultimately, these studies will also provide us with clues on the role of mergers as triggers for nuclear activity. UV spectroscopy will also make possible to explore the properties of the Broad Line Region, produced by gas located in the immediate vicinity of the SMBH. Among the different areas of interest concerning the BLR, reverberation mapping appears to be the best tool to estimate SMBH masses out to large redshift (by measuring the lag between continuum and line variability, e.g. Peterson & Bentz 2006). However, this technique relies on measurements of rest frame UV lines in high-z AGN despite the fact that it is not properly calibrated from UV observations of nearby objects.

## 2.1.4 Can we observe strong gravity in action?

Other objects observable in the local Universe and that are the product of gravitational interactions are binary stars. In particular interacting binaries (IB) can experience strong gravitational interactions.

IBs are the most powerful high energy sources in our Galaxy and Local Group. They emit strong X-ray and UV radiation arising from the release of gravitational energy of accretion material onto compact objects.

Some of these systems may be rather intriguing and exotic objects consisting of a white dwarf, a neutron star or a black hole accreting from a late type companion with luminosity class ranging from main sequence (CVs) to red giants (Symbiotics), as well as ultracompact systems in which both components are degenerate stars.

### 2.1.4.1 Compact binaries in clusters

#### Scientific Background

There are many products of the interaction of binaries in star clusters. Among these we mention the millisecond pulsars (MSP), the low-mass X-ray binaries (LMXBs), the cataclysmic variables (CVs). Also blue stragglers (BSS) and extremely hot HB stars may be related to the interacting binaries (we will discuss these objects in Section 2.1.4.2).

*Binary Millisecond Pulsars*

MSPs are formed in binary systems containing a neutron star (NS), which is eventually spun up through mass accretion from the evolving companion (e.g. Bhattacharya & van den Heuvel 1991).

Even though the disk of the Galaxy has a total mass 100 times larger than the GGC system, more than 50% of the entire MSP population has been found in the latter. This is not surprising because in the galactic field the only viable formation channel for MSPs is the evolution of primordial binaries. At variance, in the ultra-dense stellar environment of a GC core, dynamical interactions can promote the formation of various kinds of binaries suitable for recycling the neutron stars into MSPs (e.g. Davies & Hansen 1998, MNRAS 301, 15).

The search for the optical counterparts to MSP companions in GGCs has just begun: only 5 optical companions to MSPs have been identified up to now in GGCs. Among them, at least three objects seem to be helium WD, roughly located on the same-mass cooling sequence. If further supported by additional cases, this evidence could confirm that in GCs, like in the field, the favoured by-product of the recycling process of MSPs is a low mass (M = 0.15-0.2 Msun) helium WD orbiting a MSP. However, a noticeable exception has been recently discovered in the center of NGC 6397: it is a MSP binary system containing a tidally deformed star with mass M=0.1-0.4 Msun (Ferraro et al. 2001, ApJ, 561, L93). It could represent a newly born MSP, or the result of an exchange interaction

Since most of the MSP companions are expected to be He-WDs, deep NUV-optical observations at high resolution are crucial in order to detect them in high density clusters. The optical identification of companions gives the possibility to perform dedicated photometric and spectroscopic follow-ups, allowing a deep insight on the physical parameters and the processes occurring in these binary systems. In practice, the careful examination of the light curve produced by ellipsoidal variations of a bright bloated companion to a MSP may allow to constrain the orbital inclination of the system. This measure coupled with the mass ratio derived from the velocity curve leads to a direct estimate of the pulsar mass. This is particularly interesting since the NS in these systems is expected to be somewhat more massive (due to heavy mass accretion) than the canonical 1.35 Msun. Because of that, the predictions on the maximum NS mass by different equations of state can be better checked and constrained.

*Interacting Binaries (LMXBs AND CVs)*

The first evidence of IBs in GCs was obtained through the discovery of X-ray sources. The so-called LMXBs, characterized by Lx > $10^{34.5}$ erg/sec and by bursts in their X-ray luminosity, are thought to be binary systems with an accreting neutron star. LMXBs are much more abundant (by a factor of 100) in GGCs than in the field, presumably because of the higher stellar density that leads to more collisional binaries. Given the existence of neutron star systems in GCs, one might expect to find many more analogous systems involving WDs. Binary systems where a WD accretes material from a late type low mass companion (i.e., a MS or a subgiant star), through a hot, UV-emitting accretion disk are observed in the field as CVs.

IBs in high density GCs are thought to be formed through stellar interactions. However, a comparable number of IBs has been recently identified in low density GCs (M13, NGC6712, M13), where stellar interactions are expected to be negligible. Thus, IBs should also form from another process (the evolution of primordial binaries), and in GGCs one can expect at least two types of these objects: those created by dynamical processes in dense clusters (Hut & Verbunt 1983, Bailyn 1995), and those resulting from primordial binary systems in low-density clusters (Verbunt & Meylan 1988). An extensive search for IBs in GCs with different properties would allow answering the following questions: Which processes generate IBs in GCs? Are the IBs in high density clusters different from those in low density systems (due to different formation mechanism)? How does the dynamical state of the cluster affect the IB population? Are the IBs in GGCs similar to those in the field? For example the relative paucity of IBs in the core of high central density clusters, with respect to moderate density ones would indicate that encounters between binaries and single stars are effective in destroying IBs.

The work done with HST has shown that several optically-faint but UV-bright objects are associated with faint X-ray sources and CVs. X-ray observations with Chandra and XMM are dramatically changing the game of searching CVs in GGCs, showing that, for instance in 47Tuc, many UV objects are indeed associated to X-ray sources. While this suggestion has been further supported by X-ray observations with Chandra and XMM for 47 Tuc Grindlay et al 2001, Ferraro et al 2001) definitive confirmation depends on direct observations of diffuse Halpha emitting gas. While the latter observations are made difficult by the large crowding in the R band, the cores of even the densest clusters are relatively open in the UV. Therefore, a "calibrated" UV search scheme that will overlap the H(alpha)/R band and the Chandra surveys will dramatically increase the sample of known CVs in clusters and allow to determine the correlation (or lack thereof) between UV, X-ray, and emission line searches of CVs in GCs. The collection of a homogeneous data-set with extensive WSO-UV observations will therefore be extremely valuable also in this context.

## Key observable with WSO-UV imagers

The identification of MSP companions needs broad-band NUV/UVO high-resolution observations. Also the shape of the light curve of the companion will yield hints on the possible ellipsoidal variations (in case of a bright bloated companion) or any peculiar heating processes of the companion, due to the radiation coming from the pulsar.

Combined FUV, NUV, and UVO broad band/H-alpha imaging shall allow detecting IBs in the core of GCs. The observational properties of IBs in different environments (GCs with different central density) will allow understanding IBs formation process in GCs.

## Key observable with WSO-UV Spectrographs

Phase resolved high-resolution spectroscopy with HIRDES of MSP companions will allow determining the velocity curve of the system. From the period and the velocity curve amplitude the mass function of the system can be determined. This can be used to infer the NS mass. Direct measures of the surface abundance pattern of the brightest companion can also put into evidence anomalies like that observed in the MSP companion in NGC6397.

Phase resolved high-resolution spectroscopy (the so-called Doppler tomography) with HIRDES will yield crucial information on the structure of the accretion processes (at least for the brightest IBs in GCs).

### 2.1.4.2 Other interacting binaries

**Scientific Background**

NOVAE: One of the most intriguing problems in CV's is the evolutionary link between subclasses and in particular between classical novae, which explode every million years or so, and dwarf novae which have short term variability (hours to days). Dwarf Novae could indeed represent the hibernated status' of classical novae. Chemical abundance determinations (through high resolution spectroscopy) and appropriate analysis of the circumsystem ejected material are crucial keys to verify whether the hyphotized evolutionary link can be supported on the ground of chemical analysis and ejection dynamics.

The ejecta of classical novae have a very rich emission line spectrum in the UV range, which includes transitions from astrophysically important elements such as C, N, O, He, Ne, Al, Si. Precise determinations of chemical abundances in the ejecta are primary indicators of the internal structure and mass of the WD primary component.

Another problem concerning classical novae is the dynamics of the ejection and how mass loss decreases after outburst. These items, that are clearly inter-related, are both poorly known, with the result that, at present, there is a fundamental discrepancy between predicted and observed values of ejected masses (these latter are often 10-100 times larger). In addition, the very few studies available of time-dependent expansion velocities, are not sufficient to definitively assess the problem of the origin of the complex absorption systems.

Current theories of the thermonuclear runaway causing nova outbursts provide rough predictions of the abundance, structure and mass of the ejecta as well of the energetics of the explosive events (Shore 2002). On the other hand, firm observational constraints on these items are missing since so far only a few objects have been monitored in detail, thus leaving great uncertainties.

NON-RADIAL PULSATORS IN CVs: The possibility to open a new window in the study the physical state of white dwarf primaries in CVs has come from the detection of non-radial pulsations in low accretion rate CVs (van Zyl et al. 1998; Warner & Woudt, 2004 Gaensicke et al. 2006; Szkody et al. 2007). Whether WD CVs are g-mode pulsators when entering in the ZZ Ceti-type instability strip has to be established by systematic searches. The UV can enable searches for pulsators because amplitudes are amplified in this range and simultaneously determine the effective temperature of the primaries. For a handful of objects for which UV data are available temperatures appear to be higher than those of single ZZ Ceti type stars. The UV range has therefore the power to provide first basis of asteroseismology of degenerates in binaries.

NOVAE IN EXTERNAL GALAXIES: Beyond our Galaxy there are great challenges in the understanding Interacting Binaries and evolution. Indeed, extragalactic Novae are the tracers of interacting binary stars, and of the star formation history of a galaxy (Yungelson et al. 1997; Neill & Shara 2004).Up to now, nova statistics in external galaxies are sparse and are based on optical (H alpha) surveys. The first problem with these surveys is their incompleteness, because they allow observations of novae in far away galaxies only close to maximum, and the Halpha magnitude declines rapidly. Moreover, there are problems in imaging the most crowded regions.

SUPER-SOFT X-RAY SOURCES: Supersoft X-ray Sources (SSS) are extremely interesting as likely progenitors of type Ia supernovae (e.g. van den Heuvel et al. 1992) but they are often transient (or recurrent) in X-rays. It turns out that they are instead mostly steady bright UV sources, especially in the Far UV. In a recent projects based on Galex imaging of M31 in the regions outside the bulge, it was found that 50% of SSS are detected at FUV or NUV wavelengths, or both (Nelson & Orio 2007, preprint). The identification and monitoring of the large population of SSS in the bulge of M31, M33 and other Local Group galaxies can enable to understand the secular evolution of these binary systems and hence in turn to allow one to understand progenitor evolutionary paths towards SNIa.

*Accretion Processes*

While X-rays provide information on the hottest accretion shock regions, the UV is the range where the inner regions of accretion discs and/or of magnetically confined accretion flows dominate both the continuum and the emission lines of high ionization species. IUE first and HST later have provided important results helping to improve accretion disc models and magnetic accretion flows (e.g. Horne et al. 1994; Orosz & Wade 2003; Eriksson et al. 2004; Haswell et al. 1997; de Martino et al. 1999; Belle et al. 2003). However large discrepancies between observations and theory especially in the far UV range still exist for disc accreting systems. Indeed, our knowledge of vertical and radial temperature structure needs to be verified and improved through appropriate observations.

*Outflows*

Outflow phenomena such as those occurring in Dwarf Novae (e.g. Sion et al. 2004) can be efficiently studied in the UV because the onset of outbursts in the UV, the time-lag with respect to X-rays and optical ranges are fundamental to determine the energetics and to identify the responsible mechanisms.

Important information for Dwarf Novae can be obtained from the analysis of the P-Cygni profiles seen in the high ionization lines present in the UV range. Recent observations have indeed revealed that winds might be collimated (Knigge & Drew 1997). However, just a few observations are so far available, so that it is not possible at present to clarify whether a basic relation exists between wind features and disk luminosity (Hartley et al. 2002).

**Key Observables**

The study of Interacting Binaries then requires UV observables of different kinds.

To diagnose the physical conditions of accreting plasma and mass loss one requires broad band UV (and optimally down to optical range) SEDs as well as UV emission line profiles as HI Ly alpha, O V, CIV, NV, SiIV, and HeII to be monitored at the rotational (mins) and at the orbital (hours, days and even years for Symbiotics) periods and on timescales of days, months, years to trace spectral evolution during parossistic events.

To determine stellar properties of underlying components both FUV and NUV detailed line analysis are essential to separate the contribution of the unexposed and exposed atmospheres. Also, spectropolarimetry in the UV can allow one to identify high magnetic field systems

To obtain a statistical significant sample of systems to trace different paths of binary evolution (e.g. Thermal Time Scale Mass Trasfer (Schenker et al 2002) one need to map CNO diagnostics which are easily identified in the Nitrogen, Carbon and Oxygen lines in the FUV and unobservable in the optical range.

To allow a systematic search for non-radial pulsators in DA WD primaries one need to monitor the FUV light with high time resolution as well as to cover this range with moderate-high resolution.

Population studies in external galaxies definitively require FUV coverage to isolate the counterparts of X-ray sources to be monitored at high temporal resolution as well as wide UV spectral coverage. Monitoring at different epochs to follow the evolution of outbursts (Dwarf Novae and Novae) is essential in this respect.

WSO-UV with its multifold observing (including ToO) capabilities can allow one to achieve a great step ahead in the understanding of interacting binaries and their evolution.

**Observations with the WSO-UV imagers**

**UV Imaging**

*SSS and Novae in external galaxies*: the FUV and NUV absolute magnitude of a Nova is around value -7.5 or slightly higher for several months, during the so called constant bolometric luminosity phase (Gallagher & Code 1974, Gallagher & Starrfield 1976, 1978). The H alpha magnitude remains at comparable value for only less than a month, declining steeply in a short amount of time. WSO-UV opens therefore a completely new era for the study of extragalactic Novae, allowing rather complete statistics based on both the FUV and NUV ranges, with observations of galaxies within ~10 Mpc once every month or two. The high resolution of the cameras will allow one to image also crowded areas like the bulge of M31, in which about 15 novae per year are observed. Any extragalactic region that is not heavily affected by absorption is basically open for this new type of surveys. SSS comprise of a variety of sources ranging from Very Soft (VSS), SSS and Ultraluminous systems (ULX). WSO-UV will be able to identify and monitor the large population of SSS in the bulge of M31, in M33, and in many other galaxies even outside the Local Group by means of FUV and NUV broad band imaging.

*Imaging Galactic Novae*: It is predicted that CVs undergo Novae explosions during their lifetime (Shara 1986). The census resides on those systems which have been recorded to undergo Nova explosions. For these there are still discrepancies in the turn-off times as well as in their temperature evolution. Also, recent Galex UV observations (Shara et al. 2007) have revealed extended shell around a Dwarf Nova, which, if confirmed in other systems, would confirm theoretical predictions. A systematic imaging survey of Dwarf Novae, magnetic systems and recent Novae in FUV and NUV with broad band filters with temporal coverage of a few hours in one filter as well as one narrow band image can provide strong constraints on nova evolutionary theory.

**Fast UV photometry**

Search for non-radial pulsators: A systematic search for ZZ Ceti-type pulsators in short period (Porb<100min) low mass accretion CVs with FUV broad band fast photometry can allow a proper census of pulsating DA WDs. Expected pulsation amplitudes are of a few tens of mmag and with periods ranging from a few hundreds of seconds to a thousand of seconds. The important WSO-UV spectroscopic connection can allow determining the white dwarf temperature and hence the instability strip of WDCVs pulsators.

*QPOs in Magnetic CVs*: QPOs of a few seconds were identified in a few magnetic Polar systems in the optical. Such variations are believed to arise from shock oscillations at the strongly magnetized polar regions. High frequency QPOs were found to be blue and indeed also observed by HST in the UV (Larsson 1995). High temporal resolution UV photometry using broad band and narrow band filters centred on HeII emission line can allow one to trace oscillation through the accretion region and to identify the driving mechanis which is unknown so far.

*Dwarf Novae Oscillations*: Dwarf Novae during outbursts have been recently found to show quasi-periodic oscillations whose amplitudes and frequencies change during the evolution of the outburst (Warner et al. 2003). These DNOs are the corresponding low frequency QPOs found in LMXRBs in the X-rays. While detailed information on outburst evolution can be provided by spectroscopy, fast photometry in the continuum and emission lines in both FUV and NUV ranges can allow tracing the DNOs not only in temporal domain but also with wavelength.

## Slit-less spectroscopy

NUV spectroscopic low resolution capabilities of WSO-UV to observe X-ray source counterparts at UV wavelength represent a unique observing tool to provide alternative coverage to broad band photometry which instead will be required for the dimmest sources.

## Spectro-polarimetry

The cyclotron emission from strongly magnetized (B>20MG) white dwarfs in CVs was detected so far in the optical-nIR range. The extension to UV wavelengths of spectropolarimetry can allow one tracing high cyclotron harmonics. Also, fast photometry with polarized filters can allow one to follow the variability of circular and linear polarization along the rotational period of the accreting white dwarf. This provides strong constraints on magnetic field strength and topology especially for the highest magnetic field systems for which cyclotron at low harmonics are optically thick at long wavelengths.

## Observations with the WSO-UV Spectrographs

Both LSS and HIRDES (FUV and NUV) spectrographs will provide complementary information on interacting binaries.

## LSS observations

*Accretion flows*: The moderate resolution of LSS and fast timing capabilities will allow to study in great details the accretion flow geometry including magnetically confined flows and discs. The relative large spectral coverage is essential to identify different components which can be resolved at the spin (magnetic systems) and orbital periods and thus to trace temperature radial dependence in the accretion discs and magnetized flows. The combination of information from LSS and HIRDES (see below) will further determine the accretion geometry.

*Anomalous line ratio systems*: LSS resolution is greatly suitable to identify in FUV spectra anomalous line intensities of O V, CIV and NV, which diagnose CNO processed accretion material. To date only 11 CVs are known to display great enhancement of Nitrogen over Carbon and Oxygen (Gaensicke et al. 2005). A systematic census of CV emission line properties is required since theory predicts that about one-third of accreting white dwarf systems should have followed non-standard evolutionary paths.

*Disc winds*: To understand the physical mechanism of disc winds it is of fundamental importance to observe a large sample of systems at different inclinations as the appearance of disc dominated systems is strongly modified by inclination. Also, the onset of short time scale variability is not understood as some systems show rapid variations while others do not.

*Outbursts*: LSS can provide strong constrains on the evolution of outbursts. The observations - to be performed with tight sequence according to the type of explosion - can allow inferring velocities, abundances and energetics of the explosion in Novae. Duty cycle of Dwarf Novae and time delay from FUV to NUV provide constrains on the propagation of heating within the disc.

*Temperature determinations in accreting DA WDs*: LSS coverage down to HI Ly alpha will allow the determination of DA white dwarf atmospheric temperatures as well as to identify Zeeman components in high field systems.

*SSS in the Local Group*: UV or FUV spectroscopy of SSS has so far been possible, and with low resolution, for only the four brightest SSS in the Large Magellanic Cloud (Hutchings et al. 1995, Gaensicke et al. 1998, van Teeseling et al. 1999, Schmidtke et al. 2004). With LSS it will be possible to obtain a clear determination of the interstellar column density from the HI Ly-alpha absorption line, and OI and Si II lines, and thus clearly constrain the spectral fit of broad band CCD-type X-ray observations of SSS and to derive reliable temperatures and bolometric luminosities. It will be possible to detect emission lines intrinsic to the sources, (e.g. HeII, C IV, O IV, OV, and NV etc.). All these lines and the correlation of their variations with X-rays/optical luminosity will allow to model accretion and wind outflows.

**HIRDES observations**

*White dwarf atmospheres*: While low resolution spectroscopy is sufficient to determine the effective temperatures of DA white dwarfs, abundance determinations and rotational velocities of accreting primaries can be obtained only with high (R>20000) resolution spectroscopy.

*Accretion flow mapping*: Doppler tomography from phase-resolved spectroscopy at the primary rotational and/or orbital periods is known as one the most powerful method to infer the structure of accretion flow. HIRDES covering both FUV and NUV ranges can allow diagnosing strong emission lines of N V, He II, CIV, Si IV and Mg II. Components originating in the accretion flow but also in the irradiated atmosphere of the companion will provide crucial information on geometry and energetics. Irradiation of the companion star is important aspect to understand the physical status of donor stars and hence the binary evolution. This is especially important in the ultracompact AM CVn type binaries where the donor star is a white dwarf as these systems are could be one of the progenitor channel for SNIa.

## 2.2   How Do Galaxies Form and Evolve

We know very little about the nature of dark matter and dark energy, and we are more familiar with 'ordinary matter', made of protons, neutrons, electrons, etc. Yet, much remains to be done to fully map the evolution of this 'baryonic' component of the Universe. The finite volume accessible to astronomical observations is bounded by the sphere at a redshift of about one thousand emitting the microwave background, beyond which the Universe is fully opaque to radiation. Within this volume, three quarters of all the baryons we could, in principle, detect lie between redshifts of seven and a thousand. These are the Dark Ages, between the epoch of recombination (~400 000 years after the Big Bang) and the most distant galaxy so far detected (~750 million years after the Big Bang). No direct evidence has yet been gathered for any kind of event in the Dark Ages, in spite of density fluctuations having grown by many orders of magnitude during this critical half billion years in the life of the Universe. Nevertheless, our first glimpses at redshift seven reveal that the young Universe was by then almost completely re-ionized, while stars, galaxies, and quasars had begun to form and shine, many with the metal-rich signatures of even earlier generations of star formation. Understanding this rapid build-up of stars, metals, galaxies, and supermassive black holes, as well as the subsequent transformation of these young objects to the present-day Hubble sequence of galaxies is a major challenge. As the Universe was re-ionized, most baryons were heated by stellar radiation and mechanical energy input from exploding stars (supernovae) as well as from active galactic nuclei powered by supermassive black holes. As a result, over 90 per cent of the baryons were left in a diffuse intergalactic medium. Nearly half of even the local intergalactic baryons have yet to be detected. Understanding these cycles of formation and evolution of galaxies, and the intergalactic baryons and metals (elements heavier than Boron) that link them, will require a clever combination of observational and theoretical approaches.

The key questions are:

1) How can we peer into the Dark Ages, and map the growth of matter density fluctuations from their tiny size at redshift one thousand to the formation of the first stars and galaxies?

2) What are the dominant sources for re-ionization of the Universe: star light, black hole powered active galactic nuclei, or even decaying supersymmetric particles? How long did the process take?

3) How did the structure of the cosmic web of galaxies and intergalactic gas evolve?

4) What are the histories of the production and distribution of the metals in the universe, within and between the galaxies?

5) How was the present-day Hubble sequence of galaxies assembled, as traced by the buildup of their mass, gas, stars, metals, and magnetic fields?

6) What is the detailed history of the formation and evolution of our own Galaxy, and what lessons does it hold for the formation and evolution of galaxies generally?

While it is certain that WSO-UV can provide fundamental inputs for the answer to all of these questions, in the following we will further exploit the main contribution of WSO-UV within the 5) and 6) astronomical problems.

## 2.2.1 How were galaxies assembled

Understanding the formation and evolution of galaxies is driving most of the current efforts in observational cosmology. Galaxies are thought to form inside the dark matter (DM) halos produced by gravitational instability from the seed initial fluctuations. The evolution of the DM component is successfully predicted from first principles, because it evolves under the sole action of gravity. However, the predictive power of theory drops dramatically when it comes to the luminous, baryonic component of the universe. Complex physical processes come into play, and it is no surprise that the attempts at modelling galaxies have so far attained limited success.

A number of different investigations with WSO-UV can provide fundamental inputs to the problem of galaxy formation and evolution.

### 2.2.1.1 FUV/F150W, NUV/F220W, and UVO F300W imaging of GOODS/UDF Deep Fields

**Scientific background**

Ongoing, community wide efforts in observing selected fields aim to answer some of the most profound questions connected to these processes:

a) The mass assembly of galaxies. The full spectral energy distributions (SED) of galaxies with measured redshift (in particular the rest-frame near-IR data from Spitzer/IRAC) provides the best measure of the stellar mass of galaxies, and thus trace the mass assembly history of galaxies from z~5 to z=0. Spectroscopic line-widths provide dynamical mass estimates for galaxy sub-samples, establishing a connection between stellar and dark halo masses.

b) The onset of the morphological types of galaxies in relation to their environment. The morphologies of galaxies observed today are believed to be produced by a combination of initial local conditions and evolution processes linked to interactions and mergers of smaller building blocks. Multi-wavelength information, combined with the knowledge of the underlying large scale structure from redshift information and projected mass estimates from weak-lensing maps, allows to evaluate how galaxies build up in relation to the local matter density.

c) Cosmic history of star formation. High-redshift star formation can be traced by the rest-frame ultraviolet, far IR, mm, radio, and emission line measurements. Such multiple SFR indices for the same galaxies obviate most concerns about systematic errors in measuring SFRs at high z.

The overall SEDs also constrain extinction and stellar population ages.

*The evolution of the AGN population.* Deep X-ray surveys provide the best census of accretion onto SMBH throughout the Universe due to the penetrating nature of X-rays (which mitigates absorption bias) and the fact that AGN host galaxies provide minimal dilution in X-rays. X-ray surveys have found by far the highest density of AGN on the sky (> 5500 per sq. deg as compared to ~400 per sq. deg for the deepest optical survey). The X-ray observation in these large areas such as CDF-S, enables to pin down the slope of the AGN X-ray luminosity function (XLF).

d) Evolution of early-type galaxies. The reddest elliptical and early-type spiral galaxies are dominated by very old stellar populations, and therefore point to the earliest epochs of star formation.

Among these efforts, GOODS stands out for its depth over a moderate large area (~ 300 arcmin^2 over two separated fields). The Italian community is deeply involved in the project, in particular ground based (VLT) spectroscopy and (VLT) imaging, as witnessed by the many papers with Italian leadership.

Within the exceptional multiwavelength coverage, ranging from X to Radio, and obtained with Chandra, XMM, HST, Spitzer, VLT and Keck among others, a noticeable gap is present in the Far and Near UV.

The deep observations with GALEX, cannot fully complement the data at the other wavelengths due to the very poor PSF (~6").

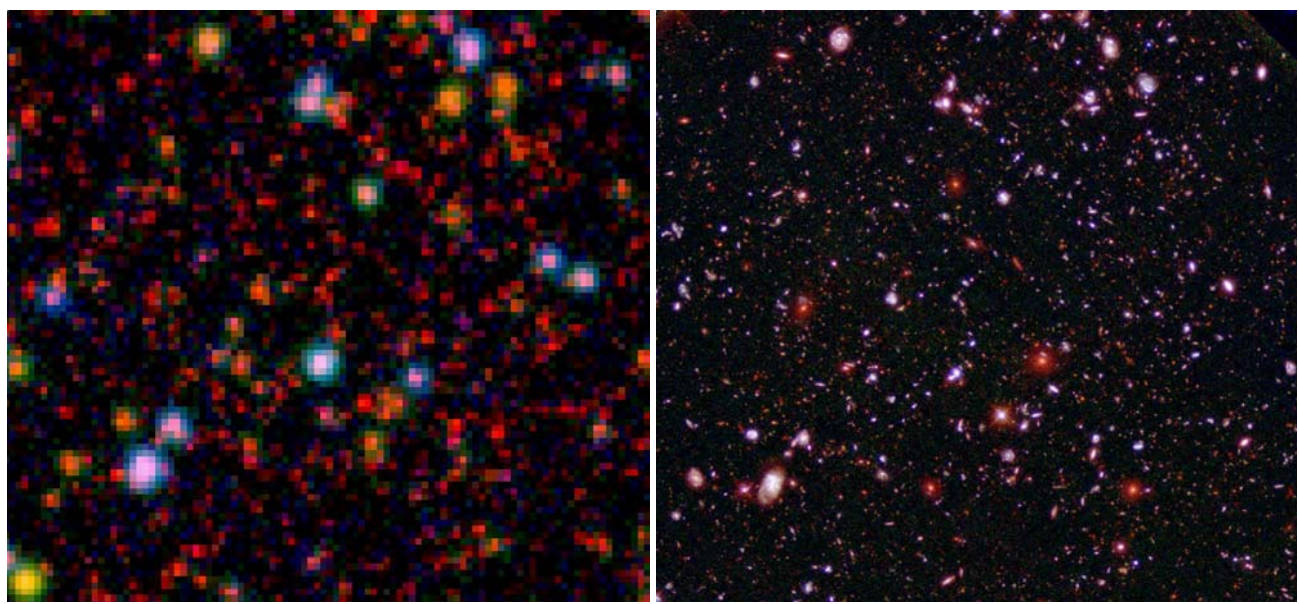

Figure 2: *Left:* A true colour image of part of the UDF: the blue and green channels are respectively GALEX FUV (~1500A, ~75Ks) and GALEX NUV (~2000A, ~76 Ks), while the red channel is the VIMOS@VLT deep imaging (~36000Ks) in band U. Most of the red "noise" are objects detected in ground based U band imaging, but not detected in GALEX data. The depth of GALEX data is 25-26 AB, with a PSF of 4". The size of the image is 2'x2'. *Right:* A true colour image with a simulation of the same field of the previous image: the blue and green channels are the WSO-FUV and WSO-NUV respectively, while the red is U band image.

**Key observables with WSO-UV imagers**

WSO-UV shall allow covering the whole GOODS fields with the FUV+150W combination and with UVO+F300W (9 pointing per field). Moreover observations of the Hubble Ultra Deep Field (UDF), which lies inside GOODS-South, should be carrier out in near UV (band F220W), see Figure 2.

The lack of deep and high resolution FUV/NUV observations still prevents detailed study of e.g.

a) Search for Lyman Break Galaxies at relatively low z. While the search for LBG galaxies is a very well established technique at z >=2-3, the interval around z~1.-1.5 is poorly explored due to the lack of deep and good angular resolution observations.

b) Study of the ultraviolet luminosity function: the ultraviolet (rest frame 1500A) luminosity function is currently poorly constrained at the faint end. The UVO/F300W observations will fill the gap in particular at z~1-1.5, where the global cosmic star formation rate stars to decline.

## 2.2.1.2 Galaxy clusters: unfolding the star formation history through FUV/NUV imaging.

### Scientific Background

Early-type (E/S0) galaxies are among the oldest and most massive galaxies over a large range of cosmic time. The scatter and slope of the colour-magnitude relation (CMR) in galaxy clusters, the evolution of the Fundamental Plane and Balmer line strengths in these same cluster galaxies all point toward an early origin for these objects. At z > 1.0, observations imply that the stellar formation redshift for cluster ellipticals is at z~3-4 (Blakeslee et al. 03, Mei et al. 06). By fitting models to the spectral energy distributions (SEDs) of these same sources the inferred stellar masses are in excess of $10^{11}$ solar masses even at this early epoch. Conversely, however, the early-type population as a whole appears to undergo non-negligible evolution between redshift one and the present epoch. The stellar mass-density in galaxies on the field CMR has almost doubled since z=1 (Bell et al. 2004, Faber et al. 2005). A viable mechanism to increase the number/mass density of the "red sequence" population is by merging faint read galaxies. However, deep optical studies of distant clusters (z>0.8) fail to find significant faint red galaxy population, and this deficit is even more pronounced in lower density environments (Tanaka et al 2005). Hence, merging of red galaxies cannot be the only mechanism responsible for the observed increase in the number of massive early-type galaxies. Both in-situ star-formation and mergers involving gas-rich galaxies must also play a role. Evidence that active star-formation moves to galaxies of lower mass over time, the so called "downsizing" scenario, may suggest that red-sequence galaxies are built-up via merging with star-forming dwarf galaxies. Somewhat surprisingly, the brightest cluster members of RDCS~1252 at z=1.24 show evidence for recent (i.e. ~1 Gyr earlier) star-formation from their H_delta absorption line strengths (even although their luminosity-weighted ages are around 3 Gyr). Current existing data place some constraints on the star-formation history (SFH) within the last few Gyrs. However, little is known about the instantaneous star-formation rates in early-type galaxies, and the recent SFR of early-types at high redshift. Therefore, we propose to study on-going and recent star-formation occurring in early-type cluster galaxies in clusters at z >= 0.8 by observing the rest-frame near and far-ultraviolet (1250-1500\AA). Studies of local early-type galaxies have revealed the existence of relatively young stellar populations. Such {\it fossil record} observations of absorption line-strengths (Trager et al. 2000, Thomas et al. 2004) find that stellar populations younger than ~5Gyr (i.e., which must have formed between z ~1 and the present-day) are relatively prevalent in early-type galaxies. Furthermore a significant fraction of z < 0.1 early-types show relatively blue near-UV-optical colours (Kaviraj et al. 2005, Kaviraj et al. 2006), indicating star-formation over the past Gyr or a substantial contribution from hot helium (super helium rich?) - burning horizontal branch (HB) stars (Brown et al 2002, Ree et al., 2007) or from Post AGB stars. In the last scenarios however the predictions are for a rather fast fading of the rest frame FUV flux with look back time.

The inferred recent star-formation amounts to only ~1\% of the total stellar mass. This cannot be responsible for the evolution of the bright end of the mass function since z=1. To definitively identify the star-formation responsible for the residual mass assembly since z=1 it is crucial to probe the 1250-1500\AA~rest frame wavelengths to constrain the on-going/recent star-formation in massive early-types at z~1.

**Key observables with WSO-UV Imagers**

WSO-UV imagers should allow to do a survey of z~1 clusters with the UVOF300W in order to estimate the 1500A rest frame luminosity, and disentangle among different assembly scenarios for the clusters early type galaxies.

The proposed GOODS-field observation will fully complement the possible WSO-UV observations, since the GOODS FUV/UVO observations will constitute the "field" control sample. This is fundamental to study the environment dependence of early-type galaxy assembly evolution.

## 2.2.1.3 UV Surface Brightness Fluctuations: An innovative tool to investigate unresolved stellar populations.

**Scientific background**

Most of our knowledge on the formation and evolution mechanisms of galaxies comes from the study of the physical and chemical properties of their stellar components. Stars play a fundamental role in understanding the properties of elliptical galaxies, as the integrated light of this morphological class is less affected by internal dust extinction, or, equivalently, the small amount of gas cannot be adopted as a primary tool to analyze their formation/evolution processes. Integrated starlight analysis is the typical tool to investigate external galaxies, since, even with the high-resolution capabilities of HST instruments, single star analysis is possible only for a few galaxies outside of the Local Group. However, integrated magnitudes and colours, together with spectroscopic and spectral index are more or less affected by the well-known age/metallicity degeneracy (Worthey 1994), which acts in such a way that differences in the measured quantity could be interpreted as due either to age or chemical composition variations.

The surface brightness fluctuations (SBFs) technique was originally invented as an extragalactic distance estimator (Figure 3). To date it has proven to be effective at distances of over 50 Mpc from the ground (e.g. Tonry et al. 2001, Liu et al. 2002) and to more than 100 Mpc from space (Jensen et al. 2003). Soon after first applications it was recognized that SBFs also serve as a powerful tool to investigate stellar population properties of early-type galaxies (e.g. Blakeslee et al. 2001, Cantiello et al. 2003, 2005, Raimondo et al. 2005) and star clusters (e.g. Ajhar & Tonry 1994, Raimondo et al. 2007). Being quantified via the ratio between the second to the first moment of the luminosity function, SBFs are very sensitive to the brightest stars of the population, therefore, they are an excellent tool to detect different population components in galaxies where episodes of star formation might be happened in a relatively recent past. Multiband SBFs involving UV range of wavelength are an extremely powerful tool for investigating the evolution of unresolved stellar populations in distant galaxies, and for deriving consistent and accurate estimations of their age and metallicity. In addition, the WSO operated in diffraction limit mode will reach the image quality required to successfully measure SBF differences inside galaxies, and to use them for studying the galaxy assembly/formation history.

**Key Observable with the WSO-UV imagers.**

*SBF in the UV range of early-type galaxies.*

Mapping the sky in the wavelength range 1100-3100 A, the WSO is the ideal telescope useful to apply the SBF technique to study the presence and properties of hot-stars in early-type galaxies. Being quantified via the ratio between the second to the first moment of the luminosity function at a given wavelength, the SBF signal is more sensitive than classical tools, as integrated magnitudes and colours, to the brightest stars in the population. Thus, the SBF signal in the UV will provide the unprecedented opportunity to detect and identify the hot-stellar component responsible for the UV upturn observed in ellipticals, where the distance hampers to resolve stars. For old stellar populations, accurate SBF models show that, moving from the optical to UV, the SBF signal becomes very sensitive to the number-ratio between post-AGB and hot-HB stars (Figure 4). Thus UV-SBF measurements give the opportunity to understand the nature of the stellar objects producing the UV flux excess. Moreover, the SBF technique applied at ultraviolet wavelengths can be a valuable method to trace the physical (evolutionary timescales, colours, and temperatures) and the rarity/abundances of hot-HB, post-HB and post-AGB stars which are still far from being properly understood.

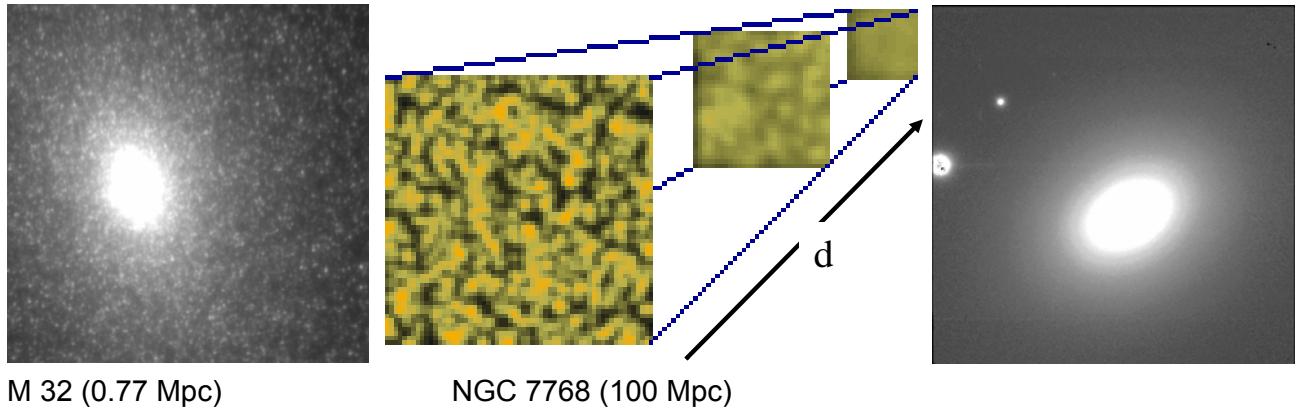

M 32 (0.77 Mpc)                    NGC 7768 (100 Mpc)

Figure 3 : Surface Brightness Fluctuations signal obtained from images of early type galaxies. Distant galaxies appear smooth compared to nearby ones.

In addition, recent SBF models show that SBF-UV are very sensitive to the temperature distribution of the HB stars disclosing the opportunity of investigating the HB morphology in unresolved stellar systems and then to derive information about their metallicity as well as to probe the occurrence of the HB-second parameter in stellar populations out of the Local Group galaxies.

*Merging and starburst galaxies.*

A further science case is related to the extremely high capability of SBF in tracing the brightest stars in an unresolved stellar population. This property of SBF coupled with the possibility of ultraviolet imaging of WSO provides the unique opportunity of precise and accurate study of star formation events in distant galaxies. In fact, the occurrence of relevant burst of star formation imply that bright hot massive main sequence stars strongly contribute to ultraviolet region of the spectral energy distribution of the galaxy. As shown in Raimondo et al. (2005), the SBF magnitudes in $U$ band are sensitive to the age of the recently born stellar population and this becomes much more effective when ultraviolet wide band SBF observations are considered. The age-metallicity dichotomy is expected to be removed by coupling UV SBF measurements with the optical $V$ or $I$ ones (and for this reason a WSO optical imager of similar field of view would be very important in this kind of research ).

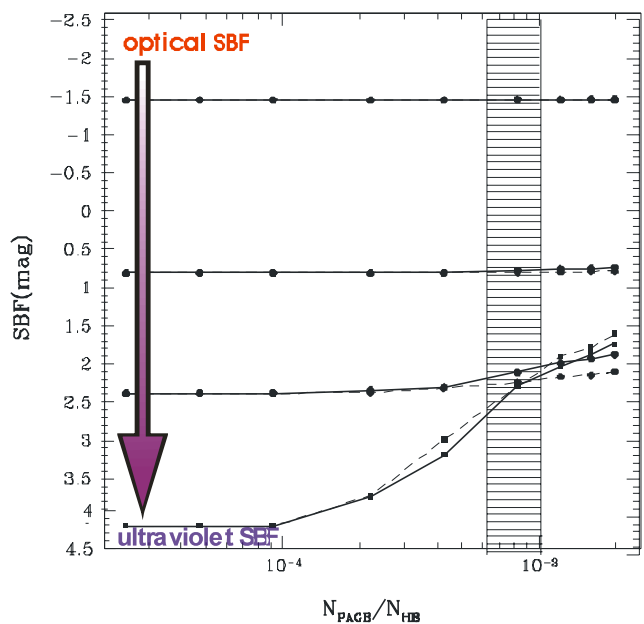

Figure 4 : The behaviour of the SBF magnitude from optical (top) toward ultraviolet (bottom) wavelengths as a function of the ratio between the number of post-AGB and hot-HB stars.

*Merging and starburst galaxies.*

A further science case is related to the extremely high capability of SBF in tracing the brightest stars in an unresolved stellar population. This property of SBF coupled with the possibility of ultraviolet imaging of WSO provides the unique opportunity of precise and accurate study of star formation events in distant galaxies. In fact, the occurrence of relevant burst of star formation imply that bright hot massive main sequence stars strongly contribute to ultraviolet region of the spectral energy distribution of the galaxy. As shown in Raimondo et al. (2005), the SBF magnitudes in $U$ band are sensitive to the age of the recently born stellar population and this becomes much more effective when ultraviolet wide band SBF observations are considered. The age-metallicity dichotomy is expected to be removed by coupling UV SBF measurements with the optical $V$ or $I$ ones (and for this reason a WSO optical imager of similar field of view would be very important in this kind of research ).

A far UV imager (1100A < λ< 2000) on WSO would provide the unique possibility of probing, with extremely high accuracy, the age of the star forming event. It allows us a very deep insight of the star formation history of elliptical galaxies which experienced recent merging events as far as of starburst galaxies.

Moreover, recent SBF spatial gradients have been detected in elliptical galaxies in the optical bands (Cantiello et al. 2005). A similar analysis with far UV imaging will detect the presence of possible spatial distribution and/or radial gradient of young stellar components in unresolved galaxies. In particular, UV-optical SBF will provide a new tool to trace the evolutionary history of the stellar populations in galaxies.

**Observations with WSO-UV Imagers**

WSO instruments for imaging are expected to provide the high quality images required by this technique. A preliminary evaluation shows that WSO should be able to provide an S/N high enough to derive reliable UV SBF measurements for elliptical with high UV emission (i.e. (m1550-V) ~ 4).

Similarly starburst galaxies and merging galaxies, in nearby cluster of galaxies, are expected to present a spectral energy distribution which allows WSO images with the S/N required for SBF measurements both in UV and optical wavelengths.

### 2.2.1.4 Onset and Evolution of the UV-upturn in Evolved Stellar Populations

**Scientific Background**

In old, quiescent systems such as elliptical galaxies and spiral bulges, the UV offers a major probe of stellar populations. Most old systems have been found to contain a very hot, low-mass stellar component with Teff > 20,000 K which dominates the far-UV light. The interest of the extragalactic community in this latter field has been persistently growing since the launch of the first space-borne, UV-sensitive observatories (e.g. OAO-2, IUE; Burstein et al., 1988). The UV-upturn probably originates from stars with very thin envelopes on the extreme horizontal branch and subsequent advanced evolutionary phases. Their UV output is predicted to be very sensitive to their envelope masses and compositions (see

Figure 5).

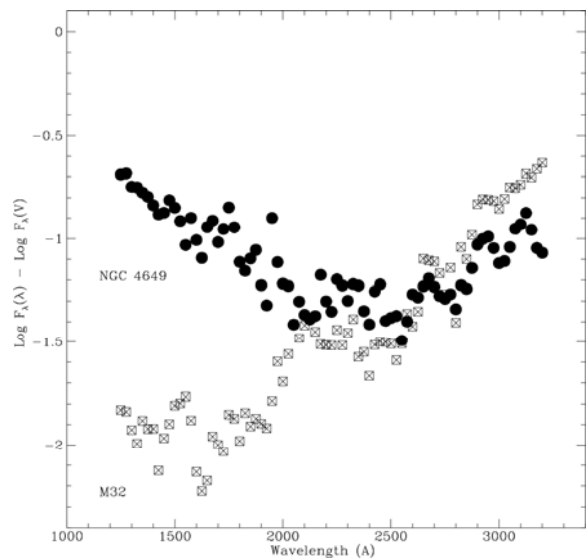

Figure 5: UV-upturn amplitude difference between the metal-rich gE NGC 4649 and the dwarf, metal poor elliptical M32. The scale is logarithmic.

What is more, it has progressively become clear that the precise knowledge of the epoch of the UV-upturn onset, is a very useful tool to constrain many cosmological crucial parameters, such as the redshift $z_f$ of galaxy formation. More precisely, recent UV observations of nearby galaxies have shown that very hot Horizontal Branch (HHB) stars are the dominant source of the UV upturn observed in evolved stellar populations (e.g. Brown et al 1998). In spite of this general result, a precise knowledge of the relative number of UV sources responsible of the UV-upturn morphology is not yet reached. HST/STIS images of M32 show a rich population of very hot HB stars but the number of post-HB, AGB-Manquè and post-AGB stars is widely smaller than the amount predicted by the theory. This means that theoretical models of evolved stellar populations concerning the UV fluxes may be overestimated if these results on M32 are confirmed for other galaxies showing a more robust UV-upturn. On the other hand, ellipticals at redshift 0.3 < z < 0.6 show that the UV-upturn fades, but not as rapidly as might be expected, suggesting either a large dispersion in the parameters that govern the formation of EHB stars, or another source of UV emission that becomes dominant at earlier ages (cf. Brown 2004).

**Key Observables**

*Individual UV-bright Stars in Local Galaxies*

A space-borne telescope with UV capabilities comparable to these of HST/ACS can identify the individual, evolved low-mass stars responsible of the UV-upturn of the closest extragalactic spheroids. This is for instance the case of the elliptical satellites of the M31 galaxy. In particular the UV flux shortward of 2000 A in M32 turned out to be dominated by EHB stars with a minor contribution of P-AGB stars. In effect, being the UV-upturn a very sensitive age indicator of evolved stellar populations, its reliability and accuracy is presently compromised by the poor knowledge of the very hot Horizontal Branch and of the hot bright stars experiencing their final evolutionary phases. In particular, multiband UV photometry of resolved stellar systems with WSO-UV is expected to provide fundamental constraint on the hot component of old stellar population in resolved stellar systems which is presently poorly understood. As an example, the relevant result about the origin of the UV flux of M32 galaxy, open new questions, like the surprising dearth of AGB-Manquè and post-AGB stars (Brown et al 2000) which discloses an unacceptable discrepancy with theoretical predictions (see Figure 6).

*UV-Optical colour as a function of Redshift*

The behaviour of the colour (1550-V) as a function of the rest frame age of a galaxy is a key issue to be compared with existing models of elliptical galaxy evolution. In this respect, the most recent attempts, allowed to constrain the level of the UV-upturn of a few distant ellipticals at redshift not higher than z=0.55 suggesting that the amplitude of the UV-upturn shows a real decline presumably beyond this latter redshift value, corresponding to approximately 5-6 billion years lookback time.

**Observations with WSO-UV Imagers**

The WSO-UV Ultraviolet-Optical Camera (UVO) shall enable to detect in the blue extended objects down to a surface brightness $m_{3600}$ ~20-21 keeps the promise of reconstructing the evolution of the UV-upturn phenomenon as a function of look-back time for objects at z > 1. This issue is, in turn, a crucial piece of information to understand the late phases of stellar evolution. As foreseen by current models, after the decline following the initial period of star formation, the (1550-V) colour suffers indeed from a sharp change, becoming bluer at a certain value of the age, while the precise value of the redshift corresponding to the ages of the UV upturn reflects, in turn, the particular model of the Universe being used.

**Observations with WSO-UV Spectrographs**

UV Spectroscopy can establish the shape and the relative amplitude (relative to the optical region) of the UV-upturn itself, thus giving basic clues to the amount and nature of the dominating hot sources (e.g. EHB vs. AGB-Manquè stars).

### 2.2.1.5 Wide-Field Investigations: UV Survey of early type galaxies in the Virgo cluster.

**Scientific Background**

In most recent years a major step in understanding the physics and evolution of early-type galaxies came from the `ACS Virgo Cluster Survey' carried out by means of Hubble Space Telescope and its follow-up studies aimed at exploring the emitting properties of this sample of galaxies from the X-ray to the infrared. In terms of depth, spatial resolution, sample size, and homogeneity, this survey represents the most comprehensive imaging survey to date of early-type galaxies in a cluster environment. Fundamental results have been reached in studying the early-type galaxies from the innermost structure up to their globular cluster systems (e.g. Cotè et al. 2004; Ferrarese et al. 2006; Peng et al 2006; Jordan et al 2007). In spite of this, an exhaustive knowledge of the properties of early-type galaxies is not yet reached. A critical key point is that no homogeneous and accurate information of a wide sample of these galaxies are available in the UV range of their spectral energy distribution (SED) which is an irreplaceable milestone to understand the state and the evolution of high redshift galaxies.

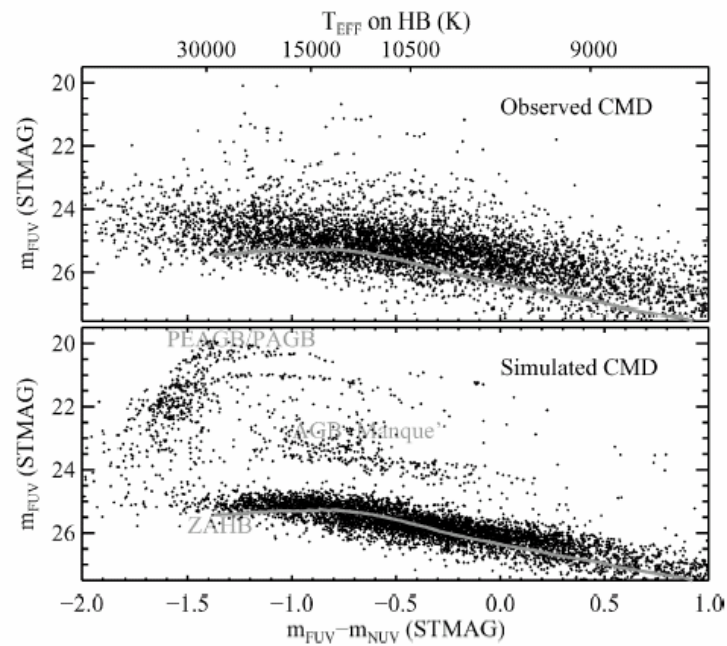

Figure 6: *Top panel*: The CMD constructed from the near UV and far UV images of M32, as observed by HST/STIS (black points), compared to the predicted location of the zero-age HB (grey curve). Although the EHB is well populated, the UV-bright stars above the EHB are under-represented. *Bottom panel*: The simulation that best reproduces the observed CMD, which assumes a bimodal HB morphology having a minority EHB population. Note the large number of UV bright stars, and the clear gap between the EHB and AGB-Manqué stars, which are not seen in the observed CMD.

**Key Observables**

*Extension of the ACS Survey*

WSO-UV shall provide the opportunity of investigating the same sample of ~100 early-type galaxies, for which the excellent optical data of the ACS Survey are already available, in the largely unexplored ultraviolet range. The UV Survey outcome shall assure valuable resources for many scientific applications, including stellar population, star formation histories as well as galaxy formation and evolution studies. Major results are expected to provide quantitative and statistically reliable new results on the 'hot' component (young stars, hot HB, post-AGB, BSS, etc) of the stellar population and their relationship with the general properties of the parent galaxy.

*Local UV Templates*

The proposed observations shall allow delivering to the scientific community a wide database showing the properties of local early-type galaxies at UV wavelengths. This knowledge does represent an irremissibly tool for researchers wanting to understand the spectral energy distribution of such galaxies observed at different redshifts (e.g. Renzini 2006).

## 2.2.2 How did our Galaxy form?

### 2.2.2.1 The Galactic globular cluster system

Globular clusters have traditionally been considered as good tracers of the process that led to the formation of their host galaxy. Absolute ages set a constraint on the epoch of the galaxy formation, with important cosmological consequences. Relative ages provide detailed information on the formation process of the host galaxy.

For the absolute age determination, accurate distances, metallicity, and reddening are mandatory.

As for the distance, while waiting for GAIA results, there is a simple, geometric method to get accurate distances of GCs: the comparison of the dispersion of internal proper motions, an angular quantity, with that of radial velocities, a linear quantity gives a distance. Radial velocities for thousands of stars in GCs, with an accuracy of a few hundred km/s are now easily attainable with multifiber facilities at the 8-10m class telescopes. As for the proper motions, already King et al. (1998) pioneeristically showed that the WFPC2 on board of HST allows astrometric position measurements with a precision of the order of a few milliarcseconds (mas) on a single image. With WSO-UV we expect to reach accuracies of <1mas/frame. This accuracy, taking into account the possibility of using the archive HST data, i.e. the fact that we have a very long temporal baseline (20-30 years), allows to obtain relative proper motions measurements with accuracies of <10 microarcsec/yr.

With the high accuracy proper motion and radial velocity measurement, the distance uncertainty mainly depends on the sampling error, which goes with the $(2n)^{1/2}$, where n is the number of measured stars (typically a few thousands).

Using background nucleated galaxies (or, even better, background point like sources like QSO) it is also possible to have absolute proper motions with an accuracy which is mainly related to the ability to measure the position of the reference galaxies. This accuracy can be estimated to be of the order of a 30 microarcsec (better in the case that the reference objects are background QSOs).

With these absolute proper motions it is possible to:

- use the 3-D kinematics of GCs as a probe of the Galactic potential;
- use the 3-D kinematics, coupled with information on the GC ages and metal content, to detect Galactic streams that are important information to understand the role of the accretion from small size satellites in the assembly of our Galaxy.

**Key observables with the WSO-UV imagers**

The accurate measurement of relative and absolute proper motions requires multiepoch near-UV or visual observations of GCs. In this case, the use of old, archive WFPC2 and ACS images shall provide a large temporal baseline, which shall increase the proper motion accuracy.

Accurate age measurements require multiband (also for an accurate reddening determination) images to build CMDs with large colour baselines and from very accurate photometry, as shown in Rosenberg et al. (1999) and De Angeli et al. (2006).

## 2.3 What is the origin and evolution of stars and planets

The life cycle of stars is a fundamental topic in astrophysics which is expected to be one of the most active in the coming decades, and one to which WSO-UV can give a breakthrough contribution. The cycle of matter from the interstellar medium into stars and then back is the basic engine that drives the evolution of the baryons across the age of the Universe. Planetary systems like our own will form during the early phases of stellar evolution and the complex chemistry of the dense interstellar medium around newly formed stars may be a necessary process to produce the complex molecules that are the building blocks for life. Understanding the life cycle of stars is thus a fundamental step to find answers on the origins of our own Solar System and life on Earth as well as for other habitable planets and life elsewhere in the Universe.

The discovery of exoplanets just over a decade ago has opened a new and fascinating front in astrophysics, also for the philosophical implications of the existence of planetary systems outside our Solar System. All extrasolar planetary systems discovered so far are very different from our own, due to the limitations of the techniques used so far for planet detection. Nevertheless, the discovery of such unexpected configurations has provoked a profound revision of our views of the formation and evolution of planetary systems. The search for extrasolar planetary systems similar to our own Solar System, of exo-Earths and of the signatures of extraterrestrial life is a long term goal for the astronomy of the 21st century.

The key questions that will have to be addressed in the coming decades span a broad range of topics: the role of magnetic fields, environment, multiple systems; star formation in clusters; interplay between the dynamical evolution and the stellar evolution in clusters; the detailed shape and possible variations of the Initial Mass Function for stars; the mysteries of the internal structure of stars; the chemical processes in the interstellar medium; the influence of stellar populations on galactic structure; the final stages of stellar evolution and the feedback to the interstellar medium; the evolution of circumstellar disks leading to the formation of planetary systems; the diversity of planetary systems and the search for Solar System analogues.

WSO-UV will contribute to answer to most of these questions. In the following, we will present just a few examples of the possible contribution by WSO-UV in this field.

## 2.3.1 How do star forms?

### 2.3.1.1 Accretion and Outflows in Young Stellar Objects (YSOs)

#### Scientific Background

The formation of stars is accompanied by many dynamically complex phenomena as, for example, the collapse of the parent magnetized molecular core, the mass accretion from the circumstellar disk, and the generation of outflows. Although scientists have tried to explain the global process of star formation in terms of basic physical ingredients (gravity, opacity to radiation, magnetic field, rotation etc. see e.g. Shu et al., 2000, Konigl & Pudritz, 2000), there are aspects of the standard theory that still pose problems of difficult solution. For example, can the mechanism envisaged for low mass stars be generalised to stars of all masses?

What is exactly the nature of the accretion/ejection engine operating in young stars? Are outflows capable of extracting the excess angular momentum from the disk/star system, and of injecting turbulence in the medium? How are planets generated, and what is the influence of energetic radiation from the central star on the occurrence of primordial life?

To answer these questions, one needs to obtain from the observations a good knowledge of the physical conditions of the immediate environment of young stars. Although YSOs are often embedded in the dense molecular material, the late stages of the formation process (as in T Tauri and HerbigAeBe stars) occur when the outflows have already cleared the stars' immediate environment, and the central object becomes naked and visible at opt/UV wavelengths. The emission lines in these ranges trace the energetic phenomena in the inner 50 AUs from the star, a region that sees in action the accretion through funnel flows, the acceleration and collimation of stellar jets, the formation of planets. Even at earlier stages, however, the UV and optical observations of the jets at a distance from the star can give precious information on the accretion/ejection history of the central object, as well as on the nature of the ISM surrounding the system. Observing young stars and their surroundings at UV and optical wavelengths is therefore of fundamental importance. Precedent studies with, e.g., IUE and HST, have shown that an adequate spectral mapping of the UV and optical features in YSO spectra is feasible (e.g., Ardila et al. 2002, Gomez de Castro et al, 1999, 2001), and also that high angular (sub-arcsecond) resolution is necessary to adequately investigate the physics of these objects, even for the closest star formation regions (e.g. Bacciotti et al., 2000, 2002, Hartigan et al. 2004). Observing with WSO-UV, therefore, will be a unique opportunity to proceed in this direction, and the instrumentation on-board is expected to provide great advances in the understanding of the formation of nearby young stars, and of the nature of their outflows. In turn, this will provide clues to the occurrence of similar accretion/ejection phenomena at many different scales, including AGNs.

**Key Observables**

The mass transfer between the young star and its surroundings occurs at highly supersonic velocities. As a consequence, shocks of various effective strengths form in the flows, heating the gas that subsequently cools producing many emission lines in a wide range of wavelengths. Among these, UV and optical lines are of extreme importance, because they trace the most energetic parts of the mass flows, and the accretion shocks in the regions closest to the star, where the accretion/ejection mechanism is at work.

Previous studies have provided extremely rich spectra of low mass T Tauri stars and intermediate mass Herbig AeBe stars that possess disks and powerful outflows. Many permitted, forbidden and semi-forbidden lines are observed, tracing very different regions in terms of density, temperature and excitation conditions (e.g. Gomes de Castro et al., 1999 Hartigan et al. 1999, Devine et al. 2000).

To mention only a few, in winds and outflows one finds powerful emission from lines as Si III 1206, HI Ly alpha, CIII] 1907,1909, FeII $\lambda$ 2383,2626 and the MgII doublet at 280 nm, whose intensity is often a factor 3-10 stronger than the more classical optical lines used to identify Herbig-Haro jets, namely H-alpha, [O I] $\lambda$ 630.2 nm, the [S II] doublet at 673 nm and [N II] $\lambda$ 658.5 nm.

On the other hand, accretion can be studied through higher excitation and density tracers like O I$\lambda$ 130.2 nm, C IV $\lambda$ 155 nm, HeII at 164 nm, OIII] nm 166.6, Si III] $\lambda$ 188.3 nm.

Temporal variability in both accretion and wind spectral signatures is an important piece in the formation puzzle, as it gives an idea on how much intermittent pulsating phenomena are important in the accretion process, an aspect too often neglected in formation model. Finally, the UV continuum bears of course important information also. For example, the FUV and UV excess of the central stars leads to estimates of the mass accretion rate, while the UV continuum in the regions distant from the sources maps propagating shocks.

**Observations with WSO-UV Imagers**

*Imaging of star formation regions.*

The cameras on board the WSO-UV, shall allow mapping in the UV relatively large areas in the known star formation regions, taking advantage of the different filters available. Associated optical imaging shall assure a proper identification of the various features, allowing high angular resolution observations (Figure 7) after Hubble will have terminated its operations.

Using UV wide-band filters one shall be able identify energetic phenomena influencing the star formation in the clouds, such as the propagation of ionization fronts and high velocity shocks, the presence of HII regions and photo dissociation regions. The Orion and Eta Carinae regions are a good example, as testified by the presence of proplyds and 'irradiated jets' (Bally et al, 1998). Imaging sites of multiple star formation with wide band UV filters can also give clues to the unexplained relationship between mass accretion rate and mass of the central object, as found recently for T Tauri and lower mass stars by Natta et al. (2004). The UV excess will allow to determine the accretion rates in higher mass Herbig Ae stars, with and without outflowing activity, and so test if the trend is continued at higher masses. The application of narrow band filters in the high density line tracer mentioned above can then help to identify the most active accretors in a star formation region, to be further examined with the WSO-UV spectrographs.

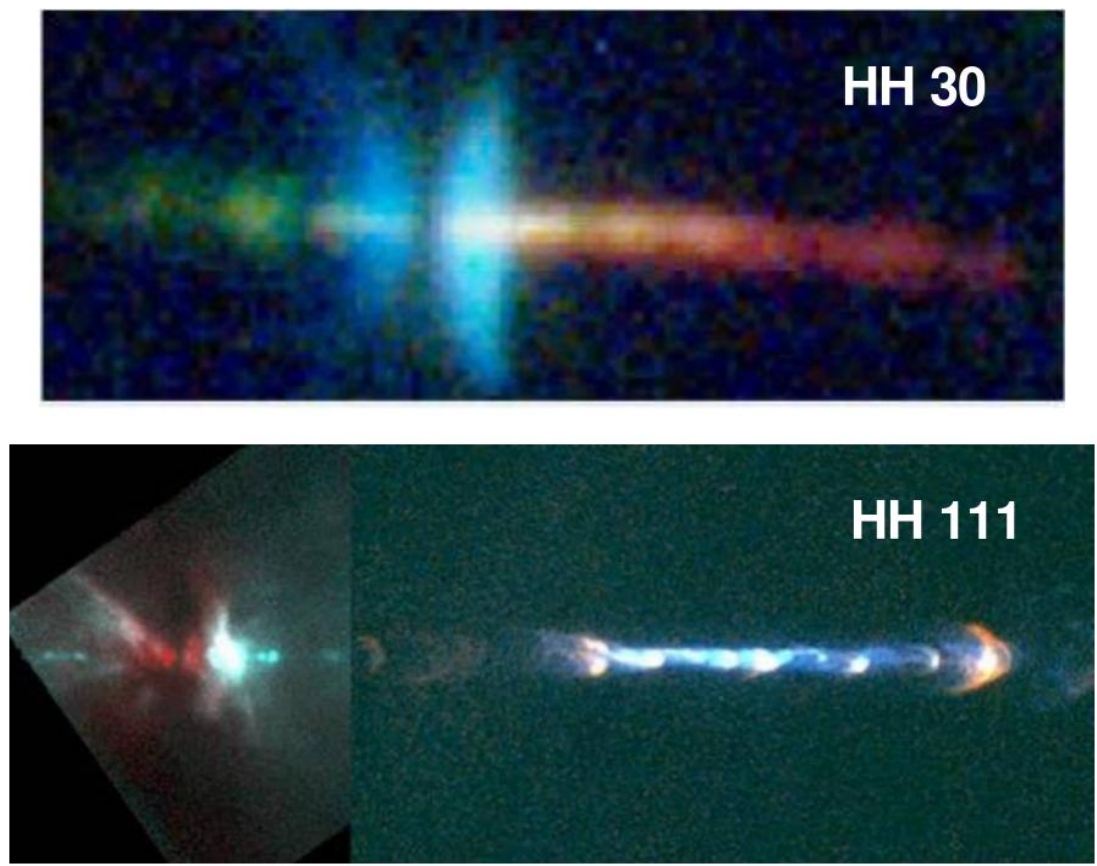

Figure 7 : Hubble Space Telescope images of the bipolar HH30 and HH111 stellar jets (Ray et al. 1996, Reipurth 1989). In both images, the two cusp-shaped nebulae on the left hand side with the dark lane in between indicate the position of the accretion disk. In both images, one lobe of the jet is brighter because it is coming out of the dark cloud from which the star formed.

*Narrowband imaging of YSO jets*

One very promising opportunity with WSO-UV shall be the possibility of forming narrowband images of stellar jets in FUV and NUV lines. High angular resolution imaging of jets in the past has been done with HST but only in optical and infrared lines. Shock models, however, have shown that FUV and UV lines can be a factor up to 10 stronger than the optical lines 'historically' associated to jet emission (Hartigan et al. 1987). Although interstellar extinction can preclude far-ultraviolet studies of the outflows, early-type young stars that have shed most of their surrounding gas and dust can offer a good opportunity. For these stars, the relative faintness of photosphere near H I Ly alpha and the low line-of-sight extinction make Ly alpha an ideal tracer of outflow activity, especially near the central object. This has been demonstrated to be feasible by the HST/STIS observations of Devine et al. (2000).

Strong lines of  Mg II at 280 nm are also produced in the jet beams, as well as in the large bow-shocks at a considerable distance form the star, as shown,  e.g.,  by HST and FUSE observations by Hartigan et al. 1999, Coffey et al. 2007.

WSO-UV shall offer the opportunity to approach the nearly unexplored field of UV imaging of outflows, which will allow us to clarify the structure of the most energetic parts of the flows at high angular resolution.  Comparison of images in distant lines useful for extinction diagnostic (e.g.

Halpha/Hbeta) shall also provide de-reddening and absolute calibration. Comparison of images taken in the strong UV lines with others taken in fainter UV and optical outflow tracers such as Si III 120.6 nm, C III] 190.7,190.9 nm, [O I] 630 nm, [S II] 671.6, 673.1 nm will allow a mapping of the relative excitation of the various jet structures.

It is particularly interesting to investigate the emission properties of the outflows in continuum-subtracted narrowband images at high angular resolution close to the star (e.g. Bacciotti et al. 2000, Woitas et al. 2002, Dougados 2000), for a better understanding of the acceleration mechanism, and its relationships with the accretion of mass onto the star.

*Slitless spectroscopy with grisms*

It is interesting to note that being stellar jets elongated line emission objects, one can fruitfully perform 'slitless spectroscopy' of these objects taking advantage of the moderate spectral resolution prisms/grisms associated to the NUV and Optical cameras. Orienting the dispersion direction transverse to the jet axis, one can form separated multiple images of the flow in the various lines of interest in the same exposure. This allows, first of all, a very accurate subtraction of the stellar continuum, unveiling the structure of the flow down to the spatial resolution limit close to the star. On the other hand, line ratios of the whole flow can be easily formed in one go and give diagnostic information on the physical conditions of the gas.

This technique has been applied to YSO jets using HST/STIS at optical wavelengths (see Hartigan et al. 2004), and it has proven to be successful provided that the jet under study is well collimated, and that the grism spectral resolution is neither too low (to allow a complete separation of the spectral images in the emission lines of interest) nor too high (in order not to generate confusion between spectral/spatial features in one single line). The grisms provided with the FCU, which have R ~ 300 are ideally suited to this scope.

It should be stressed that the availability of narrowband filters and of a grism is particularly useful in the visible range (longward than 320 nm) which is not covered by the WSO-UV spectrographs.

In fact many important lines for the diagnostics of star formation nebulae, and in particular of stellar outflows are located between 400 and 700 nm, like [O III] 436.9 nm, [N I] 520 nm, [O I] 630.2 nm, Halpha, [N II] 658.5 nm and the [S II] doublet at 673 nm.

*Spectro-polarimetry*

Finally, spectro-polarimetry operated at NUV wavelengths, can give information about the geometry and structure of the condensations around the star. Moderate-resolution spectropolarimetry provides in fact tight constraints for the theoretical models of disks and outflows, since it helps understanding the distribution of matter in the different regions of density and excitation traced by the different lines. Where the polarization mechanism is wavelength-dependent (e.g. dust scattering in an equatorial disk plus electron scattering in polar regions), one can obtain evidence of unresolved bipolar microjets, and infer the presence and orientation of circumstellar disks, as well as the presence of disk inhomogeneities leading to proto-planetary systems. The variation of position angles from the optical to the UV can also give clues to the nature of the dust in the disk (e.g. the grain size).UV spectro-polarimetry can also be used to infer the direction of magnetic field at the surface of young stars. The technique allows one to detect weak (microGauss) magnetic fields, and provide information about the 3D topology of magnetic field in the magnetosphere. This is very useful if one has to investigate the accretion of matter channelled from the disk to the star through the funnel flows.

**Observations with WSO-UV Spectrographs**

**LSS:** The moderate spectral resolution of LSS is ideally suited to identify global spectral properties of the accretion region around the YSO, and of the flows at small and large distance from the star.

The line tracers of outflows and accretion have been discussed above, as well as diagnostic possibilities highlighted by the use of available narrowband filters and grisms, thus this part is not repeated here.

The resolution of LSS will in addition allow separating lines of doublets useful for the diagnostics, like for example the two lines of MgII at 2769 and 2803 Angstroms. Their ratio indicates the optical thickness of the examined region, thus giving very useful hints about the concentration of the gas.

Similar cases can be found among the other lines emitted by these objects.

One limitation of the LSS is the fact that the slit width is quite large, (contrary to STIS, for example) thus limiting the possibility of combining spectral and spatial resolution to obtain crucial information about the physics of the flows across their width. This would be especially useful in the zone near the accelerating source (see also below). The spatial resolution will however be retained along the slit, allowing, for example, to map accurately the variation of the emission in the lines along the flow and the values of the physical quantities derivable from spectral diagnostics with distance from the source (see e.g. Bacciotti Eisloffel & Ray 1999, Podio et al. 2006).

   One powerful technique that is more and more applied to the study of accretion/ejection engines around YSOs is the so-called `spectro-astrometry', which could be successfully attempted also with spectra of WSO-UV/LSS. It consists in the combination of astrometry and spectral principles, to find the displacement of spatially unresolved velocity features associated to accretion and/or outflows with respect to the position of the source, down to sub-AU scale. With this technique, for example, the first jet from a Brown Dwarf could be recently identified (Whelan et al, 2005). Spectro-astrometry has usually been performed at optical and IR wavelengths, and there is no reason to think that it couldn't be applied to UV spectra. This would be particularly useful for the study of magnetospheric flows.

**HIRDES**: The combination of high angular and very high spectral resolution offered by HIRDES gives unique insights to the problem of accretion and ejection of matter around young stars.

With this instrument, in particular, it will be possible to resolve spectrally in a very accurate way the profile of lines tracers of accretion, for the identification of infalling gas, possibly combined with winds and collimated flow in the magnetospheric region associated with P-Cygni profiles in the observed line. This will greatly help understanding the complex physics of the star/disk interface, identifying ways to validate the current accretion/ejection magnetohydrodynamic models (e.g. Shu et al. 2000).

Regarding stellar jets, a very accurate mapping of the kinematics will be possible in each position from the source. On would like to study in detail, for example, phenomena like the acceleration of the flow close to the source, the internal shock structure and efficiency, the propagation of large bow shocks in the parent cloud and the consequent transfer of linear momentum. We stress that these analyses, common in the visible wavelength range, have been performed in the UV range only for a few jets (see e.g. Devine et al. 2000, Coffey et al. 2007).

The high spectral resolution of the instrument will in addition allow one to resolve the different velocity components of the flow (as for example the faster portion closer to the axis and the slower one closer to the jet borders). This will allow testing MHD models for the jet launching and propagation, elucidating the role of the different components in the jet physics and in the interaction with the ambient or with coaxial outflows.

For those jets that have a large enough extension transverse to the axis with respect to the available slit widths, as well as for extended H2 regions and PDRs, one should be able to perform multiple slit spectroscopy by stepping the slit across the flow, to form 3D datacubes (2D spatial and 1D velocity) similar to those obtained by Integral Field Units. This allows one to construct 2D images of the region in separated velocity bins. With such a technique it has been possible to recognise with HST/STIS, for example, the onion-like kinematic structure of the jets close to their origin, as predicted by the models (Bacciotti et al. 2000).

Finally, if the borders of the flow can be resolved spatially, one can attempt to ascertain if asymmetries of the radial velocity, consistent with rotation of the jet around its central axis, are seen also in UV lines. The confirmation of the detection of jet rotation (Bacciotti et al. 2000, Woitas et al. 2005, Coffey et al. 2004) would be quite important, to prove that the outflows are capable of carrying away the excess angular momentum from the YSO star/disk system. A positive detection of rotation signatures in UV lines, in agreement with the indications in optical lines for the same flow, has been obtained recently for two jets before the loss of STIS (Coffey et al. 2007). The present statistics is however insufficient to definitely prove the presence of jet rotation at UV wavelengths, and WSO-UV/HIRDES may contribute significantly in this direction.

### 2.3.1.2 Star formation history in Local Group Dwarf Galaxies

There are ~50 dwarf galaxies making the Local Group (systems within ~<1 Mpc). Although, they are bound to be considered "simple stellar populations", dwarf galaxies display such a wide range of star formation histories and chemical enrichment that we still lack a comprehensive scenario explaining what triggers and regulates star formation in them. Indeed, dwarf spheroidals galaxies are usually gas-poor systems with little, if any, recent star formation. The star formation history of a typical dwarf spheroidal (e.g. Ursa Minor) resembles very much that displayed by Milky Way halo globular clusters: a metal-poor stellar population formed via a single star formation episode that took place some >10 Gyr ago. On the other hand, dwarf irregular galaxies (e.g. SagDIG) are gas-rich systems that have experienced a continuous star formation history and are currently forming stars. The sub-division is however not sharp, and many "transition" systems (e.g. Leo A and Carina) show signs of discrete and multiple star formation episodes that are separated by quiescent periods.

The variety of star formation history, metallicity and distance(from the Milky Way and M31) and the availability of a statistically representative sample make the Local Group dwarf galaxies a unique laboratory where one can address the properties of their more distant counterparts at higher red-shifts.

The UV-imaging performance of the WVO cameras will allow detailed studies of the following issues:

1) Propagation of star formation in dwarf irregulars: spaced-based UV deep imaging of star forming regions in dwarf irregular galaxies is ideal in deriving a detailed view of their recent star formation history (a look-back time ~1-2 Gyr). Indeed, colour-magnitude diagrams show that Blue and red super-giants (~<1 Gyr) do not overlap in magnitude, and older generations are simply shifted towards fainter magnitudes. Thus, with the help of stellar evolutionary models one can disentangled young stellar population in terms of their age. This will allow a reconstruction of how star-formation proceeds and propagates (both in time and space) within a dwarf galaxy (e.g., Figure11 in Momany et al. 2005).

2) UV-upturn: extreme blue HB stars are considered the dominant source of the UV emission in elliptical galaxies. The blue HB of a dwarf spheroidal galaxy sample seems however to be truncated at temperatures of ~12000 K (Momany et al. 2004), thereby lacking the HB stars in the so-called extreme-HB regime (>20000 K). This puzzle can be solved by a systematic UV-survey of nearby dwarf spheroidals.

Blue stragglers in Dwarf galaxies: Blue stragglers are usually studied in globular and open clusters, little is known about Momany et al. presented a study of Local Group dwarf spheroidals and derived a significant anti-correlation between the BSS frequency and the galaxy Mv that resembles that found by Piotto et al. for Galactic globular clusters. It was argued that BSS in dwarf galaxies are mostly primordial binaries and little if any have a collisional binary origin. Thus, Dwarf galaxies offer a loose and different environment where the properties of BSS can be investigated. An UV-survey of the Local Group spheroidals is necessary to provide a fundamental Observational constraint on the frequency of the BSS in dwarf galaxies.

## Observables with WSO-UV

### Dwarf Galaxies in the Local Group

There are ~50 dwarf galaxies making the Local Group (systems within ~<1 Mpc). Dwarf galaxies display such a wide range of star formation histories and chemical enrichment that we still lack a comprehensive scenario explaining what triggers and regulates star formation in them. Indeed, dwarf spheroidal galaxies are usually gas-poor systems with little, if any, recent star formation. The star formation history of a typical dwarf spheroidal (e.g. Ursa Minor) resembles very much that displayed by Milky Way halo globular clusters: a metal-poor stellar population formed via a single star formation episode that took place some >10 Gyr ago. On the other hand, dwarf irregular galaxies (e.g. SagDIG) are gas-rich systems that have experienced a continuous star formation history and are currently forming stars. The classification is however not sharp, and many "transition" systems show signs of discrete and multiple star formation episodes that are separated by quiescent periods.

The variety of star formation histories, metallicities and distances from the Milky Way and M31, and the availability of a statistically representative galaxy sample, make the Local Group dwarf galaxies a unique laboratory where one can address the properties of their more distant counterparts at high redshift.

### Nearby star-forming galaxies: pinpointing the recent star formation

The ultraviolet bands of WSO-UV imagers will represent a breakthrough to observe star forming galaxies in the Local Group. Regions of star formation will be observed at high spatial resolution in both the FUV and NUV bands to obtain photometry of individual stars and place the young, hot stellar population in the HR diagram. The far-ultraviolet colors will be essential to study the properties of the most massive, hottest stars thus characterizing the most recent episodes of star formation in the field and star clusters of nearby dwarf irregulars. In fact, redder bands and colors either completely miss the most massive stars or are unable to characterize their effective temperatures and location in the HR diagram. WSO-UV imaging will allow us to study the star formation process in environments of different metallicity, approaching - as it is the case for Leo A - the lowest metallicities known for distant dwarf star-forming galaxies (known as Blue Compact Dwarfs). As such, it will help to interpret the integrated properties of more distant galaxies using star-formation indicators.

### Measuring proper motions of LG galaxies with WSO-UV

Also, WSO-UV will allow measuring absolute proper motions for nearby dwarf spheroidal galaxies in the Local Group with unprecedented accuracy, by comparison with astrometric observations obtained over many years with HST. In several galaxies, fields containing background QSO's have been observed to provide a zero-velocity reference frame. Assuming we can reach absolute proper motion accuracy as discussed in Section 5.1.2, it will be possible to infer the orbits of individual dwarf galaxies, test if they belong to large-scale *streams* in the LG. The ultimately goal is to model the *dynamical history of the Local Group*.

Moreover, with astrometry based on observations on a wide temporal baseline, it will become possible to measure the internal motions of stars, i.e. the tangential-velocity dispersion of individual stars, at least in nearby dwarf spheroidal satellites of the Milky Way. Models suggest that comparing the dispersion in the radial and tangential velocities across the galaxy body may overcome the well-known difficulties in disentangling the mass distribution and velocity anisotropy, thus allowing one to derive the distribution of dark matter halos (Strigari et al. 2007).

### Hot stars in old stellar populations

There are UV-bright stellar populations in dwarf spheroidal galaxies that still lack unambiguous interpretation. One major puzzle is the nature of blue stars located above and blue-ward of the old main-sequence turnoff in the HR diagram (e.g., Momany et al. 2007; Mapelli et al. 2007). Are these "blue-straggler stars" (in the case of dSph, the product of the evolution of mass-transfer binary systems) as in Galactic globular clusters, or rather the result of a prolonged star formation? Answering this question is crucial to correctly reconstruct the star formation history of dSph. Characterizing the UV photometric properties of BSS candidates and other UV-bright stars, such as blue HB stars, in dwarf spheroidals with multi-band WSO imaging will be of help for our understanding of UV observations of the integrated light of early-type galaxies.

## 2.3.1.3 UV Observations of Young Populations

### Scientific Background

Unlike the optical band light (coming basically from a mix of warm and cool stellar components in which older populations of galaxies have significant influence), the ultraviolet region shortward of 3200A is more sensitive to the hottest stellar components (both young massive stars and advanced evolutionary phases of low-mass stars)as well to the presence of dust than are either the optical itself or the near-IR radiation (for instance, stars with surface temperatures above ~10,000 K are brighter in the UV than at any longer wavelengths). In this respect, one should be aware that the sources dominating the UV emission and those governing the optical/IR light are so physically separate that the appearance of galaxies in the UV can be dramatically different from that in the above wavebands. In general terms, the UV has the highest sensitivity of any spectral region to stellar temperature and metal abundance, implying that it is especially valuable as a means of characterizing stellar populations, current star formation rates (SFRs), and star formation histories.

### Key Observables

#### The Stellar Content of the Nearest Galaxies

Since the time it was recognized our Milky Way system belongs to a loose association of galaxies dubbed "The Local Group" (LG), observers realized that the existence of such a few, nearby systems (spiral and irregulars) offer the opportunity of calibrating the luminosities of several kinds of secondary standard candles, such as stellar aggregations and/or massive individual stars. In particular, the best direct detection of hot objects in galaxies (i.e. the sources responsible for relevant astrophysical phenomena such as ionization, photodissociation, kinetic energy input, and element synthesis) is assured by space-borne UV imagers. In effect, the study of stellar cluster systems and compact associations provides us with insight into a galaxy's star formation history and stellar content, while far- (FUV) and near-UV (NUV) bands (as those recently obtained by means of GALEX), are a sensitive probe for measurement of the physical parameters of young clusters and --when combined with optical data -- of their extinction.

#### Integrated UV colour of stellar populations: Magellanic Clouds clusters.

It is worthwhile to emphasize that before facing the problem of understanding the young stellar population in distant galaxies, it is mandatory to have a precise and accurate knowledge of nearby (resolved) young single-burst stellar populations. For instance, there is no doubt that WSO-UV images of Magellanic Clouds (MC) clusters are excellent candidates for these studies, because: i) their distance is well established, ii) they can be spatially resolved by the above imagers, and iii) they span a wide range of ages (from a few million years to a few billion years).

Multicolour UV imaging of unresolved distant stellar systems as obtained in the range of wavelength available with WSO-UV discloses the possibility of deriving age evaluation of young stellar populations. The contribution to the integrated UV light of the main sequence turn off stars make possible the use of UV colours for age determination. For example, in the range of 1100-3100 A it has been proved that a two colour diagram, which is free by uncertainties on the distance determination, can be calibrated in terms of age when dealing with stellar population younger than 1 Gyr (Figure 8, Barbero et al. 1990). Moreover, the same two colour diagram makes possible the investigation of the HB morphology in non resolved stellar systems.

*The Missing Satellite Problem in Local Group Analogs*

The expected number of dark matter clumps around galactic, Milky Way-sized halos exceeds the number of smaller satellite galaxies actually observed by at least one order of magnitude. Verifying whether other LG analogs within reach of WSO-UV share the same paucity of luminous satellite galaxies orbiting dominant (spiral) galaxies is thus a key issue to address the so-called "missing satellite" problem. If this is the case, unless gas accretion and star formation are suppressed in dwarf dark matter clumps by some unknown mechanism, this investigation could ensure that the amplitude of the small-scale primordial density fluctuations is considerably smaller than expected in the cold dark matter scenario, thus affecting cosmological studies themselves.

**Observations with WSO-UV Imagers**

An UV-optical imager in a 2-m class space telescope, capable of assuring a subarcsec resolution within a FOV as large as 5 arcmin, is ideally suited for identifying any compact, UV-bright star-forming region of local, individual galaxies, down to a surface brightness of about $m_{UV}$ ~15-16. For more distant galaxies one can simply resort to the UV luminosity, being the flux around 1500 A proportional to the current star formation rate (SFR); in this latter view, the combination of WSO-UV Ultraviolet and Optical imaging will provide a highly reliable calibration of the UV-SFR relationship to be used also for the interpreting the observation of the rest UV emission of high-redshift young galaxies. As far as the `missing-satellite' problem is concerned, the same WSO-UV imaging capabilities will make easier to detect low-surface, dwarf star-forming objects in the surroundings of the dominant galaxies of closeby LG analogs, so as to get a more reliable view of the primordial formation processes for such systems, including our own LG.

**Observations with WSO-UV Spectrographs**

Thanks to the initial fast evolution of the UV SED shortward of 3000A (cf. Figure 9; Marcum et al. 2001), WSO-UV low-res spectrograph observations (LSS) do represent the best suited tool for dating young (sub)systems from their formation epoch to the age of ~ 3 Gyr. More generally, UV spectroscopic observations offer a unique probe of the star formation history (i.e. their star formation rate as a function of time) of galaxies on intermediate timescales of 10-1,000 Myr (though always potentially affected by extinction).

## *2.3.1.4 Multiple stellar populations in star clusters*

**Scientific Background**

Star clusters have been always considered ideal laboratories for the study of stellar populations, because of the general idea that they host simple stellar populations, with all stars at the same distance, formed at the same time, with the same metal content. However, there is a growing body of evidence, both from spectroscopic and photometric investigations, that some clusters host more than one generation of stars. Very recent observations have shown that younger generations may form from material strongly polluted from the material ejected by former generation. For example, recently discovered multiple main sequences in globular star clusters could be interpreted with the presence of stars with very high (Y=0.40) He content (Bedin et al. 2004, Piotto et al. 2005, Piotto et al. 2007). Multiple generations of stars have been identified also in Magellanic clusters (Mackey & 2007).

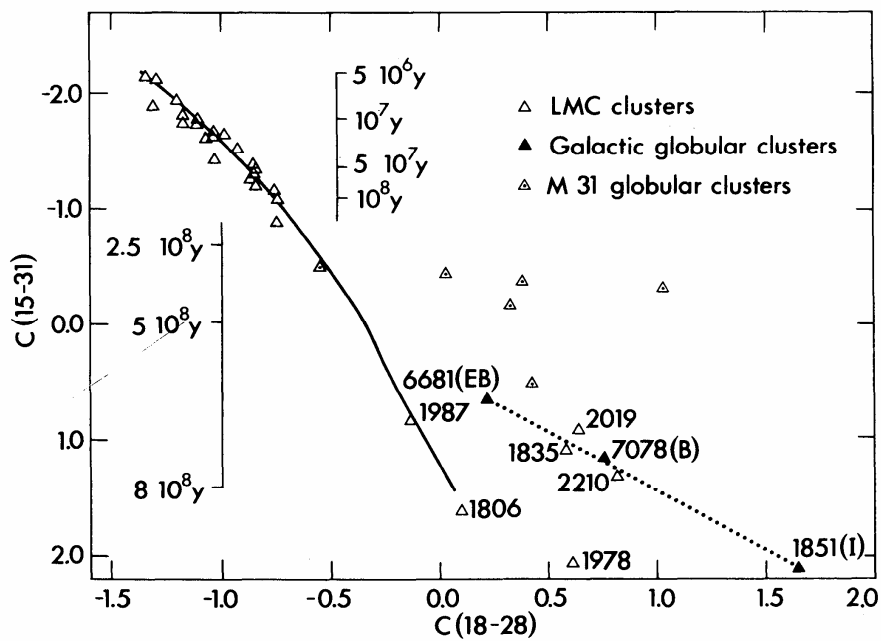

Figure 8: Data from IUE observations of young LMC star clusters lye on the solid line, while the old galactic globular clusters are located on the lower-right side part of the diagram according with the different morphology of their HB.

## Key Observables

High accuracy, multiwavelength photometry, with a large color baseline and in crowded field is mandatory to identify multiple populations in stars clusters. The possibility to remove field stars (e.g. by proper motion membership estimate) is of fundamental importance in order to extract a pure sample of cluster stars.

## Observations with WSO-UV imagers

We need imaging observations with near UV and optical broad-band filters (including Strömgren filters) of star cluster fields. It would be important to repeat observations at distance of years for proper motion measurements. In many cases, HST archive images can provide the needed first epoch data.

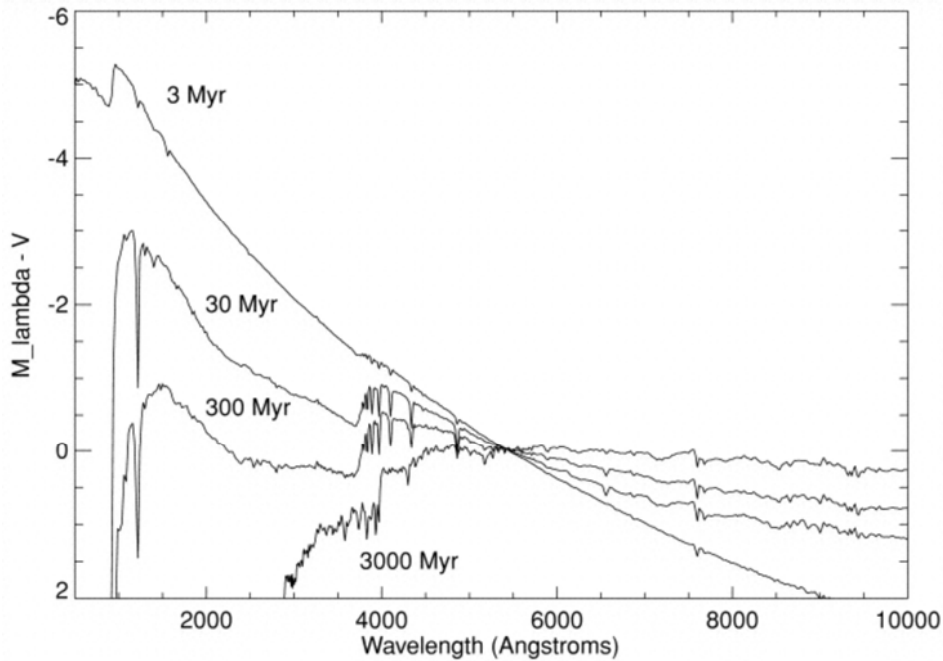

Figure 9: Synthetic spectral energy distributions of single generation stellar populations having Salpeter IMFs and solar abundances for ages 3-3000 Myr. The shape of the near-IR spectrum is much less sensitive to age.

## 2.3.2 Do we understand stellar structure and evolution?

### 2.3.2.1 Asteroseismology

**Scientific background**

Seismic investigations of stellar non-radial pulsations, based on the analysis of oscillation frequencies, provide the tool to probe the inner parts of stars, enabling us to study their internal structure, evolutionary state and the physical processes occurring in the stellar interiors. Furthermore star parameters inaccessible to classical observation or that can not be determined with an accuracy that is sufficient for many purposes could be determined with this technique. Examples of these parameters are the mass and angular momentum of stars. Currently, accurate masses of stars can only be determined if they are members of close binary systems. Even for the best spectroscopic data and stellar atmospheres modelling currently available, for single stars their mass is not known to better than 20%. So, if one wishes to determine the mass function of stars (e.g. to study the star formation process) the current data are heavily biased to stars that have formed in binary systems, which may not reflect the variety of conditions under which stars can form. As a consequence of the fact that forming stars contract from molecular clouds many orders of magnitude larger than their main sequence stellar radius, a substantial amount of angular momentum must be shed during the contraction. An important constraint for theories of mechanisms mediating this angular momentum loss would be the measurement of the angular momentum of stars for a range of masses. Also, for the post-main sequence evolution of in particular high-mass stars the loss of angular momentum during evolution is similarly important. The oscillation frequencies are dependent on the internal structure and on the physical properties of the stellar layers, like pressure, temperature and density through the inverse dependence by the sound speed in the gas. Sound speed and masses are tied together because the hydrostatic equilibrium. Furthermore the sound speed underwent to a secondary dependence on hydrogen abundance which arises through the change in the mean atomic mass on which the sound speed depends. Thus if a sufficient number of modes is present and identified in the time series of an oscillating star, it should in principle be possible to measure its mass, its core hydrogen abundance ("age"), and its rotation rate. The power of such analysis has been demonstrated in the solar case, with Helioseismology, and with the seismic analysis of solar-like oscillations in other stars than the Sun, see Bedding and Kjeldsen, 2003.While Helioseismology can use also higher degree modes, seismology of distant stars is restricted to deepest modes with lower degree, because they are the only ones that don't cancel out when integrated over the visible hemisphere.

Asteroseismic analyses have already provided significant constraints on stellar modelling for a broad range of stars, including Solar-like stars, massive main-sequence stars and white dwarfs. The coming years, with the possibility to make observation from space, will show increasingly reliable asteroseismic determination of stellar ages, of great value to investigations of the evolution of the Galaxy. Utter information will be obtained on the extent of convective envelopes and cores, and on the importance of mixing processes which affect the composition and hence the oscillation frequencies. In addition, constraints will be placed on the internal rotation of stars from observations of rotational splitting of oscillation frequencies, and hence on the modelling of the evolution of stellar rotation.

Space based observations can solve the problems due to the aliases, thanks to the higher duty cycle. Photometry is then favoured by observations from space, since the atmospheric scintillation is null.

WSO-UV represents the possibility to obtain very accurate asteroseismological data for the object that cannot be observed by optical satellites such as COROT, MOST or Kepler, because the ultra-violet range allow observing even faint stars as White Dwarfs.

**Key observables**

*Solar-like oscillations*

The oscillations in the Sun are stochastically excited by the outer convection zone, so generally all the oscillations driven by this process are called "solar-like". Therefore all the stars having an outer convection could present this kind of pulsations. In the H-R diagram they belong to the region extending from the cool edge of the classical instability strip to the red giant.

In this case the so called *p-modes*, generated by pressure fluctuations, are the more superficial waves, while the *g-modes*, driven by gravity pressure, move only in the central part of the star.

These modes are unfortunately less numerous and have extremely low amplitudes in comparison to the solar ones. Nevertheless, the improvement of instruments and techniques, has made possible their detection and their identification. Asteroseismic observations of solar-like oscillations produced very encouraging results which have brought to conceive and realize massive ground-based campaigns and spatial missions. Thanks to this continuous development of instruments and computational techniques, asteroseismology has reached a valuable degree of precision and accuracy in its measurements.

From earth the main detection technique of these oscillations is the high-precision Doppler measurement, performed with the high-resolution spectrographs, while photometric observations are penalized by the atmospheric scintillation phenomenon.

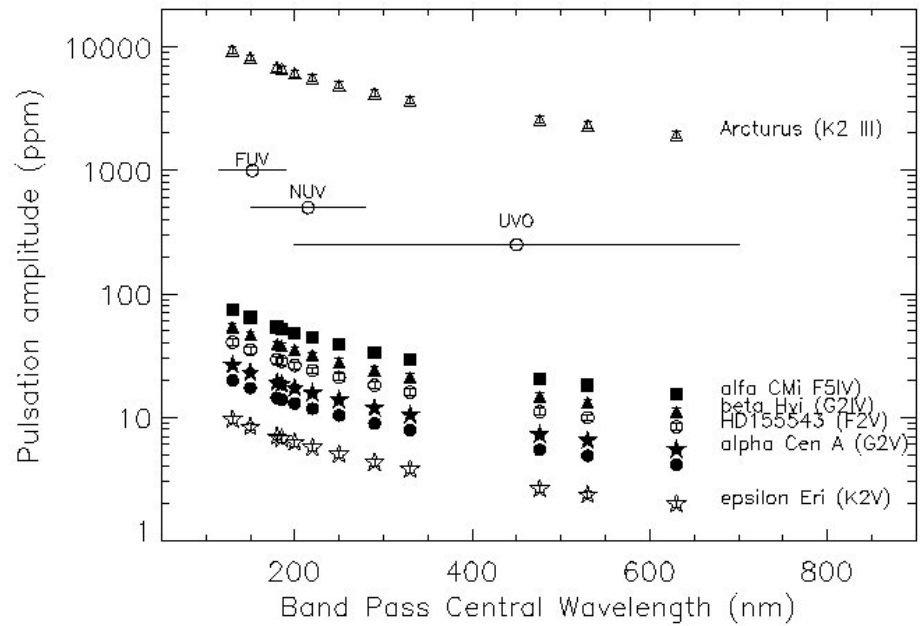

Figure 10: Variation of oscillations amplitude with band. In the plot the wavelength ranges of each FCU camera is also shown.

Figure 10 shows the trend of the oscillation amplitudes as function of wavelength. We can see that they increase towards the ultraviolet domain, so space observation with WSO-UV could provide very interesting and useful data.

*White Dwarfs*

The White Dwarfs represent the last phase of the life of stars with initial masses ranging between about 1 and 6 Msun. They occupy the region of the H-R diagram called *Cooling Sequence*, where instability phenomena occur.

In particular we define three variability episodes in these stars: *a*) during the Planetary Nebula or in the immediately subsequent phases, *b*) during the DBV phase (high Helium abundance), with effective temperature ranging between 23.000 - 25.000 K, *c*) in the DAV phase (high Hydrogen abundance), characterised by effective temperature ranging between 11.000 and 12.000 K.

Since their relatively high oscillation amplitudes, short periods and the great number of excited modes, the White Dwarfs are very interesting targets of Asteroseismology. In these objects the non-radial pulsations are gravity waves, or g-modes, due to $k$ and $g$ mechanisms in the ionization zone near the stellar surface. In spite of their behaviour in the solar-like stars, the g-modes are the more superficial modes, because their cannot pass through the extreme conditions in the degenerate core, while the p-modes show very short periods, of about 1 second.

The characteristic of the g-modes is to be equally spaced in period, according to the asymptotic theory. A variation in this regular pattern is index of chemical gradient in the surface layers.

White Dwarfs are too faint objects for the satellites (dedicated to asteroseismology) MOST, COROT and Kepler. As pointed out by Kepler et al. 2000, analysing HST data from the FOS spectrograph (see Figure 11) the amplitude of the pulsations increases towards lower wavelengths, allowing to ultra-violet instrumentation a major possibility to detect them.

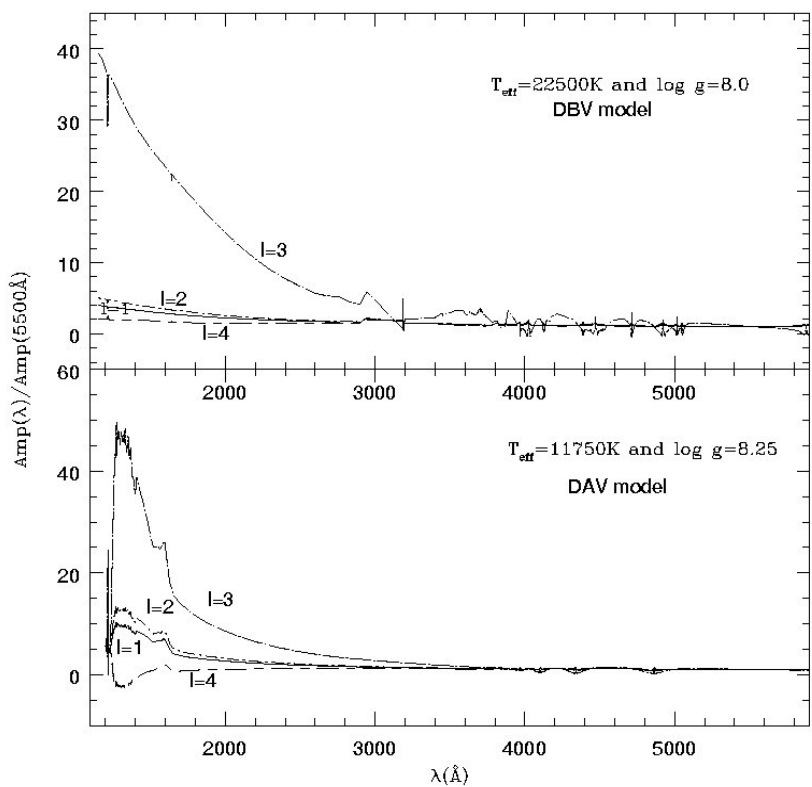

Figure 11: Variation of White dwarf pulsation amplitude with wavelength (by Kepler et al. 2000)

In particular are the modes with higher degree *l* which take more profit from this fact. They become also free from the limb-darkening effect, which is responsible for the undetection of the higher modes in distant stars. For these reasons the Asteroseismology in the UV range could provide interesting results as in the optic.

Through this kind of analysis we should obtain very important parameters of the structure of the White Dwarfs, and probe the models of stellar evolution since we could have at our disposal a posteriori investigation. In particular we could achieve:

- the total mass of the star, from the period spacing;

- the mass of the outer layers and information's on their chemical abundance, from the deviation in the period spacing;

- the rotation period, from the rotational splitting of the multiplets in the power spectrum;

- the magnetic field from the magnetic splitting of the multiplets in the power spectrum;

- the evolutionary time scale, achievable quantifying the crystallization effect, which slows down - the cooling time. This information gives also a useful estimate of the age of the galactic disk in the solar vicinity;

- test the physics of the neutrino production in dense plasma. Even in this case it gives information's on the cooling time of the White Dwarf.

The power spectrum of these objects is often complex, since the Fourier analysis carried out for two basic frequencies $w\_1$ and $w\_2$ produces peaks at the combined frequencies $kw\_1+jw\_2$.

*B spectral type stars*

B stars represent another type of objects that shows a variability characterised by non radial pulsations. Near or above the Main Sequence, they are subdivided in several classes: β Cep stars, slowly pulsating B stars, pulsating Be stars, pulsating supergiant B stars and subdwarf B stars (see Aerts 2006). All of these stars present both p- and g-modes of oscillations.

Even if the driving mechanism of the oscillations is known (the κ mechanism acting in the partial ionization zone of the iron-like elements), we yet cannot explain the reason for which only some modes are excited compared to the expected. These stars show oscillation periods ranging from few hours to few days, so it is necessary to get long run observations, as provided by space observations.

Comparison between theory and observations of some β Cep stars show incompatibilities between the number of expected modes and the observed ones. Thanks to new observations it could be possible to confirm the hypothesis of new models on the role played by the radiative diffusion, that in the particular case of ν Eri solved the problem.

A complete modelling of these stars is not yet achieved because of the lack of data, in particular for the pulsating Be and supergiant B stars. In the latter case, since they are evolved stars, we have to face the effect of the frequency spacing variation due to the avoided crossing phenomenon, but as in Garrido 2004, the mode identification could be achieve using different filters. We can apply this technique to identify the modes also when the rotation produce asymmetries in the regular pattern of frequencies.

Observations of Subdwarf B stars can help us to understand their evolution, and perhaps clarify the nature of the strong mass loss mechanism which is responsible for their very tiny inert hydrogen envelope. As White Dwarfs, sdB stars show low amplitudes, multiperiodicity and small pulsations periods, even if there are some particular cases of subdwarf B stars with longer periods and g- mode pulsations. This last category of B stars shows iron abundance anomalies (responsible for the κ mechanism) probably due to a weak surface wind, as explained in Billeres et al. 2004. Since this wind has not been yet observed, we could have the possibility to detect it.

Another problem is in the estimation of atmospheric parameters, in particular the effective temperature. Asteroseismology can set a good estimation of the Teff scale which could help to compare theoretical models and observations.

Since subdwarf B stars present extremely low amplitude, to clearly identify the period in the Fourier transform, it is necessary to obtain long baseline observations.

## 2.3.2.2 Stellar Magnetic Activity

### Scientific Background

Turbulent convection in the layers under the photosphere and differential rotation, both present in late type stars, generate strong magnetic flux ropes by dynamo action. The tubes bounce and, arising from the stellar surface, become braided and twisted by surface velocity fields. Eventually, magnetic free energy, released by magnetic field reconnection, heats the outer atmospheric layers. This hot plasma can be studied in a single ultraviolet spectrum, providing simultaneous information on thermal structures of a wide range of atmospheric components. In fact, UV continuum together with UV emission lines of, e.g., C I, Mg II and Si I, are formed at temperatures between 7000 and 10000 K which are typical of the chromospheric plasma, while Ly-alpha, He II, C II, Si IV, C IV, O IV, OV, O VI, and higher ionized UV lines form from 10000 K to million degrees allowing to finely sample the transition region till the corona (see, e.g., the HST/STIS spectrum of the dM1e AU Mic in Figure 12).

We will focus in the remainder of this section on a few selected open issues in cool-star physics that can be addressed by new observations in the ultraviolet spectral range. These issues are, in our opinion, hot topics in stellar physics and key ingredients to progress in our understanding of space weather in our solar system and in other planetary systems because the stellar activity (irradiance cycles, flare, CMEs, winds) directly affect planetary environments on a range of timescales (Pagano et al. 2006).

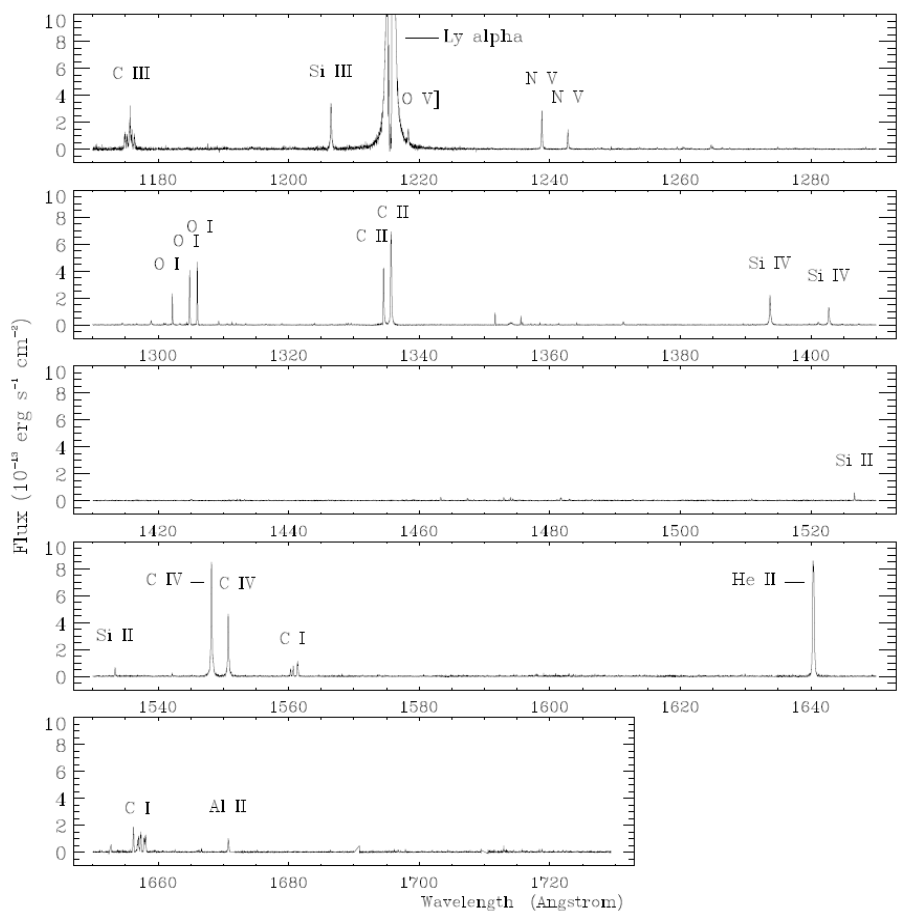

Figure 12: The spectrum of AU Mic (dM1e) – a typical magnetic active star - acquired at 45000 resolution by HST/STIS. Only major emission lines are labelled. They probe the plasma from $10^4$ to $10^6$ K, from the chromosphere, to the transition region and low corona (Pagano et al. 2000).

*Theories on the dynamo and the transport of magnetic energy in plasmas*

Butler (1994) and Pallé Bagó & Butler (2001) have found quantitative evidence that much of the warming of the past century can be accounted for by the direct and indirect effects of solar activity. However, at the moment we cannot forecast long and short-term solar magnetic variability because we do not have a comprehensive understanding of their root causes: the dynamo mechanism and the physical processes that shape plasma structure and control dynamics in the solar outer atmosphere.

One way to progress in this area is to step back from the Sun, and instead evaluate how magnetic activity phenomena depend on fundamental stellar parameters: mass, rotation rate, chemical composition, binarity, and so forth. The diverse stellar examples of activity might reveal insights into the underlying physical processes that observations of the singular example of our Sun cannot, no matter how detailed.

Information on the spatial distribution of active regions in the different layers of the stellar atmosphere, on their temporal evolution (activity cycles), on the velocity fields in the quiet and active regions, and finally on the impulsive events (microflares, flares, CMEs) due to magnetic field flux tubes reconnections are fundamental to progress in this field.

For instance, data on the surface organization of magnetic activity in large stellar samples and on stellar cycles constraint the stellar dynamo theories. We know from optical observations that magnetic active stars have large dark starspots, site of kiloGauss order magnetic fields. The optical observations of stellar-spots provide gross information on the way the magnetic fields are spatially organized. Polar spots, migration of active regions toward the poles, preferred active longitudes, "flip-flop" effect (e.g., Rodonò et al. 2000) are phenomena observed for highly active stars. However, when looking at photospheric spots only we have an incomplete description. In fact, as we see on the Sun, a large fraction of the global magnetic flux is carried by plages and supergranulation. These features are present even at cycle minimum when few or no spots are on the disk and show very large intensity contrasts in the ultraviolet while are not easily distinguishable from the surrounding quiet photosphere at visible wavelengths. Hence, to derive information on a large fraction of the global magnetic flux, UV emission must be investigated.

The understanding of chromospheric and coronal heating is one of the major unsolved problems in solar and stellar physics. Plasma dynamics is an important by-product of the heating mechanisms, and thus a potential window into their nature. In fact, turbulence in the convection zone transforms gas kinetic energy into sound waves and propagating electrodynamic disturbances. The latter are thought to provide the bulk of coronal heating. However, the shock dissipation of pure acoustic waves -- mostly damped in the TR because of the steep temperature gradient -- may play also a role in the chromosphere and the lower transition region heating. Quantitative details of the energy transport and dissipation processes are poorly known (c.f., Judge et al. 2004 and references therein). Hence, information on the plasma dynamics is fundamental to assess the role of sound waves shock dissipation with respect to the magnetic mechanisms. Such information can be derived by the inspection of resolved emission line profiles of ion transitions originating at different temperatures from the chromosphere to the corona.

*Winds from cool stars*

The winds of low-mass cool stars feed back on their coronal evolution owing to the significant angular momentum carried away by a fast, magnetized outflow. In the evolved red giants, winds can potentially directly affect the nuclear evolution of the star by removing significant mass from the surface layers. Coronal winds can erode volatiles from primitive planetary atmospheres by sweeping up charged molecules and ions from exospheric regions ionized by the stellar UV radiation field. Moreover, stellar winds control the flow of material and flux of cosmic rays from the galactic environment.

Hence, exploring the winds of low-mass and evolved stars is of fundamental importance from a number of standpoints.

By modelling H I Lyman-alpha profiles obtained at high resolution (R~45000) from HST/ GHRS and STIS, Wood et al (2002) performed the first quantitative measurements of mass loss rates for G and K dwarf stars. These authors found that the mass loss rates increase with activity - with a saturation at very high activity level - and thus with decreasing stellar age. By implication, the solar wind might have been 1000 times stronger when the Sun was very young, and thereby likely played a major role in the evolution of planetary atmospheres, particularly the stripping of volatiles from primitive Mars.

*Activity in young galactic clusters and star-forming regions: How stellar activity affects planets*

How stellar activity affects planets, especially habitable ones? The evolution of a planetary atmosphere under the joint erosive impacts of coronal ionizing radiations and wind from the parent star is a hot and open scientific issue, with important ramifications for understanding the evolution of Earth's paleoclimate and the birth of life (see Lammer et al 2003).

The high energy emission, both as radiation and particles, resulting from magnetic activity on the central star influences the thermal structure of young star disks, the formation process of planetesimals, and the photoexcitation and photoionization of protoplanets and young planetary atmospheres.

The study of late-type stars in young (~50 -100 Myr) galactic clusters, and even younger star-forming regions (1-10 Myr), is a valuable window into not only the evolution of magnetic activity, but also its basic characteristics. In particular, within a given cluster or other coeval group of objects, the relationships between, for example, activity indicators and rotation periods for stars of different spectral types will not suffer a hidden age bias.

It is generally accepted that late-type stars undergo spin-down due to magnetic braking throughout their main-sequence phase, especially near the beginning where their angular momentum is highest thanks to spin-up by their natal disks (see, e.g., Collier Cameron and Jianke 1994 and references therein).

Studies of the "Sun in time" (see Ribas et al 2005, Guinan et al 2007 and reference therein) have made use of observations of stars with different rotation periods and ages, but spectral types around the solar one, to build a scenario of what might have happened to the Sun at different stages along its evolutionary path. The several efforts have used observations of activity tracers in the optical, ultraviolet and X-rays. In particular, in the ultraviolet domain *IUE* spectra were intensively exploited to extract information on chromospheric and transition region fluxes; while more recently *FUSE* observations were used to study the whole outer. The general activity-rotation-age relationship in stars also has been the subject of many studies. The first efforts showed that the activity-age relation for main sequence stars older than about 100 Myr can be modelled by an exponential law whose rate of falloff depends on surface temperature.

Unfortunately, only few subsequent extensive studies in this direction have been carried out using the much more sensitive contemporary instruments of *HST* or other missions. Ayres (1999) obtained GHRS spectra (115 - 67 nm) of three solar-type dwarfs in the young clusters *alpha* Per (85 Myr) and the Pleiades (125 Myr) to complement an earlier FOS study of 10 cluster stars, including 5 in the Hyades (625 Myr); with the aim to investigate the behaviour of the stellar activity and the dynamo at early ages. Although the study drew important conclusions on the activity-age relationship for TR lines like C IV, these referred specifically to early G-type dwarfs, and generalization to other spectral types and luminosity classes was not possible. The lack of modern UV work on the age-activity relation contrasts with the very extensive surveys accomplished in the optical (see e.g. Soderblom et al. 2001) and X-ray domains. In the latter, for example, Pizzolato et al 2003 investigated the relationship between coronal X-ray emission and stellar rotation in a sample of 259 dwarfs in the B-V range 0.5-2.0 observed with *ROSAT*, including 110 field stars and 149 members of the Pleiades, Hyades, alpha Per, IC 2602 (30 Myr) and IC 2391 (30 Myr) open clusters. The "missing" UV part of the puzzle is extremely important to problems such as the radiative erosion of primitive planetary atmospheres by young hyperactive parent stars because the dominant ionizing radiations for abundant atmospheric molecules like $N_2$ fall in the Lyman continuum (see Ayres 1997); the crucial bright O V 62.9 nm feature, for example, cannot be observed directly in stars, but its strength can be inferred through measurements of O IV, O V, and O VI UV lines longward of the 91.2 nm H I edge.

At even younger ages, are pre-main sequence objects in star-forming regions. The UV is extremely valuable in studies of the less-obscured of the classical and "naked'" T Tauri stars. For example, fluorescent excitation of the Lyman bands of $H_2$ lines by H I Ly-alpha can be used to probe the physical conditions in the accretion disk. UV signatures of the hot splashdown point of the accretion stream can be exploited to estimate the accretion rate and geometry. Furthermore, these very young objects, not surprisingly, are hyperactive in terms of the usual UV and X-ray indicators, and thus provide a laboratory for "magnetic activity in extreme environments". The dissection of the highly complex TRs and coronae of these quite exotic objects is an important challenge for future emission-line Doppler imaging efforts.

*Magnetic activity of stars hosting planets*

Theoretical models suggest - and first observations confirm - that giant planets orbiting close to their parent star – the so-called hot-Jupiters – magnetically and tidally interact with the star inducing magnetic activity on it. Understanding the nature of this process is relevant to predict the evolution of the stellar-planets systems. This topic has a more extensive description in Section 2.3.4.1, so we will not discuss it further.

**Key observables**

*3D maps from the chromosphere to the corona*

Direct stellar images of active stars hot layers are not obtainable at moment. In fact, UV maps of the closest solar-like dwarf stars -- with good enough resolution to record small active regions and crudely image the supergranulation network -- require a spatial resolution of about a tenth of a milliarcsecond. Extending the sample to dwarfs within 100 pc would require microarcsec imaging.

However, maps of active regions can be reconstructed by photometric and/or spectroscopic methods. Both methods work when the spatial distribution of active regions is not uniform and the stars are not seen pole-on. In all other conditions, the visibility of active regions changes as the star rotates producing rotationally modulated variability of the stellar light curve observable by photometric measurements and rotationally modulated distortions of line profiles observable spectroscopically.

Measurements made in narrow band filters around intense chromospheric (e.g. Mg II 280 nm) and transition region (e.g. C IV 150 nm) emission lines allow the reconstruction of rotationally modulated light curves and to study the stochastic variability (flares and microflares) due to magnetic field reconnection events.

High-resolution UV spectra repeated during the stellar rotation allow to measure Doppler shifted signatures of active regions in emission line profiles (Doppler imaging), useful to map the surface structure of chromosphere and TR (see Pagano et al. 2001). Doppler imaging requires spectral resolution better than 30000 and the ability to obtain time series of spectra with good enough cadence to avoid smearing the Doppler information, and long enough coverage to distinguish between persistent surface features and transient flare activity. In fact, the best targets for emission-line Doppler imaging are fast-rotating stars, and these by their nature are highly active and flare frequently.

Also, fluorescent lines of molecules (e.g., H2 and CO) and atomic species (e.g., Fe I, Fe II, O I, Cl I) falling in the UV spectrum can be useful to map the plasma temperature structure in the outer atmospheres of active stars. These lines are produced thanks to the excitation of specific low-excitation lines by photons originating in hotter regions of the stellar (or disk) atmosphere. The presence in the spectrum of such fluorescent lines is and indication of co-presence of hot and cold gas in the outer atmospheres of these stars. In short, fluorescence processes can be a guide to the geometrical organization of a stellar atmosphere (or accretion disk) on small physical length scales not accessible to direct telescope observations.

*Velocity fields in plasma from $10^4$ to $10^7$ K*

*Line shift vs. temperature of line formation:* Observations of the Sun show that transition region emission lines are, on average, redshifted, and that redshifts increase with increasing formation temperature up to about $10^5$K. The maximum redshift (~15 km/s) is reached at about $1$-$1.2 \times 10^5$K (active regions and quiet Sun, respectively). At higher temperatures, the centroid velocities decrease, crossing over to blueshifts at $T \sim 10^6$ K, reaching about -6 km/s (see e.g. Peter & Judge 1999). This behaviour is not anticipated by models of upward propagating acoustic waves, for which both optically thin and optically thick lines are predicted to be blueshifted. On the stellar side, a solar-like behaviour is seen in most of the investigated late-type stars even though the maximum red-shift is sometimes reached at higher or lower temperatures than in the solar, however very active dMe stars show hardly any redshifts, and certainly show no conspicuous trend of line shift versus formation temperature (see Pagano 2004 and reference therein).

Broad wings of transition region lines*:* In addition to redshifts, stellar transition region emission lines also show a curious bimodal structure: broad wings superimposed on a narrower central peak. The broad wings have widths comparable to, or wider than, the broadened C IV profiles observed in solar transition-region explosive events (see Dere et al 1989), small bursts thought to be associated with reconnections of emerging magnetic flux to overlying pre-existing canopy fields. Wood et al (1997) showed that the narrow components can be produced by turbulent wave dissipation or Alfvèn wave-heating mechanisms, while the broad components---whose strength correlates with activity indicators like the X-ray surface flux---can be interpreted as a signature of ``microflare'' heating. Peter et al. (2001) showed that broad components are a common feature in the thermal regime 40000 K to $10^6$ K above the magnetically dominated chromospheric network on the Sun, and presented evidence that the narrow line core and broad wings are formed in radically different physical settings: small closed loops and coronal funnels, respectively, the latter being the footpoints of large coronal loops. Non-thermal widths of the broad components follow a power-law distribution with respect to line-formation temperature, a signature of upward propagating magneto-acoustic waves. In stars other than the Sun, the broad components have been observed over formation temperatures of 40000 K to $2 \times 10^5$ K. Extending the stellar coverage to higher temperatures, as in the solar work, is a challenge for the future.

*Dynamics of the corona by means of UV forbidden lines:* Extending the observational side of plasma dynamics to higher, coronal temperatures is relatively straightforward on the Sun, because many suitable strong coronal permitted lines (e.g., Mg X 61,025 nm) fall in the Lyman continuum region immediately below 91,2 nm, where high-resolution far-UV spectroscopy still is practical. Unfortunately, these key features are not accessible even in the nearest stars owing to interstellar extinction. Observing permitted coronal X-ray lines, say in the important iron-L shell band at 1 keV, is not feasible at present, because contemporary high-energy missions like *Chandra* and *XMM-Newton* fall short in spectral resolution by an order of magnitude, and there are no planned future missions that will push that limit. However, a number of coronal *forbidden* lines in the UV range longward of the Lyman continuum edge have been observed in the Sun (see for example Doschek et al. 1975, Feldman et al. 2000). On the stellar side, *HST*'s GHRS and STIS instruments, and *FUSE* have been able to reach highly ionized iron forbidden lines in many cool stars (see for example Maran et al. 1949, Pagano et al. 2000, Ayres et al 2003a, Redfield et al. 2003) with high spectral resolution (up to R = 40000).

The UV coronal forbidden lines detected in cool stars for the most part show negligible Doppler shifts, suggesting that the emissions arise mainly from confined structures, analogous to magnetic loops on the Sun, rather than, say, in a hot wind. Moreover, the Fe XII and Fe XXI line widths generally are close to their thermal values (FWHM 40-90 km/s at $T \sim 10^{6.2}$-$10^7$ K), except for the Hertzsprung-gap giants 31 Com (G0 III) and Capella (G1 III) and the K1 IV component of the RS CVn type system HR 1099, all of which display excess broadening in Fe XXI. If the additional broadening is rotational, it would imply that the hot coronae of "X-ray deficient" 31 Com and Capella are highly extended, in opposition to the compact structures suggested by recent density estimates in a number of active coronal sources. On the other hand, the more common case of purely thermal line widths imply that supersonic turbulent motions are absent in the coronal plasma, eliminating shock waves as an important heating mechanism.

*The physics of impulsive heating: stellar flares and microflares*

Flares, lasting from a few minutes to several days, are the most dramatic examples of transient energy release in solar and late-type stellar atmospheres. A magnetic reconnection process is though to power these events: magnetic free energy is converted---in thin current sheets---into thermal heating, Alfvén waves, and the acceleration of relativistic particles. Accordingly, the relaxation phenomenon is complex, fast, and radiates across the whole electromagnetic spectrum from non-thermal radio synchrotron, to thermal emissions in the UV and X-ray ranges. In fact, the flare phenomenon involves the whole atmosphere, from the corona down to the lower chromosphere and photosphere, blasted by hard radiations and particle beams from the high-altitude flare kernel. The investigation of detailed flare physics historically has relied on multi-wavelength simultaneous observations, involving both spectroscopy and photometry.

Ultraviolet emission lines provide an important source of plasma cooling, as well as lower-atmospheric heating, during the gradual phase of stellar flares (e.g., Hawley et al 2003), and thus are valuable diagnostics of the flare evolution. Although large flares are a conspicuous contributor to transient heating of stellar coronae, smaller scale events---so-called micro- and nano-flares---might play a key role in the "steady" heating of the outer layers of cool stars. Robinson et al (2001) studied the statistics of transient bursts in high time resolution UV observations of a typical flare star (AU Mic) with *HST* STIS, and concluded that the power-law slope of the occurrence rate versus time-integrated flux was considerably steeper for low-energy flares than for the rarer high-energy ones; implying that microflares potentially can account for a significant portion, if not all, of the coronal heating. However, this is such an important issue that more conclusive evidence is needed: the investigation should be extended to a larger sample of flare stars, collecting a significant number of events in each target.

Despite the broad band photometry delivered by GALEX does not provide optimized information, Robinson et al. 2005, and Welsh et al 2006 analyzed GALEX serendipitously observed flares from dMe stars thanks to time-tagged data (resolution time <0.01 sec) provided by this instrument. These authors found evidence of two distinct classes of flares: a first type - showing a flux increase while the temperature stays constant – that can be modelled by avalanche of nanoflares in a plage, supporting the view that plages are heated by microflares, as first proposed by Parker (1988); a second type – showing both flux and temperature increase - that are instead compatible with classical explosive events (solar like two ribbon flares). The latter flare type is very important for the influence they have on the planet atmospheres.

Moreover thanks to time-tagged data, Welsh et al 2006 found evidence of coronal loop oscillation during flares (30–40 sec range) that are signature of acoustic waves in coronal loops.

The objective t study flares and microflares is achievable with UV imaging time-tagged photometry in narrow band filters tuned on strong chromospheric and transition region emission lines or time-tagged low resolution field spectroscopy (R≥100) possibly in stellar clusters.

*UV emissions from very late M dwarfs and brown dwarfs*

Magnetic activity decreases for spectral types later than approximately M7. However, flares have been observed in optical, UV and X-rays from very late dM stars and brown dwarfs (e.g., Linsky et al 1995, Rutledge et al. 2000). Several authors have suggested that hot gas in the atmospheres of very low mass main-sequence stars and brown dwarfs is present only during flares. However, by recording UV spectra of a sample of these stars, Hawley & Johns-Krull (2003) showed that a persistent quiescent chromosphere and transition region, similar to those observed in earlier type dMe's, is present at least through spectral type M9. The existence of persistent magnetic activity in these fully convective stars poses challenges for contemporary dynamo models that require a shear layer between the convective outer envelope and the radiative interior. The current thinking is that a ``distributed dynamo'' might be in play, one that operates directly on convective turbulence and does not require the catalyzing agency of differential rotation. Incidentally, this same "alpha-2" dynamo process probably also is operating in the slowly rotating red giants, to account for the feeble, but nonetheless present, coronal activity of the inhabitants of the so-called "coronal graveyard" (Ayres et al. 2003b). Extending this work -transient vs. persistent hot plasma - to the even lower mass range, and much cooler atmospheres, of the brown dwarfs will require significantly more sensitive UV spectroscopy than was available, for example, from *HST* STIS. Moreover, UV imaging time-tagged photometry narrow band filters tuned on strong chromospheric and transition region emission lines or time-tagged low resolution field spectroscopy (R≥100) in stellar clusters will result in a much convincing statistics on flare occurrence in very low-mass stars and BDs.

*Evolved stars winds and solar-like stellar winds*

The velocity structure, mass loss rate, and ionization of the outflowing gas in the low-temperature ($\sim 10^4$ K) winds of late-type giants and supergiants can be inferred from high resolution spectra of optically thick UV resonance lines. In fact, the UV provides unique access to the dominant circumstellar wind absorption species in the red giants: H I, O I, Mg II, and C II

The winds of main sequence stars like the Sun are too hot and too thin to provide detectable UV or X-ray absorption (or emission) signatures. Fortunately, however, in a few favourable cases of nearby stars, coronal winds can be studied by subtle distortions of H I Ly-alpha due to the "hydrogen walls" produced by the interaction of the stellar outflow with inflowing interstellar gas. The stellar "astrosphere" causes an absorption feature on the blue side of the interstellar absorption. For studies of cool winds and astrospheres, higher resolutions are needed: R= 100000 is considered a goal for such studies, but previous HST STIS work have showed that is possible to get results also at R~45000.

*UV radiation fields in the stellar system*

The characterizations of the UV photoionizing and photoexciting radiation in the surrounding of the stars can be obtained by means of imaging photometry in narrow band filters tuned on strong chromospheric and transition region emission lines or low resolution field spectroscopy (R≥100) of stellar populations with known age, of both old and young kinematic groups. Young stars have very high level of stellar activity (e.g. UV and X-ray emission) that can strongly affect the formation and evolution of their circumstellar disks and protoplanetary systems. The dwarfs of the Local Associations, whose ages range from 7 to 200 Myr, are a remarkable sample of young stars very close to the Sun (typically near than 100pc) that permit the detailed study of early evolution of stellar activity. The crucial radiation field that photoionizes protoplanetary disks and atmospheres is dominated by transition region and low corona emission lines emitted in the EUV (30-90 nm) region, that are unobservable because of strong interstellar absorption. However, their strength can be estimated by UV (110-300 nm) observations of other emission lines from the same ions or ions formed at similar temperatures. Similarly, UV data can provide detailed information on the magnetic activity level of old stars.

## The WSO-UV key contribution

The study of the outer atmospheres of cool late-type stars (isolated and in binaries) and their magnetism – that affect the evolution of protoplanetary systems -  can benefit from WSO-UV photometry and spectroscopy.

Magnetic activity phenomena – spots, faculae (plages), flares, loops – are highly variable in time on different time scale from fraction of seconds (microflares), to seconds/hours (flares), days (rotational modulation), months/years (magnetic cycles). This field of research highly benefit from detectors providing time tagged data, and from telescopes capable of monitoring observations. Up to date only the sensitivity limited spectrographs on *IUE* have allowed monitoring studies for stellar activity. While the possibility to study the stellar transition region variability on short time scale has been provided only by EUVE on few bright targets With *HST*, the small field of view of UV imagers, together with the strong pressure to this project (certainly not UV dedicated) and the difficulties to get monitoring observations have strongly limited the study of stellar chromospheric and transition region emission in stellar clusters in contrasts with the very extensive surveys of photospheric and coronal activity accomplished in the optical (see e.g. Soderblom et al. 2001) and X-ray domains (Pillitteri et al. 2005, Rebull et al. 2006, Stelzer et al 2007). Also the beautiful spectra obtained by *HST* GHRS and STIS (e.g., Linsky et al. 1998, Pagano et al. 2000, 2004) gave us only snapshots information. Imagine studying the solar activity just taking only one spectrum of the Sun at a random epoch.

We need now to have access to an instrument dedicated to UV astronomy, capable to monitor stellar activity in cluster of different ages to allow studies on time scales from hours – days – months and possibly years. Hence, a main requirement to the telescope is the capability to observe from an orbit with few operational constraints in order to allow long-duration uninterrupted observations. Also it should be planned to provide observations for at least half a decade (i.e., half of the solar cycle, since we have indication of stellar cycles on time scale similar to the solar one). These requirements are satisfied by the presently choice of a geostationary high altitude orbit for WSO-UV, and from its planned life of 5(+5 years). Moreover, the focal plane instruments – UV spectrographs and imagers - must provide time-tagged data in order to allow investigations on time scale from fraction of seconds to hours.

**Observations with WSO-UV imagers**

The most important lines longward of the 91.2 nm H I edge that are diagnostic for magnetic activity are Ly-alpha 121.6 nm, O I 130 nm, Si IV 140 nm, C IV doublet 150 nm, He II 164 nm, and Mg II 280 nm. Narrow-band filters tuned to these important lines sampling plasma from $10^4$ to $10^6$ K (He II is also sensitive to coronal radiation fields) coupled with detectors capable to deliver time-tagged data and to an optical design that provides a large enough field of view (let say, $\geq 6 \times 6$ arcmic$^2$) will allow us to investigate microvariability, flares, rotational modulations, and active cycles in the chromosphere and transition region of large sample of late-type stars at the same time, for example, members of a compact galactic cluster or PMS candidates in a star-forming regions, thus complementing and adding value to works made in the optical and X-ray which describe the magnetic activity phenomena in the photosphere and in the corona.

The possibility to get UV low dispersion spectra, let say, R$\geq$100, especially in the region 100-200 nm, of all the objects in the field should also be explored, being in such a case a single image capable to provide contemporary information on structures at different temperature in the vertical atmosphere.

**Observations with WSO-UV spectrographs**

A resolution of R about 30,000 is sufficient to resolve most of the narrow chromospheric emission lines seen in dwarf stars, which typically have FWHM >10 km/s. The hotter TR lines are broader owing to higher thermal velocities, but they also benefit from good resolution for deblending purposes, dynamical studies, and Doppler imaging. The WSO-UV/HIRDES (UVES & VUVES) sensitivity allow us to reach the faint, interesting objects beyond the solar neighbourhood, out to at least 150 pc to include, for example, the important young galactic clusters alpha Per and the Pleiades, as well as the key TW Hya star-forming region. With the LSS spectrograph (R~ 2500) we can study flares and other types of variability (i.e., velocity resolution ~120 km/s) because then one can have cleaner separation of close lines, some velocity discrimination in short-period binary systems, and the possibility of measuring hypersonic dynamics in large flare events. The adoption of photon counting detectors for the three spectrographs on board WSO-UV is perfect to allow time variability studies of line profiles, investigation on flares and microflare statistics.

### 2.3.2.3 Globular Clusters

*Interplay between stellar evolution and dynamical evolution*

Although the bolometric light from old stellar populations (as the galactic globular clusters, GGCs) is dominated by cool bright giants, these stellar systems harbour a significant amount of hot objects that are best observable at ultraviolet (UV) wavelengths Figure 13: extreme horizontal branch (EHB) stars, blue stragglers stars, cataclysmic binaries, binary millisecond pulsar, etc.

The origin and the formation mechanisms of these hot stellar populations are not fully understood yet. Most of them cannot be interpreted in the framework of the passive stellar evolution of a single star, and are thought to be produced by the evolution of primordial binaries, and/or by dynamical processes involving the formation, evolution, and interaction of binary systems in the over-crowded regions of GC cores. The significant radiation of these populations in the UV domain produces key observational signatures that make them stand out of the ordinary cluster stars at these wavelengths. Moreover, the exotic objects preferentially populate the central regions of GCs, that are usually very congested in the optical bands, but relatively uncrowded in the UV (see Figure 14). Hence, UV observations are **key observables** to identify and fully characterize the properties of these puzzling populations.

*HB morphology and the nature of the EHB stars*

The colour (temperature) distribution of stars along the Horizontal Branch (HB; hereafter the HB morphology) is due to a distribution of the stellar envelope mass (see Rood 1973): stars with large envelopes populate the red (cool) side of the HB (redder than the RR Lyrae strip), while stars with thin envelopes populate the blue (hot) extension of the HB. The HB morphology can differ significantly from cluster to cluster, with the stellar metallicity being the primary driving parameter: metal poor GCs tend to show essentially blue HB, while metal rich ones tend to have stubby red HB. Notable exceptions are however well known (see classical case of M3 and M13), and the origin of the differences in the HB morphology from cluster to cluster is not yet understood. Age is a popular "second parameter" (see Rood 1973; Lee et al. 1994; Catelan et al. 2001), but evidences have been found that clusters with the same age and metallicity can have quite different HB morphologies (see the example of the M3/M80/M13 triplet; Ferraro et al. 1998). The amount of mass loss during the red giant branch (RGB) phase certainly plays a major role in determining the envelope mass (and hence the temperature) of a star on the HB. A growing number of GCs have been found to show a prominent hot extension (the "blue tail", BT) of the HB, that defines an almost vertical sequence extending down to visual magnitudes similar to (or fainter than) the Main Sequence (MS) turnoff in the (V, B-V) colour-magnitude diagram (CMD). Therefore, studying the distribution of BT stars along the HB is crucial to get information about the most extreme mass loss activity, and to provide valuable constraints to the still poorly understood physics of mass loss. Moreover, the HB stellar distribution in some GCs is interrupted by underpopulated regions or "gaps", that can explained by assuming the existence of "multiple" mass loss "drivers" (Ferraro et al. 1998, Piotto et al. 1999). The gaps might also be related to phenomena like the helium dredge-up (Sweigart et al. 2002, "Omega Cen: A Unique Window into Astrophysics"), or primordial helium abundance variations (D'Antona et al. 2002; D'Antona et al. 2005; Piotto et al. 2005, 2007). Even if these mechanisms have been suggested a while ago (see Fig.16 of Rood & Crocker 1989), a deep understanding of these issues has not been reached yet, and the lack of a high quality, homogeneous data-base has precluded definitive answers to several important questions, as: "Are the gaps always located at the same effective temperature in all clusters?", or "Do the gaps correspond to HB regions completely devoid or only poorly populated of stars?". Another vexing question is the origin of the "extreme HB" (EHB) stars. These objects populate the extreme extension of the BT of the HB, and have suffered significant mass loss during their late red giant evolution (Dorman et al. 1993). EHB stars and their progeny are the brightest UV objects in GGCs, and are found even in metal rich ([Fe/H]=-0.4) systems, like NGC6388 and NGC6441 (Rich et al 1997). They are also observed as hot subdwarf (sdO/sdB) stars in the galactic field, and have been suggested to be at the origin of the UV upturn in elliptical galaxies and spiral bulges (Buzzoni, 1989; Greggio & Renzini 1990). The primary issue in understanding their origin is why some stars loose much more mass than others during their RGB stage. For instance, are stellar interactions or binarity required to produce EHB stars? Preliminary results suggested that this is not the case, since the EHB population does not vary with the cluster radius (Whitney et al. 1998; Ferraro et al. 1999; Bedin et al. 2000), and Moni Bidin et al. (2006) did not find any close binary in the BT of NGC 6752. Recio-Blanco et al. (2006) has recently found an intriguing correlation between the maximum extension of the EHBs and the cluster total mass. This result might indicate that indeed the cluster dynamical evolution has something to do with the presence of EHBs.

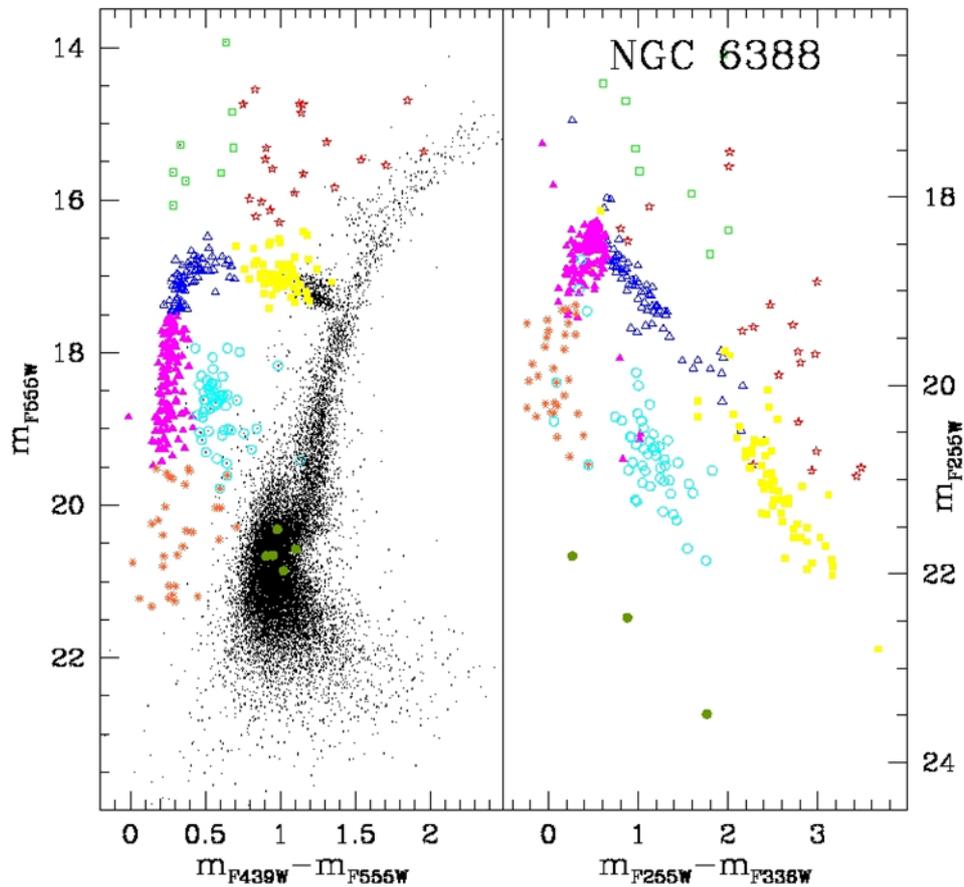

Figure 13: BSS and EHB stars dominate the flux in the UV (from Busso et al. 2007).

*Exotic Objects*

GCs are also very efficient "kilns" for forming exotic objects, such as low-mass X-ray binaries, cataclysmic variables, millisecond pulsars, blue straggler stars, etc. Most of these types of stars cannot be interpreted within the standard contest of single mass stellar evolution, and are thought to be produced by the evolution of primordial binaries, and/or by the effect of some dynamical processes. They have been found to be best studied in the UV band, and UV observations are therefore crucial for a deeper understanding of binary evolution, cluster dynamics, and of the complex interplay between stellar and dynamical evolution in dense stellar systems. Possible WSO-UV observations of MSP, LMXRB, and CVs are described in Section 5.1.4.1. Here we describe the other kind of exotica.

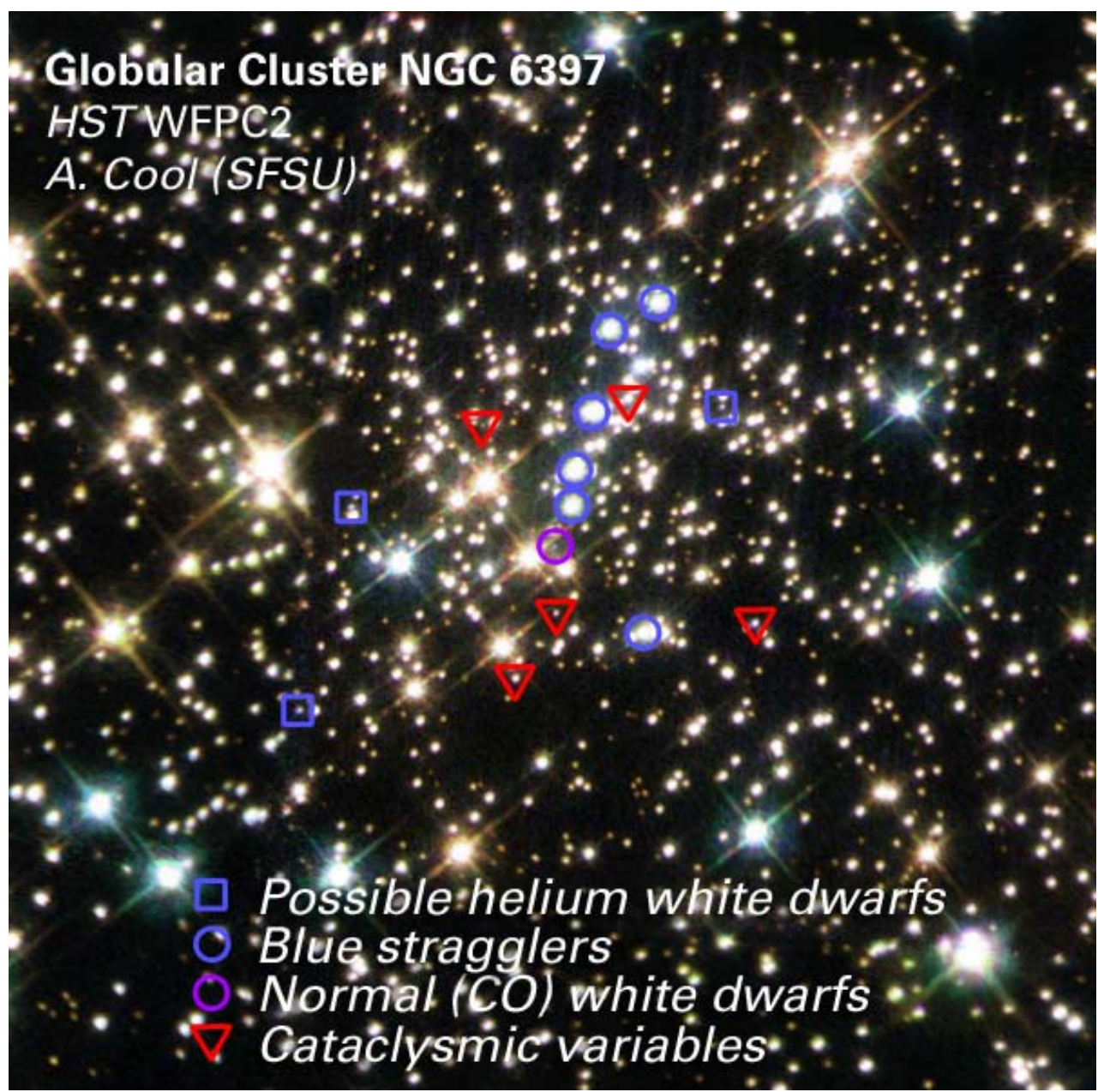

Figure 14: Globular clusters images in UV are not dominated by the red giant light, and therefore significantly less crowded.

*Blue Stragglers*

Commonly defined as those stars brighter and bluer (hotter) than the MS turnoff stars, BSS lie along an extrapolation of the MS, thus mimicking a rejuvenated stellar population. First discovered by Sandage (1953) in M3, their nature has been a puzzle for many years, and their formation mechanism is still not completely understood. BSS are more massive than the normal MS stars (Shara et al. 1997), thus indicating that some process which increases the initial mass of single stars must be at work. Such processes could be related either to mass transfer between binary companions, the coalescence of a binary system, or the merger of two single or binary stars driven by stellar collisions.

According to Fusi Pecci et al. (1992; see also Davies, Piotto \& de Angeli 2004), BSS in different environments could have different origins. In particular, BSS in loose GCs might be produced by mass-transfer/coalescence of primordial binaries (hereafter MT-BSS), whereas in high density GCs (depending on survival-destruction rates for primordial binaries) BSS might arise mostly from stellar collisions (COL-BSS), particularly those that involve binaries. This might explain the strong anticorrelation between the BSS frequency and the parent cluster absolute luminosity found by Piotto et al. (2004). As shown by Ferraro et al. (2003), the two formation channels can have comparable efficiency in producing BSS in their respective typical environment (see the case of M80 and NGC288; Ferraro et al. 1999; Bellazzini et al. 2002). Moreover, these formation mechanisms could also act simultaneously within the same cluster, with efficiencies that depend on the radial regions, corresponding to widely different stellar densities. This is suggested by the bimodality of the BSS radial distribution observed in a few clusters (M3, 47 Tuc, NGC6752, M5, and M55), where the BSS specific frequency has been found to be highly peaked in the cluster center, rapidly decreasing at intermediate radii and rising again outward (see Ferraro 2006, astro-ph/0601217, for a review). A substantial contribution to these kinds of studies has been provided by UV observations from the space, that have made possible to obtain complete BSS samples even in the densest cores of GCs, and thus turned the BSS issue into a more quantitative basis than ever before (e.g., Ferraro et al. 2003). Theoretical models still predict conflicting results on the expected properties of BSS generated by different production channels. For instance, Benz & Hills (1987) predict high rotational velocities for COL-BSS, whereas Leonard & Livio (1995) have shown that a substantial magnetic braking could occur, and the resulting BSS are NOT fast rotators. In the case of BSS formed through the mass transfer production channel, rotational velocities larger than those of typical MS stars are predicted (Sarna & de Greve 1996). Concerning the chemical surface abundances, hydrodynamic simulations (Lombardi et al. 1995) have shown that very little mixing is expected to occur between the inner cores and the outer envelopes of the colliding stars. On the other hand, signatures of mixing with incomplete CN-burning products are expected at the surface of BSS formed via the mass-transfer channel, since the gas from the donor star is expected to come from deep regions where the CNO burning was occurring (Sarna & de Greve 1996). Systematic spectroscopic observations have recently begun to provide the first set of basic properties (mass, rotation velocities, etc.; see the recent work by De Marco et al. 2005). However, with the exception of a few bright BSS in the open cluster M67 (Mathys 1991; Shetrone & Sandquist 2000), and a restricted sample in 47 Tuc (Ferraro et al. 2006) an extensive survey of BSS surface abundance patterns is still lacking, particularly in GCs. Despite all the most recent observational efforts, one can hardly say that the BSS problem is solved; indeed, in some ways it is more puzzling than ever. Extending the UV searches for BSS to a larger sample of GGCs, characterized by different structural and dynamical properties is therefore crucial to finally unveil the nature and the formation mechanisms of these peculiar objects.

*Interacting binaries*

When mass transfer processes take place in close binary systems containing a compact object (e.g. a neutron star or a white dwarf), the streaming gas, its impact on the compact object, or the presence of an accretion disk can produce observational signatures that make these systems stand out of the ordinary cluster stars. Signatures include significant radiation in the UV, X-ray emission, rapid luminosity variations, or emission lines (such as in Halpha); exotic objects include millisecond pulsars (MSPs), low-mass X-ray binaries (LMXBs), and cataclysmic variables (CVs). These objects and the possible observations with WSO-UV have already been described in Section 5.1.4.1.

**Observations with WSO-UV imagers**

The collection of complete and unbiased samples of hot HB stars in the central regions of GCs definitively requires FUV and NUV broad band imaging. A proper combination of multi-band observations (in at least two filters of the NUV/FUV) will allow to best determine hot HB star temperature, addressing specific issues as for example the temperature and the extension of the HB gaps.

The collection of complete and unbiased sample of BSS in the core of GCs requires NUV/UVO broad-band imaging. The proper mapping of GC cores will allow deriving the specific frequency of these stars with respect to "normal" cluster stars. This will yield hints on the cluster dynamical evolution, the evolution of binary systems, and the complex interplay between stellar and dynamical evolution in dense stellar systems.

Combined FUV, NUV, and UVO broad band/H-alpha imaging will allow detecting IBs in the core of GCs. The observational properties of IBs in different environments (GCs with different central density) will allow understanding IBs formation process in GCs.

FUV, NUV and UVO broad-band imaging of fully resolved stellar populations in GCs represents a formidable data-base for empirically studying the dependence of colour indexes on the stellar population properties (as metallicity, age, etc).

**Observations with WSO-UV spectrographs**

High resolution spectroscopy with HIRDES will allow determining abundance patterns and rotation velocity for BSS in different environments: this will assess physical and chemical properties of BSS, thus helping distinguishing among the different formation processes.

Phase resolved high-resolution spectroscopy (the so-called Doppler tomography) with HIRDES will yield crucial information on the structure of the accretion processes (at least for the brightest IBs in GCs).

## *2.3.2.4 Variable Stars in globular clusters*

**Scientific Background**

GCs of stellar population(s) presenting a large spread in He content (Omega Cen - Piotto et al. 2005; NGC6388 and NGC6441 - Busso et al. 2007; NGC2808 - Piotto et al. 2007). All these clusters show an extended blue HB, thus suggesting a strong correlation between the presence of a He-rich stellar population and the occurrence of a blue HB tail.

The initial He content has a significant impact on the UV flux emitted by a given stellar population, since it strongly affects the evolutionary lifetimes, the stellar luminosities and the range in mass that might evolve as strong UV emitters. Therefore, if current findings concerning the spread in He content are confirmed, then the current explanations for the UV excess phenomenon in ellipticals and the use of mid-UV to constrain age and metallicity of stellar populations in high redshifts "red envelope" systems (Bernardi et al. 2002, AJ, 123, 2990) should be revised.

Moreover, recent near and far UV data collected with GALEX opened new opportunities to measure the UV flux in several ellipticals, and in turn to constrain the UV emission as a function of the look-back time (Lee et al. 2005, ApJ, 619, L103). Moreover, data collected with GALEX disclosed that Galactic field RR Lyrae are among the brightest UV sources (Welsh et al. 2005, AJ, 130, 825). In particular, these low-mass helium burning variables present in the UV a luminosity amplitude of the order of 6-9 magnitudes (Wheatley et al. 2005, ApJ, 619, L123). This feature is caused by the significant effective temperature variation during the pulsation cycle (roughly 1400 K) and the possible occurrence of sonic shocks in their atmospheres (Bono & Stellingwerf 1994, ApJS, 93, 223). The same outcome, but to a less extent, applies to classical Cepheids (Bono, Marconi & Stellingwerf 1999, ApJS, 122, 167). This circumstantial empirical evidence indicates that radial variables might play a crucial role in the UV emission of both old and intermediate-mass stars.

**Key Observable with WSO-UV**

The WSO-UV will have a significant impact on these long-standing astrophysical problems. The reasons are manifold.

A large field of view of UV cameras together with a high-speed shall allow us to derive accurate UV light curves not only for RR Lyrae and Cepheids, but also for the short period SX Phoenicis (variable Blue Strugglers). The comparison between pulsation and evolutionary predictions with observations will provide firm constraints on the fraction of UV flux contributed by static and variable HB stars and their progeny. This also means that we can constrain the role that the HB morphology plays in the UV integrated emission of simple and complex stellar populations.

The high resolution spectrographs will provide the unique opportunity to measure the radial velocity curves of cluster variables in the UV region with unprecedented accuracy. The new data will allow us to trace the formation and the propagation of sonic shocks in RR Lyrae stars, and in turn to constrain the efficiency of convective transport in the outermost layers along the entire pulsation cycle.

### 2.3.2.5 Open Clusters

#### Scientific Background

Galactic Open Clusters (OCs) are ideal tracers of the general properties of the Galactic disk, being observable at all positions, and covering the entire metallicity and age distribution. Even if the situation is steadily improving, the largest part of catalogued clusters still miss accurate, modern studies.

The study of OCs will benefit of the WSO-UV performances, also given the quite large field of view (about 5') that will permit to observe the entire cluster with some mosaicing (this is unpractical with the HST).

These objects are located near the Galactic plane. This often complicates the derivation of fundamental cluster parameters like age and distance, because key features in the Colour-Magnitude Diagrams (e.g., turn-off from the main sequence, RGB, red clump) are heavily contaminated by field stars.

However, given the astrometric precision of WSO-UV in the near UV/optical regions and the temporal baseline (5 years or more), membership could be determined  with proper motion measurements also in those clusters not having HST first-epoch images.

Since OCs are distributed all over the disk, are very numerous and their distances can be determined with precision (at variance with isolated field stars), the OC population can contribute to the knowledge of the 3-d reddening distribution near the galactic plane.

#### Key Observables

#### WSO-UV Observations

For all clusters where the cluster/field separation is less critical single  epoch observations will be sufficient to derive the fundamental parameters obtaining photometry in at least three filters: e.g., using F339W, F435W, F555W we would derive the reddening from the two-colour diagram, plus distance and age from the CMDs (and a good estimate of the metallicity). Furthermore, the reddening could also be derived using LSS, obtaining spectra near the 220nm feature.

Two epoch observations will be needed for all clusters for which separation from field stars is necessary.

Deep observations in visual bands (possibly in two epochs for proper motion measurement for membership) will allow studying the mass function down to very small masses.

### 2.3.2.6 Stellar Initial Mass Function in star clusters

The stellar initial mass function (IMF) of any stellar system is important tool for understanding the star formation processes as well as the dynamical evolution of the system itself. In particular, the mass function of star clusters is relatively easy to obtain, in view of the fact that all stars are located at the same distance, have the same metal content and age. A still debated issue concerns the universality of the mass function: do different environments form stars with similar or different IMFs?

*Young clusters*

The true mass function of young star clusters is a milestone in the knowledge of the star-forming process, as well as on the formation and evolution of their host galaxies over cosmic time, since they are one of the most relevant products of star-forming episodes triggered by galaxy mergers, collisions, and close encounters.

Unfortunately, we still do not know if blue massive clusters observed in starburst galaxies will survive and evolve in globular cluster type objects. A crucial role to proceed along this evolution toward old age is played by their IMF. For example, the star clusters with a shallow IMF slope (e.g., compared to solar neighbourhood) are expected to be dispersing before they reach an age of few Gyr (Chernoff & Shapiro 1987, Mengel et al. 2002).

This means that the need to observe the luminosity function (i.e., to study the IMF) of the nearby young star cluster to understand their formation, evolution and destruction is still urgent.

In fact, fundamental questions still require to be adequately addressed: do recent events of star formations show similar IMF? Is the IMF universal? What is their IMF slope? What is the percentage of binary stars? How efficient is the mass segregation of stars in recently formed clusters?

*Open clusters*

To obtain the IMF of open cluster may be a challenging problem. In fact, due to the gravitational interaction with the Galaxy, open cluster stars tend to evaporate. The star loss is selective, lower mass stars being favoured with respect to the more massive ones, therefore altering the mass function shape.

As open clusters are typically located on the Galactic plane, strong contamination due to field stars is usually present. In this case, proper motion membership could give an important contribution to the derivation of the present day mass function.

In older clusters, mass segregation and evaporation might have so much affected the mass function that only a dynamical model can recover the IMF. In this case, coverage of different fields, located at different distances from the cluster center can help the construction of the model.

*Globular Clusters*

Being on average more massive than open clusters, globular clusters are less affected by star loss. Still, internal dynamical evolution, and loss of stars from the envelope of the cluster, alter the initial mass function, and make the mass function strongly dependent from the location inside the cluster, and from the cluster orbit in the Galaxy. A dynamical model can help to recover the initial mass function from the local mass function observed at different distances from the cluster center. This is a time consuming project, already started with HST. The main scientific target (Piotto and Zoccali 1999) is to understand whether the globular cluster stellar IMF is universal, or rather it depends on some cluster parameter (e.g. metal content).

A challenging, but very important program possible only with space-based facilities is the study of the mass function down to the minimum mass needed to ignite hydrogen burning (hydrogen burning limit, HBL) in the stellar core (approximately 0.1 solar masses). Because of the sharp steepening of the mass-luminosity relation at those small masses, the number of stars per unit magnitude tends to vanish, and the few observable cluster stars are completely mixed with the field stars, even for clusters with small field contamination. In this case, the possibility to distinguish cluster stars from field objects using proper motions become of fundamental importance (King et al. 1998, Bedin et al. 2001).

**Observables with WSO-UV imagers**

WSO-UV imagers shall provide the unique possibility of deriving accurate luminosity function (and then, precise constraints on the actual mass function) of any star clusters in the Local Group.

The UV imaging will make possible to explore the brightest and hottest part of the luminosity function in young systems. The optical camera will allow investigating the distribution of low mass star, eventually down to the HBL for the closest systems.

Multiepoch images (possibly of fields already observed in the past with HST) will allow proper motions. Proper motions can be used for field star cleaning, and for the measurement of the stellar motion within the cluster for proper dynamical model development.

Multiband photometry is very important in order to test the radius-luminosity theoretical relations in colour-magnitude diagrams. The capability of models to reproduce the radius-luminosity relation allow us to establish the reliability of the corresponding theoretical mass-luminosity relations, essential to transform luminosity functions into mass functions, in particular for the old stellar populations (Bedin et al. 2001).

## 2.3.3  What is the life cycle of the Interstellar Medium and Stars?

### 2.3.3.1 Chemistry of the interstellar medium (ISM)

**Scientific Background**

The interstellar medium (ISM) plays a crucial role in the evolution of our own and other galaxies: it is the amniotic liquid in which (and from which) new stars form and, at the same time, the collector of the spent ashes of previous generation of stars. Dying stars provide metals to the ISM either through SNe explosions or stellar winds, which at the same time inject both mechanical energy and fast, energetic particles into it. The ISM is heated by the radiation field of stars (mainly in the UV-vis) and emits this energy, reprocessed, in the IR. The ISM exists in a number of physical (and chemical) states, roughly in pressure equilibrium, with H atom number densities ranging from $\sim 10^{-3}$ cm$^{-3}$ for the hot intercloud (HIC) medium on one extreme, to more than $10^3$ cm$^{-3}$ in molecular clouds (MCs) on the other extreme, and kinetic temperatures correspondingly ranging from $\sim 10^6$ (HIC) to $\sim 10$ K (MCs). A thorough description of the structure, physics and chemistry of the ISM can be found in Tielens (2005).

With the exception of the hottest phases, a more or less rich chemistry takes place, producing a large number of molecules which in turn affect the energetics (and thus the physics) of the ambient interstellar cloud (see e.g. Williams et al., 2007). Electronic, vibrational and rotational transitions of such molecules provide tracers amenable to observation from the microwave to the far-UV range of the spectrum[1]. While the chemistry of the ISM is qualitatively understood in general, many crucial unsolved problems remain. It remains unclear how the diffuse medium, despite being pervaded by harsh UV radiation and its low density, can host molecules previously thought to pertain only to regions where dense chemistry becomes efficient (Snow & McCall, 2006); moreover, the formation in the diffuse medium of the observed amounts of CH+ (which occurs through an endothermic reaction) and the population of highly excited rotational states of molecular H2 require some transient heating process (to be determined) to be at work.

Direct observations of H2 from its ground state can only be performed in the far-UV, and indeed has been performed very successfully, to some extent, with Copernicus (see Shull & Beckwidth, 1982, and references therein) and FUSE (Shull et al., 2000, Snow et al., 2000, Richter et al., 2001,

---

[1] See http://www.cv.nrao.edu/~awootten/allmols.html or http://www.astrochymist.org/ for up to date lists of the molecules detected in space.

Rachford at al., 2001, 2002, Tumlinson et al., 2002); the superior sensitivity and resolution of WSO-UV will enable us to perform much deeper surveys of the weak rovibronic lines from highly excited rotational states, which will help to discriminate among the different excitation mechanisms proposed, such as, e.g., intermittent dissipation of turbulence (Falgarone & Philips, 1990), slow magnetic waves (Viti et al., 2007), residual formation energy, etc.. WSO-UV will also make feasible direct measurements of the Lyman and Werner bands of the HD isotopomer of H2 in relatively dense clouds, where both H2 and HD are self-shielded and their ratio is a direct proxy for deuterium abundance (Ferlet et al., 2000).

The second most abundant interstellar molecule, CO, is also observable in the far-UV via several intense vibronic bands in the far UV.

The spectral resolution of the spectrometers on board WSO-UV will be able to resolve their rotational structure, meaning that a large number of individual transitions will be simultaneously measurable in a single spectrum. Moreover, WSO-UV, thanks to its superior sensitivity, will enable us to directly measure the relative abundances of rare CO isotopomers, as was previously done only for Ophiuchus using HST (Federman et al., 2003).

Another molecule which has never been observed until recently is N2: even if it is believed to contain a large fraction of the available interstellar nitrogen, its first observable transition is in the far UV, and was first barely detected with FUSE by Knauth et al. (2004). Unfortunately, this molecule will probably be just below the low wavelength limit of WSO-UV.

Another diatomic molecule which will be detectable by WSO-UV will be HCl, which was detected in the UV in Ophiuchus using HST (Federman et al, 1995).

Going up in molecular size, H2O and CO2 gas phase absorption bands in the far UV were searched for in diffuse clouds using Copernicus and HST (Snow & Smith, 1981, and Spaans et al., 1998), but only upper limits were obtained. WSO-UV will be able to push these limits down by about one order of magnitude.

To go from relatively small to the largest molecules in the ISM, a whole class of organic compounds, polycyclic aromatic hydrocarbons (PAHs), are widely believed to be ubiquitously present in the ISM, being thought to be indeed the most abundant class of molecules after H2 and CO. PAHs appear to be able to resist very harsh conditions indeed: Tappe et al. (2006) recently detected their aromatic infrared bands immediately behind a SN blast wave, using Spitzer observations. Such molecules, albeit not directly of biological relevance, must be formed along chemical reaction paths similar to those leading to the formation of sugars and amminoacids, hence their alleged ubiquitous presence is extremely relevant for astrobiology. PAHs are detected via their characteristic vibrational emission spectrum in the mid-IR, but such features are completely non-specific, meaning that despite their incredible (supposed) overall abundance, no single, specific PAH has yet been identified in space, in the face of more than twenty years of intensive laboratory, theoretical and observational effort. All PAHs, on the other hand, show intense, specific transitions in the near UV (see e.g. Malloci et al., 2007), which can be used for an unambiguous identification. Indeed, a few specific PAH bands were unsuccessfully searched for using HST by Clayton et al. (2003), which established some loose upper limits. WSO-UV will enable us to perform a much more extensive and sensitive search for PAHs (and indeed other large organic molecules, such as long carbon chains, which have strong UV signatures), thanks to its higher efficiency and resolution.

**Key Observables**

The Lyman and Werner bands of H2 have band origins below 112 nm, with lines from highly excited rotational states extending at slightly longer wavelengths and bands from excited vibrational states shifted in the mid UV. CO has its strongest bands bluewards of ~150 nm and a weaker intersystem band system at somewhat longer wavelengths. Molecular nitrogen has its first bands at about ~96 nm, hence just out of reach with present WSO-UV specifications. HCl will be observable at 129 nm, H2O at 124 nm, CO2 at 109 nm. As to PAHs, each of them shows several very strong pi-pi* band systems in the interval between ~150 and ~300 nm, each with a characteristic vibronic structure. PAH bands become broader with increasing energy, increasing the likelihood of spectral confusion, reducing the possibility to disentangle individual bands in the vibronic structure and, in turn, making them less suitable for an unambiguous identification. The intrinsic width of electronic transitions of ionised PAHs is usually at least one order of magnitude larger than that of neutral PAHs, meaning that the odds are much more favourable for identifying neutrals instead of ionised PAHs.

Neutral and (possibly) doubly ionised PAHs may be strongly luminescent, showing broad fluorescence in the blue (peaking from 300 to 400 nm) and phosphorescence in the red (peaking from 500 to 800 nm).

**Observations with WSO-UV Imagers**

*Spectroimaging of reflection nebulae*

The UVO camera on board the WSO-UV, if capable of spectroimaging, will enable us to map blue luminescence (BL) and extended red emission (ERE) phenomena (see e.g. Vijh et al., 2005, 2006, Witt et al., 2006, Mulas et al., 2006) in reflection nebulae, such as the Red Rectangle, the Orion Bar, the so called newly discovered "Red Square" (around MWC 922), as well as in PNe.

The high spatial resolution will enable us to apply blind signal separation(BSS) techniques (see e.g. Bernè et al., 2007) to unravel the individual contributions to BL and ERE as well as their spatial distributions in well-defined environments; the spatial distributions will be directly checked for correlations with the spatial distribution of the mid-IR emission in aromatic bands in the same sources, available from Spitzer observations; individual spectra will be directly comparable with theoretical predictions and laboratory data for specific PAHs; tentative matches will be cross-checked by searching for the far-IR bands expected for the same molecule in Herschel data (deep spectral surveys of both the Orion Bar and the Red Rectangle are already scheduled for Herschel).

Polarimetric capability, if implemented, will supply an independent constraint for the BSS, enabling us to unambiguously distinguish linearly polarised scattered light from unpolarised luminescence.

**Observations with WSO-UV Spectrographs**

*LSS:*

The moderate spectral resolution of LSS will be appropriate for deep spectral surveys searching for relatively broad PAH absorption bands, which, if detected, will appear as fine structure on top of the wide UV extinction bump at ~220 nm.

*HIRDES:*

The high spectral resolution of HIRDES will enable us to disentangle and resolve the individual rovibronic lines of the vibronic transitions of $H_2$, $CO$, $H_2O$, $HCl$, $CO_2$ and any of their rare isotopomers, if detected. In the case of $H_2$, this will greatly enhance the detectability of weak transitions from highly excited rotational levels, which was hindered, in FUSE observations, also by spectral confusion. In the case of heavier molecules, tens of rovibronic transitions of each detected band will be measured at once, providing a wealth of information with a single observation.

## 2.3.4  What is the diversity of planetary systems in the Galaxy?

### 2.3.4.1 Properties of extrasolar planet atmospheres

**Scientific background**

Since 1995, the year of the discovery of the first Jupiter-sized object orbiting a star other than the Sun, the number of detected extrasolar planets has steadily increased each year. The vast majority of these discoveries have been accomplished by various high precision Doppler surveys on samples of more than a thousand nearby stars. The unexpected properties of the extrasolar planets found so far have sparked much new theoretical work, with the aim to move from a set of models describing separate aspects of the physics of the formation and evolution of planetary systems to a plausible, unified theory, capable of making robust and testable predictions. After a decade of extrasolar giant planet discoveries, the only idea that has not yet undergone significant revision is the paradigm that planets form within gaseous discs around young T Tauri stars. Many old ideas were revisited or revived, and a number of new ones were proposed in an attempt to explain the observational data on extrasolar planets.

Also the existence of life outside the Earth, that has been one of the major interest of mankind for several thousand years, gained a renew interest. The study of planetary systems around other stars in search for life signatures is a prerequisite if we want to understand how life is distributed in the Universe and what its origin on Earth is. In particular, we need both to measure the distribution of planet sizes, masses and orbits, in order to determine which of these exoplanets are likely to be habitable and the to determine the characteristics of their host stars, such as element abundances and age. In this way, we could relate the characteristics of the planets to the chemical composition and age of their parental clouds and determine the initial conditions and at what stage of their evolution planets can be provided the necessary environment for life. Even if telluric planets in the habitable zone are privileged sites for development of life as they recall the Earth, the complete picture of exoplanet population we need requires a full study of the whole planet mass and orbit spectrum.

The extrasolar planet sample exhibits many interesting and surprising orbital characteristics compared to the giant planets of our Solar system. The most striking feature is the presence of giant planets on very short orbits. This has profound implications for our understanding of planet formation. Interactions between the planet and the disc at the early stage of formation can affect the orbit of the planet. Resonant interactions of a planet with a disc of planetesimals inside its orbit, dynamical friction between a planet and a planetesimal disc as well as tidal interaction between a gaseous disc and an embedded planet can lead to the migration of the planet up to very short orbital distances of its star. However, the scenarios for giant protoplanet migration in gaseous discs are not without problems. Timescales for migration are very short, much shorter than typical disc and planet formation lifetimes. Furthermore, a stopping mechanism must be devised in order to prevent the migrating protoplanets from plunging into the central star, and to reproduce the observed pile-up of planets on few-day orbits.

The typical planet frequency in the giant planet mass range and with semi major axis shorter than 3 AU, is estimated to be about 5–7 per cent. The planet detection rate increases strongly for stars with metallicity higher than the Sun. This is likely due to the initial enrichment of the nebulae but its precise link to planet formation mechanisms is a matter of debate.

The current and future missions, CoRoT and Kepler, will certainly discover the first telluric exoplanets, allowing us to start studying the distribution of planet sizes and orbits down to earth-sized bodies. But both missions will have their limitations in terms of minimum planet size, maximum orbital period and number of detected exoplanets. Clearly, a further mission with the objective of increasing these factors is highly desirable. FCU on WSO-UV-UV is the opportunity to explore extrasolar planet systems in an utterly new wavelength range for this kind of science. This new approach in exoplanetary science, providing a full analysis of exoplanetary systems such as magnetic activity linked to planetary auroral emissions, compositional analysis of planetary atmospheres through detection of evaporating phenomena.

## Key Observables

*Extrasolar planet atmospheres*

The possibility of detection of UV emission from hot Jupiters is still debated, but it is very promising. It may happen that the star/planet contrast in the UV is better than in the VIS-NIR. The region of interest is: 100-130nm for the spectroscopy of $H_2$ Werner bands and H Lyman-alpha emission (121.6nm); in addition with spectroscopy of $H_2$ Lyman bands longward of 130nm.

## Evaporation

Among the more than two hundreds extrasolar planets identified so far, there are massive hot-Jupiters. Since the discovery of the first of them, 51 Peg b (Mayor & Queloz 1995), the issue of their evaporation status has been raised. The discovery of transiting extrasolar planets has led to direct detections and characterization of their atmospheres. Charbonneau et al. (2002) measured the relative change in eclipse depth for HD 209458b across a sodium doublet (5893 Å), resulting in the first detection of atomic absorption in an EGP atmosphere. Vidal-Madjar et al. (2003) discovered an extended hydrogen-rich atmosphere surrounding HD 209458b in their UV measurement, finding that at Lyα wavelengths, the planet is ~3 times larger than in the optical. The far UV HI Lyα line is the strongest absorption line of atomic hydrogen. Far UV HST observation of HD209458b revealed a hydrogen obscuring disk which is about 15% of the size of the stellar disk, a factor ten of the about 1.5 % planetary disk seen in the optical. The material encompasses an enormous obscuring area larger than that defined by the Roche lobe outside of which the gravitational pull of the parent star becomes important. This effect is caused by the wavelength-dependence of opacities in the planets atmosphere which obscure stellar light at different planet radii, leading to a wavelength-dependent depth of the light curve during primary eclipse. Consequently, searching for relative changes in eclipse depth as a function of wavelength directly probes the absorption properties of the planet's atmosphere, with the potential to reveal the presence (or absence) of specific chemical species. If this effect is real and present in other extrasolar planets among the 20 transiting planets discovered up to day, it is possible detect their extended atmosphere searching also for heavier species dragged in the outflow by the lighter gas. Vidal- Madjar et al (2004) have in fact shown that far UV absorption of OI and CII are present in the upper atmosphere of HD209458b pointing out that Hot Jupiters can lose a significant fraction of their atmosphere. Since these planets exist and orbit stars that are not particularly young, the evaporation rate of the observed massive hot-Jupiters should be modest enough to not dramatically impact their evolution (Lecavelier des Etangs, 2007, A&A, 461, 1185). Thus, the discovery of a large number of massive hot-Jupiters led to the conclusion that the evaporation of massive planets has to be modest. In this scenario, the discovery that the transiting extrasolar planet HD 209458b is indeed losing mass was unexpected. On the other hand, such a evaporation process due to strong insolation by the parent star could lead to a new type of planets with hydrogen-poor atmosphere or even with no atmosphere at all (see Trilling et al.1998) and justifies the recent discoveries of Hot Neptune extrasolar planets (e.g. μ Arae, Santos et al., 2004).

## Auroral emission by extrasolar giant planets

Many advantages are offered by searching for and studying the ultraviolet auroral emission generated by extrasolar giant planets: i) it allows a direct detection of a planet instead of the indirect methods employed to date (radial velocity and pulsar timing); ii) the presence of an auroral signature is strictly linked to the presence of a planetary magnetic field and the evidence of this effect can not be observed with any other detection method; iii) ultraviolet auroral could contribute to characterize the near space environment around planet, so that their study could provide information about both basic atmospheric composition and the energies of the impacting particles. In general, UV wavelengths provide observational advantages compared to the optical. In fact, higher contrast ratios can be achieved in UV and the UV diffraction limit allows planets to be detected at smaller angular separations to their host stars.

## Magnetic activity of stars hosting planets

More than 200 extrasolar planets have been detected so far, mainly around solar-type stars. Most of these planets are massive Jovian--types, observationally favoured by the currently popular

Doppler-reflex techniques. However, space missions like CoRoT, and will discover – and eventually allow us to characterize - Earth-sized extrasolar planets.

How a planet might directly interact with its parent star is a new field of research, motivated by the tight orbits of some of the extreme "roasters". Cuntz et al. (2000) predicted that a giant planet orbiting close-by a star incites increased stellar activity by means of tidal and magnetospheric interactions. Preliminary observational support for planet-induced enhancements of chromospheric activity was reported by Shkolnik et al. (2001, 2005), who observed the optical Ca II H&K lines of stars hosting planets. The effects should be exaggerated in the upper chromosphere, transition region, and corona; thus UV light curves and/or monitoring spectroscopy will certainly contribute importantly to exploring such planet-star interactions.

UV spectroscopy of parent stars of planetary systems would be feasible with a 2-m class telescope for most of the relatively nearby stars in contemporary planet searches by the Doppler reflex technique.

*Biomarkers in the UV optical spectrum*

The terrestrial atmosphere is unique with abundant $O_2$ and $O_3$ produced by biological activity. Also terrestrial atmosphere has significant amount of water. Biomarkers like ozone ($O_3$) have very strong transitions in the ultraviolet. These are electronic molecular transitions, hence several orders of magnitude stronger than the vibrational or rotational transitions observed in the infrared or radio range. The Hartley bands of $O_3$ are the main absorbers at 200-300 nm. $O_2$ has strong bands in the range 150-200 nm. CO has strong band below 180 nm and weaker Cameron bands from 180 to 260 nm. The $CO^+$ first negative bands are located in the 210-280 nm range, while $CO_2$ can be detected through the $CO_2^+$ Fox-Duffenback-Barker bands from 300 to 450 nm. All these transitions can be observed when the planet transits in front of a strong parent star UV background. The atmosphere of HD 209458b is the first one ever observed in such a way, thanks to UV-optical spectroscopy obtained by HST (Charbonneau et al. 2002, Vidal-Madjar et al 2003, 2004)

The spectral resolving power required to detect biomarkers in the atmosphere of exoplanets is not a crucial capability. According to Gòmez de Castro et al. (2006) a resolution R~10000 is adequate to these investigations, and even R≤ 1000 could be enough to detect the broad band signatures of many molecules. The main limit is instead represented by the sensitivity: Gòmez de Castro et al. (2006) have computed the number of possible detections of UV atmospheric signatures as a function of telescope sensitivity in unit of the *HST*/STIS sensitivity, finding that the presence of biomarkers and other constituents in the atmospheres can be searched by WSO-UV high resolution spectrographs - whose sensitivity is about 10 times the *HST*/STIS one - for about 100 exoplanets orbiting K, G and F-type main sequence stars.

## 2.3.4.2 Properties of Solar System planet atmospheres and magnetospheres

### Scientific Background

In the solar systems there are a number of astrophysical processes which may be investigated in the UV domain. Interestingly, the UV spectral window may provide invaluable information for almost all classes of bodies -independently of their physical sizes-: from giant gaseous planets, to minor bodies (e.g. comets), passing from rocky planets and natural satellites. Roughly speaking such astrophysical processes can be divided into two main groups: on one side, there are energetic interactions between high velocity particles and planetary atmospheres which are responsible of bright UV emissions such, for instance, the aurorae; on the other side, there are UV emissions stimulated by the interaction between solar radiation and atoms/molecules, as in planetary atmospheres and cometary comae. Therefore UV observations are crucial to investigate gaseous environments and provide insights on a number of interesting processes. For instance, apart for the obvious reason that UV observations provide information on the composition and the chemistry of those environments, we recall that they provide the best remote method for studying planetary magnetospheres and thus for obtaining information on planetary interiors.

## Key Observables

One of the most interesting aspects related to UV observations is certainly the auroral emissions of gaseous planets, in particular Jupiter and Saturn. The presence of aurora for such planets has been recognized since long time, and recently the Hubble Space Telescope has provided important data toward a correct interpretation of such processes. However, along with this data many new questions arose. For instance, it turned out that Jupiter and Saturn aurorae are very complex and different from the best-known Earth's aurora. Jupiter aurora is the results of a mixing of solar wind particles and 'local' plasma, mostly released from its natural satellites. It is not clear however, how these particles interact with Jupiter magnetic field and what are the specific mechanisms responsible for those auroral UV emissions. Similar considerations also apply to Saturn's aurora, and detailed studies concerning Uranus and Neptune are still lacking.

Another important application of UV observations of giant planets regards the spatial distributions of certain molecules, like the anomalous $H_2$ bulge of Jupiter detected via Ly-alpha emission. Dedicated observations are needed in order to understand the origin and the evolution of such anomaly in the atmosphere of Jupiter.

It is now established that satellitary systems are complex environments, much more then being just a simple list of satellites. In particular, some of them fill the surrounding space with emission of gas and/or grains of a wide span of compositions. Such processes have been observed in all satellitary systems. The most spectacular one is the Jovian system with the emissions from Europa, Ganymede and in particular Io. The latter is responsible of a large torus of plasma, composed mainly of sulfur and oxygen (all having emission lines in the UV). There is evidence that Io's torus is refilled by outgassing material triggered by Io volcanic activity. Moreover, this plasma may be responsible for the auroral emission from Ganymede atmosphere (notice that Ganymede is the only satellites with a significant intrinsic magnetosphere).

Concerning Saturnian system, it has been discovered recently the vents of gas and grains outgassing from south pole of Enceladus. These vents are composed mainly of water ice with additional volatiles ($N_2$, $CO_2$, $CH_4$). There is evidence that such vents maybe the source for the large Saturnian OH torus, nevertheless a global view -involving the kinematic, the composition and the energetics of such vents- still need to be figured out. Very important -but still largely unknown- are also the evolution over time of such particles along with the refilling mechanisms.

Much of the upper atmospheres of rocky planets may also learn in the UV. We can study again auroral phenomena, but also the chemistry of the upper atmosphere induced by the ionizing solar radiation and impinging solar electrons. Processes such atmospheric airglows belong to this category. A corollary of such studies is to determine the composition of the upper atmosphere and trace the abundances of elements like H, He, C, O and so on. These measurements are important, for instance, in order to determine atmospheric escape rates. Moreover, the nature of the emissions constrains the chemistry of these regions. This is of particular interest for understanding the evolution of Mars and Venus atmospheres, and to study the difference with respect to Earth's atmosphere.

Moreover, in the UV we have the possibility to detect the major constituents of the exospheres of airless rocky bodies, like the Moon, Mercury, and possibly asteroids. These exospheres are thought to be composed mainly of He, Ar (from the solar wind) and Mg, Al, OH (from the soils). However these species can only be detected in the UV, and no firm detection has obtained yet. This data are essential to obtain information of the loss mechanisms (e.g. solar wind sputtering vs micrometeorite impacts) and also to provide indirect information on surface composition, precious where it is difficult to observe directly the surface, like in the case of Mercury.

Concerning minor bodies, UV observations are invaluable for studying comae of comets. Only a hand full list of comets has been observed in the UV (mainly by IUE and HST). Many important molecules-atoms may be investigated in this spectral range, providing information which is complementary to other spectral regions, like the visible. Among them we have OH, CO, CS, $C_2$, CN, OI, CI, SI etc. The comparative study of such emission provides indication of the molecules production rates (which vary with the heliocentric distance) and the photodissociation in the coma (notice many molecules are the product of the dissociation, such as OH and $H_2O$). This also gives an indication on the nature of the ices in the nucleus (intimate vs segregated mixtures) and thus on the primordial conditions where these materials formed. Moreover, the study of the time variability of the coma provides information on the 'activity' of the surface.

It is important to stress that a versatile facility is needed in order to study all the processes indicated, however such facility is neither presently available nor planned for the near future, in spite of the several space-born facilities presently available (HST, Rosetta, Cassini, New Horizon etc) and a few others planned for the future (JUNO, BepiColombo etc) thus the importance of WSO-UV. Moreover, WSO-UV will provide a great step ahead in comparison to past satellites and missions, thanks to its sensitivity and both spectral and spatial resolution.

## Observations with WSO-UV Imagers

In particular, WSO-UV shall allow to study in details the structure of the aurora of the major planets, along with the surroundings. To give an idea, the angular diameters of Jupiter, Saturn, Uranus and Neptune are 47", 20", 4" and 2", respectively (computed at opposition), therefore fine structures can be imaged even for distant planets wit diffraction limited cameras.

Moreover, the extension of the Io torus is of the order of 5' (maximum projected Io-Jupiter distance), while the Saturnian OH torus extends for some arcmin[2].

Thus WSO-UV shall provide the possibility of mapping such environments. This is fundamental for understanding the time evolution of plasma environments and their interaction with the planetary magnetic fields and the auroral emissions. We stress that our present understanding of this complex problem is made basically by 'snapshots'; in other words, we miss completely the dynamic, which is the link between the release of the gas and the resulting effects on the atmospheres. WSO-UV will be able to fill this gap.

The diffraction limited imaging of the UV channels along with appropriate narrow band filters shall give the possibility of a close look to the satellites and their outgassing mechanisms. For instance, Io and Titan have diameters of 1.1" and 0.8", respectively, thus we expect to be able to resolve fine details in their outgassing mechanisms (see for instance the spectacular eruption imaged by New Horizon).

As for planetary exospheres and cometary comae WSO-UV imaging shall provide important data for unveiling their composition and source mechanisms. In the case of cometary coma, the WSO-UV cameras shall provide a very detailed view of the spatial structures of the delivery from the nucleus of a good number of molecules (e.g. CO, OH, and CN, if appropriate narrow filters will be provided) and of the dust.

WSO-UV narrow band imaging shall provide an important contribution to the study of the atmospheres of Mars and Venus (angular diameters of about 16" and 19", respectively), and in particular the airglows (e.g. from OI) in the upper layers. Notice also, that although WSO-UV cannot certainly compete with the dedicated in-situ missions, it can still obtain useful data. As for instance, the scale height of Mars' atmosphere (about 15 km) at opposition corresponds to about three pixels of the NUV channel.

## Observations with WSO Spectrographs

Most of the topics mentioned above will also take advantage of observations with both low- and high spectral resolution.

Concerning the auroral emission of giant planets the region of interest is: 100-130nm for the spectroscopy of $H_2$ Werner bands and H Lyman-alpha emission (121.6nm); in addition with spectroscopy of $H_2$ Lyman bands longward of 130nm. Spectroscopy is of particular interest for the possibility of detecting the $H_2$ (a major constituent of the gaseous planets), and thus may be used to map compositional anomalies and to track the dynamics of the atmospheres.

Another very important application is the study of the abundances and kinematics of satellites' plasma tori. The main subject will be the investigation of Io production of S and O and their ionization states (in particular OII, SII, SIII and SIV). The detection of the ionized species is very important because allows tracking the interactions between the neutral gases and the solar radiation and can only be done via spectroscopy. Other applications may be the HI torus of Triton and the OH torus of Enceladus. Moreover, precise emission line measurements may be used to study the dynamics of such environments.

Concerning comets, the spectroscopy is required for abundances and compositional determinations. The region 100nm-320nm is useful not only for studying several major molecules-atoms, like CO, CS, CI, OI, SI and OH but also for studying minor constituents not easily detectable with other techniques (like $S_2$ and $H_2$). Moreover, spectroscopy provides important data for disentangling the contribution of the gas emissions from the dust continuum.

Moreover the region 100-320nm can be used for spectroscopy of the exospheres of Moon and Mercury (in particular for the detection and production rates of new species like MgI at 285.3nm and 202.6nm; and OI at 130.2-130.6nm) and low resolution spectroscopy of asteroids in the region 100-320nm is useful for studying the "spectral reversal anomaly" observed on silicate-rich bodies and on the Moon.

# 3. HIGHLIGHTS OF THE FCU CAPABILITIES

## 3.1 FCU Imaging Photometry: criteria for filters and detectors

The chosen set of broad and narrow band filters shall assure continuity with the HST photometric bands, and with some other standard photometric system (e.g. SDSS, Johnson-Cousins, Strömgren).

Narrow band filters are chosen according to the key science drivers outlined in Section 2, but also taking in mind that WSO-UV is a general observatory. General purpose choices have been preferred to specific requests satisfying only a restricted field of research. The selected filters and an additional list of high priority filters are listed in RD4.

The filter and detector optical quality shall guarantee high accuracy photometry, as close as possible to the Poisson noise limit.

Near-UV and optical photometry shall be possible also in crowded environments.

The near-UV optical images shall have a dynamical range up to 65000.

Time tagged observations down to the millisecond temporal sampling shall be possible in the FUV and NUV channels.

In order to reach the scientific objectives, it must be guaranteed that the total system throughput and the filter transmission efficiency allow reaching an S/N=10 in one hour of exposure for targets as in Table 1.

Table 1: Magnitude V of targets observable in 1 hr with S/N=10

| Band (a) | Sp. type | | |
|----------|----------|----------|----------|
|          | O3 V | A0 V | G0 V |
| F150W    | 26.2 | 21.5 | 11.1 |
| F250W    | 24.6 | 21.9 | 19.5 |
| F555W    | 26.3 | 26.3 | 26.3 |
| a) Bands are defined in Chapter II. | | | |

## 3.2  FCU Astrometry: complementarities to GAIA and JWST

### 3.2.1  Expected astrometric performances

The optical design of the NUV and UVO channels shall be optimized in order to allow imaging quality as close as possible to diffraction limit, small geometrical distortion (less than 10%, corner to corner), and, most importantly, high geometrical distortion stability (at the level of 3 mas, corner to corner, for UVO).  All this in order to optimize the astrometric performances.

In the UVO channel, we should reach an angular resolution of the order of 0.07 arcsec or better, with a well sampled PSF.

As fully demonstrated on WFPC2 and ACS/HST images (Anderson and King 2000, 2003, 2006, Bedin et al. 2004) it is possible to reach an astrometric accuracy of the order of 1/100 of the resolution element on well exposed (S/N>100) undersampled images, and up to 1/150 of the resolution in well sampled images (as shown by Anderson et al. 2006, on groundbased images).

This implies that we shall expect to have in near-UV-optical resolution of the order of 0.7 milliarcsec (mas) on single, well exposed point sources, close to the astrometric accuracy reached with the ACS camera (~0.5 mas on single, well exposed stars).

As shown by the experience on HST images, (Anderson and King 2000, 2002, 2006), the astrometric accuracy scales with the square root number of the number of images used to measure point source positions. Therefore, with a sample of 25-30 well exposed images, we shall expect to have an astrometric accuracy in near-UV-optical of the order of 0.10-0.15 mas.

This astrometric accuracy becomes of great importance in the measurement of proper motions (both relative and absolute) for which we can take advantage also of the huge amount of high angular resolution images in the HST archive for real breakthrough science. The WSO-UV observations, expected in the 2012-2020 for a huge amount of Galactic and extragalactic fields, coupled with the HST dataset will provide multiepoch observations spanning a temporal interval of up to 25-30 years. This immediately translates into a capability of measuring **relative proper motions with an error of the order of 5 microarcsec** (in the best cases).

For absolute proper motions, we are limited by the accuracy in the astrometric position of the inertial reference points. In most cases, these will be faint galaxies. Even compact or nucleated galaxies will allow an astrometric measurement from 2 to 5 less accurate than what is reachable on point sources (Milone et al. 2006), which translate in an **absolute proper motion accuracy of the order of 10-20 microarcsec**, at best (the highest accuracy can be reached when in the observed field there will be one or more bright, well exposed point like inertial reference source, like a QSO, Bedin et al. 2003).

One important aspect is that these proper motions can be measured for stars up to 5 magnitudes fainter than GAIA, and, most importantly also in crowded fields, unreachable by GAIA.

It is therefore a matter of fact that WSO-UV can be fully complementary, and in some cases competitive, with GAIA, and this, long before (10 years at least) the full GAIA catalog will be available.

Because of the wavelength coverage, and because of the detector quality (for UVO) WSO-UV will have astrometric performances very similar to what is expected for JWST. WSO-UV will be complementary (in wavelength coverage) to JWST, and it will investigate a region of the electromagnetic spectrum more appropriate for the study of high energy phenomena, and of the local Universe stellar populations.

Below we briefly list a number of science topics which can be carried out taking advantage of WSO-UV capability to perform high accuracy astrometry and proper motion measurements. This list should just give the flavour of the impressive quantity (and quality) of the science that astrometry based on UV-optical images in space allows. Astrometry based on large size UV-optical telescope is a very new science, and many investigation fields have just been started, thanks to HST. On this respect, WSO-UV is an absolutely mandatory instrument, which will be able to fully exploit a research field just started by HST.

The science which should be allowed by WSO-UV astrometric performances and the consequent ability to obtain high precision proper motions has been fully exploited in Section 2. Here we just summarize some of the most exciting expected main applications.

## 3.2.2 Possible scientific applications

*Galactic star clusters*

*separation field/clusters in any Milky Way region,* also in crowded environments (galactic disk, spiral arms, bulge, central regions of the Galaxy);

measurement of *internal proper motions in all open and Galactic globular clusters*, from the crowded (unreachable by GAIA) centers to the outskirts; this measurement will allow: (1) a measurement of accurate distances (independently from GAIA, e.g. by comparing internal proper motion dispersion with radial velocity dispersion in the central parts), and (2) the development of realistic dynamical models of the cluster, able to follow the cluster dynamical evolution and the evolution of its stellar population (e.g. formation of exotic objects as a consequence of dynamical interactions, estimate of the cluster stellar initial mass function accounting for the loss of stars due to various dynamical processes, etc.);

measurement of the *cluster absolute proper motions* using background QSOs and galaxies as reference frame; estimate of their orbit in the Galaxy; identification of Galactic streams, relics of mergers.

particularly in globular clusters, determination of *the fraction of clusters which host central intermediate mass black holes (IMBH)*, and *measurement of the IMBH mass* by mapping the proper motion and the acceleration of the stars in the very inner core (as done for the Galactic center);

measurement of the fraction and the properties (period, orbit separation, masses) of binaries with massive (black hole, neutron star, white dwarf) components in the cluster central core. This study is of great importance for the development of cluster dynamical models (binaries are the most important energy sources in the cluster cores), but they are also important for stellar evolution, and for the estimate the amount of dark matter in the cluster cores.

*Extragalactic systems*

*separation field/clusters in the Magellanic Clouds and other Local Group galaxies,* including M31 and M33; this separation is of great importance for the study of the cluster stellar population, also in regions where field contamination is a severe limitation;

internal proper motions in Magellanic Clouds and other Milky Way satellites and their clusters (assuming a 20 yr temporal baseline);

*absolute motions of stars and stellar systems in Milky Way satellites and in M31.* Note that at the distance of M31, 50km/s correspond to 15 mas/yr, 3 times our expected accuracy in proper motion measurement with a >20 yr temporal baseline (using HST images as first epoch data).This kind of measurement becomes important for the mapping of the gravitational potential and the measurement of the dark matter content.

# 3.3 FCU slitless spectroscopy, imaging polarimetry and spectropolarimetry

The use of three additional modes of observations, namely slitless spectroscopy, imaging polarimetry and spectropolarimetry are required by science issues illustrated in Section 2.

## 3.3.1 Slitless spectroscopy

The goal of the slitless spectroscopic capabilities of WSO-UV is to obtain low-to-moderate resolution (R≥100) spectra of all objects in the field of view. This clearly represents a very useful tool when combined with UV imaging. It is also envisaged that, when implemented on the UVO channel, it can provide (almost) simultaneous optical spectra for UV sources observed with HIRDES and/or LSS. Furthermore, the requested low resolution will enable one to extend the feasibility of UV spectroscopic observations to fainter objects than expected for the high resolution spectrographs. This mode of observations is also very useful to explore the properties of individual sources in moderately crowded fields. In fact, it will be possible to separate the spectra of objects falling into the 1" slit planned for the LSS.

Possible scientific applications

Among the many possible applications, slitless spectroscopy can be used to identify hot stellar counterparts of X-ray sources in crowded fields much more efficiently against the time consuming single-slit single object observations. The opportunity to uncover UV identified blue stars from UV broad band photometry can allow one to easily pick-up those objects with signatures of accretion (i.e. showing emission lines of high ionization species). This tool is the alternative coverage to broad band multi-wavelength photometry which might be used for the dimmest sources.

One can fruitfully perform 'slitless spectroscopy' also of stellar jets taking advantage of moderate spectral resolution prisms/grisms associated to the NUV and Optical cameras. Orienting the dispersion direction transverse to the jet axis, one can form separated multiple images of the flow in the various lines of interest in the same exposure. This technique has proven to be successful provided that the jet under study is well collimated, and with a properly selected grism spectral resolution, of the order of R~250. The same line of reasoning applies to observations of the emission line regions surrounding AGN. Here the emission line has a highly structured morphology, dominated by compact knots. Slitless spectroscopy can provide simultaneously morphology in different emission lines, kinematical information as well as spatially resolved diagnostic emission line ratios.

## 3.3.2 Imaging polarimetry

Polarized emission can arise due to three main mechanisms: scattering, dichroic transmission through aligned dust grains, and it can also be intrinsic to the emission process, such as in the case of cyclo- and synchrotron radiation. The study of several scientific issues can then take significant advantage from the possibility of obtaining broad-band imaging polarimetry, leading to the indication from the Science Team to implement a mode of polarimetric observations.

The goal of the WSO-UV polarimetric observations is to allow obtaining imaging polarimetry over the largest possible field-of-view, optimally the whole field of the cameras.

Possible scientific applications

For example, it has a wide application in the context of AGN. In fact, imaging polarimetry gives detailed information on the structure of the magnetic field in extragalactic jets and on the processes of particle acceleration in AGN. This observing mode represents also the only way to disentangle scattered light from in situ emission and it plays a crucial role to test the unified models of AGN, based on orientation and scattering of otherwise hidden sources (Antonucci 1993). The same technique will allow the separation of the genuine host galaxy emission in QSO from the nuclear light scattered into our line of sight.

Asphericity in SNe can be probed via polarization because electron scattering causes the electric field vectors of scattered photons to take on a preferential orientation. For core--collapse SNe there are indications that the degree of asphericity is inversely correlated with the amount of envelope material present at the time of explosion, so that SNe Ib/c and SNe IIb have on average higher polarization that SNe II. Only a few SNe Ia have been tested for polarization. In most cases only upper limits have been provided, in others a low degree of polarization of the order of 0.2--0.3% (corresponding to an asphericity of about 10%) has been detected before maximum light, likely due to a distorted photosphere or element distribution.

Imaging polarimetry is almost completely pioneering for the study of GRB, and would represent an invaluable progress here, because it is very poorly known in optical and totally unexplored in FUV-NUV. UV polarimetry is an excellent diagnostic of the non-thermal phenomena active in the early phases.

Concerning the applications to stellar physics, dusty shells with both jet and disk structures of luminous blue variables have been observed in the optical. It is possible to observe the previous stages of the star's activity through the decomposition of scattered light in the environment at different distances. Having been successfully applied in the optical and near IR to sources such as Eta Carinae, the method could be extended to a large class of sources (including P Cygni and less massive members of the class). The observations are similar to those used for planetary nebulae but are an independent check on the dust properties derived from the IR.

A map of the sky can provide an integrated view of the magnetic field distribution and the correlation with jets and flows in evolving objects. The polarized flux images taken at different epochs will result in maps of time varying and probably exotic objects. In regions like Lockman's window the extragalactic background can studied. UV polarization provides two important advantages, namely reduced galactic starlight background (both intensity and polarization) and reduced interstellar polarization.

### 3.3.3 Spectropolarimetry

As explained above, it is recommended that the UVO and NUV channels of the FCU on board WSO-UV will have slitless spectral and polarimetric capabilities that can be combined for imaging spectro-polarimetry. During the last decades, only a few experiments from space, such as WUPPE, FUSP and FOS/HST have obtained UV spectro-polarimetric observations (down to 115 nm). UV spectro-polarimetry has been under-exploited in the study of astrophysical objects, in part because of the limitations of optical designs (for instance, COSTAR and its effect on the polarization properties of extended sources).

Possible scientific applications

Spectropolarimetry has been demonstrated to be an essential tool for revealing the structure of deeply imbedded sources (e.g. AGN, especially Sy 2) and for studying structure of extended objects in dusty environments (e.g. shells of planetary nebulae). UV is ideal to study regions of relatively low optical depth, because it is possible to sample dust properties without the contaminating background of thermal emission or emission lines.

Information about geometry and structure of the condensations around the star can be obtained. One feature of polarization measurements is the sensitivity of the scattering to absorption ratio to grain size and mean orientation, and with multi-wavelength observations imaging spectro-polarimetry provides detailed geometric information and intrinsic dust properties unobtainable from other measurements but complementary to optical and IR spectroscopic imaging.

For planetary nebulae and AGB stars, where specific emission lines from the star dominate the scattered light, the separation of the wind and interaction regions should be possible using polarimetric methods independent of line profile modelling.

For variable stars, especially the Mira types that have resolvable photospheres, and the extreme late type objects such as Betelgeuse, the variation of polarization and effective radius provides an additional probe of the geometry and structure of the stellar envelope. This has been used in the optical for some time, however it has yet to be exploited in the UV.

UV spectro-polarimetry can be used also to infer the direction of magnetic field at the surface of young stars. Weak magnetic fields can be detected and information about the 3D topology of magnetic field in the magnetosphere provided. This technique is very useful to investigate matter accretion channelled from the disk to the star through the funnel flows. Extending the detection of cyclotron emission from strongly magnetized white dwarfs in CVs to UV wavelengths to search for polarized emission in the NUV continuum can allow tracing of high cyclotron harmonics. Also, fast photometry with polarized filters can allow following the variability of circular and linear polarization along the rotational period of the accreting white dwarf on timescales of minutes to hours. This will provide constraints on magnetic field strength and field topology especially for the high magnetic field systems for which cyclotron at low harmonics are optically thick at long wavelengths.

During the course of the year to maintain nominal solar pointing the "effective slit", i.e. dispersion orientation, will rotate through about 47°. This is a major advantage that allows orientation of the spectrum without slits to reconstruct spectro-polarimetric images of extended objects (i.e. jets).

For all the above reasons, the WSO-UV Italian Science Team has required the FCU team to explore the possibility to have the spectropolarimetric mode in the NUV channel. It is contemporary required that its implementation does not affect the performances of the NUV channel as imagers, for throughput, FoV and spatial resolution.

## 3.4   FCU Throughput and Discovery efficiency

The expected system throughputs are compared in Figure 15 to the ones of the cameras that have flown or will fly on board the Hubble Space Telescope i.e. WFPC2, ACS/WFC, ACS/HRC and WFC3/UVIS (Bond et al. 2006) Throughput includes the telescope, all of the optical elements of the instruments themselves, the sensitivities of the detectors and filter transmissions, that for the FCU are estimated based on the best information currently available and are subject to change.

The throughput of the UVO channel is much better than that of WFPC2 which, by the way, is the oldest of the HST cameras, and is comparable to that of ACS/HRC and WFC3/UVIS being better in the UV range.ACS/WFC, being optimized in the visual range, has a much better throughput than the UVO channel.

The discovery efficiency, defined as system throughput times the area of the FOV as projected onto the sky, is useful when comparing different instruments in the context of wide-angle surveys. Due to its large field of view the UVO channel of the FCU has a discovery efficiency equal or greater than that of ACS/WFC, as shown in Figure 16. The performance of the FUV channel is even better when compared to HST because, in this case (by the way ACS/SBC is not present in the plot), no camera working in this wavelength range has a large field of view.

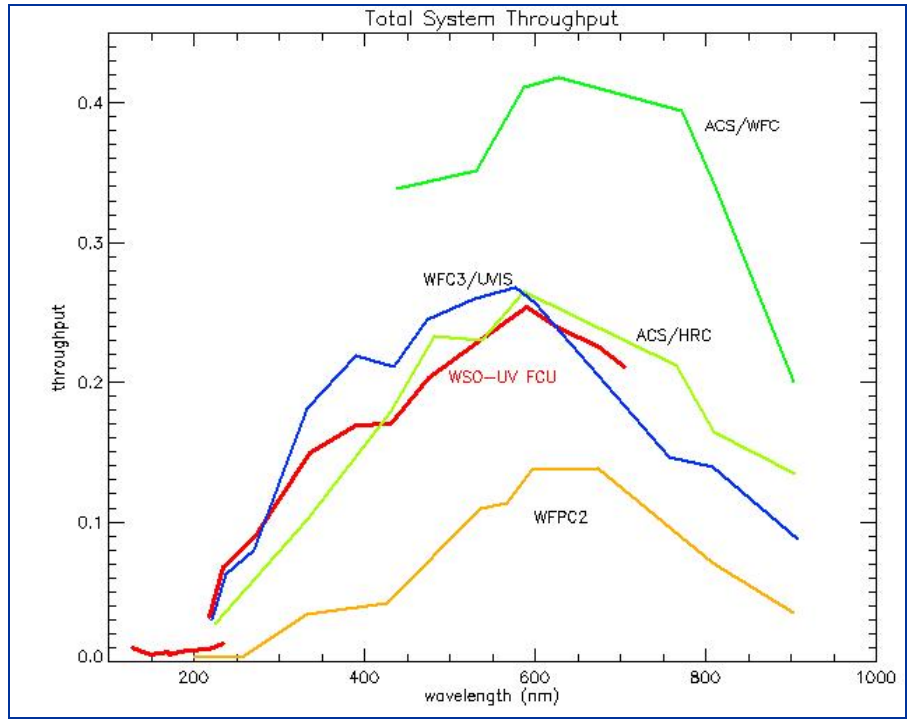

Figure 15 - System throughputs of the FCU vs. wavelength compared to that of the HST imaging instruments. Note that the y-axis is logarithmic.

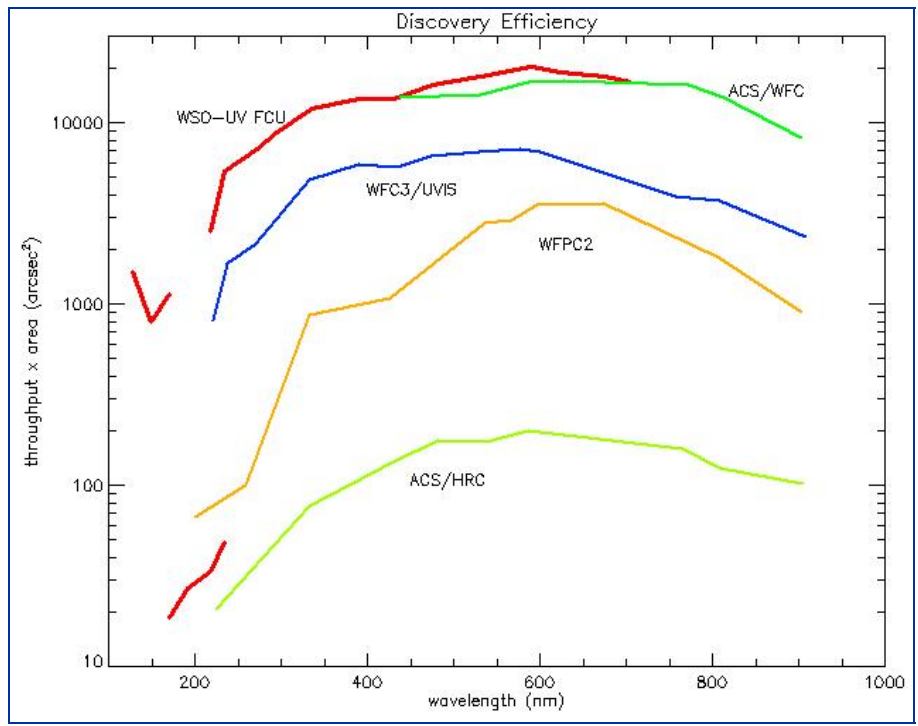

Figure 16 - System discovery efficiencies of the FCU vs. wavelength compared to that of the HST imaging instruments. Discovery efficiency is defined as the system throughput multiplied by the area of the field of view. Note that the y-axis is logarithmic.

# Chapter II.

# WSO-UV FCU Top Level Requirements

## 1.    INTRODUCTION

This chapter presents a consistent list of Top Level Requirements for the FCU instrument that will be mounted on the focal plane of the WSO-UV space telescope. The requirements described here come directly from the science case described in Chapter A. We specify that the FCU should have three different channels in order to maximize the instrument performances in three different and well defined wavelength ranges (from Visible to far UV). For each channel a set of TLR concerning general instrumental functionality, together with optics and detector technical requirements, are given in order to reach the scientific goals proposed.

This requirement list is organized according to the instrument channels:

- FUV (in its various modes): Section 3
- NUV (in its various modes): Section 4
- UVO (in its various modes): Section 5

Each Top Level Requirement (TLR) is identified by a letter (G for General, O for UVO, N for NUV and F for FUV) that associates it with the appropriate module m and a number .## within that module, e.g. TLR_m.##. Any text immediately following but not in a paragraph directly headed by a TLR_ is for clarification or comment. For particular requirements, as e.g. in Detector case, a successive level of numeration individuates a further subset of requirements.

The FCU shall include different channels to provide imaging, polarimetric, and low resolution slit-less spectroscopy facilities from far-UV to visual bands. The main purpose of the camera is direct observation of specific targets, and survey observations.

The main science targets for the imaging instruments are described in the "Key Science Drivers to the Italian participation to WSO-UV" document (Chapter I).

The FCU mainly should consist of three channels:

- A far ultraviolet high sensitivity and large field of view channel with the possibility of low resolution slitless spectroscopy (FUV)
- A near ultraviolet high angular resolution imager, with the possibility of low resolution slitless spectroscopy and polarimetric facilities (NUV)

- A near ultraviolet-visual diffraction limit imager, with the possibility of low resolution slitless spectroscopy and polarimetric facilities (UVO)

Being one of the focal plane instrument of a general-purpose orbital telescope as WSO-UV is, the FCU will provide high resolution and high sensitivity imaging available to the international community in a wavelength range otherwise uncovered during the years of operation.

Detailed analyses supporting the present Top Level Requirement document are presented in Chapter III.

The **science** for FCU (Chapter I) requires:

- the possibility to obtain diffraction limited images for $\lambda > 230$ nm (goal $\lambda > 200$ nm).
- S/N=10 per pixel in one hour exposure time down to:
- V=26.2 (21.5 – 11.1) for a O3 V (A0 V - G0 V) star in the band F150W
- V=24.6 (21.9 – 19.5) for a O3 V (A0 V - G0 V) star in the band F250W
- V=26.3 (26.3 – 26.3) for a O3 V (A0 V - G0 V) star in the band F555W
- Spectral resolution R=100 at $\lambda$=150 nm, R$\geq$100 at $\lambda$=250 nm, and R=250 at $\lambda$=500 nm.
- Polarimetric filters
- High time resolution imaging
- The possibility to obtain wide field (~ 5x5 arcmin$^2$) images from Far-UV to visual wavelengths.
- Partial overlap between spectral range covered by FUV and NUV and NUV and UVO, for relative calibration purposes.

Table 2 summarizes the FCU instruments modes.

Table 2 List of instrument modes

| Channel | Modes |
|---------|-------|
| FUV | Imaging |
| | Field Spectroscopy |
| NUV | HAR imaging in the NUV |
| | Field  spectroscopy |
| | Polarimetry |
| UVO | HAR imaging in the V-NUV |
| | Field Spectroscopy |
| | Polarimetry |

# 2.    COMMON CAPABILITIES

## 2.1    Operational requirements

### 2.1.1  TLR_G.1 Efficiency

Every effort shall be made to achieve the highest possible efficiency on all instrument modes.

## 2.2    Requirements toward the WSO-UV system

### 2.2.1  TLR_G.2 Parallel Observation of Focal Plane Instruments

Every effort shall be made to allow the focal plane instruments, i.e. one or more channels of the FCU, HIRDES, and LSS spectrographs, to be operated in parallel for science observations.

### 2.2.2  TLR_G.3: Pointing stability

As described in [RD2], it has been requested to the WSO-UV PI that the telescope has a (Gaussian) pointing stability distribution with a sigma<15 mas (aim 7 mas), with systematic drifts <7mas in a 30 minute exposure.

### 2.2.3  TLR_G.4: Dithering accuracy

As described in [RD2], for an optimal recover of the PSF shapes in UVO and NUV channels, differential shifts (dither) of the telescope shall be accurate at the level of $\leq$ 7mas.

# 3.    FAR ULTRAVIOLET CHANNEL (FUV)

This camera will be used in imaging mode.

## 3.1    General Properties

### 3.1.1  TLR_F.1: Spectral domain

The useful spectral domain shall extend from 115 to 190 nm.

### 3.1.2  TLR_F.2: Spatial sampling

Images shall be sampled with a pixel scale of 200 mas/pixel

### 3.1.3  TLR_F.3: FoV

The FoV shall be $\geq$ 6.5x6.5 arcmin².

### 3.1.4 TLR_F.4: Detector

In order to meet the science goals reachable with this channel the detector shall be a sensor efficient in the working wavelength range with the following characteristics.

#### 3.1.4.1 TLR_F.4.1: Pixel size and format

The detector shall have no less than 2k×2k pixels. The pixel size shall be $\leq 20\ \mu m$ square.

#### 3.1.4.2 TLR_F.4.2: Detector DQE

The Quantum efficiency of the detector shall be

DQE$\geq$20% for λ=120 nm

DQE$\geq$20% for λ=150 nm

DQE<$10^{-5}$ % for $\lambda > 400$ nm

#### 3.1.4.3 TLR_F.4.3: Local Dynamical Range

The local dynamical range shall be >10 count/s on a PSF-size light spot

#### 3.1.4.4 TLR_F.4.4: Global Dynamical Range

The global dynamical range shall be >200000 count/s

#### 3.1.4.5 TLR_F.4.5: Time Resolution

The FUV camera detector time resolution shall be $\leq (20\ ms)^{-1}$.

#### 3.1.4.6 TLR_F.4.6: Dark Current

The dark current of the FUV Detector will be <$4\times10^{-5}$ counts/pix/hr at detector operating temperature.

#### 3.1.4.7 TLR_F.4.7: Detector Response Uniformity

The detector shall be correctable to a uniform pixel-to-pixel response to <1% at all wavelengths.

#### 3.1.4.8 TLR_F.4.8: Large scale non uniformity

Large scale non-uniformity shall be < 10%

#### 3.1.4.9 TLR_F.4.9: Detector QE Stability

The absolute QE stability shall be $\pm$0.5% (peak to peak) per hour and $\pm$1% (peak to peak) per month.

#### 3.1.4.10 TLR_F.4.10: Detector Flat Field Stability

The difference of two flat fields taken one month apart with the same instrumental configuration shall be < 1%.

### 3.1.5 TLR_F.5: Operations

It shall be possible to operate FUV alone in all its modes. The possibility to use the FUV channel in parallel with other channels is not in the baseline, but should be investigated during Phase B1.

#### 3.1.5.1 TLR_F.5.1: Time Tag

It shall be possible to operate FUV in time tag mode.

## 3.2 Imaging mode

### 3.2.1 TLR_F.6: Imaging FoV

The full FUV FoV shall be used for imaging (see TLR_F.3).

### 3.2.2 TLR_F.7: Limiting magnitude

S/N=10 per resolution element in one hour exposure time down to V=26.2 (21.5 – 11.1) for a O3 V (A0 V - G0 V) star in the band F150W (corresponding to $5.5 \times 10^{-16}$ erg cm$^{-2}$ s$^{-1}$ A$^{-1}$).

### 3.2.3 TLR_F.8: Number of available filters

Space for at least 10 different, selectable, filters shall be provided.

### 3.2.4 TLR_F.9: Required filter and characteristics

The FUV channel list of filters is given in Table 3. Filters (positions 1 to 10) are listed in order of priority.

Table 3: FUV list of filters

| N | Filter name | Central wavelength [A] | Band width [A] | Description |
|---|---|---|---|---|
| 1 | F130M | 1300 | 160 | Should exclude Ly-$\alpha$ geocoronal line |
| 2 | F150W | 1500 | 400 | Similar to HST F160BW |
| 3 | F175W | 1750 | 400 | |
| 4 | F122N | 1216 | 40 | Lyman |
| 5 | F155N | 1550 | 40 | C IV |
| 6 | F164N | 1640 | 40 | He II |
| 7 | F130N | 1304 | 40 | O I |
| 8 | F135N | 1348 | 20 | |
| 9 | F140M | 1400 | 100 | Si IV |
| 10 | F160N | 1587 | 20 | Continuum |
| 11 | GRISM | 1100-1900 | | Grism (R=100) |
| 12 | Neutral 1 | | | Neutral Density TBD |
| 13 | Neutral 2 | | | Neutral Density TBD |

No spectral element shall relatively displace the image by more than 0.5 pixels or degrade the image quality by more than 10% of the diffraction limit at any given wavelength.

#### *3.2.4.1 TLR_F.9.1: Filter Minimum transmission*

The minimum transmission inside the filter band depends by the filter under consideration. The minimum value shall be

T$\geq$ 10% for broad band filters

T≥ 10% for narrow band filters

### 3.2.4.2 TLR_F.9.2: Quality of filtered images, ghosts

The optical set-up shall avoid ghosts whenever possible; when practically unavoidable, the number and level of ghosts shall be <0.2% of the total incident light from a point source.

## 3.3    Field Spectroscopy mode

### 3.3.1  TLR_F.10: Spectral resolution and spectral coverage

A low resolution spectroscopy mode with a resolving power obtained on a diffraction-limited source R=100 @ 150 nm shall be provided covering the full FUV spectral domain simultaneously.

### 3.3.2  TLR_F.11: Spectroscopy Field of view

The FoV in the field spectroscopy mode shall be the full width of the FUV FoV (see TLR_F.3), even though the spectra of the targets in the external part of the FoV will not be fully recorded.

### 3.3.3  TLR_F.12: Disperser Efficiency

The efficiency of the dispersive element shall be > 50% @ 150 nm.

# 4.    NEAR ULTRA VIOLET CHANNEL (NUV)

## 4.1    General properties

### 4.1.1  TLR_N.1: Spectral domain

The useful spectral domain shall extend from 150 to 280 nm.

### 4.1.2  TLR_N.2: Optical efficiency

The NUV optical scheme shall have a photon transmission efficiency >70% (not including filter or polarizer transmissivity) in imaging mode and >60% in the spectroscopic mode.

### 4.1.3  TLR_N.3: Spatial sampling

The spatial resolution, including the contributions from optical aberrations and resolution losses in the detector shall be < 60 mas (goal 30 mas).

### 4.1.4  TLR_N.4: Field of view

The FoV shall be ≥1.0x1.0 arcmin²

### 4.1.5  TLR_N.5: Detector

In order to meet the science goals achievable with this channel of the FCU instrument the detector shall be a sensor efficient in the working wavelength range with the following characteristics.

### 4.1.5.1 TLR_N.5.1: Pixel size and format

The detector shall have no less than 2k × 2k pixels. The pixel size shall be 20 μm square, corresponding to 30 mas on the focal plane.

### 4.1.5.2 TLR_N.5.2: Detector DQE

The Quantum efficiency of the detector shall be

DQE≥5% for λ=150 nm

DQE≥10% for λ=200 nm

DQE≥10% for λ=250 nm

DQE<0.1% for λ=400 nm

### 4.1.5.3 TLR_N.5.3: Local Dynamical Range

The local dynamical range shall be >10 count/s on a point-like source

### 4.1.5.4 TLR_N.5.4: Global Dynamical Range

The global dynamical range shall be >200000 count/s

### 4.1.5.5 TLR_N.5.5: Time Resolution

The NUV camera detector time resolution shall be $< (20 \text{ ms})^{-1}$.

### 4.1.5.6 TLR_N.5.6: Dark Current

The dark current of the NUV detector will be $<4\times10^{-5}$ counts/pix/s at detector operating temperature

### 4.1.5.7 TLR_N.5.7: Detector Response Uniformity

The ICCD detector shall be correctable to a uniform gain per pixel to <1% at all wavelengths.

### 4.1.5.8 TLR_N.5.8: Large scale non uniformity

Large scale non-uniformity shall be < 1%

### 4.1.5.9 TLR_N.5.9: Detector QE Stability

The absolute QE stability shall be ±0.5% (peak to peak) per hour and ±1% (peak to peak) per month.

### 4.1.5.10 TLR_N.5.10: Detector Flat Field Stability

The difference of two flat fields taken one month apart with the same instrumental configuration shall be < 1%..

## 4.1.6 TLR_N.6: Operations

It shall be possible to operate NUV alone in all its modes. The possibility to use of NUV channel in parallel with other channels is not in the baseline, but should be investigated during Phase B1.

### 4.1.6.1 TLR_N.6.1: Time Tag

It shall be possible to operate NUV in time tag mode.

### 4.1.6.2 TLR_N.6.2: FGS Parameters

It shall be investigated the possibility to operate NUV making on board position correction using the FGS parameters.

### *4.1.6.3 TLR_N.6.3: Focusing*

It shall be possible to operate NUV making on board position correction using a focusing mechanism in order to compensate any changes from the optimized optical path.

## 4.2 Imaging mode

### 4.2.1 TLR_N.7: Imaging FoV

The full NUV FoV shall be used for classical imaging (see TLR_N.4).

### 4.2.2 TLR_N.8: Limiting magnitude

S/N=10 per resolution element in one hour exposure time down V=24.6 (21.9 – 19.5) for a O3 V (A0 V - G0 V) star in the band F250W (corresponding to $5.0 \times 10^{-16}$ erg cm$^{-2}$ s$^{-1}$ A$^{-1}$).

### 4.2.3 TLR_N.9: Number of available filters

Space for at least 45 different, selectable, narrow band (NB), broad band (BB) and neutral filters shall be provided on at least two filter wheels.

### 4.2.4 TLR_N.10: Required filter and characteristics

The NUV channel list of filters is given in Table 4. Filters (positions 1 to 29) are listed in order of priority.

Table 4: NUV List of Filters

| N | Filter name | Central wavelength [A] | Band width [A] | Description |
|---|---|---|---|---|
| 1 | F175W | 1750 | 400 | |
| 2 | F200W | 2000 | 400 | |
| 3 | F218W | 2175 | 400 | Similar to HST/F218W |
| 4 | F250W | 2500 | 400 | Similar to HST/F255W |
| 5 | F175N | 1750 | 20 | NIII] |
| 6 | F185N | 1855 | 50 | Al III |
| 7 | F189N | 1890 | 20 | SiIII] |
| 8 | F190N | 1906 | 20 | CIII] |
| 9 | F280N | 2800 | 50 | Mg II |
| 10 | F141N | 1410 | 20 | S IV |
| 11 | F152N | 1520 | 50 | CO |
| 12 | F164N | 1640 | 20 | He II |
| 13 | F165N | 1657 | 40 | C I |
| 14 | F172N | 1720 | 40 | S III |

| N | Filter name | Central wavelength [A] | Band width [A] | Description |
|---|---|---|---|---|
| 15 | F181N | 1810 | 40 | S I |
| 16 | F183N | 1830 | 20 | Continuum |
| 17 | F188N | 1883 | 10 | Si III] |
| 18 | F189N | 1892 | 10 | Si III] |
| 19 | F190N | 1908 | 20 | C III] |
| 20 | F200N | 2000 | 40 | Continuum |
| 21 | F212N | 2123 | 20 | Continuum |
| 22 | F230N | 2300 | 20 | CII] |
| 23 | F244N | 2466 | 20 | Continuum |
| 24 | F250N | 2500 | 20 | Continuum |
| 25 | F257N | 2576 | 50 | CS |
| 26 | F264N | 2646 | 20 | Continuum |
| 27 | F279N | 2796 | 10 | Mg II |
| 28 | F238N | 2383 | 50 | Fe II |
| 29 | F262M | 2626 | 100 | Fe II |
| 30 | POL0UV | | | Polarizer |
| 31 | POL60UV | | | Polarizer |
| 32 | POL120UV | | | Polarizer |
| | Wollastone Prism | | | Alternative to Polarizers (30-32) |
| 33 | GRISM | 1500-2800 | | GRISM (R≥100) |
| 34 | Neutral 1 | | | Neutral Density TBD |
| 35 | Neutral 2 | | | Neutral Density TBD |
| 36 | Neutral 3 | | | Neutral Density TBD |
| 37 | Neutral 4 | | | Neutral Density TBD |

No spectral element shall relatively displace the image by more than 0.5 pixels or degrade the image quality by more than 10% of the diffraction limit at any given wavelength.

### 4.2.4.1 TLR_N.10.1: Filter Minimum transmission

The minimum transmission inside the filter band depends by the filter under consideration. The minimum value shall be

T≥ 20% for broad band filters

T≥ 10% for narrow band filters

### *4.2.4.2 TLR_N.10.2: Quality of filtered images, ghosts*

The optical set-up shall avoid ghosts whenever possible; when practically unavoidable, the number and level of ghosts shall be <0.2% of the total incident light within a discrete ghost image.

## 4.3    Field Spectroscopy mode

### 4.3.1  TLR_N.11: Spectral resolution and spectral coverage

A low resolution spectroscopy mode with a resolving power obtained on a diffraction-limited source R≥100 @ 200 nm shall be provided covering the full NUV spectral domain simultaneously.

### 4.3.2  TLR_N.12: Spectroscopy Field of view

The FoV in the field spectroscopy mode shall be the full width of the FUV FoV, even though the spectra of the targets in the external part of the FoV will not be fully recorded

### 4.3.3  TLR_N.13: Disperser Efficiency

The efficiency of the dispersive element shall be > 50%.

## 4.4    Polarimetric mode

### 4.4.1  TLR_N.14: Polarimetric mode

NUV shall provide a polarization imaging mode

### 4.4.2  TLR_N.15: Polarization state

Images of the FCU field can be obtained in at least 2 linear polarization state (Extraordinary beam, E, and Ordinary beam, O).

### 4.4.3  TLR_N.16: Transmission Polarization Analyser (TPA)

A transmission polarization analyzer (TPA) shall be introduced into the light beam to produce twin images of the source, with orthogonal senses of polarization (E and O beams), at the detector.

### 4.4.4  TLR_N.17: Throughput Efficiency

The throughput efficiency of the TPA shall be > 20% for each of the E and O beams in the full NUV spectral domain.

### 4.4.5  TLR_N.18: Polarimetry mode FoV

The FoV for the beam of the polarimeter shall be > 30x60 arcsec$^2$ (goal: full NUV FOV)

### 4.4.6  TLR_N.19: Polarizers transmission uniformity

The polarizers transmission shall be uniform over the field of view to better than 10%.

### 4.4.7  TLR_N.20: Instrumental Polarization

The instrumental polarization shall be < 1%.

### 4.4.8 TLR_N.21: Spectropolarimetry

Slitless spectroscopy and polarimetric modes can be optionally combined in order to achieve a low spectral resolution NUV spectropolarimetry mode. Both the TPA and the dispersive element shall be introduced into the light beam to produce twin dispersed images of the source (E and O beams), at the detector.

### 4.4.9 TLR_N.22: Spectropolarimetry FoV

The spectropolarimetry mode is slitless. The FOV shall be 15 x 60 arcsec$^2$ on the detector.

# 5. UV-OPTICAL CHANNEL (UVO)

## 5.1 General Properties

### 5.1.1 TLR_O.1: Spectral domain

The useful spectral domain shall extend from 200 nm to the red limit of the detector sensitivity. In the definition of the detector properties, performances for λ<500 nm shall be maximized.

### 5.1.2 TLR_O.2: Spatial sampling

In order to optimize the field coverage, and still allow to take full advantage of the diffraction limit angular resolution, the space sampling shall be 0.07 arcsec/pixel.

### 5.1.3 TLR_O.3: Field of view

FoV shall be ≥ 4.5 x 4.5 arcmin²

### 5.1.4 TLR_O.4: Field Distortion

Field distortion shall be smaller than ≤10%, edge to edge

### 5.1.5 TLR_O.5: Field Distortion Stability

Field distortion shall be as stable as possible. In any case, the field distortion shall be stable at the level of 3 mas.

### 5.1.6 TLR_O.6: Operations

It shall be possible to operate UVO alone in all its modes. The possibility to use of UVO channel in parallel with other channels is not in the baseline, but should be investigated during Phase B1.

### 5.1.7 TLR_O.7: Detector

In order to meet the science goals reachable with this channel of the FCU instrument the detector shall be a sensor efficient in the working wavelength range with the following characteristics.

#### *5.1.7.1 TLR_O.7.1: Pixel size and format*

The detector shall have 4k × 4k pixels. The pixel size shall be ≤15 μm square, corresponding to ≤0.07 arcsec on the focal plane.

### 5.1.7.2 TLR_O.7.2: Detector QE

The Quantum efficiency of the detector shall be:

QE$\geq$40% for $\lambda$ = 200 nm

QE$\geq$60% for $\lambda$ = 300 nm

QE$\geq$60% for $\lambda$ = 400 nm

QE$\geq$60% for $\lambda$ = 500 nm

QE$\geq$60% for $\lambda$ = 700 nm

### 5.1.7.3 TLR_O.7.3: Dark Current

The dark current of the UVO detector will be <18 e-/pix/hr at detector operating temperature.

### 5.1.7.4 TLR_O.7.4: Read Out Noise

The read out noise shall be <3 e$^-$.

### 5.1.7.5 TLR_O.7.5: Black and white spots and columns defects

Total number of black and white spots shall be < 800. Columns defects should be less than 5.

### 5.1.7.6 TLR_O.7.6: Readout Time

Minimum read out time per frame should be < 90 seconds.

### 5.1.7.7 TLR_O.7.7: Detector Response Uniformity

The CCD detector shall be correctable to a uniform response per pixel to <1% at all wavelengths. No more than 5% (goal <1%) of the pixels shall have response outside 90 – 110% of the mean response (goal 95 – 105%).

### 5.1.7.8 TLR_O.7.8: Large scale non uniformity

Large scale non-uniformity shall not exceed 3% (peak to peak), and shall be correctable to <1%.

### 5.1.7.9 TLR_O.7.9: Detector QE Stability

The absolute QE stability shall be $\pm$0.5% (peak to peak) per hour and $\pm$1% (peak to peak) per month.

### 5.1.7.10 TLR_O.7.10: Detector Flat Field Stability

The difference of two flat fields taken one month apart with the same instrumental configuration shall not exceed 1% (0.5% goal); no more than 5% of the FoV shall exceed 5% variation.

### 5.1.7.11 TLR_O.7.11: Detector Overlight response

Blooming for exposures exceeding the Full Well capacity shall be only along the columns containing the overlight pixels. Electron loss for blooming shall be <1%. Residual images from 100 times full well overlight conditions shall not exceed 1 electron/second after 2 hours.

### 5.1.7.12 TLR_O.7.12: Dynamic Range

A dynamic range up to 65536 shall be guaranteed.

### 5.1.7.13 TLR_O.7.13: Gain

It shall be possible to operate the camera with different gains. Gains 1 and 2 shall be supported for the UVO, since gain=1 provides the smallest readout noise, while gain=2 (or higher) is needed to sample the available full well depth. Gain 4 and 8 should also be available (though not calibrated).

## 5.1.8 TLR_O.8: Shutter fly time

Opening and closing timing as well as differences in the shutter coverage of the observed field shall be smaller than 0.01%. For any exposure time from 0.5s to 60 minutes the shutter shall guarantee a uniform exposure time across the detector such that no two pixels differ in exposure time by more than 0.01 sec.

## 5.2 Imaging Mode

## 5.2.1 TLR_O.9: Imaging FoV

The full UVO FoV shall be used for imaging.

## 5.2.2 TLR_O.10: Limiting Magnitude

S/N=10 per resolution element in one hour exposure time down V=26.3 (26.3 – 26.3) for a O3 V (A0 V - G0 V) star in the band F555W (corresponding to $1.8 \times 10^{-16}$ erg cm$^{-2}$ s$^{-1}$ A$^{-1}$).

## 5.2.3 TLR_O.11: Number of available filters

Space for at least 21 different, selectable, filters shall be provided. All filters should have the same optical thickness in order to not require a focus change at changing filters.

## 5.2.4 TLR_O.12: Required filter and characteristics

The UVO channel list of filters is given in Table 5. Filters (positions 1 to 20) are listed in order of priority.

Table 5: UVO List of Filters

| n | Filter name | Central wavelength [A] | Band width [A] | Description |
|---|-------------|------------------------|----------------|-------------|
| 1 | F250W | 2500 | 400 | Similar to HST./F255W |
| 2 | F300W | 2900 | 400 | Similar to HST/F300W |
| 3 | F345W | 3450 | 350 | Stromgren u |
| 4 | F430W | 4300 | 1000 | Johnson B - HST/439W |
| 5 | F555W | 5300 | 1000 | Johnson V - HST/555W |
| 6 | F606W | 5800 | 1500 | Similar to HST/F606W |
| 7 | F625W | 6300 | 1000 | Similar to HST/F625W |
| 8 | F814W | 8200 | 1700 | Similar to HST/F814W |
| 9 | F411M | 4110 | 190 | Stromgren v – HST/F410N |
| 10 | F468M | 4680 | 180 | Stromgren b – HST/F467N |

| n | Filter name | Central wavelength [A] | Band width [A] | Description |
|---|---|---|---|---|
| 11 | F547M | 5470 | 230 | Stromgren y |
| 12 | F280N | 2800 | 20 | Mg II |
| 13 | F468N | 4686 | 20 | He II |
| 14 | F486N | 4862 | 20 | H $_.$ |
| 15 | F500N | 5007 | 20 | O III |
| 16 | F587N | 5875 | 20 | He I |
| 17 | F630N | 6300 | 20 | O I |
| 18 | F656N | 6562 | 20 | H |
| 19 | F672N | 6720 | 50 | S II |
| 20 | F658N | 6585 | 50 | N II |
| 21 | GRISM1 | | | GRISM1 R=250 |
| 22 | GRISM2 | | | GRISM2 R=250 |
| 23 | POL | | | 1 slot - 3 polarizer |
| 24 | RAMP | >3000 | | Ramp filter |

No spectral element shall relatively displace the image by more than 0.5 pixels or degrade the image quality by more than 10% of the diffraction limit at any given wavelength.

### 5.2.4.1 TLR_O.12.1: Filter Minimum transmission

The minimum transmission inside the filter band depends by the filter under consideration. The minimum value shall be

$T \geq 30\%$ for broad band filters

$T \geq 10\%$ for narrow band filters

$T \geq 30\%$ for ramp filters

### 5.2.4.2 TLR_O.12.2: Quality of filtered images, ghosts

The optical set-up shall avoid ghosts whenever possible; when practically unavoidable, the number and level of ghosts shall be <0.2% of the total incident light from a point source.

## 5.3    Field Spectroscopy mode

## 5.3.1  TLR_O.13: Spectral resolution and spectral coverage

A low resolution spectroscopy mode with a resolving power R=250 @ $\lambda$ 500 nm obtained on a diffraction-limited source shall be provided covering the as much as possible the range 200 – 700 nm.

### 5.3.2 TLR_O.14: Spectroscopy Field of view

The FoV in the field spectroscopy mode shall be the full width of the UVO FoV, even though the spectra of the targets in the external part of the FoV will not fully recorded.

## 5.4    Polarimetry Mode

### 5.4.1  TLR_O.15: Polarimetry mode FoV

The FoV for the beam of the polarimeter shall be > 60x60 arcsec$^2$ (goal: full UVO FOV)

### 5.4.2  TLR_O.16: Minimum polarizers transmission

The minimum polarizers transmission shall be greater than 60%.

### 5.4.3  TLR_O.17: Polarizers transmission uniformity

The polarizers transmission shall be uniform over the field of view to better than 10%.

### 5.4.4  TLR_O.18: $T_{//}$ to $T_{\perp}$ ratio

The minimum ratio between the transmission parallel to the polarizer axis with respect to the perpendicular transmission shall be greater than 100.

### 5.4.5  TLR_O.19: Instrumental polarization

The instrumental polarization shall be < 1%.

### 5.4.6  TLR_O.20: Spectropolarimetry

Slitless spectroscopy and polarimetric modes can be optionally combined in order to achieve a low spectral resolution UVO spectropolarimetry mode. To do so, both the polarizers and the dispersive element shall be introduced into the light beam.

### 5.4.7  TLR_O.21: Spectropolarimetry FoV

The spectropolarimetry mode is slitless. The FoV width shall be  60x60 arcsec$^2$ on the detector.

# Chapter III.

# FCU phase A report – Optical, mechanical and electronics configurations

## 1. INTRUMENT CONCEPT

### 1.1 Mission Contest

The World Space Observatory-UV is an international collaboration led by Russia to build a space telescope dedicated mainly to UV astrophysics. The WSO-UV telescope, will be an UV optimized instrument that will investigate numerous astrophysical phenomena from planetary science to cosmology. Table 6 summarizes WSO-UV spacecraft characteristics while Figure 17 shows the satellite in flight configuration.

Table 6. Main characteristics of WSO-UV space telescope

| | |
|---|---|
| **Spacecraft mass with propellant** | 2900 kg |
| **Payload mass** | 1600 kg |
| **Instrumentation Compartment power consumption** | 750 W |
| **Data transmission rate (S-band)** | 1 Mb/s |
| **Service telemetry data transmission rate** | 32 kb/s |
| **Platform star tracker pointing accuracy** | 30 arc sec |
| **Stabilization and pointing accuracy** | 0.03 arc sec |
| **Spacecraft angular rate in stabilization mode** | $2 \times 10^{-5}$ degree/sec |
| **Spacecraft slew rate** | 0.1 degree/sec |
| **Maximum duration of scientific observation in continuous** | 30 hr |

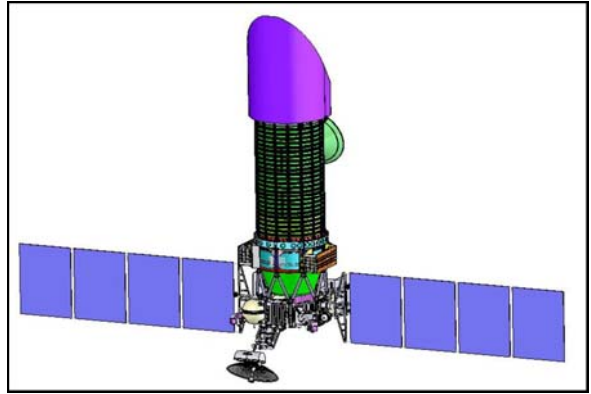

Figure 17. WSO-UV spacecraft

Telescope, launcher (Zenith 2SB) and platform (Navigator) will be developed in Russia. The spacecraft will be in a geosynchronous orbit at a height of 35800 km and an inclination of 51.8 degrees. The mission life time will be longer than 5 years. The telescope will be a Ritchey-Chretien with a diameter of 1.7m with a focal ratio of F/10 and a corrected field of view of 0.5 degrees. The primary wavelength range is 100-350 nm with an extension into the visible range. The telescope will host two spectrographs and one imager. There will be a high resolution echelle spectrograph (HIRDES) developed by Germany with a resolution of about 50000 and two channels (UVES and VUVES) to cover more efficiently the entire wavelength range. The second spectrograph, developed by China, will be a long slit low resolution (R ~ 1000-2500) spectrograph (LSS). This instrument will have also two channels. The imager, Field Camera Unit, will be developed in Italy and will provide the WSO-UV satellite with imaging capabilities allowing to obtain diffraction limited, deep UV and optical images.

## 1.2  Instrument concept

The Field Camera Unit (FCU) is a multi-spectral radial instrument on the focal plane of WSO-UV. Being a space telescope imaging camera, the highest priorities of the FCU instrument will be to guarantee high spatial resolution and high UV sensitivity while trying to maximize the wavelength coverage and the size of the field of view (see Table 7). To meet this challenge we are designing an instrument that will perform deep UV and diffraction limited wide and narrow band imaging.

Table 7. Science requirements for the FCU camera

| Parameter | Channel | | |
|---|---|---|---|
| | **Far-UV** | **Near-UV** | **UV-Optical** |
| **Spectral Range** | 115-190 nm | 150-280 nm | 200-700 nm |
| **Field of View** | 6.6'x6.6' | 1'x1' | 4.7'x4.7' |
| **Scale** | 0.2"/pix | ≤0.06"/pix | 0.07"/pix |
| **Pixel Size** | 20μm | 20μm | 15μm |
| **Array Size** | 2kx2k | 2kx2k | 4kx4k |
| **Detector** | MCP (CsI) | MCP (CsTe) | CCD (UV optimized) |

The instrument design concept is illustrated in Figure 18. Three wide band channels (FUV, NUV and UVO) are provided, covering a wavelength range from 115 nm to at least 700 nm. Each channel is specialized in a specific wavelength range and will have also spectroscopic, polarimetric and spectropolarimetric capabilities. Figure 19 shows the overall layout of the FCU instrument.

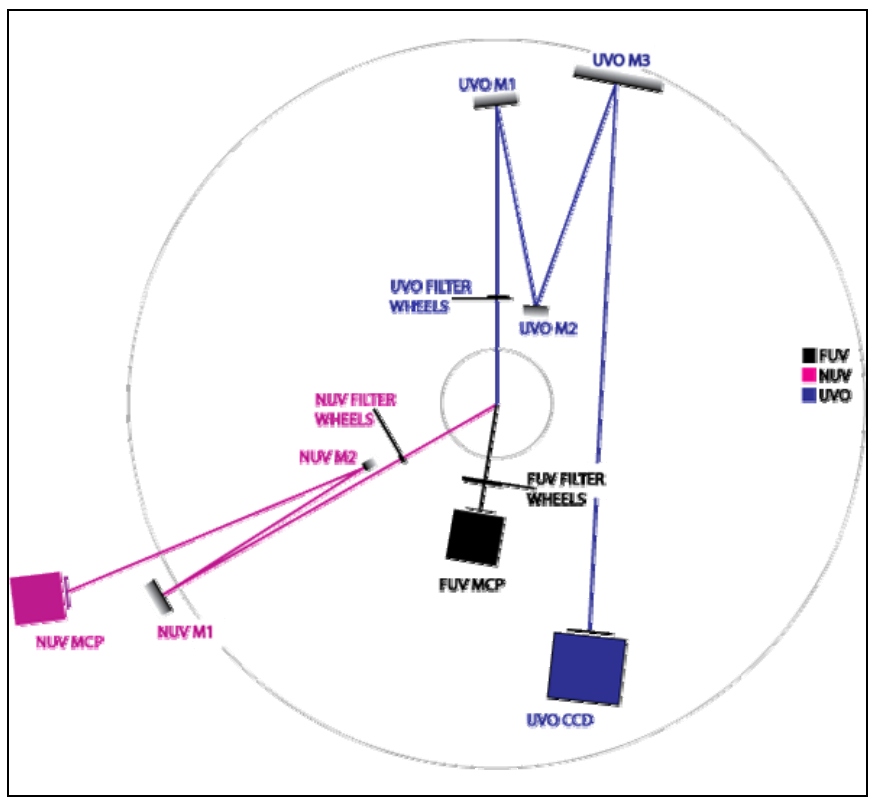

Figure 18. FCU Schematic design.

The Far-UV channel will cover the 115-190 nm range. To reduce losses in the throughput, this channel will not have any optics but the mirror to feed it. The scale of the telescope will be unchanged yielding a large field of view at expenses of spatial resolution. So, for this channel, we favour sensitivity over resolution. The FUV channel filters wheels will accommodate narrow and wide band filters and a disperser which will allow low resolution (R ~ 100) slitless spectroscopy. The FUV channel will use a photon counting detector based on a MCP detector with a 2kx2k format. It will have a CsI photocathode directly deposited on the MCP. The read-out system will be based on a CCD detector.

The Near-UV channel will cover the 150-280 nm range overlapping the FUV range on the shorter wavelengths and the UVO range on the longer ones. To exploit the diffraction limited optical quality of the telescope in this wavelength range the NUV channel has the highest spatial resolution. Its filters wheels will accommodate wide and narrow band filters and polarizers for imaging and imaging polarimetry. A grating will allow low resolution (R >= 100) slitless imaging spectroscopy. A spectro-polarimetric observational mode is being investigated. The NUV detector will differ from the FUV detector only for the photocathode ($Cs_2Te$), optimized for this wavelength range and deposited on the detector window.

Finally the UVO channel will extend to visual wavelengths to exploit the wide spectral sensitivity of its CCD detector. The pixel scale of this channel is a compromise between the need of a large field of view and high spatial resolution. Filters, dispersers and polarizers will allow to have narrow and broad band imaging, low resolution (R ~ 100-500) slitless spectroscopy and imaging polarimetry. The UVO channel will use a single 4kx4k CCD detector optimized for the UV.

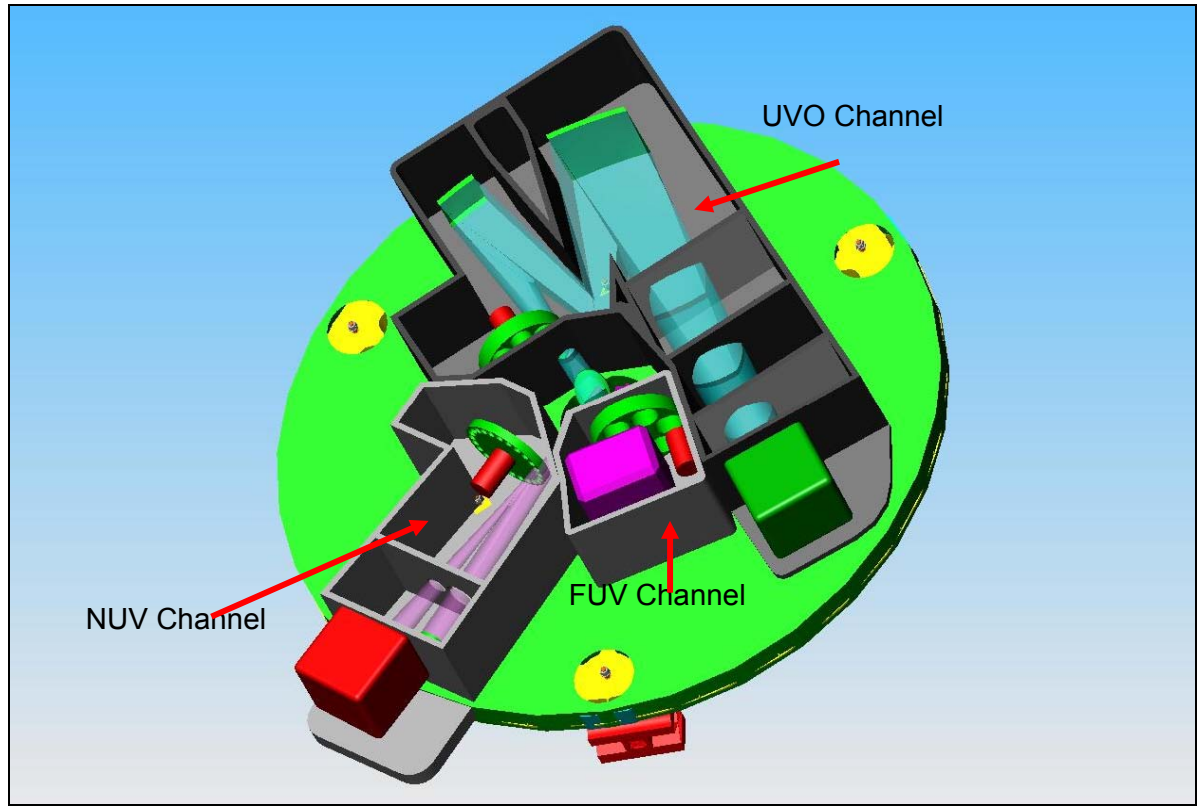

Figure 19. Preliminary overall layout of FCU

Table 8 summarizes the preliminary fundamental parameters for the three channels obtained as results of this feasibility study.

Table 8. FCU preliminary performances characteristics summary

| Channel | Detector | Wavelength region (nm) | Resolution element width (μm) | Resolution element width (mas) | Effective focal length (mm) | Internal Magnification | F/# | Array size | Field of view (arcsec) | Array width (mm) |
|---------|----------|------------------------|-------------------------------|--------------------------------|-----------------------------|------------------------|-----|------------|------------------------|------------------|
| FUV | MCP/CsI | 115-190 | 20 | 0.20 | 17000 | 1.0 | 10 | 2048 | 409.6 | 40.0 |
| NUV | MCP/Cs$_2$Te | 150-280 | 20 | 0.03 | 137510 | 8.1 | 81 | 2048 | 61.44 | 40.0 |
| UVO | CCD | 200-700 | 15 | 0.07 | 44200 | 2.6 | 26 | 4096 | 286.7 | 61.44 |

In Table 9 the preliminary list of mechanisms present in the FCU instrument is shown.

Table 9. FCU preliminary list of mechanisms

| Mechanism | Positions | Driven Size | Optical Element Size | Positioning Tolerance | Stability Tolerance | Motion |
|-----------|-----------|-------------|----------------------|-----------------------|---------------------|--------|
| Rotating Mirror | 3 | 56x56mm mirror | 56x56mm mirror | 1.2 arcsec | 0.10 arcsec | Full rotation |
| FUV Filter Wheel #1 | 6 | 150mm dia. | 45mm dia. filter | 0.35 degree | 0.15 degree | Full rotation |
| FUV Filter Wheel #2 | 6 | 150mm dia. | 45mm dia. filter | 0.35 degree | 0.15 degree | Full rotation |

FCU phase A report – Optical, mechanical and electronics configurations

| Mechanism | Positions | Driven Size | Optical Element Size | Positioning Tolerance | Stability Tolerance | Motion |
|---|---|---|---|---|---|---|
| NUV Filter Wheel #1 | 24 | 150mm dia. | 5mm dia. filter | 0.35 degree | 0.15 degree | Full rotation |
| NUV Filter Wheel #2 | 24 | 150mm dia. | 5mm dia. filter | 0.35 degree | 0.15 degree | Full rotation |
| NUV Grating Wheel | 2 | 150mm dia | TBD | 0.35 degree | 0.15 degree | Full rotation |
| NUV Tip/Tilt & Focus | Multiple | | | TBD | TBD | Focus |
| UVO Filter Wheel #1 | 9 | 150mm dia. | 24x24mm filter | 0.1 degree | 0.05 degree | Full rotation |
| UVO Filter Wheel #2 | 9 | 150mm dia. | 24x24mm filter | 0.1 degree | 0.05 degree | Full rotation |
| UVO Filter Wheel #3 | 9 | 150mm dia. | 24x24mm filter | 0.1 degree | 0.05 degree | Full rotation |
| UVO Tip/Tilt & Focus | Multiple | 150mm dia. | | TBD | TBD | Focus |
| UVO CCD Shutter | 4 | | | 2 degree | 0.15 degree | Full rotation |

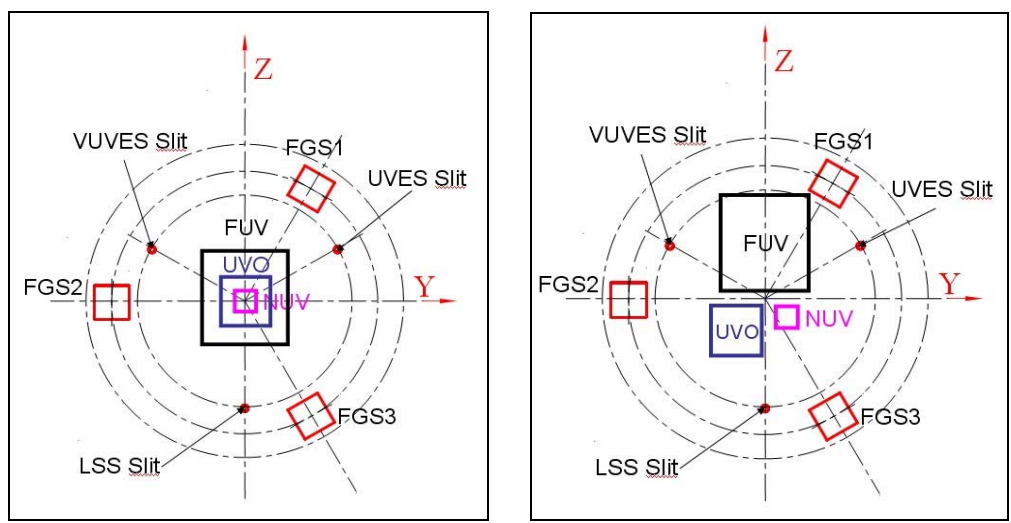

Figure 20. The FCU fields of view for the two optical layouts. Left: rotating pick up mirror. Right: mosaic pick up mirror.

Figure 20 shows the layout of the focal surface of the telescope with superimposed the fields of view of WSO-UV instruments and FGS. This area has a diameter of 150mm and as it can be deduced from the figure is rather crowded. The geometry of the focal plane of the telescope makes necessary to use a mirror to fold the optical beam coming from the telescope in a direction parallel to the optical bench where the three channels of the FCU will be deployed.

After a preliminary analysis we decided to develop an all reflective design to maximize throughput and to stay as much as possible close to the optical axis to simplify the design.

We have studied two alternative optical layouts (see Figure 20):

Rotating pick up mirror: in this case the mirror will be flat and using a mechanism will rotate feeding one of the three channels at a time.

Mosaic pick mirror: this layout foresee to use three fixed mirrors which will be oriented to feed all three channels at the same time.

Table 5 summarizes the scientific investigation parameters for each channel of FCU and for the different layouts.

## 1.3    Mechanical Layouts

Figure 21 shows orthogonal views of the FCU instrument.

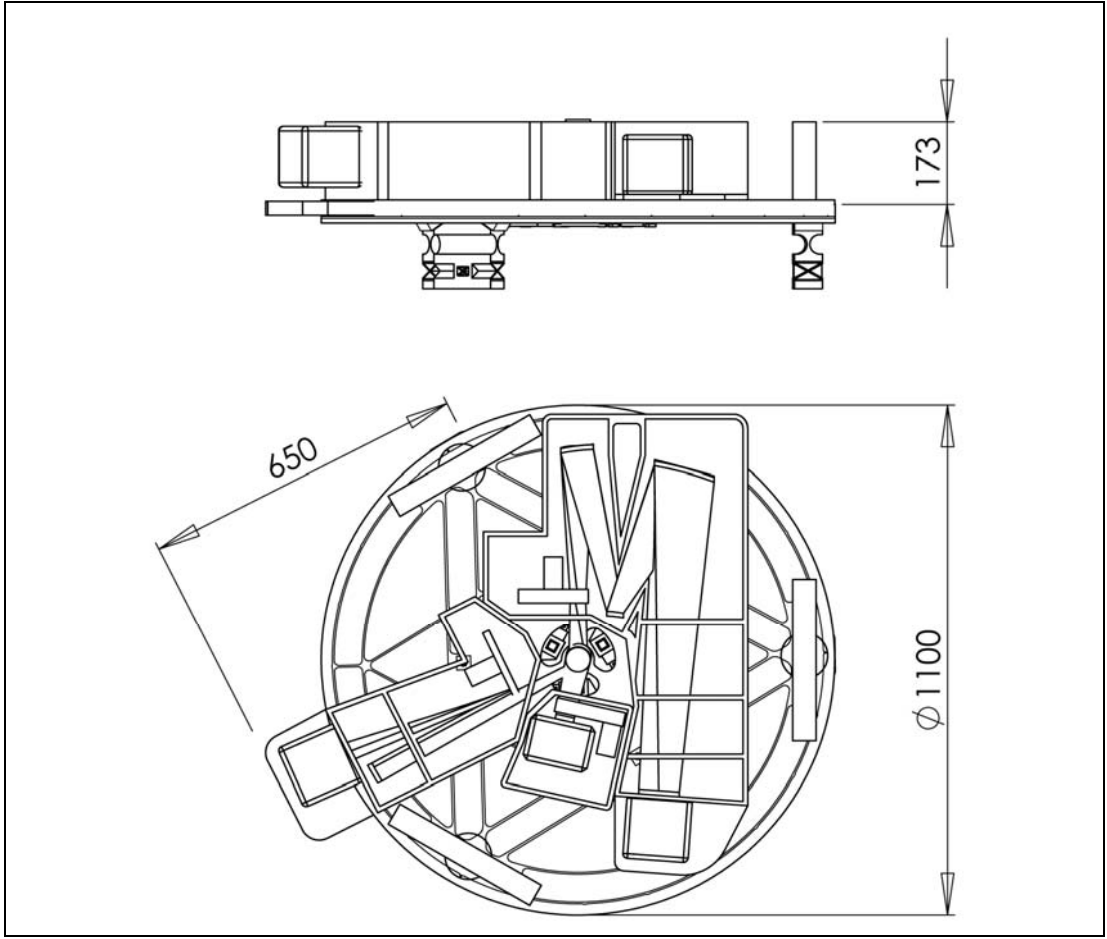

Figure 21. Top and lateral view of FCU instrument.

Table 10. FCU expected performance characteristics

| Parameter | FUV | NUV | UVO |
|---|---|---|---|
| Detector | ICCD: CsI | ICCD: Cs2Te | CCD-UV Optimized |
| Detector Array Size | 2048 x 2048 | 2048 x 2048 | 4096 x 4096 |
| Pixel Size | 20 microns | 20 microns | 15 microns |
| Spectral Range | 115 – 190 nm | 150 – 280 nm | 200 – 700 nm |
| Field of View | 409.6 x 409.6 arc sec | 61.4 x 61.4 arc sec | 286.7 x 286.7 arc sec |
| Pixel Width | 0.20 arc sec | ≤0.06 arc sec | 0.07 arc sec |
| Encircled Energy at Best Field-Nominal (On axis Design) | 0.95 @ 150nm in 0.2 arc sec diameter | 0.54 @ 200nm in 0.03 arc sec diameter | 0.45 @ 550nm in 0.07 arc sec diameter |

| Parameter | FUV | NUV | UVO |
|---|---|---|---|
| Encircled Energy at Worst Field-Nominal Design (On axis Design) | 0.51 @ 150nm in 0.2 arc sec diameter | 0.50 @ 200nm in 0.03 arc sec diameter | 0.43 @ 550nm in 0.07 arc sec diameter |
| Encircled Energy at Best Field-Nominal (Off axis Design) | 0.90 @ 150nm in 0.2 arc sec diameter | 0.47 @ 200nm in 0.03 arc sec diameter | 0.46 @ 550nm in 0.07 arc sec diameter |
| Encircled Energy at Worst Field-Nominal Design (Off axis Design) | 0.37 @ 150nm in 0.2 arc sec diameter | 0.29 @ 200nm in 0.03 arc sec diameter | 0.44 @ 550nm in 0.07 arc sec diameter |
| Enpixeled Energy at Best Field-Nominal Design (On axis Design) | 0.97 @ 150nm in 2x2 pixel | 0.71 @ 200nm in 2x2 pixel | 0.67 @ 550nm in 2x2 pixel |
| Encircled Energy at Worst Field-Nominal Design (On axis Design) | 0.93 @ 150nm in 2x2 pixel | 0.68 @ 200nm in 2x2 pixel | 0.66 @ 550nm in 2x2 pixel |
| Enpixeled Energy at Best Field-Nominal Design (Off axis Design) | 0.96 @ 150nm in 2x2 pixel | 0.67 @ 200nm in 2x2 pixel | 0.67 @ 550nm in 2x2 pixel |
| Enpixeled Energy at Worst Field-Nominal Design (Off axis Design) | 0.87 @ 150nm in 2x2 pixel | 0.55 @ 200nm in 2x2 pixel | 0.66 @ 550nm in 2x2 pixel |
| Maximum Induced Polarization | <1% (TBC) | <1% (TBC) | 1% (TBC) |
| Temporal Resolution, Typical | 10 msec | 10 msec | 500 msec |
| Typical Exposure Time | - | - | 5 to 30 minutes |
| Minimum Exposure Time | NA | NA | 0.5 sec |
| Dynamic Range | LDR > 10 GDR > 2x105 | LDR > 10 GDR > 2x105 | 65535 (16 bit) |
| Readout Times | | | 84 sec |
| RMS Read Noise | 0 | 0 | <3 electrons |
| Limiting Flux (3600 s, SNR=10) | 5.5x10-16 erg cm-2 s-1 A-1 @ 150 nm | 5.0x10-16 erg cm-2 s-1 A-1 @ 255 nm | 1.8x10-16 erg cm-2 s-1 A-1 @ 550 nm |
| Flat Field Non Uniformity | <10% | <10% | 3% |
| Detector Absolute Photometric Stability/Hour | <0.5% | <0.5% | 0.5% |
| Detector Absolute Photometric Stability/Month | <1% | <1% | 1% |
| Detector Visible Light DQE | < 0.0001% above 400 nm | < 0.0001% above 400 nm | > 60% from 400 to 700nm |
| Detector Ultraviolet Light DQE | > 20% @ 120nm | > 10% @ 180nm | > 40% @ 200nm |

## 1.4  Operating Modes and Relation to Scientific Objectives

We are planning to implement the following operating modes for the FCU instrument:

### 1.4.1 Optical Modes

All the three channels will allow to have narrow, medium and wide band imaging and also low resolution slitless spectroscopy. Polarimetric modes are foreseen for the NUV and UVO channels and a spectropolarimetric mode in under study for the NUV channel.

### 1.4.2 Parallel Observing Modes

In case the pyramid pick up mirror will be the selected solution for the opto-mechanical layout two channels of the FCU could be used at the same time looking at different parts of the sky. The implementation of this mode will be subject to thermal and power budget analysis. Furthermore one channel of the camera can be used when one of the spectrographs is working as primary instrument. In this case the only restriction could come from limitation of data transfer rate.

### 1.4.3 High Temporal Resolution Modes

The FUV, NUV and UVO channels can all be operated in a mode where subarrays of the detector can be read-out, thereby reducing the readout time and permitting a more rapid observing sequence. The FUV and NUV detectors can also operate in a time tag mode whereby the image location and the time of arrival of the detected photons are recorded permitting a few ms temporal resolution at the expense of increased data volume.

### 1.4.4 Calibration Modes

The FCU optical design will include a calibration unit with a set of continuum source lamps covering the full wavelength ranges of the three detectors. These lamps will be used periodically to perform flat field calibrations.

## 1.5 Model Philosophy

The FCU preliminary model philosophy together with the items that will be produced is given in Table 11. A more detailed discussion on this subject is to be found in [RD4].

Table 11. FCU preliminary model philosophy

| Item Name | Quantity | Deliverable | Description |
|---|---|---|---|
| NUV Bread Board | 1 | No | |
| FUV Bread Board | 1 | | |
| UVO Bread Board | 1 | | |
| FCU STM | 1 | Yes | FCU optical bench, dummy masses to simulate the mirror and mechanisms, dummy masses to simulate the three Detector assemblies, and the Electronics (ICU, PSU, harness), heaters to simulate the thermal behavior |
| FCU EQM | 1 | Yes | The EQM model will be fully equivalent to the Flight Model, as required in Moisheev et al. ( 2006) section 17 |
| FCU FM | 1 | Yes | |
| FCU EGSE + Science Console | 1 | Yes | |

| Item Name | Quantity | Deliverable | Description |
|-----------|----------|-------------|-------------|
| FCU MGSE | 1 | Yes | |
| FCU OGSE | 1 | Yes | |

## 1.6   FCU Budgets

The FCU mass and power budgets are listed in Table 12 and Table 13, respectively. The two numbers in the ICU & PSU row in Table 13 refer to power consumption during optical and calibration modes respectively. The two numbers in the total row give minimum and maximum power budget.

Table 12. FCU preliminary mass budget

| SUBSYSTEM | Weight (kg) |
|-----------|-------------|
| Optics & Mechanisms | 10.5 |
| Detectors | 14.2 |
| Mechanical Structure | 15.5 |
| Electronics | 10.5 |
| **Total** | 50.7 |
| Contingency (20%) | 10.1 |
| **Grand Total** | 60.8 |

Table 13. FCU preliminary power budget

| SUBSYSTEM | Power (W) |
|-----------|-----------|
| FUV Channel | 14.0 |
| NUV Channel | 14.0 |
| UVO Channel | 35.0 |
| ICU & PSU (Opt. /Cal.) | 29.0 / 40.5 |
| **Total** (2 channels in parallel) | 57.0 / 78.0 |

## 1.7   Environmental Requirements

### 1.7.1  Vibro–Acoustic requirements

The FCU instrument should not have resonance frequencies lower than 40Hz unless agreed with the Lavochkin Association (WSO-UV manufacturer).

The FCU instrument design will have to comply with stationary, impact, and acoustic vibration environment as defined in sections 8.1.3, 8.1.4, 8.1.5 in Moisheev et al. ( 2006).

## 1.7.2 Operational, non-operational and storage temperature requirements

During its operative life on board the WSO-UV telescope the FCU instrument operational temperature will be 20±5 C (Moisheev et al. 2006).  Non-operational temperature should be in the range −40 C / +50 C (TBC).

The conditions during the assembling and development at Lavochkin Association will be the following: temperature from 15ºC up to 35ºC, relative humidity from 40% up to 90% at temperature 20ºC (Moisheev et al. 2006). Additional requirements should be agreed with Lavochkin Association.

The FCU instrument will kept at Lavochkin Association ground facilities at temperature ranging from 15ºC up to 35ºC, and relative humidity no more than 85% at temperature 20ºC (Moisheev et al. 2006). Additional requirements should be agreed with Lavochkin Association.

## 1.7.3 EMC requirements

All FCU components will be selected to comply with EMC requirements at spacecraft level, see section 9 of Moisheev et al. ( 2006) for details.

## 1.8    Reliability/Operative life

The reliability factor of FCU operations during its operations on WSO-UV Spacecraft should be not less than 0.9 (Moisheev et al. 2006).

The operative life of FCU instruments should be no less than nine years (Moisheev et al. 2006) and should include the following phases:

maintenance and storage in the ground conditions for four years; this includes time for the production and all the tests performed on the instrument before launch on the ground;

operation in flight for five years.

## 1.9    Handling/Transportability Requirements

The FCU will be transported at the Lavochkin Association (Russia) inside a dust and waterproof container with thermal control from 5ºC up to 35ºC, relative humidity no more than 35% at temperature 20ºC (Moisheev et al. 2006).

## 1.10  Cleanliness

Air clearness at spacecraft level will be no worst than ISO class 8 (100000 on FED-STD-209E) (Moisheev et al. 2006). During the development and assembling phases of FCU instrument great care will be put on the cleanliness. The assembling of the detector head will be done in class 11 (100 on FED-STD-209E) clean room. To prevent degradation of the optical coatings, all the FM optics must be kept in class 9 (TBC) environment.

# 2. OPTICAL DESIGN

## 2.1 Introduction

The overall concept for the FCU is a functional split between three imager channels: the FUV Channel, covering the spectral range 115 – 190 nm; the NUV Channel, covering the spectral range 150 – 280 nm; and the UVO channel, covering the spectral range 200 – 700 nm.

This functional split was chosen because of the different science requirements (spatial resolution, temporal resolution, field of view) on the covered spectral range and because of the different detector technology (CCD for the UVO, MCPs for the NUV and FUV) adopted to optimize the efficiency in the spectral intervals.

The three imager channels are fed through a dedicated fore-optics system able to select each field of view from the telescope focal plane and to send it into the corresponding channel optical path.

A functional block diagram of the FCU is shown in figure Figure 22.

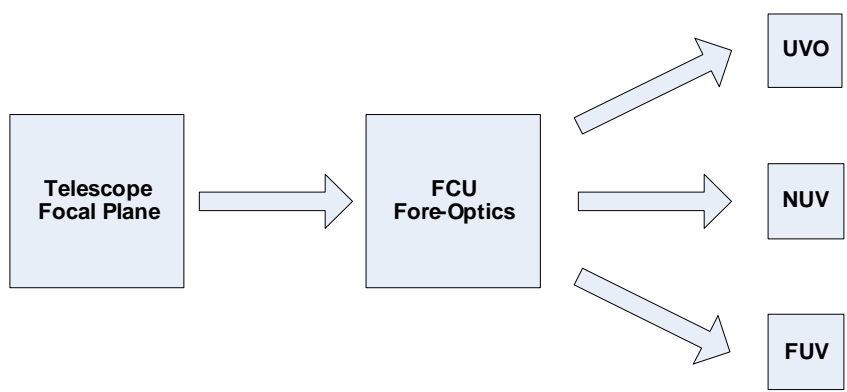

Figure 22. Functional block diagram of FCU.

### 2.1.1 WSO-UV telescope

The reference telescope data used during this study have been taken from Moisheev et al. ( 2006) and from the optical design of the T-170M telescope ZEMAX file.

The WSO-UV telescope consists of an F/10 Ritchey-Chretien aplanatic two mirrors system having a primary mirror diameter of 1.7 m and providing an accessible field of view of 30 arcmin on the telescope focal plane.

The relevant telescope data are given in Table 14 while the telescope layout is shown in Figure 23.

Table 14. Telescope parameters relevant for FCU optical design.

| Parameter | Value |
|---|---|
| Telescope entrance pupil diameter | 1.7 m |
| Effective focal length | 17 m |
| F/ratio | 10 |
| FoV diameter | 30 arcmin (148,48 mm) |
| Scale | 12,13 arcsec/mm |

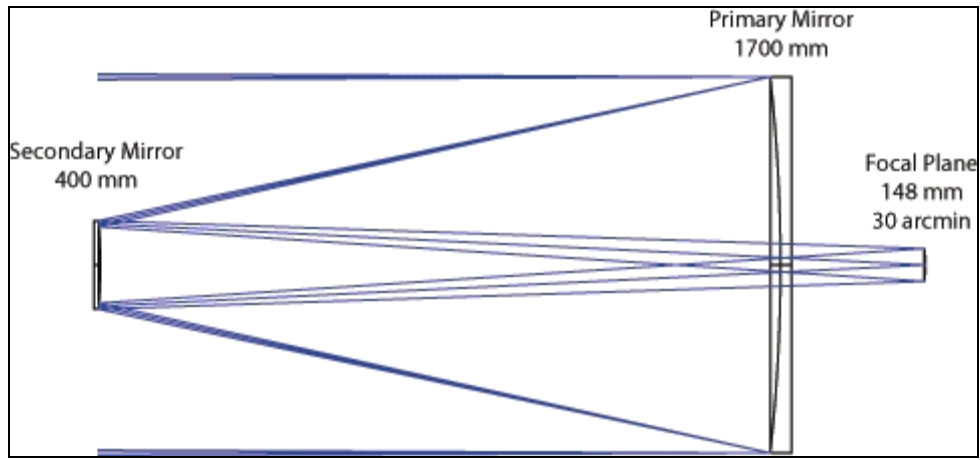

Figure 23. Telescope layout.

At the telescope FP the S/C optical bench is located and it is used as reference plane for all the onboard instrumentation. The S/C OB is aligned and maintained in the correct position with respect to the primary mirror unit (PMU) using a three rods system.

The FCU is mounted on the upper basis of the S/C OB, in the space available between the PMU and the SOB itself, while the three spectrographs UVES, VUVES and LSS are mounted to the S/C OB bottom basis. The FGS location will be nearby the spectrographs entrance slits in a TBD position.

The space available for the FCU in the direction perpendicular to the telescope optical axis is determined mainly by the dimension and the location of the other instruments entrance FP, by the dimension of the FCU optical bench (OB) and by the location of the rods supporting the S/C OB. A scheme is shown in  Figure 24.

The spectrographs and FGS entrance FP are located inside a ring centred on the telescope optical axis having a minimum radius of 45 mm and a maximum radius of 75 mm at 120 degree with respect to each others; the diameters of the spectrographs entrance FP are 10 mm, while the sizes of the FGS entrance FP have been assumed to be 10 mm (TBC).

They represent a limitation on the space available for the fore-optics of the FCU. In order to avoid any vignetting effects on the spectrographs and on the FGS, it should be considered that the forbidden area due to the spectrographs and FGS FP increases with the height with respect to the telescope FP. As an example, in Figure 24 it is shown the projection of the spectrographs and FGS beam area at a height of 210 mm upon the telescope FP.

The three rods are located at 120 degree along a circle having a radius of 485 mm (TBC), covering a rectangle area having a size of 50 mm × 250 mm. This limitation on the available space for the FCU has been taken into account.

Actually the design of the LSS is still in a trade off process; one of the possible configuration is a Roland layout. In this case, the LSS detector should be positioned on the focal plane and should occupy a segment of 119 degree (TBC) of the available S/C OB area.

The space available for the FCU in the direction along  the telescope optical axis has been determined as follows. The distance between the PMU and the telescope FP along the optical axis is 210 mm as shown in the scheme in Figure 25. The S/C OB has a thickness of 40 mm over the telescope focal FP. Moreover, it has been assumed that the FCU will be located on a dedicated instrument OB, representing the FCU reference plane, with a thickness of 20 mm. At this work level, for the diameter of the OB it has been considered a minimum value of 1000 mm and a maximum value of 1300 mm. A layer 20 mm thick has been also considered for the optics and detectors mounting supports.

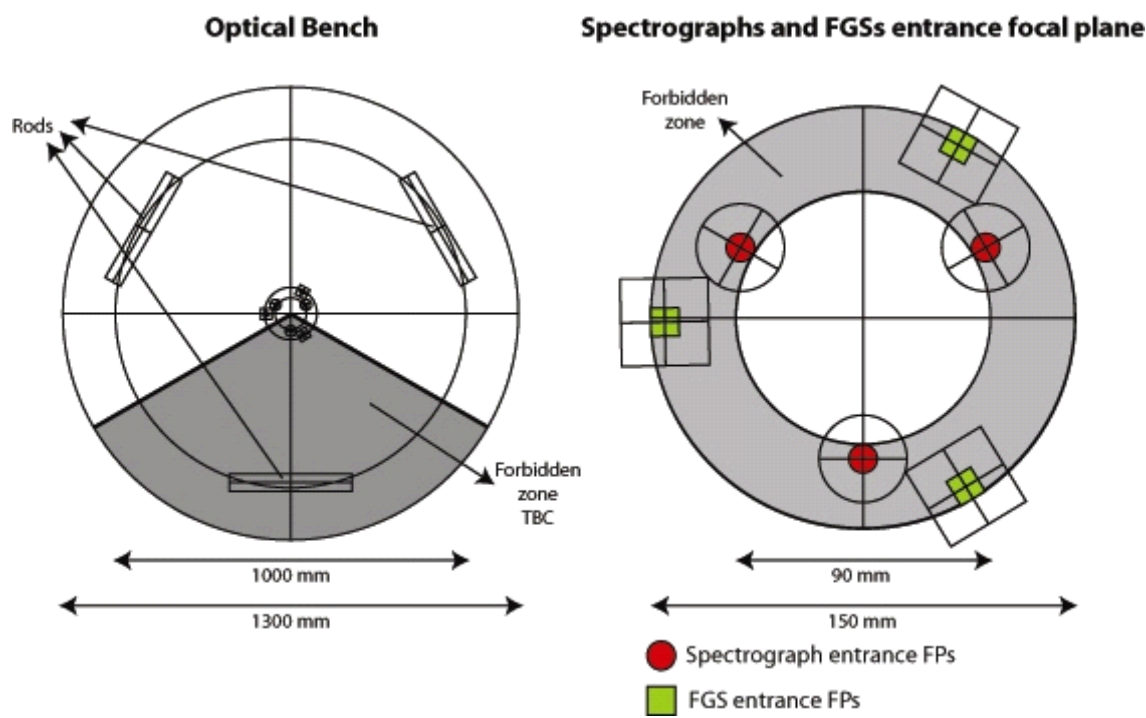

Figure 24. Available space for the FCU perpendicular to the telescope optical axis.

Tacking into account of all these terms the height from the PMU available for the FCU is 130 mm. The fore-optics is placed in such a way that the reference optical path plane lays at 65 mm from the PMU. The resulting focal extraction, i.e. the distance between the center of the fore-optics system and the FCU entrance FP, is 145 mm and then the entrance FP for each camera channels lays outside the forbidden zone represented by the spectrographs and FGSs entrance FPs.

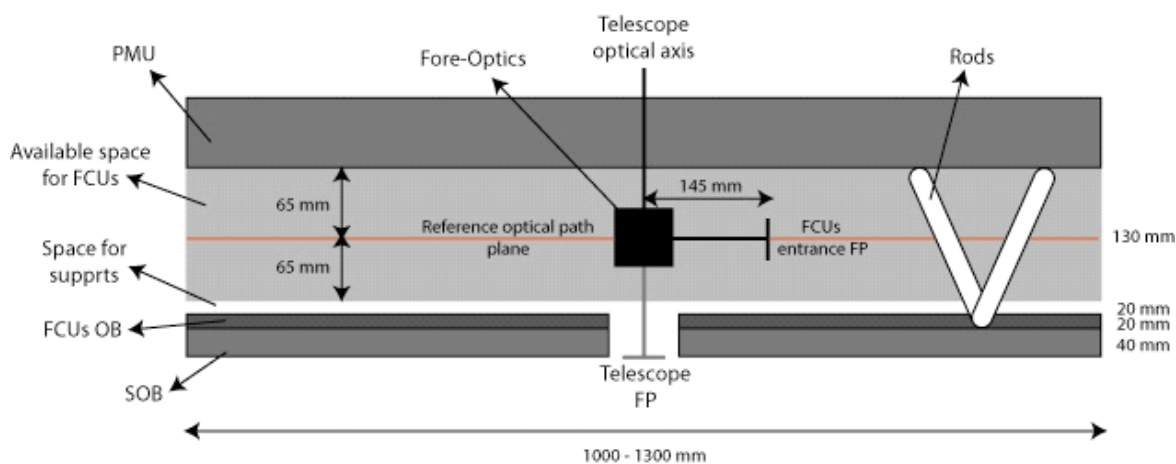

Figure 25. Available space for the FCU along the telescope optical axis.

## 2.1.2 FCU optical reference frame

For the optical design a coordinate reference frame has been used, defined as in Moisheev et al. (2006):

- the origin of the FCU optical reference frame is in the center of the telescope FOV;
- the X axis is along the telescope optical axis, orientated from the OB to toward the PMU;
- the Z axis is along a OB radius;
- the Y axis is perpendicular to the other two axis.

## 2.1.3  FUV Channel

The FUV channel covers the far ultraviolet spectral range providing medium resolution images over a wide field of view. The acquisition device is based on MCP technology.

The main FUV requirements data are given in Table 15.

Table 15. FUV requirements data.

| Parameter | Value |
|---|---|
| MCP diameter | 40 mm |
| MCP format | 2K × 2K (equivalent) |
| Equivalent pixel size | 16.5 $\mu$m |
| Instrumental PSF | 20 $\mu$m |
| Equivalent pixel sampling | 0.2 arcsec/pixel |
| Field of view diameter | 6.6 arcmin |
| Spectral range | 115 – 190 nm |

Due to the low reflective efficiency of the available coating in the FUV spectral range, the number of reflections has been minimized, introducing only the reflection of the fore-optics system.

## 2.1.4  NUV Channel

The NUV channel covers the ultraviolet spectral range providing high resolution images over a small field of view. The acquisition device is based on MCP technology.

The main NUV requirements data are given in Table 16.

Table 16. NUV requirements data.

| Parameter | Value |
|---|---|
| MCP diameter | 40 mm |
| MCP format | 2K × 2K (equivalent) |
| Equivalent pixel size | 20 $\mu$m |
| Instrumental PSF (FWHM) | 35 $\mu$m |
| Equivalent pixel sampling | 0.03 arcsec/pixel |
| Field of view diameter | 1.0 arcmin |
| Spectral range | 150 – 280 nm |

In order to achieve the highest efficiency and minimise chromatic aberrations in the whole spectral range, only fully reflective optical designs have been considered.

A pixel sampling of 0.03 arcsec/pixel has been selected tacking into account the diffraction-limited PSF at 200 nm. The telescope pointing stability and the MCP detector give actually a convolved instrumental PSF of 0.06 arcsec (FWHM). Therefore, a point-like image is sampled by 2 pixels.

If an improved telescope pointing stability and a better proximity focusing in the MCP detector will be implemented in the future, the instrumental PSF can be reduced to 0.03 arcsec and sampled by one pixel. The PSF sampling will be reduced but the spatial resolution improved.

Introducing a reflective dispersive element (a grating) to give imaging spectroscopic capabilities do not reduce the spatial resolution.

## 2.1.5 UVO Channel

The UVO channel covers the spectral range from the near ultraviolet to the visible bands, providing high-spatial resolution images over a wide field, using a CCD detector.

The main UVO requirements data are given in Table 17.

Table 17. UVO requirements data.

| Parameter | Value |
|---|---|
| CCD format | 4096 × 4096 |
| Pixel size | 15 $\mu$m |
| Pixel sampling | 0.07 arcsec/pixel |
| Field of view | 4.6 × 4.6 arcmin$^2$ |
| Field distortion | < 10% |
| Field distortion stability | < 3 mas |
| Spectral range | 200 – 700 nm |

The pixel sampling has been chosen by taking into account the diffraction limited PSF at 500 nm and the estimated telescope pointing stability (0.03 arcsec 1 sigma, Moisheev et al. 2006) over a typical exposure time, giving a convolved PSF with a FWHM of about 0.07 arcsec. Placing a single pixel on the FWHM, we maximized the FoV at the price of an under-sampled image; nevertheless, exploiting dithering techniques, full astrometric information can be recovered. The use of the dithering techniques translates on the optical design as the requirement to maximize the PSF optical quality as much as possible (goal = diffraction limited) independently by the PSF sampling.

It should be underlined that if a better telescope pointing stability will be implemented in the future, the pixel sampling can be reduced (reducing the FoV) to improve a better recovery of the astrometric information also at shorter wavelengths.

While the required field distortion is relaxed, its stability represents a challenge not only for the entire instrument thermal stability but also for the precision of the placement of any movable optic. In particular, the re-positioning accuracy of any rotating mirror mechanism must meet the field distortion tolerance.

In order to minimize the chromatic aberrations over the whole spectral range and to allow a possible extension to higher wavelengths maintaining the same optical quality, only full reflective optical designs have been considered.

## 2.2 Fore-Optics

The fore-optics system task is to pick up the optical beam from the telescope focal plane and to redirect it into each channels focal plane. The available space for the fore-optics is limited by the position of the spectrographs entrance slits and FGS focal planes, i.e. inside a circle of radius 50 mm around the telescope optical axis.

In order to illuminate the three FCU channels, two possible fore-optics solutions have been considered:

1. Single rotating pick-up mirror
2. Three mosaic pick-up mirrors

## 2.2.1 Rotating pick-up mirror

In this configuration a single mirror is placed at the center of the telescope optical axis at 45 degree with respect to the OB and deviates the optical beam towards a camera channel at a time rotating along the telescope optical axis. A scheme of the system is shown in Figure 26.

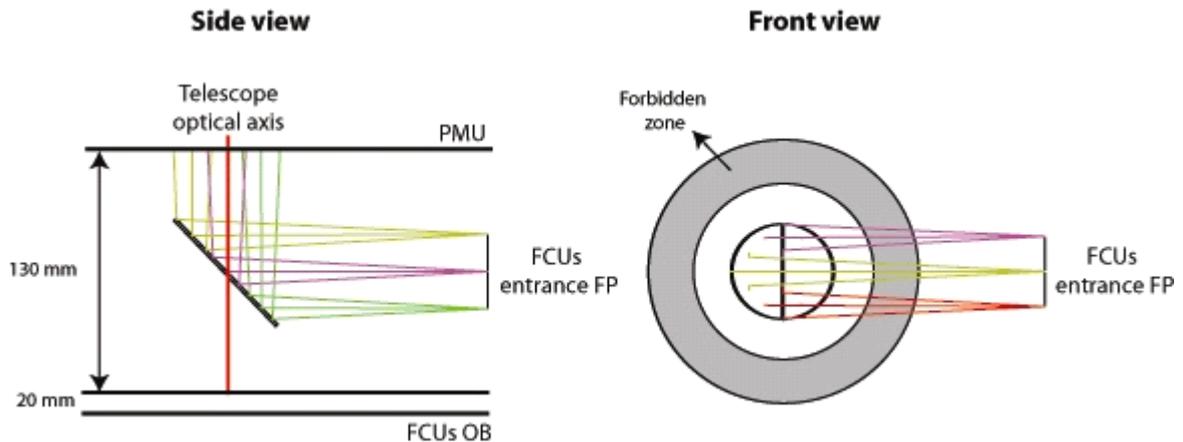

Figure 26. Central rotating mirror.

The mirror height with respect to the FCU OB is 85 mm (65mm+20mm). The mirror dimensions are determined by the FUV FoV, which has the greatest linear dimensions among the three channels. It has an elliptical shape with a major axis of 80 mm and a minor axis of 58 mm. The entrance FP lays at 145 mm from the optical axis.

In this configuration it is possible to exploit the central part of the telescope FoV for each camera channels, i.e. the less aberrated part, simplifying the optical design of the camera channels. It allows a great flexibility in the disposition of the three channels over the OB. The FoV of the three channels are co-centered so that it is possible to observe the same FoV in the all the available wavelength bands by only rotating the mirror (no telescope re-pointing is required). Of course, no parallel observations of different camera channels are allowed.

In order to simplify alignment procedures (the mirror has to rotate), no power or aspheric terms have been added to the pick-up mirror. The precision of the repositioning of the pickup mirror is actually determined by the requirement on the field distortion stability of the UVO channel.

## 2.2.2 Mosaic pick-up mirrors

In this configuration each channel has its own dedicated pick-up mirror. The three mirrors are located near the telescope optical axis at 45 degree with respect to the OB and re-directs the optical beam towards the correspondent channel entrance FP.

Being dedicated to each channel, the coating and the optical design of the various pick-up mirrors can be optimized.

On the other side in this configuration each channel uses a part of the telescope FOV that is off axis, i.e. more aberrated, making more difficult to obtain the desired optical quality. Moreover the crowding of the FP could make difficult (once frozen the positions of the FGS and the spectrographs slits) to accommodate all the mirrors.

Parallel observations of different camera channels are allowed, but the observation of the same field require a re-pointing of the telescope.

Three possible solutions for the disposition of the three mirrors have been investigated (see Figure 27, Figure 28 and Figure 29).

In each configuration the mirrors are placed leaving a security border of 2 mm to avoid mutual vignetting effects.

The position of the FUV pick-up mirror is the same in all the configurations, being the only one possible given the size of the FUV FoV, and occupies two of the quadrants of the telescope FP.

The three configurations are different for the positioning of the UVO and the NUV mirrors: in configuration 1 each of the two channels is positioned in one of the quadrant of the focal plane not occupied by the FUV mirror; configuration 2 is such that the centers of NUV and UVO pick-up mirrors have the same distance from the focal axis of the telescope, in configuration 3 the external border of the fields of the two channels has the same distance from the telescope axis.

These configurations have been investigated being representative of all the possible off axis configurations; the final mirror positions (if the solution with mosaic pick-up mirror will be chosen) will be fixed through a trade-off process between the channels and the positions of the FGS and the spectrographs slits.

The positions of the centers of the pick up mirrors in the mosaic configuration are given in Table 18.

Table 18. Positions of pick up mirrors for the mosaic configurations

| Configuration | Channel | Mirror Center (y,z)(mm) | Mirror Center (y,z) (deg) |
|---|---|---|---|
| 1 | UVO | (- 22.07,- 22.07) | (-0.0744, -0.0744) |
| | NUV | (-12.23,12.23) | (-0.0412,0.0412) |
| | FUV | (30.68,0) | (0.1034,0) |
| 2 | UVO | (-22.07, -17.15) | (-0.0744, -0.0578) |
| | NUV | (-12.23, 17.15) | (-0.0412,0.0578) |
| | FUV | (30.68,0) | (0.1034,0) |
| 3 | UVO | (-22.07, -12.48) | (-0.0744,-0.0421) |
| | NUV | (-12.23, 21.83) | (-0.0412,0.0735) |
| | FUV | (30.68,0) | (0.1034,0) |

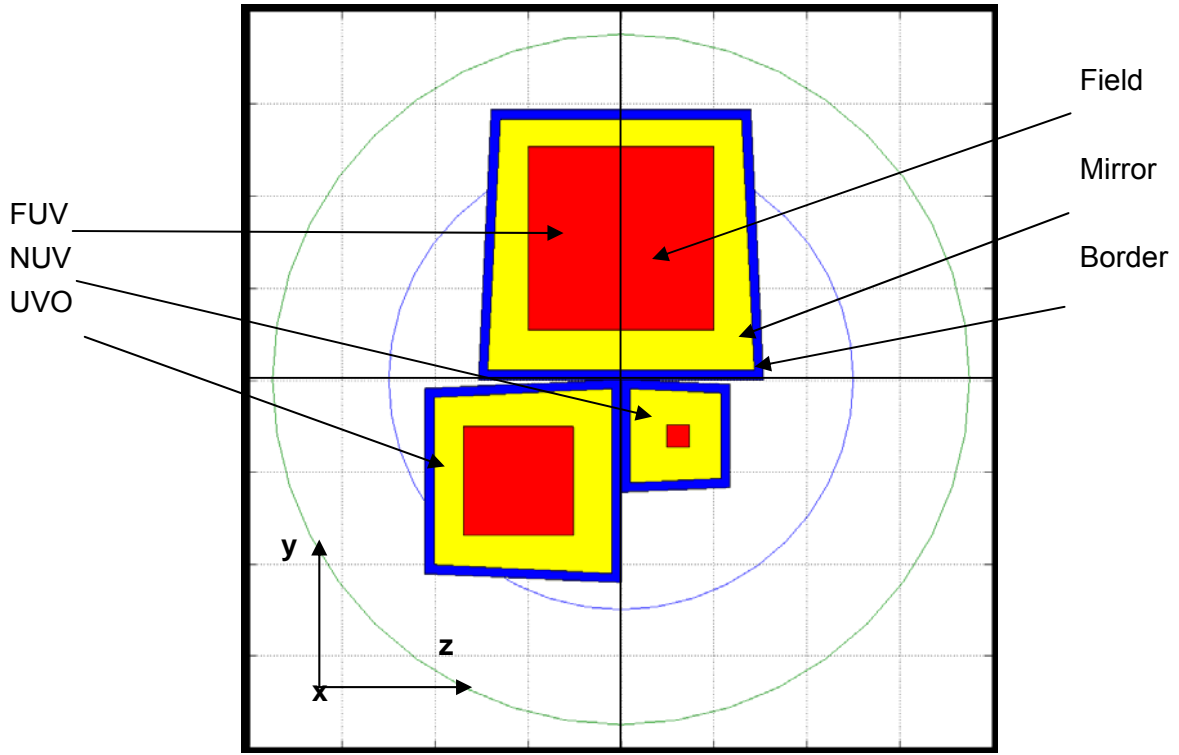

Figure 27. Mosaic configuration 1.

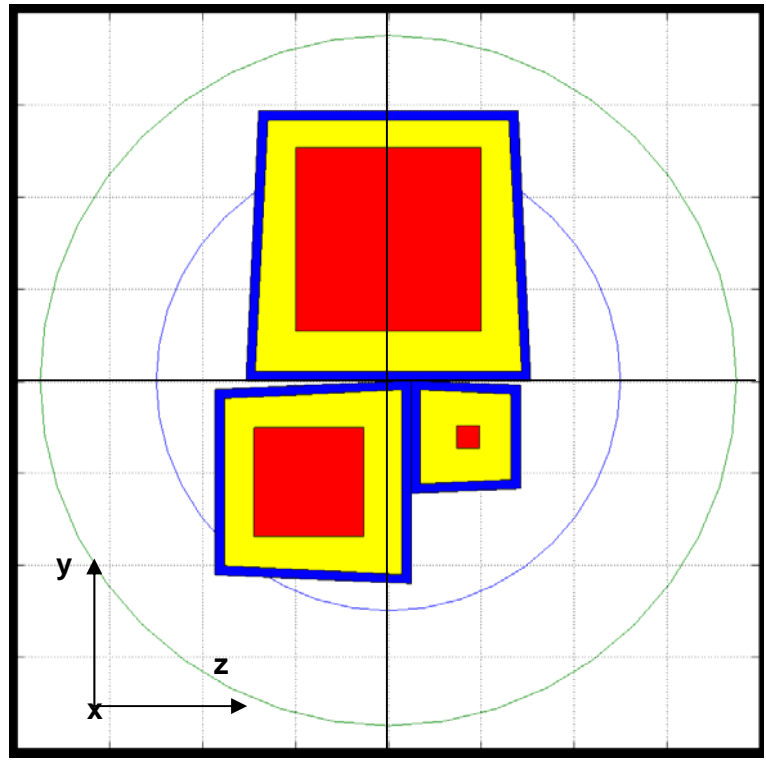

Figure 28. Mosaic configuration 2.

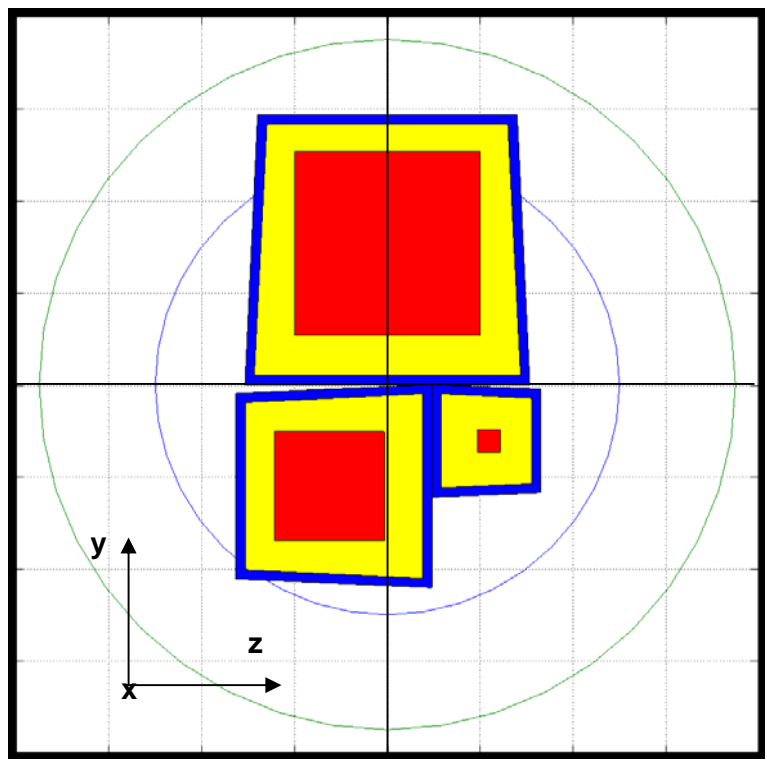

Figure 29. Mosaic configuration 3.

### 2.2.3 Reported results

In this document will be reported the results for the mosaic configuration 1 only, being the results similar for all the three configurations for all the channels.

Besides, all optical design are reported placing the entrance FP along the Z axis; this is obviously not a constraint, given the rotational symmetry of the telescope FP. The relative positioning of the channels (and a possible reflection of 180° along the Z axis of the entire configuration of a channel) will be decided on a basis of an iterative process with the mechanical design.

## 2.3 Far-UV Channel

### 2.3.1 On-Axis Layout (rotating pick-up mirror)

In this configuration the FUV channel is composed by a single rotating fold mirror deviating the optical beam towards the detector. The optical layout is shown in Figure 30.

The optical design consists only on a fold pick-up mirror. No power or aspheric terms have been added to the pick-up mirror in order to allow a precise alignment and re-positioning of the mirror itself.

The main parameters for all the mirrors and FP are given in Table 19.

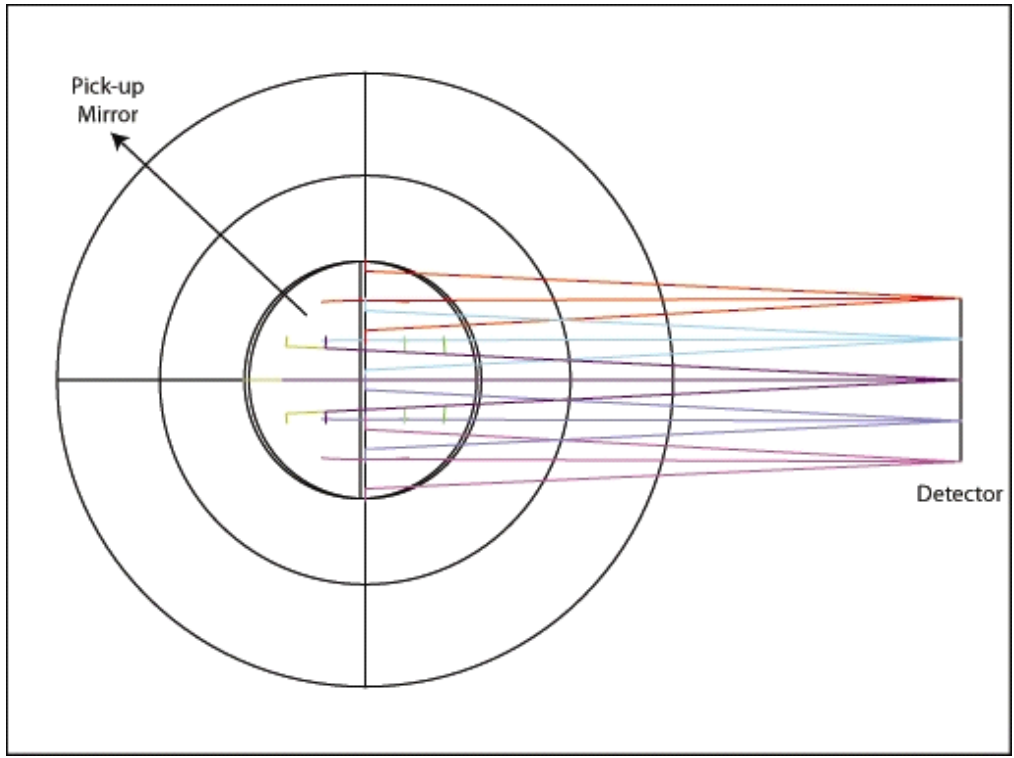

Figure 30. Optical layout of the FUV channel for the rotating pick-up mirror configuration.

Table 19. Main parameters of the FUV channel mirrors for the rotating pick-up mirror configuration.

|         | R<br>(mm) | K | Shape      | Size<br>(mm×mm) | Vertex-Center<br>(mm) |
|---------|-----------|---|------------|-----------------|------------------------|
| Pick-up | plane     | 0 | elliptical | 58×80           | 0                      |
| FP      | plane     | 0 | circular   | 40              | 0                      |

The location of the mirrors with respect to the entrance FP are given in Table 20.

Table 20. Global vertex in the coordinate system linked to FCU optical reference frame of the FUV channel mirrors for the rotating pick-up mirror configuration.

|         | Tilt y<br>(degree) | Tilt z<br>(degree) | Tilt x<br>(degree) | y<br>(mm) | z<br>(mm) | x<br>(mm) |
|---------|--------------------|--------------------|--------------------|-----------|-----------|-----------|
| Pick-up | -45                | 0                  | 0                  | 0         | 0         | 145       |
| FP      | 0                  | 0                  | 0                  | 0         | 145       | 145       |

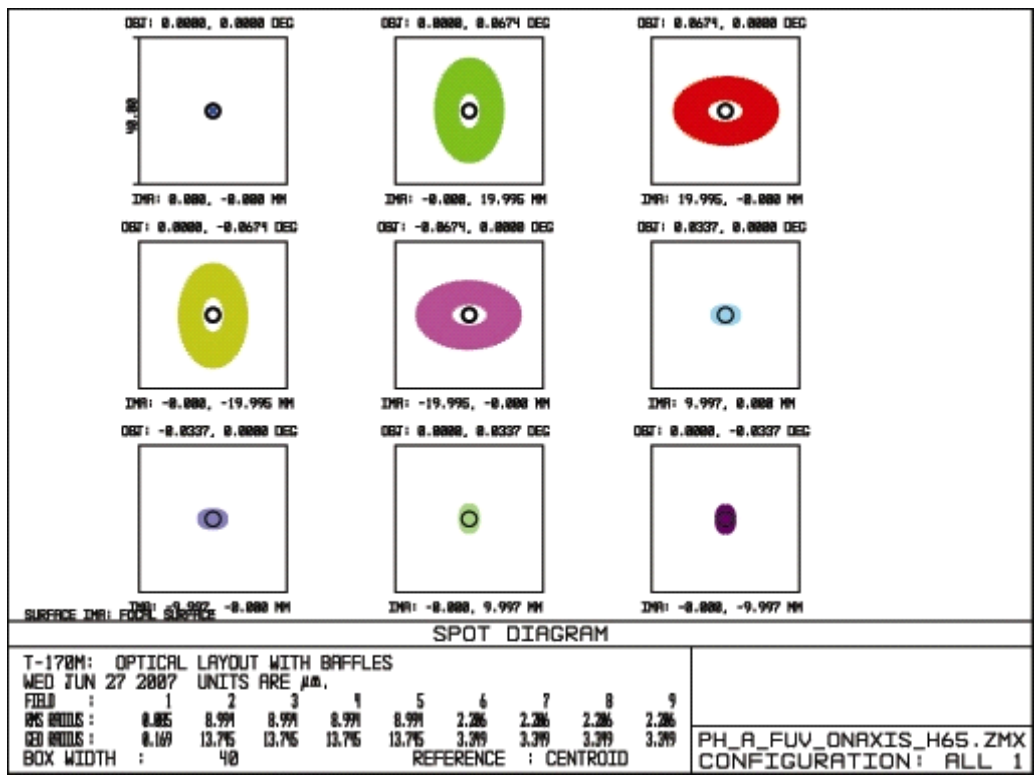

Figure 31. Spot diagrams of the FUV channel for the rotating pick-up mirror configuration.

The spot diagrams are shown in Figure 31. The spots image are given for a FoV diameter of 40 mm in the FP. The square boxes have the dimensions of 40×40 µm². The two circles have the dimension of the Airy disk at 130nm (diameter 1.59 µm) and at 165nm (diameter 2.01 µm).

The distortion grid is shown in Figure 32.

The rate of distortion is defined as the ratio $(r_{real} - r_{predicted})/ r_{predicted}$ , r being the distance of the chief ray intersection on the FP to the optical axis. The maximum distortion is less than 0.001 % over the whole FoV.

The polychromatic encircled energy is shown in Figure 33. For each fields of view, it is always less than 10.6 µm.

The Strehl ratios for all the wavelengths over the whole field of view are shown in Figure 34. The optical system has diffraction limited quality (SR>80) only within ±1.8 arcmin.

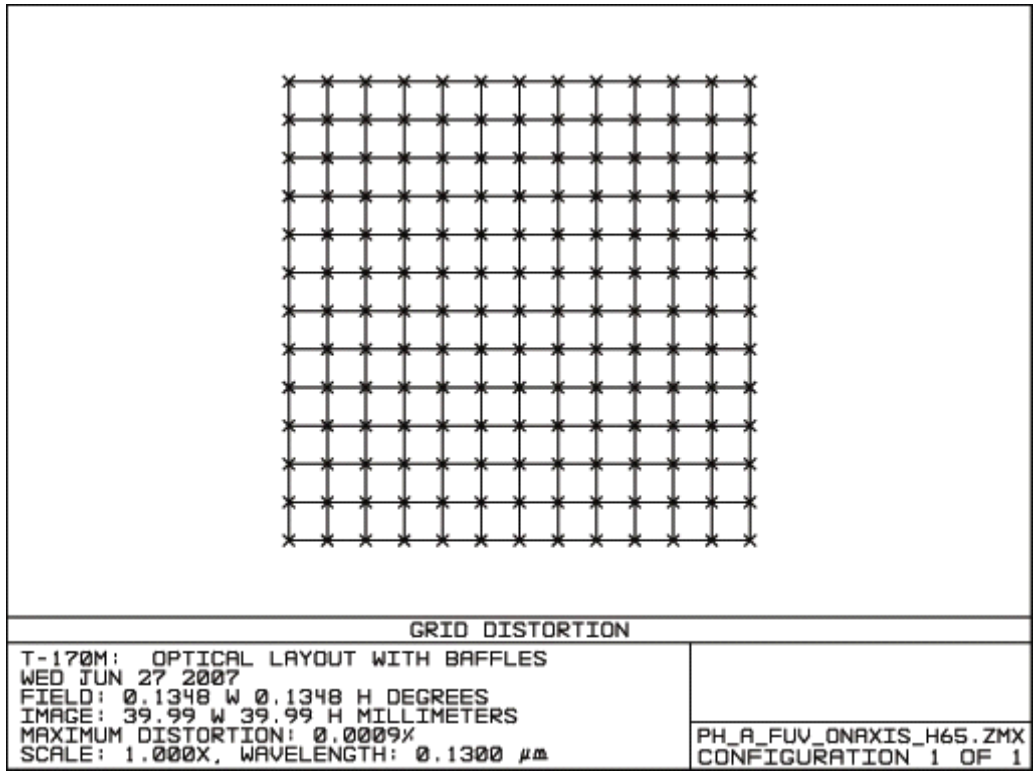

Figure 32. Distortion Grid of the FUV channel for the rotating pick-up mirror configuration.

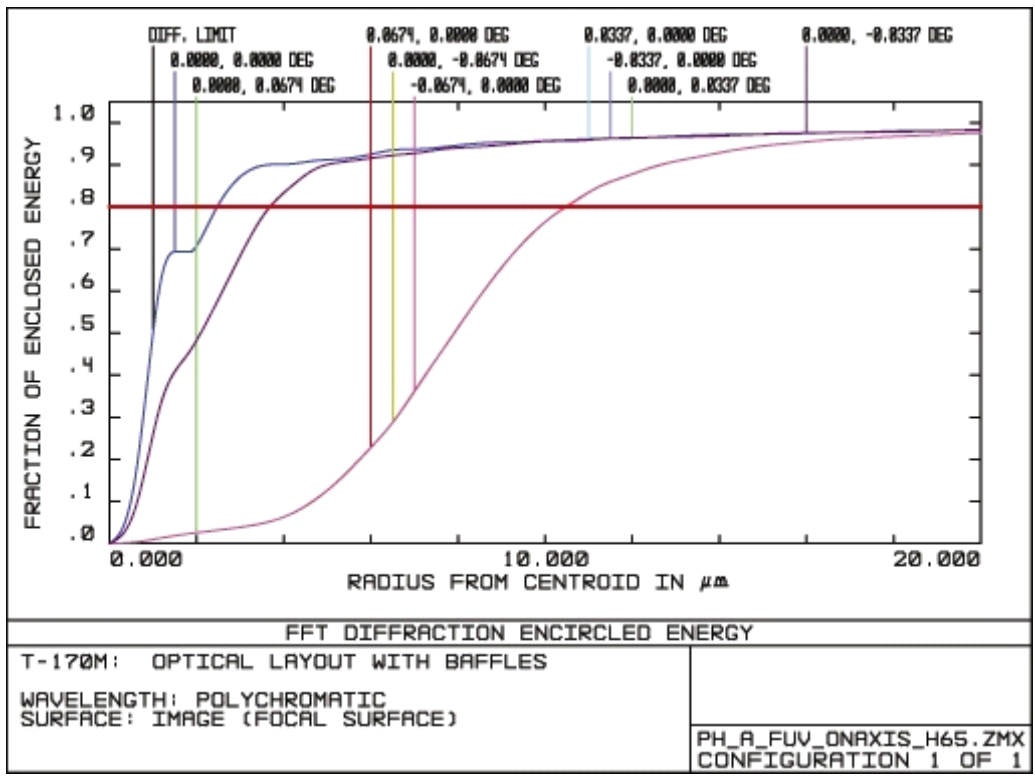

Figure 33. Encircled energy of the FUV channel for the rotating pick-up mirror configuration.

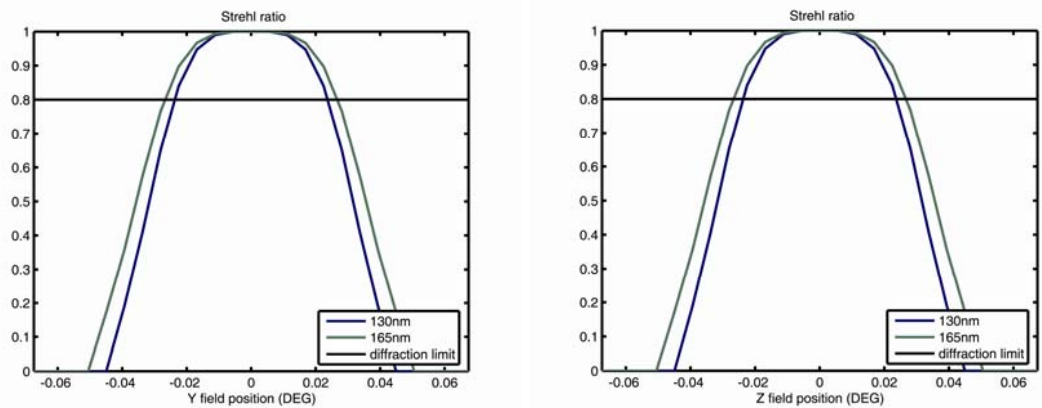

Figure 34. Strehl ratio of the FUV channel for the rotating pick-up mirror configuration.

## 2.3.2 Off-Axis Layout (mosaic pick-up mirror)

In this configuration the FUV channel is composed of a single aspheric mirror deviating the optical beam towards the detector. The pick-up mirror is de-centered with respect to the telescope optical axis as reported in Table 18 (configuration 1).

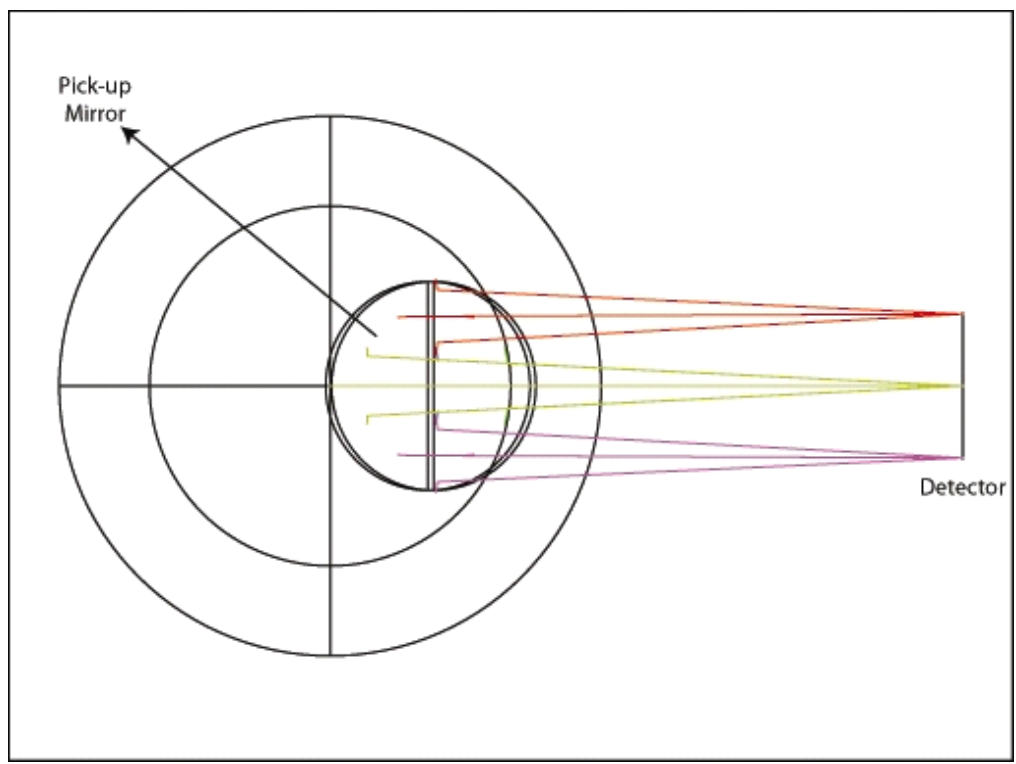

Figure 35. Optical layout of the FUV channel for the mosaic pick-up mirror configuration.

The main parameters for all the mirrors and FP are given in Table 21.

The spot diagrams are shown in Figure 36. The spots image are given for a FoV diameter of 40 mm in the FP. The square boxes have the dimensions of 40×40 $\mu m^2$. The two circles have the dimension of the Airy disk at 130nm (diameter 1.59 $\mu m$) and at 165nm (diameter 2.01 $\mu m$).

Table 21. Main parameters of the FUV channel mirrors for the mosaic pick-up mirror configuration.

| | R (mm) | K | $A_2$ | $A_4$ | $A_6$ | $A_8$ | Shape | Size (mm×mm) | Vertex-Center (mm) |
|---|---|---|---|---|---|---|---|---|---|
| Pick-up | plane | 0 | $-3.2726836\ 10^{-6}$ | $2.051402\ 10^{-9}$ | $-1.884329\ 10^{-13}$ | $7.051445\ 10^{-18}$ | elliptical | 58×80 | 21.537 |
| FP | plane | 0 | 0 | 0 | 0 | 0 | circular | 40 | 0 |

The location of the mirrors with respect to the entrance FP are given in Table 22.

Table 22. Global vertex in the coordinate system linked to FCU optical reference frame of the FUV channel mirrors for the mosaic pick-up mirror configuration.

| | Tilt y (degree) | Tilt z (degree) | Tilt x (degree) | y (mm) | z (mm) | x (mm) |
|---|---|---|---|---|---|---|
| Pick-up | -45 | 0 | 0 | 0 | 29.6895 | 145 |
| FP | 0 | 0 | 0 | 0 | 175.089 | 145 |

The distortion grid is shown in Figure 37.

The rate of distortion is defined as the ratio $(r_{real} - r_{predicted})/\ r_{predicted}$ , r being the distance of the chief ray intersection on the FP to the optical axis. The maximum distortion is less than 0.4 % over the whole FoV.

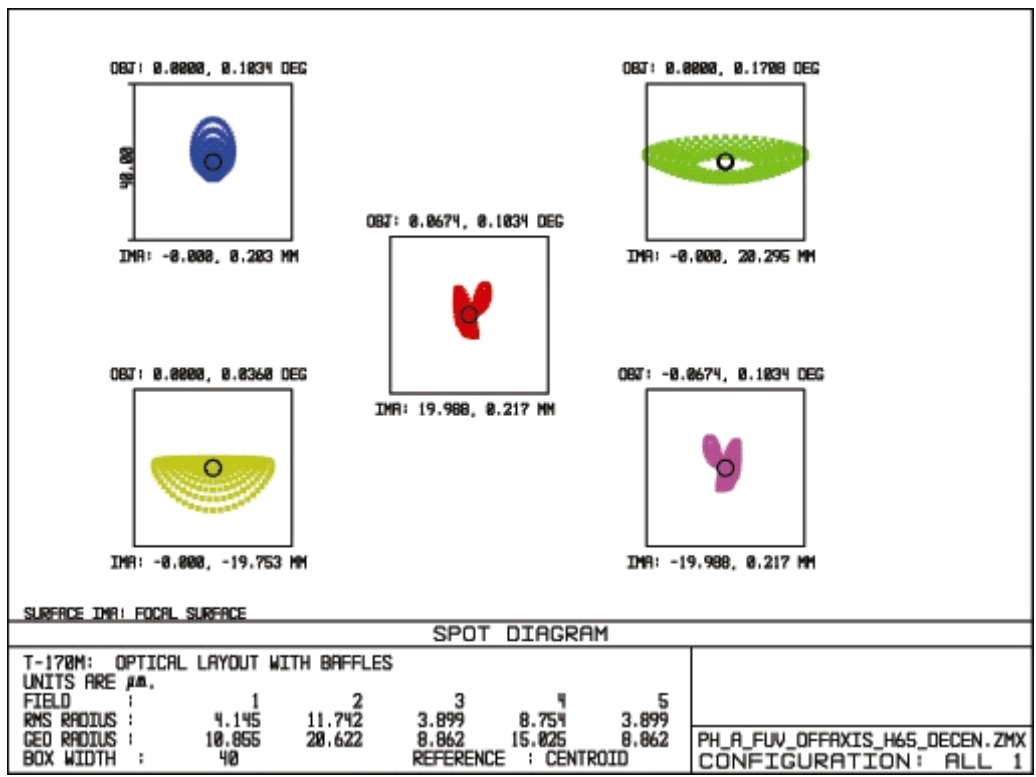

Figure 36. Spot diagrams of the FUV channel for the mosaic pick-up mirror configuration.

The polychromatic encircled energy is shown in Figure 38. For each fields of view, it is always less than 14.6 μm.

The Strehl ratios for all the wavelengths over the whole field of view are shown in Figure 39. The optical system has no diffraction limited quality (SR>80).

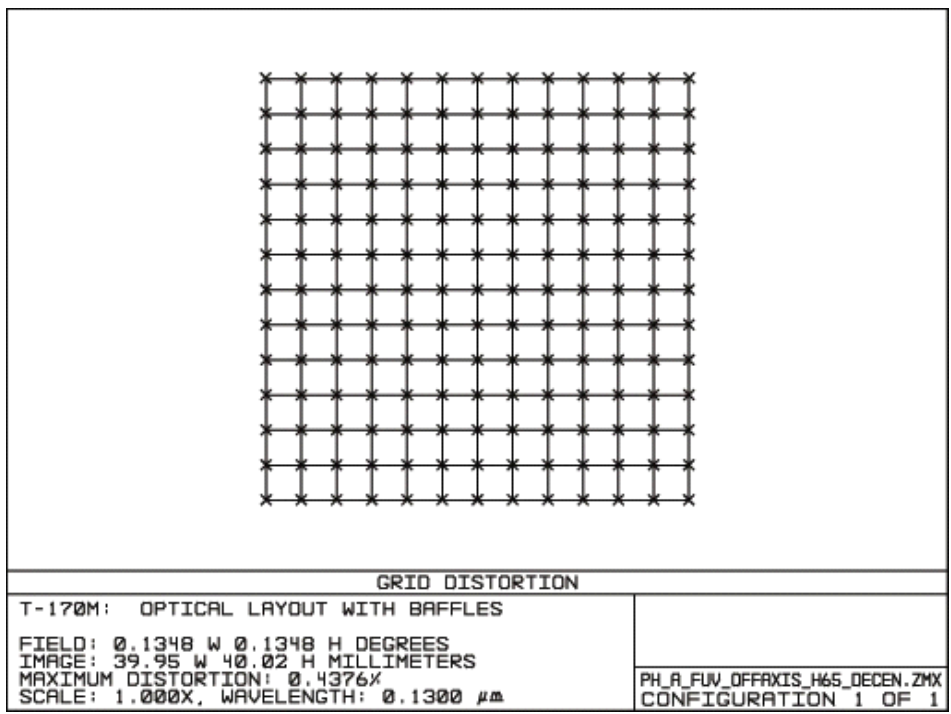

Figure 37. Distortion Grid of the FUV channel for the mosaic pick-up mirror configuration.

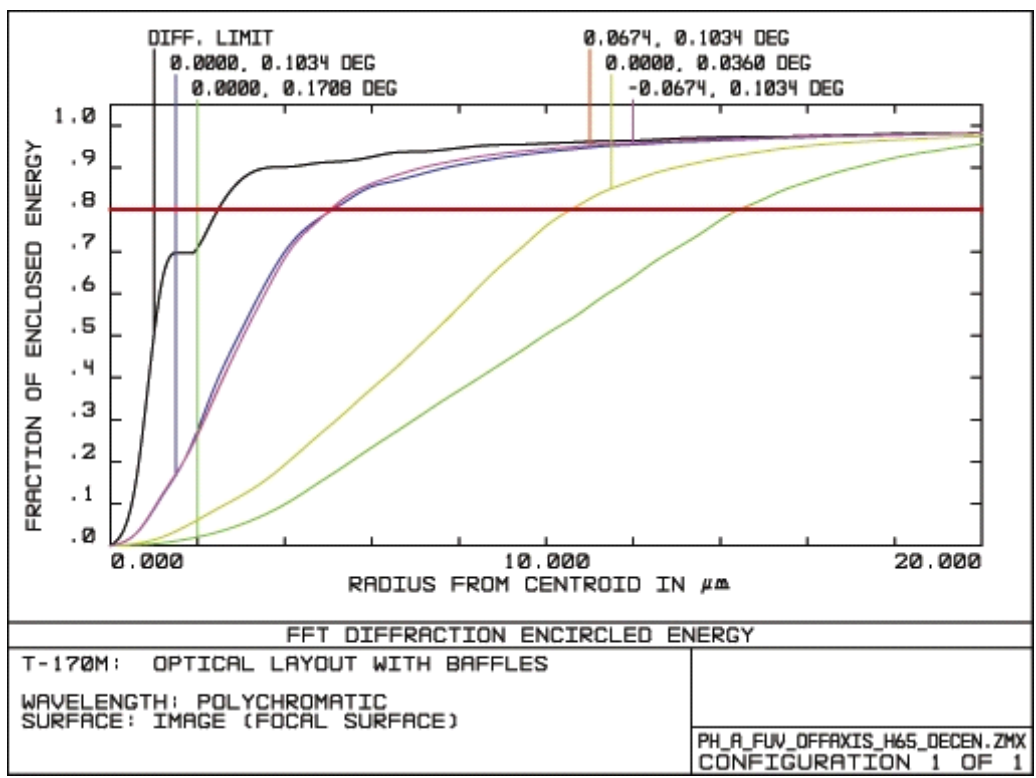

Figure 38. Encircled energy of the FUV channel for the mosaic pick-up mirror configuration.

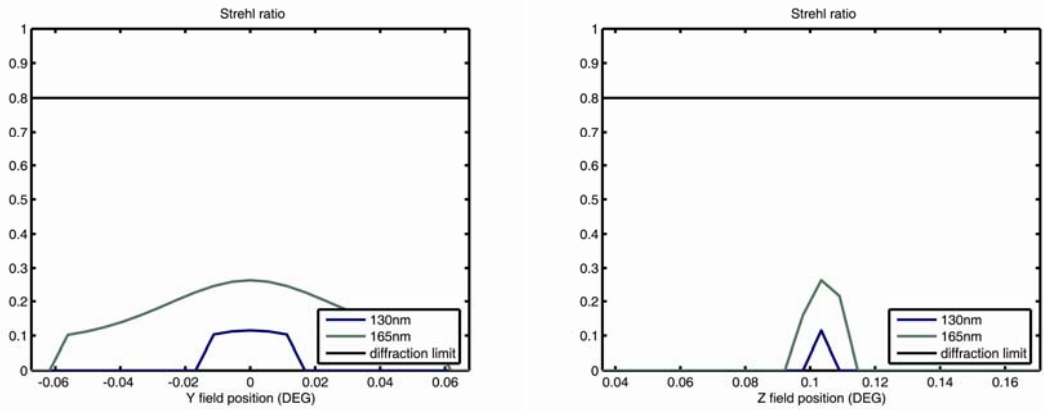

Figure 39. Strehl ratio of the FUV channel for the mosaic pick-up mirror configuration.

## 2.4    Near-UV Channel

### 2.4.1  On-Axis Layout (rotating pick-up mirror)

In this configuration the NUV channel is composed of a rotating fold mirror deviating the optical beam towards a two mirrors re-imaging system. The optical layout is shown in Figure 40.

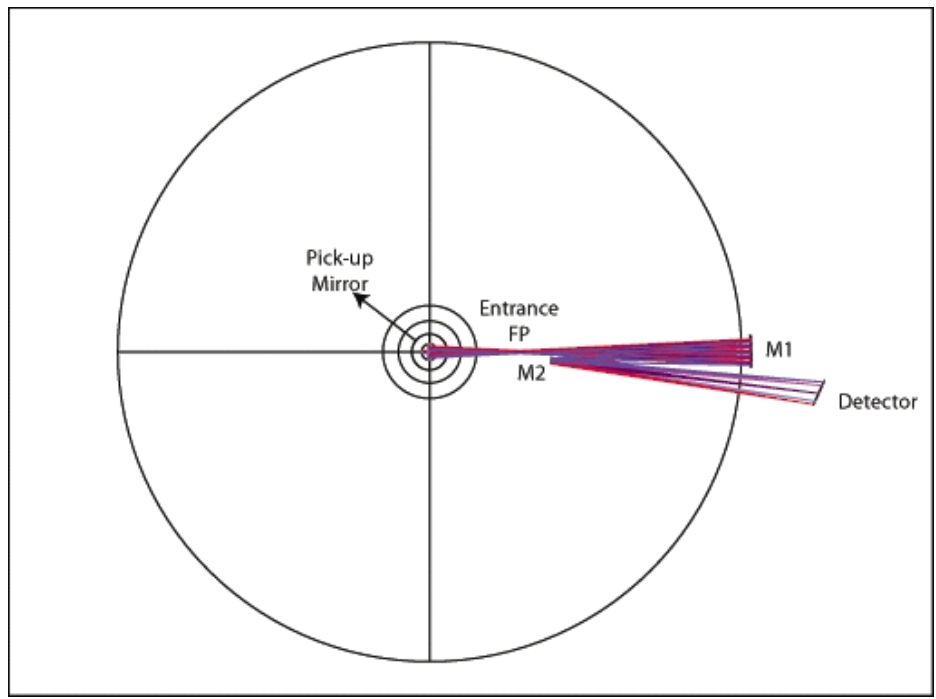

Figure 40. Optical layout of the NUV channel for the rotating pick-up mirror configuration.

The optical design characteristics are the following:

   Pick-up Mirror: flat mirror

M1: elliptical concave mirror

M2: elliptical convex mirror

The M1 and M2 optical axis are slightly tilted one with respect to the other. All the mirrors and the detector lay inside the OB having a diameter value of 1300 mm.

The pixel FoV is 0.03 arcsec/pixel. Having an equivalent pixel size of 20 μm and a detector diameter of 40 mm, the FoV diameter is 1.00 arcmin. The NUV channel magnification is 8.1.

The filter wheels are placed nearby the entrance FP, being the accessible area where the beam has the smallest dimension. The exact positioning of the filter wheels will be fixed when the opto-mechanical design will be completed.

The main parameters for all the mirrors and FP are given inTable 23.

Table 23. Main parameters of the NUV channel mirrors for the rotating pick-up mirror configuration.

|  | R (mm) | K | Shape | Size (mm×mm) | Vertex-Center (mm) |
|---|---|---|---|---|---|
| Pick-up | plane | 0 | elliptical | 58×80 | 0 |
| Entrance FP | plane | 0 | circular | 4.95 | 0 |
| M1 | 372.059 | 0.080667 | circular | 46 | 9 |
| M2 | 123.763 | 0.821435 | circular | 10 | -6.2 |
| FP | plane | 0 | circular | 40 | 0 |

The location of the mirrors with respect to the FCU optical reference frame are given in Table 24.

Table 24. Global vertex in the coordinate system linked to FCU optical reference frame of the NUV channel mirrors for the rotating pick-up mirror configuration.

|  | Tilt y (degree) | Tilt z (degree) | Tilt x (degree) | y (mm) | z (mm) | x (mm) |
|---|---|---|---|---|---|---|
| Pick-up | 45 | 0 | 0 | 0 | 0 | 145 |
| Entrance FP | 0 | 0 | 0 | 0 | 145 | 145 |
| M1 | 0 | 0 | 0 | 0 | 514.898 | 145 |
| M2 | 0 | 0 | 0.8 | -14.968 | 195.113 | 145 |
| FP | 0 | 0 | -24.2 | -66.749 | 623.610 | 145 |

The spot diagrams are shown in Figure 41. The spots image are given for a FoV diameter of 40 mm in the FP. The square boxes have dimensions of 4×4 pixels. The two circles have the dimension of the Airy disk at 150 nm (diameter 29.6 μm) and at 280 nm (diameter 55.28 μm).

The distortion grid is shown in Figure 42. The rate of distortion is defined as the ratio ($r_{real}$ − $r_{predicted}$)/ $r_{predicted}$ , r being the distance of the chief ray intersection on the FP to the optical axis. The maximum distortion is less than 1.0 % over the whole FoV.

The polychromatic encircled energy is shown in Figure 43. For each fields of view, it is always less than 33 μm.

The Strehl ratios for all the wavelengths over the whole field of view are shown in Figure 44. The optical system has diffraction limited quality (SR>80).

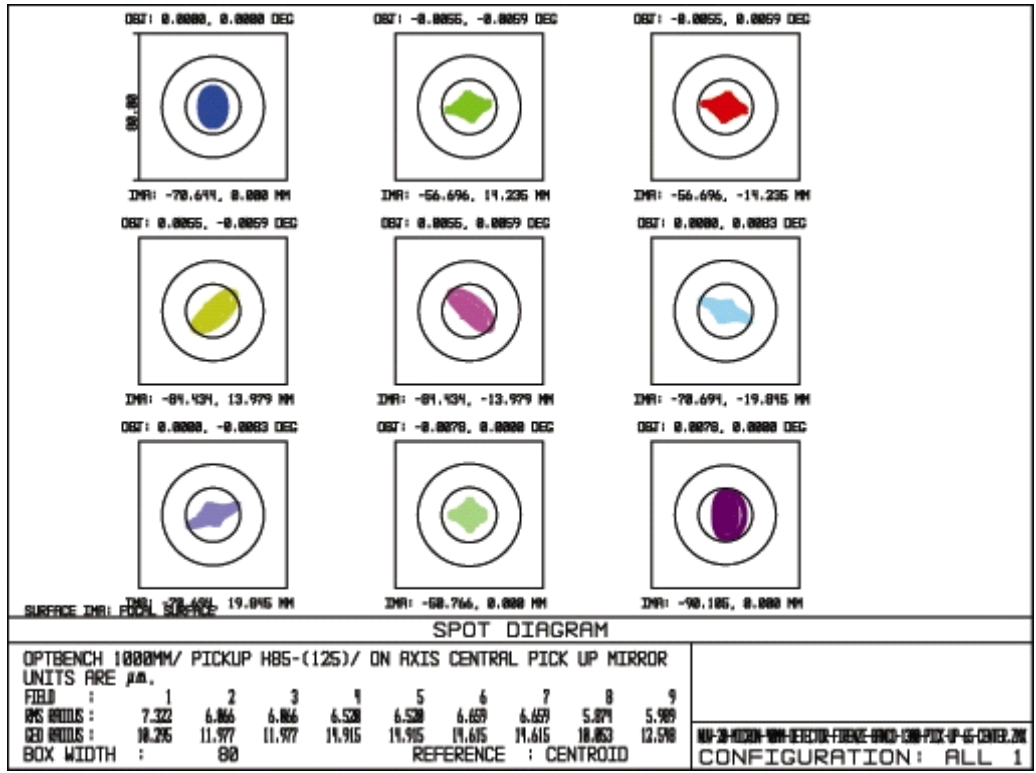

Figure 41. Spot diagrams of the NUV channel for the rotating pick-up mirror configuration.

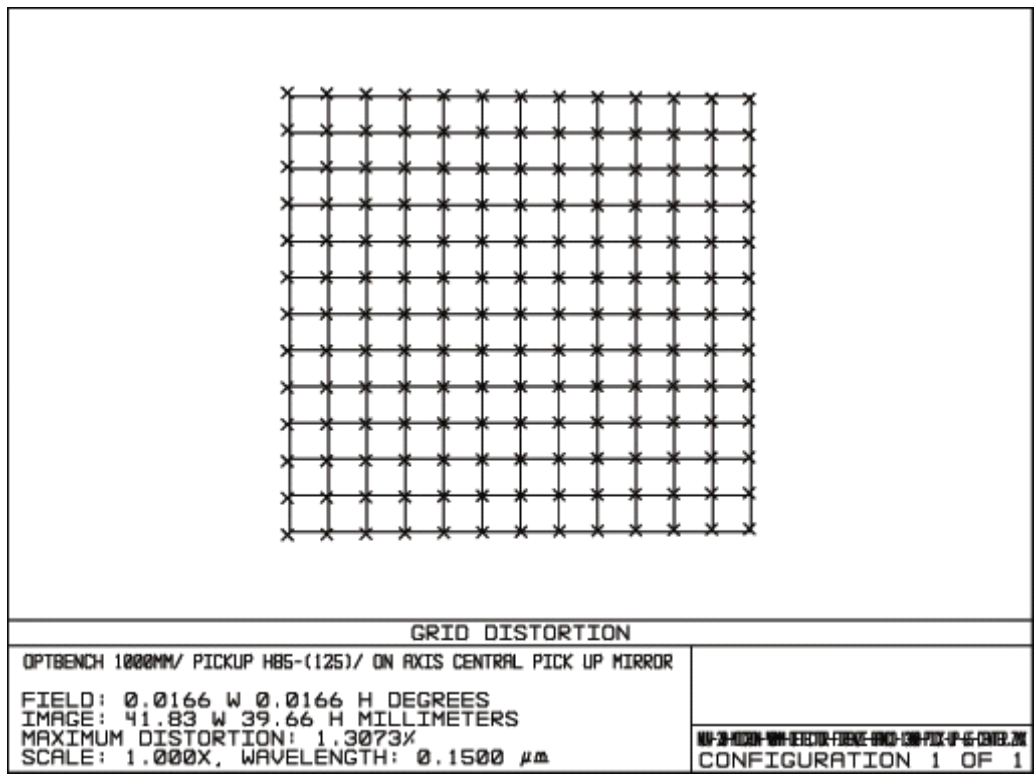

Figure 42. Distortion Grid of the NUV channel for the rotating pick-up mirror configuration.

FCU phase A report – Optical, mechanical and electronics configurations

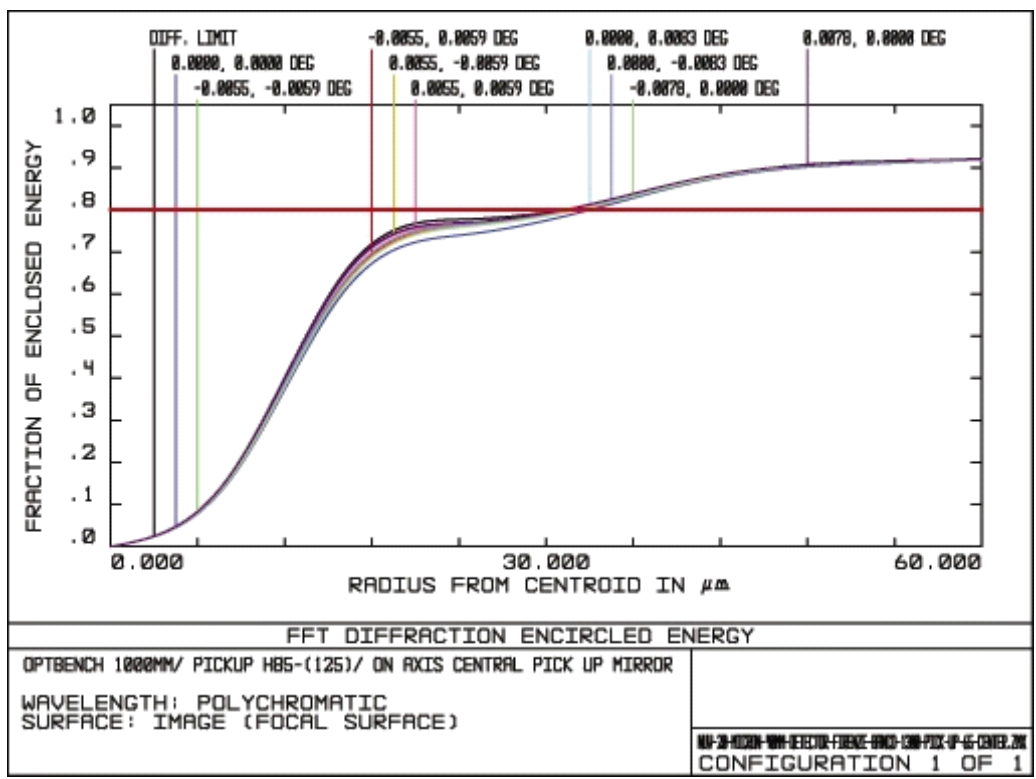

Figure 43. Encircled energy of the NUV channel for the rotating pick-up mirror configuration.

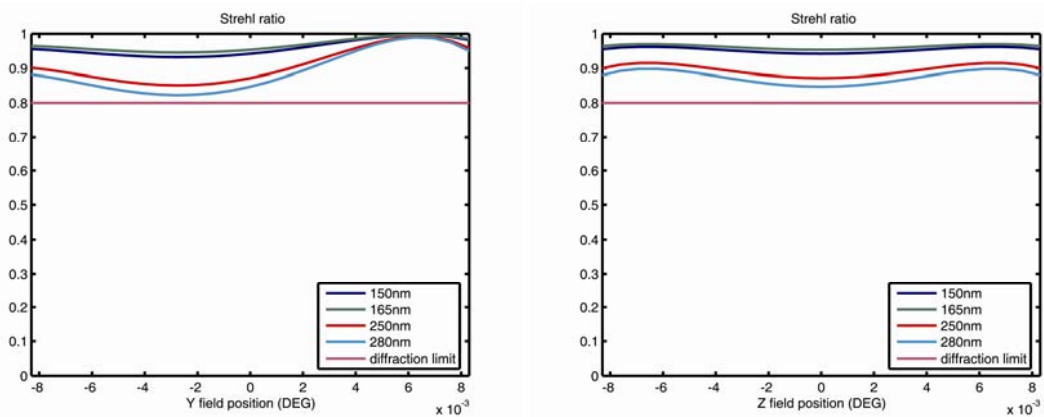

Figure 44. Strehl ratio of the NUV channel for the rotating pick-up mirror configuration.

## 2.4.2  Off-Axis Layout (mosaic pick-up mirror)

In this configuration the NUV channel is composed of a single fold mirror deviating the optical beam towards a two mirrors re-imaging system with an addition of a fold mirror. The pick-up mirror is de-centered  with respect to the telescope optical axis as reported in Table 18 (configuration 1).

The optical layout is shown in Figure 45.

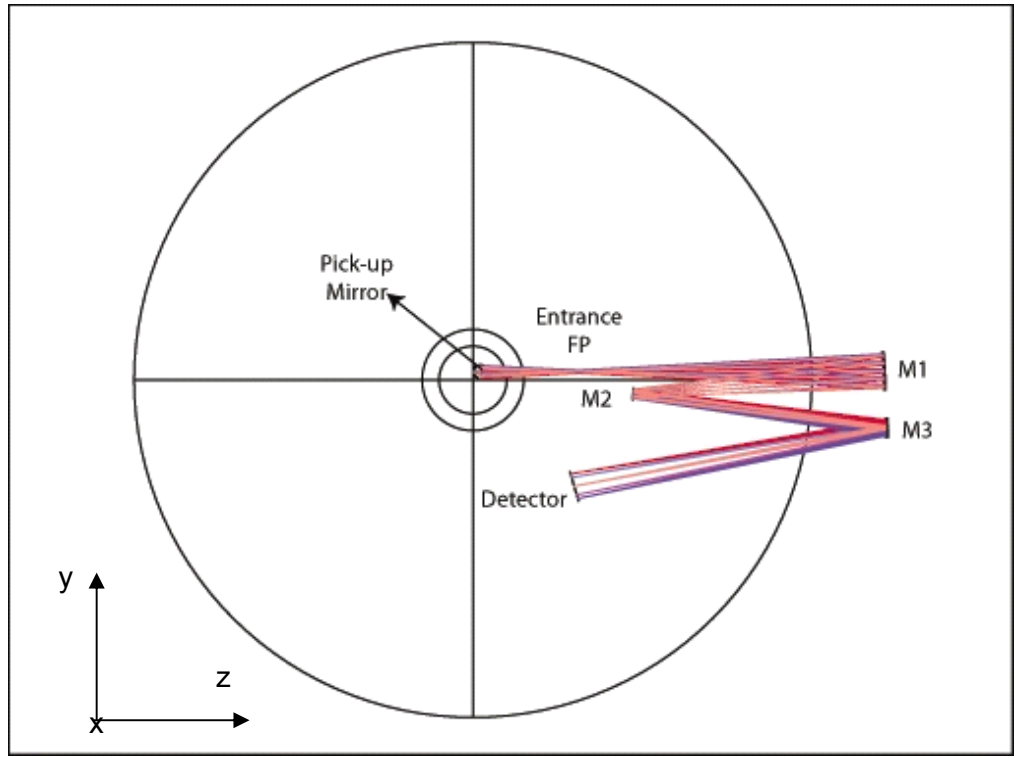

Figure 45. Optical layout of the NUV channel for the mosaic pick-up mirror configuration.

The optical design characteristics are the following:

Pick-up Mirror: fold mirror

M1: elliptical concave mirror

M2: elliptical convex mirror

M3: fold mirror

All the mirrors and the detector lay inside the OB having a diameter value of 1300 mm.

The pixel FoV is 0.03 arcsec/pixel. Having an equivalent pixel size of 20 μm and a detector diameter of 40 mm, the FoV diameter is 1.00 arcmin. The NUV channel magnification is 8.1.

The filter wheels are placed nearby the entrance FP, being the accessible area where the beam has the smallest dimension. The exact positioning of the filter wheels will be fixed when the opto-mechanical design will be completed.

The main parameters for all the mirrors and FPs are given in Table 25.

Table 25. Main parameters of the NUV channel mirrors for the mosaic pick-up mirror configuration.

| | R (mm) | K | $A_2$ | $A_4$ | $A_6$ | $A_8$ | Shape | Size (mm×mm) | Vertex-Center (mm) |
|---|---|---|---|---|---|---|---|---|---|
| **Pick-up** | plane | 0 | 0 | 0 | 0 | 0 | elliptical | 22×30 | 0 |
| **Entrance FP** | plane | 0 | 0 | 0 | 0 | 0 | circular | 4.95 | 0 |
| **M1** | 465.570 | 0.014562 | 0 | 0 | 0 | 0 | circular | 56 | 22 |
| **M2** | 259.500 | 3.635808 | 0 | 0 | 0 | 0 | circular | 16 | -12 |
| **M3** | plane | 0 | 0 | 0 | 0 | 0 | circular | 28 | 0 |
| **FP** | plane | 0 | 0 | 0 | 0 | 0 | circular | 40 | 0 |

The location of the mirrors with respect to the FCU optical reference frame are given in Table 26.

Table 26. Global vertex in the coordinate system linked to FCU optical reference frame of the NUV channel mirrors for the mosaic pick-up mirror configuration.

|  | Tilt y (degree) | Tilt z (degree) | Tilt x (degree) | y (mm) | z (mm) | x (mm) |
|---|---|---|---|---|---|---|
| **Pick-up** | -45 | 0 | 0 | 11.100 | 11.100 | 145 |
| **Entrance FP** | 0 | 0 | 0 | 11.470 | 156.832 | 145 |
| **M1** | 0 | 0 | 0 | 12.609 | 606.344 | 145 |
| **M2** | 0 | 0 | 1.5 | -20.375 | 236.875 | 145 |
| **M3** | 0 | 0 | 1.5 | -67.828 | 608.508 | 145 |
| **FP** | 0 | 0 | 16.5 | -151.396 | 147.593 | 145 |

The spot diagrams are shown in Figure 46. The spots image are given for a FoV diameter of 40 mm in the FP. The square boxes have the dimensions of 4×4 pixels. The two circles have the dimension of the Airy disk at 150nm (diameter 29.6 µm) and at 280nm (diameter 55.28 µm).

The distortion grid is shown in Figure 47. The rate of distortion is defined as the ratio ($r_{real}$ − $r_{predicted}$)/ $r_{predicted}$ , r being the distance of the chief ray intersection on the FP to the optical axis. The maximum distortion is less than 0.13 % over the whole FoV.

The polychromatic encircled energy is shown in Figure 48. For each fields of view, it is always less than 39 µm.

The Strehl ratios for all the wavelengths over the whole field of view are shown in Figure 49.

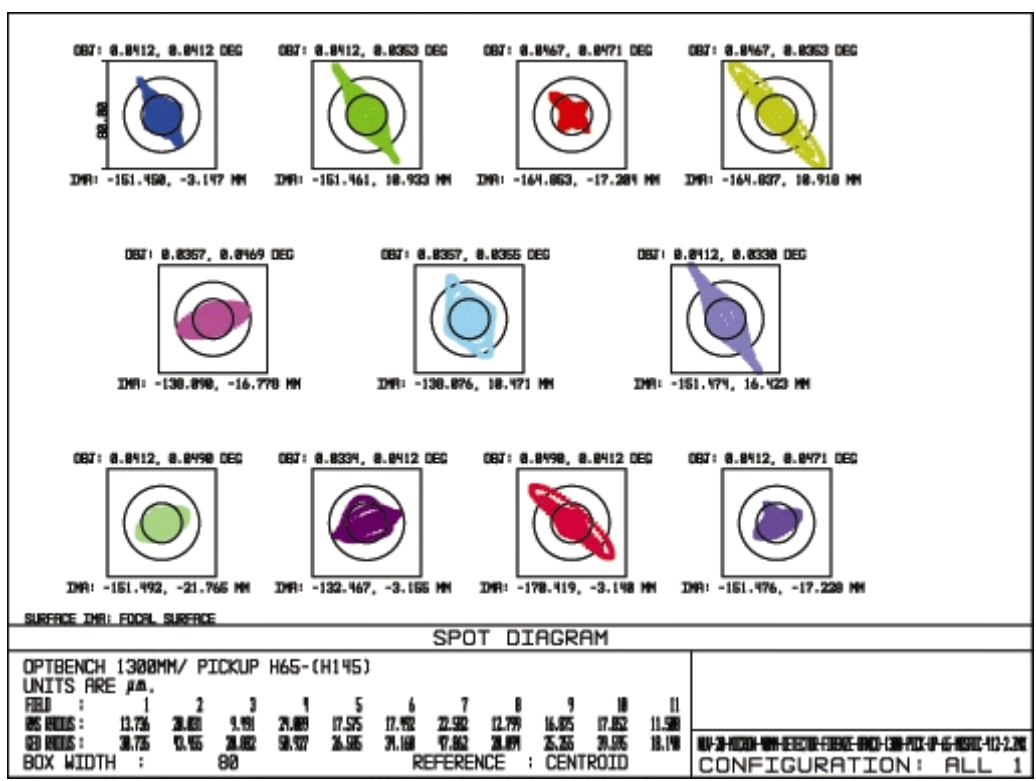

Figure 46. Spot diagrams of the NUV channel for the mosaic pick-up mirror configuration.

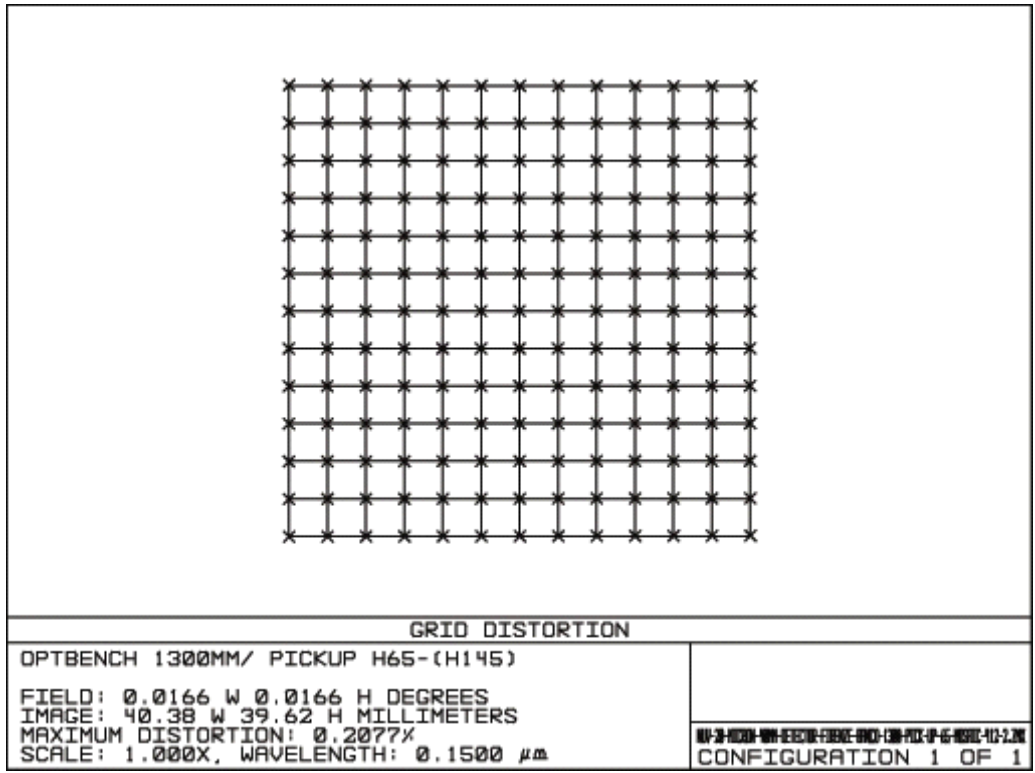

Figure 47. Distortion Grid of the NUV channel for the mosaic pick-up mirror configuration.

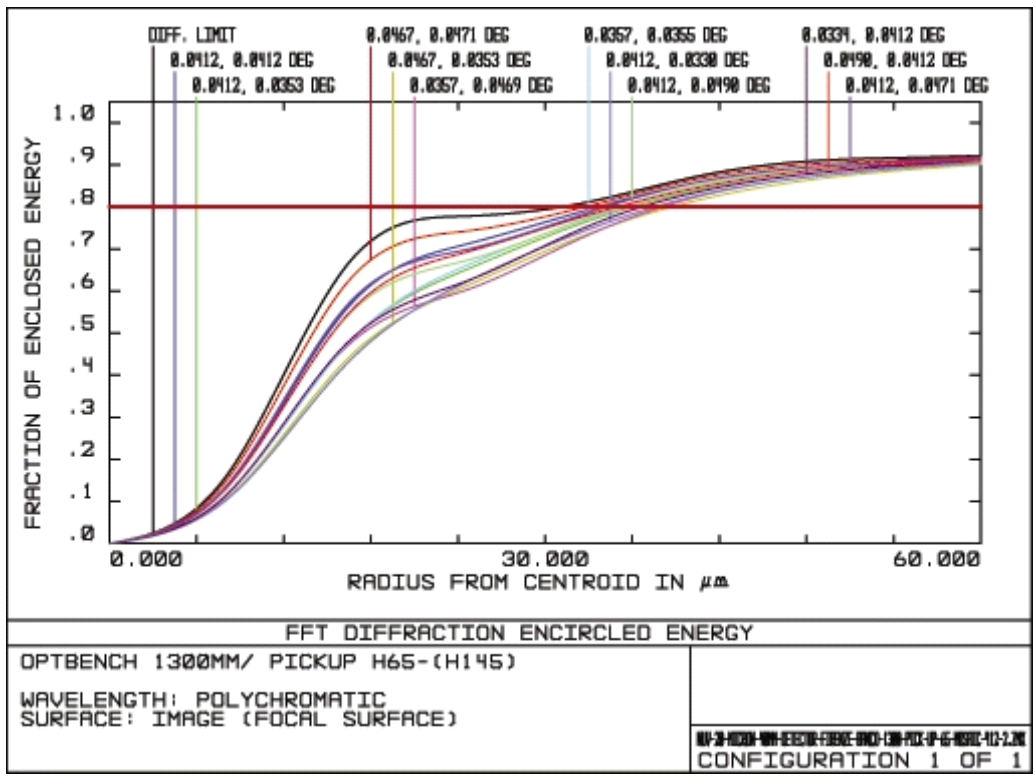

Figure 48. Encircled energy of the NUV channel for the mosaic pick-up mirror configuration.

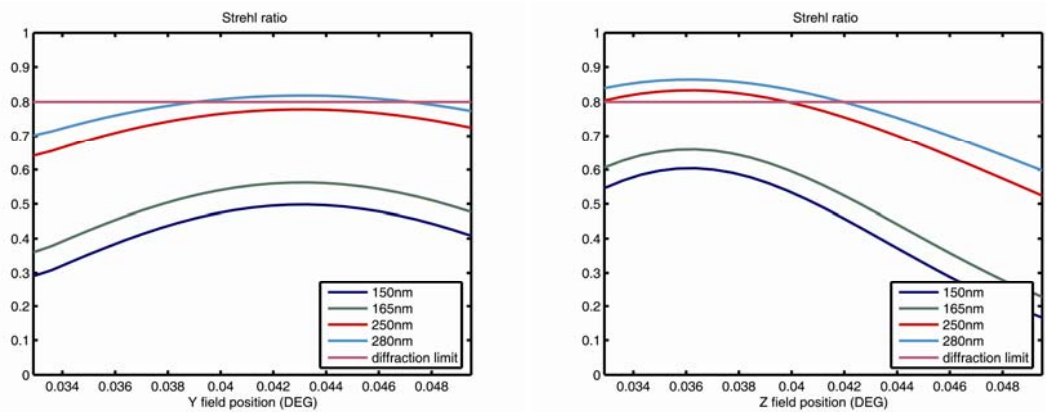

Figure 49. Strehl ratio of the NUV channel for the mosaic pick-up mirror configuration.

## 2.4.3  Slitless Spectroscopy Optimized Layout

The optical design characteristics are the following:

> Pick-up Mirror: flat mirror
>
> M1: elliptical concave mirror
>
> M3: elliptical convex mirror with grating

This configuration is the same described in section 2.4.1 with no difference in imaging capability and optical performances. The M1 and M2 optical axis are slightly tilted one with respect to the other. All the mirrors and the detector lay inside a 1300 mm OB.

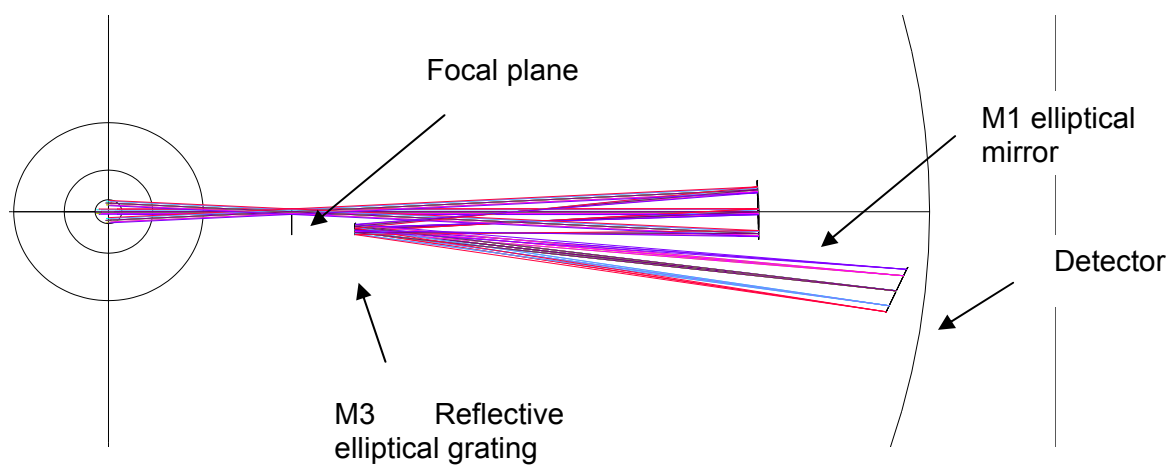

Figure 50. Layout of the NUV with a reflective convex grating replacing the elliptical mirror.

The pixel FoV is 0.03 arcsec/pixel. Having an equivalent pixel size of 20 μm and a detector diameter of 40 mm, the FoV diameter is 1.00 arcmin. The NUV channel magnification is 8.1.

The difference with the configuration of section 2.4.1 is the reflective grating on elliptical convex mirror that assures low-resolution slitless imaging spectroscopy capability.

In the spectroscopic configuration the M2 mirror is placed in a wheel similar to that of the filters. Such wheel is used to replace the M2 mirror with M3, that is an elliptical convex mirror with grating. M3 has imaging capabilities and through the grating spectroscopic capabilities with R=100.

The main parameters for all the mirrors and FPs are given in Table 27.

Table 27. Main parameters of the NUV channel mirrors for the rotating pick-up mirror configuration.

| | R (mm) | K | Shape | Size (mm×mm) | Vertex-Center (mm) |
|---|---|---|---|---|---|
| **Pick-up** | *plane* | *0* | *elliptical* | *58×80* | *0* |
| **Entrance FP** | *plane* | *0* | *circular* | *4.95* | *0* |
| **M1** | *372.059* | *0.080667* | *circular* | *46* | *9* |
| **M3** | *123.763* | *0.821435* | *circular* | *10* | *-6.2* |
| **FP** | *plane* | *0* | *circular* | *40* | *0* |

The location of the mirrors with respect to the FCU optical reference frame are given in Table 28.

Table 28. Global vertex in the coordinate system linked to FCU optical reference frame of the NUV channel mirrors for the rotating pick-up mirror configuration.

| | Tilt y (degree) | Tilt z (degree) | Tilt x (degree) | y (mm) | z (mm) | x (mm) |
|---|---|---|---|---|---|---|
| **Pick-up** | 45 | 0 | 0 | 0 | 0 | 145 |
| **Entrance FP** | 0 | 0 | 0 | 0 | 145 | 145 |
| **M1** | 0 | 0 | 0 | 0 | 514.898 | 145 |
| **M3** | 0 | 0 | 0.8 | -14.968 | 195.113 | 145 |
| **FP** | 0 | 0 | -24.2 | -66.749 | 623.610 | 145 |

The optical imaging performances are not changed with the M3 elliptical convex mirror-grating that substitutes the elliptical convex mirror M2. Those performances are shown in section 2.4.1.

Figure 51 show the spectroscopy capability with R=100 of the NUV range (150-280 nm), that covers 174 pixels on detector surface (130 nm, 1.5 nm in 2 pixels) corresponding to 3.48 mm and Figure 52 shows the dispersion in 150-280 nm for each field considered.

Figure 53 and Figure 54show the spot diagram obtained using the M3 grating mirror, at 150 nm and 280 nm respectively, of central field. In all spectral range and for each field the spot diameter is contained in Airy disk and the configuration is diffraction limited.

FCU phase A report – Optical, mechanical and electronics configurations

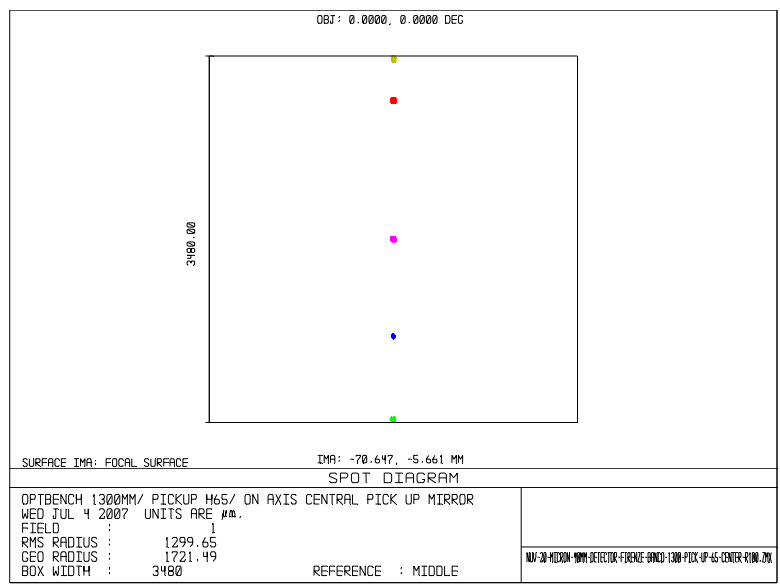

Figure 51. Dispersion for R=100 of the NUV range (150-280 nm), covering 174 pixels (130 nm, 1.5 nm in 2 pixels) corresponding to 3.48 mm for central field.

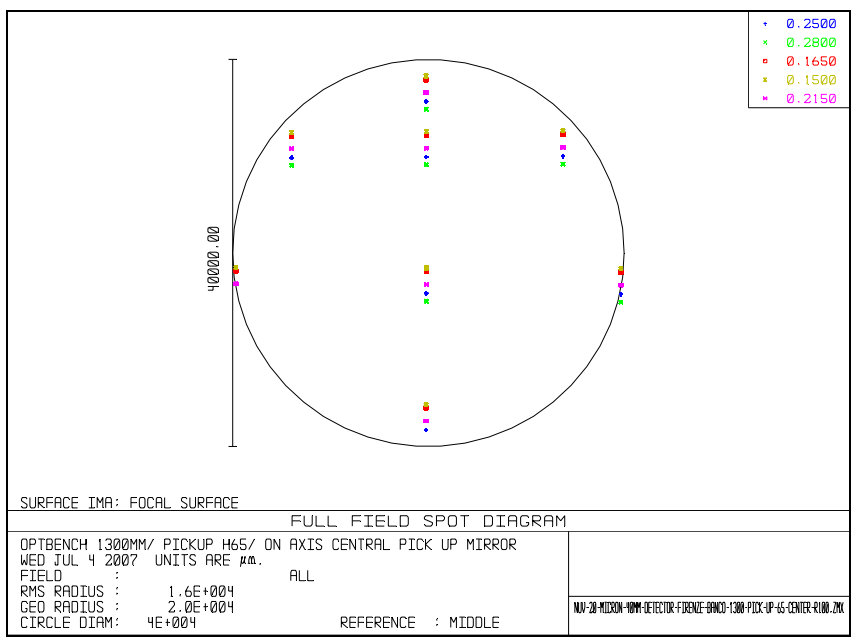

Figure 52. Dispersion of 130 nm on the detector for each field at different wavelengths represented with different colors.

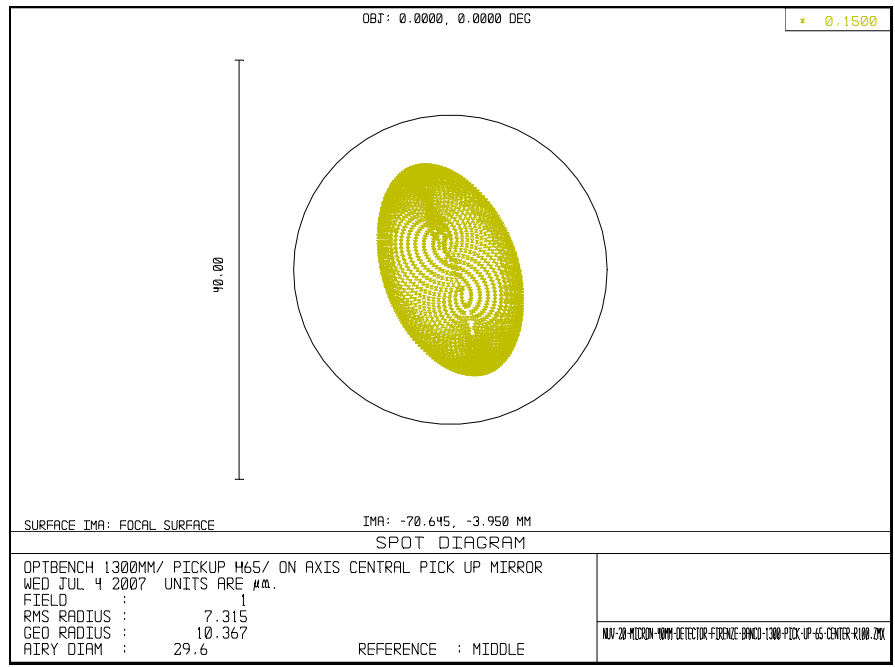

Figure 53 Spot diagram at 150 nm

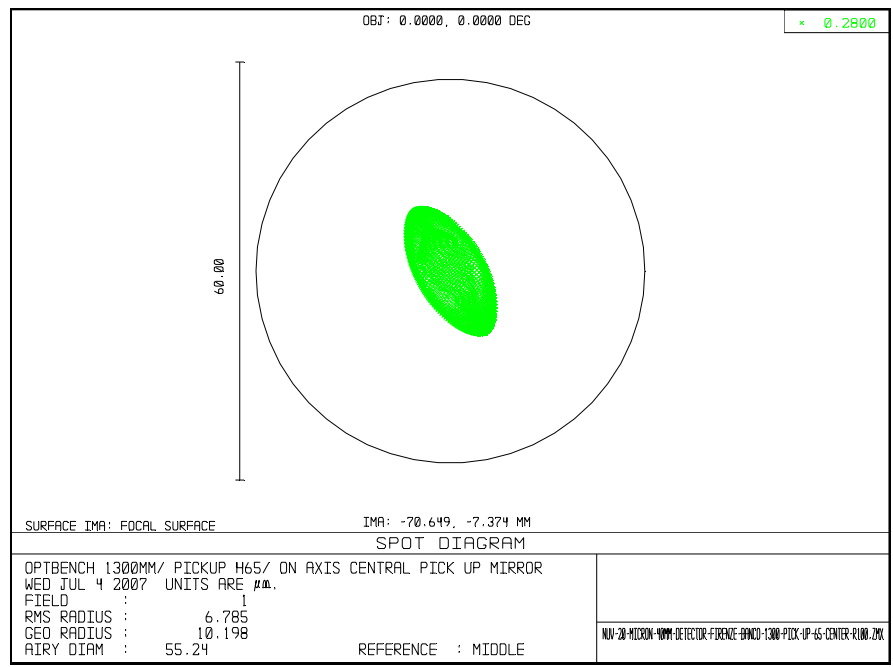

Figure 54  Spot diagram at 280 nm.

## 2.5   UV-Optical Channel

### 2.5.1  On-Axis Layout (rotating pick-up mirror)

In this configuration the UVO channel is composed of a rotating fold mirror deviating the optical beam towards a three mirrors anastigmatic system. The optical layout is shown in Figure 55.

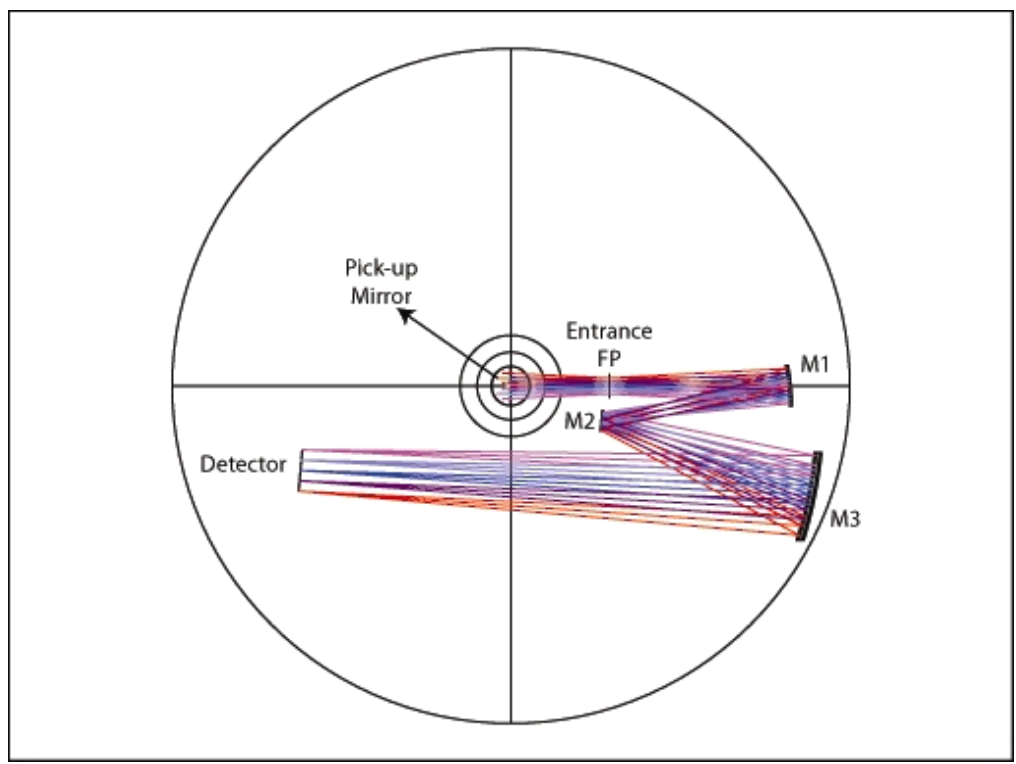

Figure 55. Optical layout of the UVO channel for the rotating pick-up mirror configuration.

The optical design characteristics are the following:

> Pick-up Mirror: flat mirror
>
> M1: elliptical concave mirror
>
> M2: spherical convex mirror
>
> M3: elliptical concave mirror

M1 and M3 have the same optical axis while M2 is slightly de-centered. All the mirrors and the detector lay inside a circle having a diameter of 1000 mm.

The pixel FoV is 0.07 arcsec/pixel. Having a pixel size of 15 $\mu$m and a detector format of 4096×4096 pixel$^2$, the FoV is 4.78×4.78 arcmin$^2$ with a detector size of 61.44 mm. The UVO channel magnification is 2.8.

The filter wheels are placed nearby the entrance FP, being the accessible area where the beam has the smallest dimension. The exact positioning of the filter wheels will be fixed when the opto-mechanical design will be completed.

The main parameters for all the mirrors and FP are given in Table 29.

Table 29. Main parameters of the UVO channel mirrors for the rotating pick-up mirror configuration.

|  | R (mm) | K | Shape | Size (mm×mm) | Vertex-Center (mm) |
|---|---|---|---|---|---|
| Pick-up | plane | 0 | elliptical | 58×80 | 0 |
| Entrance FP | plane | 0 | square | 23.62×23.62 | 0 |
| M1 | 359.115 | 0.327309 | rectangular | 60×60 | -34 |
| M2 | 265.994 | 0 | rectangular | 30×30 | 20 |
| M3 | 658.232 | -0.114339 | rectangular | 124×124 | 130 |
| FP | plane | 0 | square | 61.44×61.44 | 0 |

The location of the mirrors with respect to the FCU optical reference frame are given in Table 30.

Table 30. Global vertex in the coordinate system linked to FCU optical reference frame of the UVO channel mirrors for the rotating pick-up mirror configuration.

|  | Tilt y (degree) | Tilt z (degree) | Tilt x (degree) | y (mm) | z (mm) | x (mm) |
|---|---|---|---|---|---|---|
| Pick-up | 45 | 0 | 0 | 0 | 0 | 145 |
| Entrance FP | 0 | 0 | 0 | 0 | 145 | 145 |
| M1 | 0 | 0 | 0 | 0 | 410.746 | 145 |
| M2 | 0 | 0 | 0 | -52.169 | 135.214 | 145 |
| M3 | 0 | 0 | 0 | -161.059 | 441.543 | 145 |
| FP | 0 | 0 | 5.098 | -125.182 | -310.688 | 145 |

The spot diagrams are shown in Figure 56. The spots image are given for a FoV of ±30.61× 30.61 mm in the FP. The square boxes have the dimensions of 4×4 pixels. The two circles have the dimensions of the Airy disk at 200 nm (diameter 12.682 μm) and at 700 nm (diameter 44.380 μm).

The distortion grid is shown in Figure 57. The rate of distortion is defined as the ratio ($r_{real} - r_{predicted}$)/ $r_{predicted}$ , r being the distance of the chief ray intersection on the FP to the optical axis. The maximum distortion happens at the FoV border and has a value of 0.0711%.

In order to maintain the distortion grid stable with a tolerance of 3 mas, it has been estimated that the accuracy in re-positioning the rotating pick-up mirror must be ≤ 3 arcsec. However, it must be noted that the 3 mas tolerance takes into account all possible effects (e.g. thermal instabilities).

The polychromatic encircled energy is shown in Figure 58. For each field of view, 80% of the encircled energy always falls within a circle with radius less than 20 μm.

The Strehl ratios for all the wavelengths over the whole field of view are shown in Figure 59. The optical system has diffraction limited quality (Strehl Ratio >80%) always but at the shortest wavelengths (<230 nm).

FCU phase A report – Optical, mechanical and electronics configurations

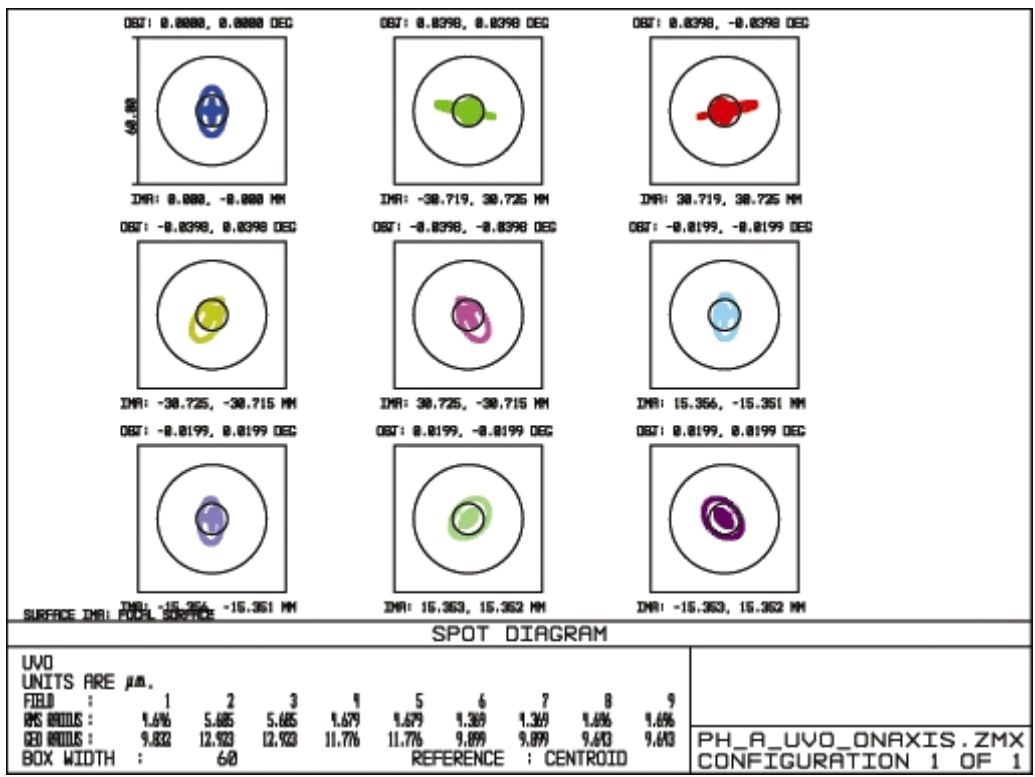

Figure 56. Spot diagrams of the UVO channel for the rotating pick-up mirror configuration.

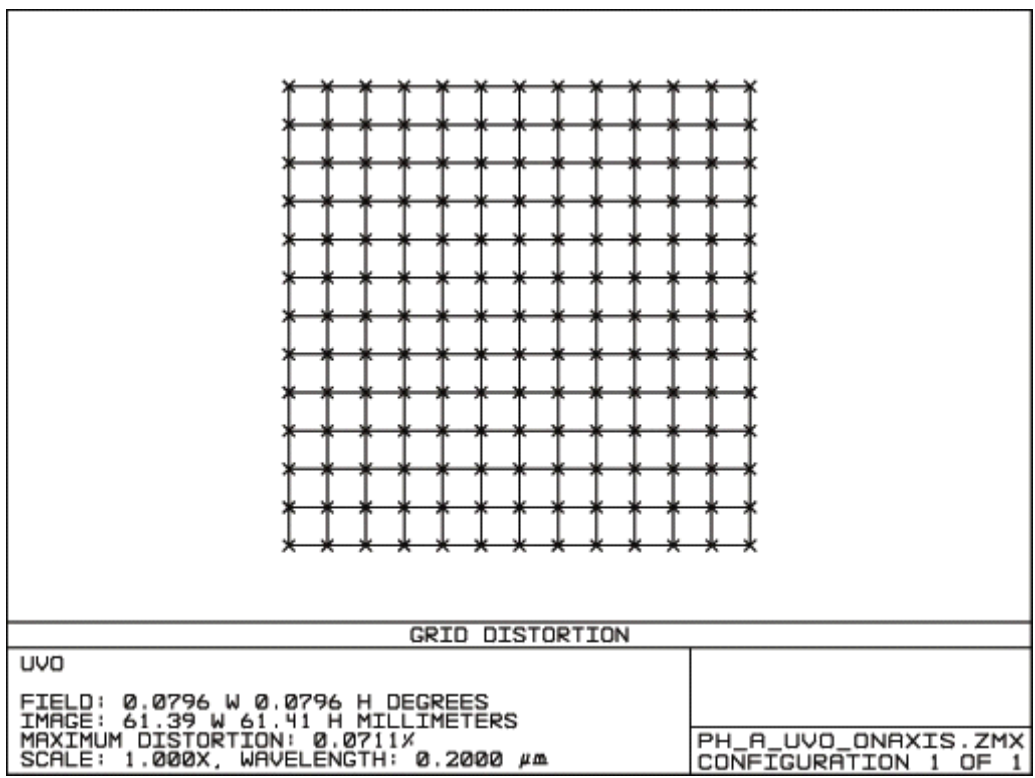

Figure 57. Distortion Grid of the UVO channel for the rotating pick-up mirror configuration.

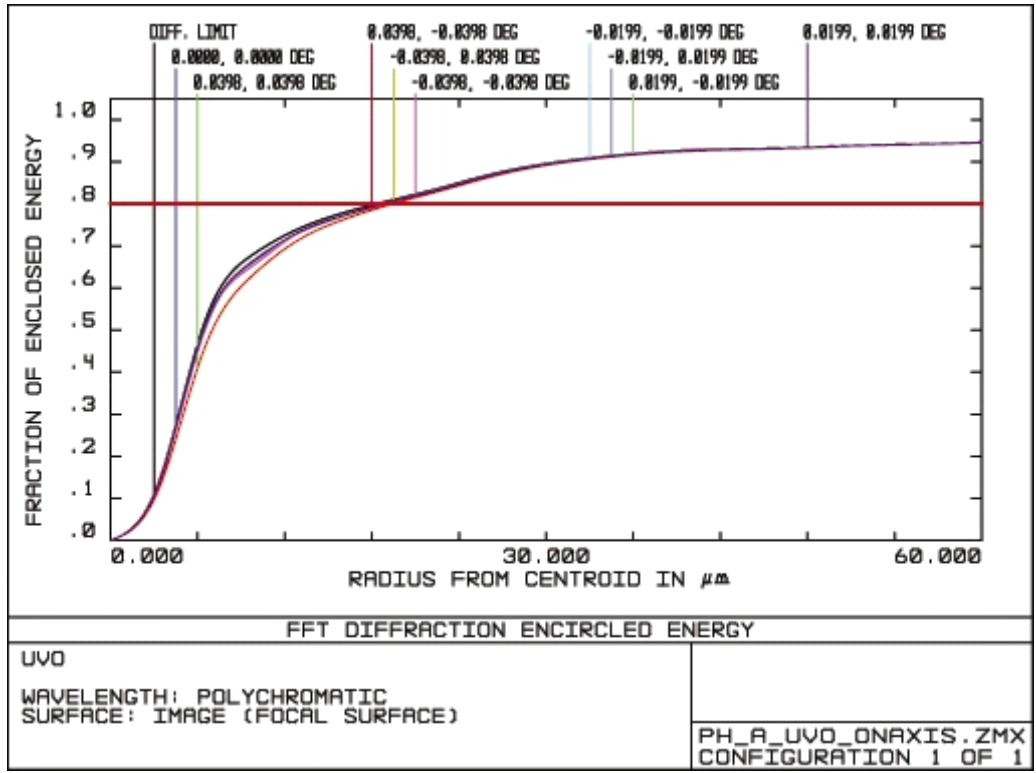

Figure 58. Encircled energy of the UVO channel for the rotating pick-up mirror configuration.

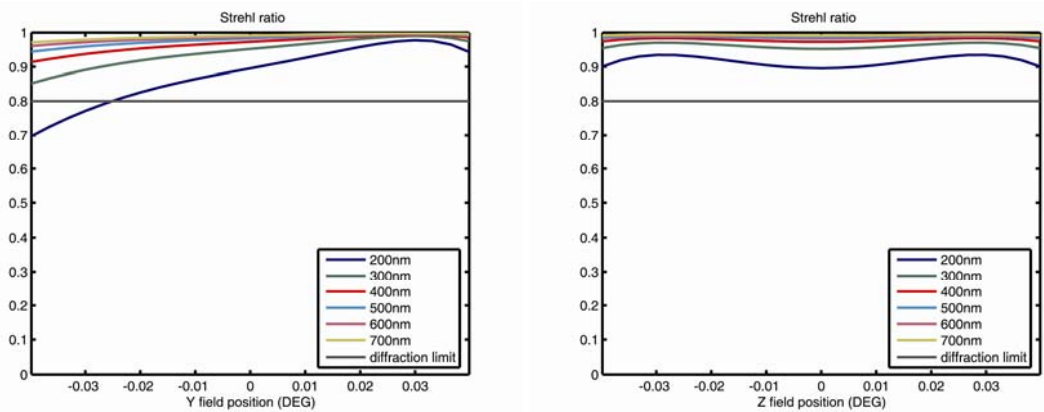

Figure 59. Strehl ratio of the UVO channel for the rotating pick-up mirror configuration.

## 2.5.2 Off-Axis Layout (mosaic pick-up mirror)

In this configuration the UVO channel is compound by a single aspheric mirror deviating the optical beam towards a three mirrors anastigmatic system. The pick-up mirror is de-centered with respect to the telescope optical axis as reported in Table 18 (configuration 1).

The optical layout is shown in Figure 60.

FCU phase A report – Optical, mechanical and electronics configurations

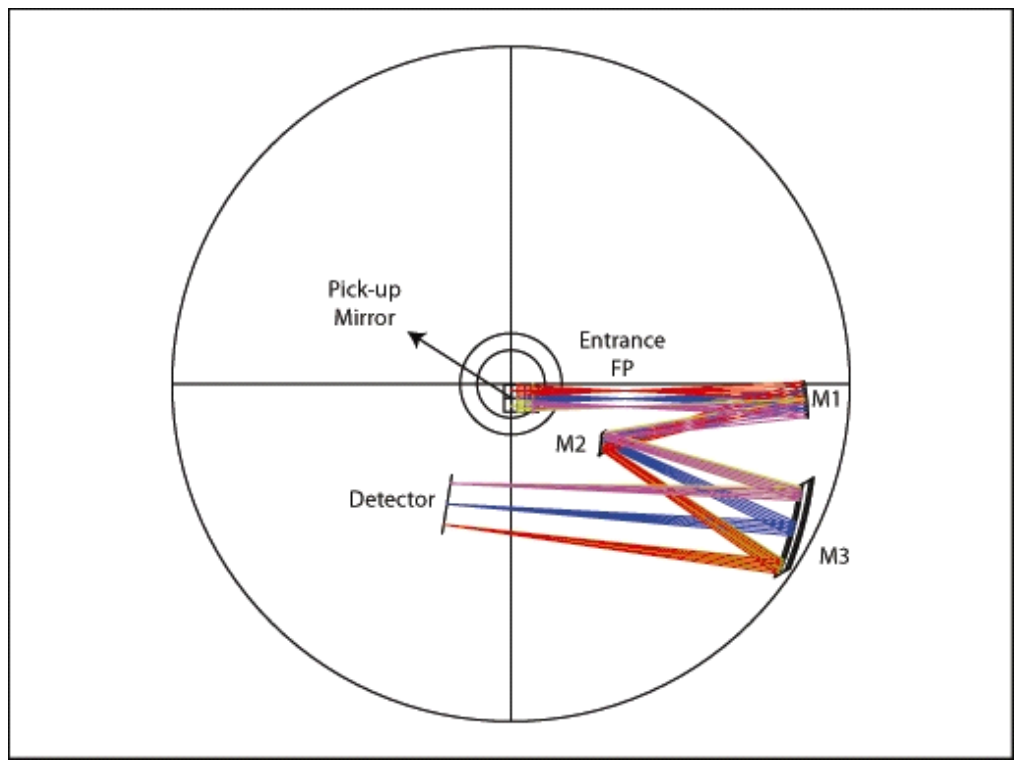

Figure 60. Optical layout of the UVO channel for the mosaic pick-up mirror configuration.

The optical design characteristics are the following:

      Pick-up Mirror: aspheric mirror

      M1: elliptical concave mirror

      M2: elliptical covex mirror

      M3: spherical concave mirror

All the mirrors and the detector lay inside the OB having a diameter value of 1000 mm.

The pixel FoV is 0.07 arcsec/pixel. Having a pixel size of 15 $\mu$m and a detector format of 4096×4096 pixel$^2$, the FoV is 4.78×4.78 arcmin$^2$ with a detector size of 61.44 mm. The UVO channel magnification is 2.8.

The filter wheels are placed nearby the entrance FP, being the accessible area where the beam has the smallest dimension. The exact positioning of the filter wheels will be fixed when the opto-mechanical design will be completed.

The main parameters for all the mirrors and FPs are given in Table 31.

Table 31. Main parameters of the UVO channel mirrors for the mosaic pick-up mirror configuration.

| | R (mm) | K | $A_2$ | $A_4$ | $A_6$ | $A_8$ | Shape | Size (mm×mm) | Vertex-Center (mm) |
|---|---|---|---|---|---|---|---|---|---|
| **Pick-up** | plane | 0 | $-3.117847$ $10^{-5}$ | $1.950661$ $10^{-10}$ | $-8.507255$ $10^{-16}$ | $1.834533$ $10^{-21}$ | rectangular | 40×56 | -23.342 × 353.047 |
| **Entrance FP** | plane | 0 | 0 | 0 | 0 | 0 | square | 23.62×23.62 | 0 |
| **M1** | 345.403 | 0.3443 | 0 | 0 | 0 | 0 | rectangular | 54×54 | -5.459 × -33.115 |
| **M2** | 250.093 | 0.100396 | 0 | 0 | 0 | 0 | rectangular | 32×24 | 2.359 × 164.221 |
| **M3** | 566.859 | 0 | 0 | 0 | 0 | 0 | rectangular | 126×98 | 15.882 × 397.107 |
| **FP** | plane | 0 | 0 | 0 | 0 | 0 | square | 61.44×61.44 | 0 |

The location of the mirrors with respect to the FCU optical reference frame are given in Table 32.

Table 32. Global vertex in the coordinate system linked to FCU optical reference frame of the UVO channel mirrors for the mosaic pick-up mirror configuration.

| | Tilt y (degree) | Tilt z (degree) | Tilt x (degree) | y (mm) | z (mm) | x (mm) |
|---|---|---|---|---|---|---|
| **Pick-up** | -45 | 0 | 0 | -21.353 | 21.353 | 145 |
| **Entrance FP** | 0 | 0 | 0 | -21.985 | 166.368 | 145 |
| **M1** | 0 | 0 | 0.251087 | -23.153 | 434.255 | 145 |
| **M2** | 0 | 0 | 35.6483 | -85.045 | 138.754 | 145 |
| **M3** | 0 | 0 | 30.2864 | -214.740 | 418.075 | 145 |
| **FP** | 0 | 0 | -9.351519 | -177.868 | -94.355 | 145 |

The spot diagrams are shown in Figure 61. The spots image are given for a FoV of ±30.61× ± 30.61 mm in the FP. The boxes have the dimensions of 4×4 pixels. The two circles have the dimension of the Airy disk at 200nm (diameter 12.682 µm) and at 700nm (diameter 44.380 µm).

The distortion grid is shown in Figure 62. The rate of distortion is defined as the ratio ($r_{real}$ − $r_{predicted}$)/ $r_{predicted}$ , r being the distance of the chief ray intersection on the FP to the optical axis. The maximum distortion happens at the FoV border and has a value of 0.4105%.

The polychromatic encircled energy is shown in Figure 63. For each fields of view, it is always less than 20 µm.

The Strehl ratios for all the wavelengths over the whole field of view are shown in Figure 64. The optical system has diffraction limited quality (SR>80%).

FCU phase A report – Optical, mechanical and electronics configurations

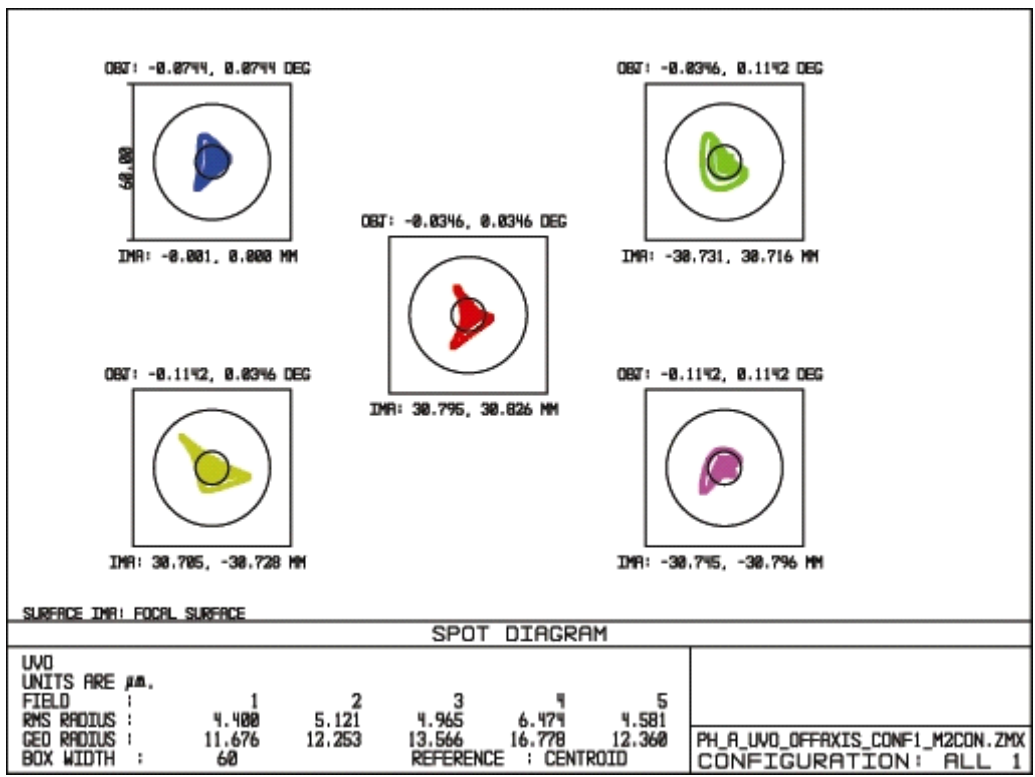

Figure 61. Spot diagrams of the UVO channel for the mosaic pick-up mirror configuration.

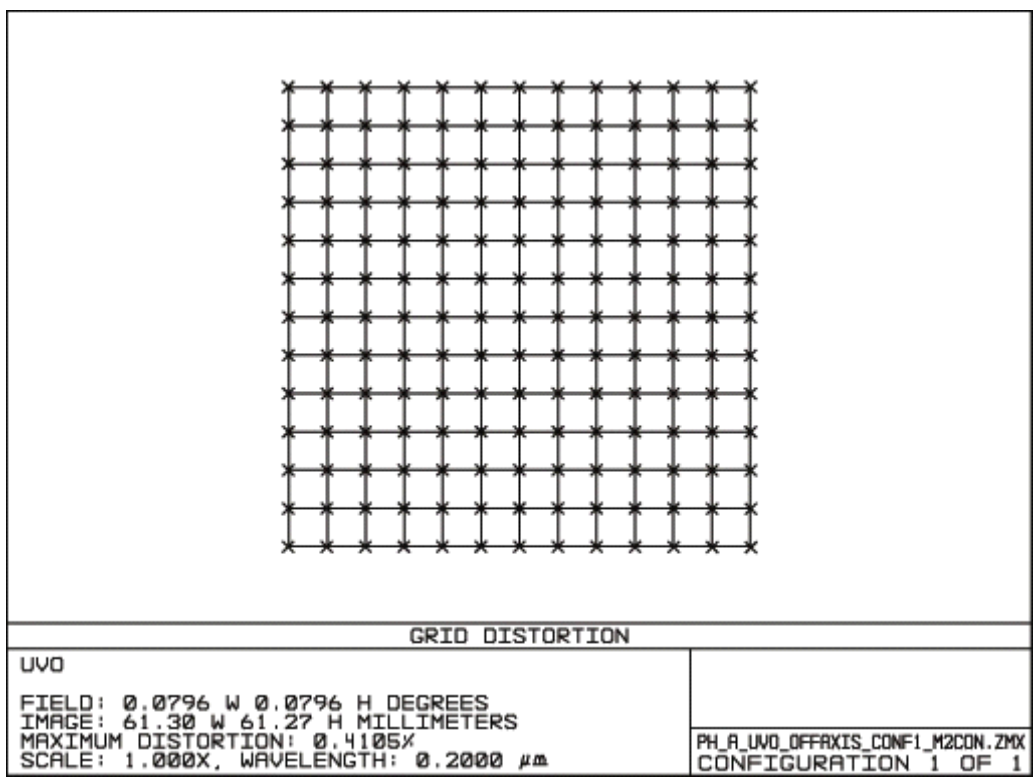

Figure 62. Distortion Grid of the UVO channel for the mosaic pick-up mirror configuration.

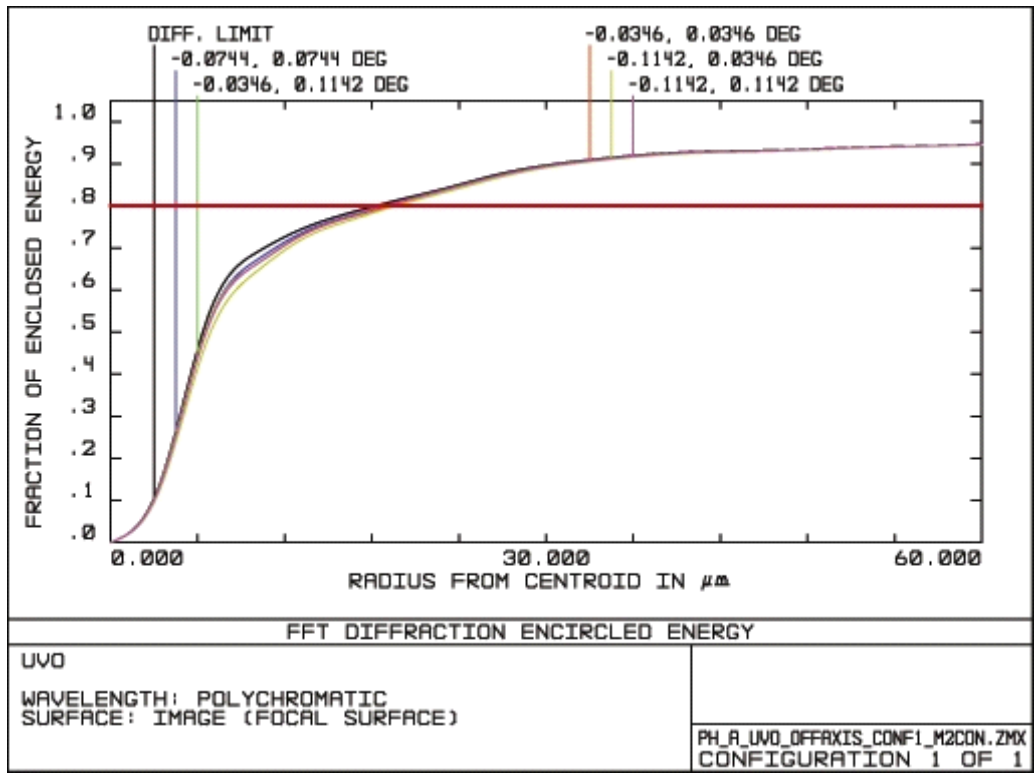

Figure 63. Encircled energy of the UVO channel for the mosaic pick-up mirror configuration.

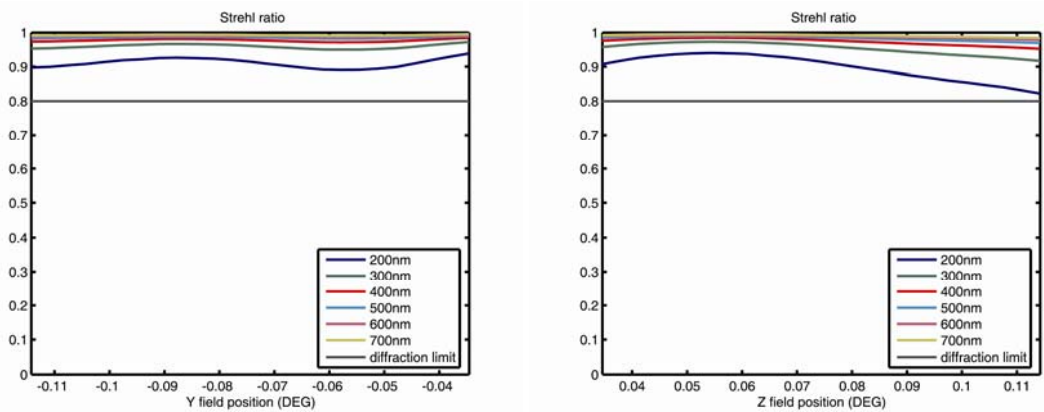

Figure 64. Strehl ratio of the UVO channel for the mosaic pick-up mirror configuration.

## 2.6 Calibration Subsystem

The FCU instrument will provide in-flight flat field calibration in all spectral bands. Wavelength calibration for the spectroscopic observation modes will be obtained using astrophysical reference sources.

Flat field calibration based on the observation of natural sources (Moon, Earth) or the use of S/C common calibration sources are being investigated but the baseline solution for the FCU calibration subsystem will be an internal calibration unit.

The optical design of the calibration unit is still TBD due to the ongoing trade off process on the optical layout of the three FCU channels. As soon as the process will be finalized the design will start.

## 2.6.1 Lamps

To cover efficiently the whole wavelength range from 115 nm to 700nm two kind of lamps will be used and if necessary combined: tungsten lamps and deuterium lamps. No wavelength calibration lamp is considered at this stage.

## 2.7 Optical layout trade-off analysis

For each of the three channels composing the FCU instrument, on-axis and off-axis layouts have been investigated.

Off-axis design can reach performances similar, from the point of view of optical quality, to that of the on-axis design except for the NUV channel. However it requires optical elements of more difficult realization and alignment and for the NUV channel, the addition of a folding mirror with consequent loss of throughput. Furthermore, the positioning of the three pick up mirrors, once known the positions of FGS and slits, may be a non trivial problem to face, due to the available space on the focal surface.

Table 33. Optical layout trade off matrix

| Optical Layout | Rotating Mirror | Mosaic |
|---|---|---|
| **Opto-Mechanical Design** | | |
| Optical Quality | Compliant to requirements | NUV worst performances |
| Field Distortion Stability | Requirement for high precision re-positioning of mirror | No specific requirement |
| Efficiency | 3 reflections in NUV channel <br><br> Mirror coating not specialized | 4 reflections in NUV channel <br><br> Mirrors coating specialized |
| Stray Light | Mirror pointing in one direction (needs detailed analysis) | 3 mirrors always illuminated (needs detailed analysis) |
| Optics Manufacturing | Simpler | More Complex |
| Alignment | Pick up mirror flat <br><br> 1 element, simpler procedure | Pick up mirrors aspheric <br><br> 3 elements, more complex procedure <br><br> Need to simulate off-axis telescope FoV aberrations |
| Mechanism | Rotating mechanism: single point failure | No mechanism |
| Accommodation | Redundant motor difficult to accommodate | Crowded central FoV |
| **FCU Operations** | One camera operating at a time | More cameras operating at the same time (subject to restrictions from telemetry and power budget) |
| **Telescope Operations** | No particular request on telescope | Require re-pointing if to observe same FoV with different channel |

On the other side, on axis solutions exploiting the central, less aberrated, part of the telescope focal surface are simpler for optical elements manufacturing and alignment procedures. The only limit for this solutions may be represented by the central mirror mechanism: it must be accurate enough so to ensure the required distortion stability.

Obviously the on-axis/off-axis designs are different from the point of view of FCU and spacecraft operations. The off-axis solution gives the possibility of contemporary observation of different fields, but observing the same field means re-pointing of the telescope; the on axis solution can provide the observation of the same field only by the rotation of the central mirror, but not the contemporary use of two or more channels.

These considerations are summarized in the trade off matrix in Table 33.

# 3. MECHANICAL DESIGN

## 3.1 Description of the design

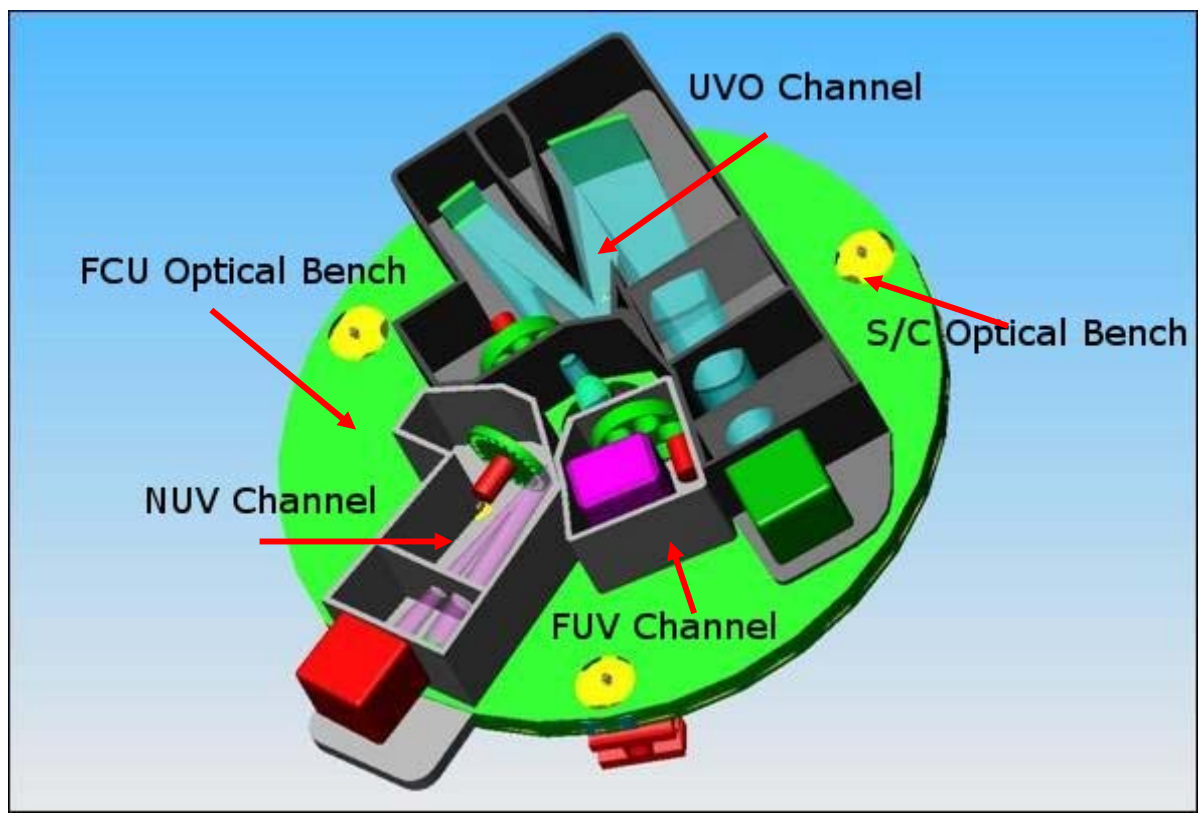

Figure 65. 3D Preliminary mechanical layout of FCU

The Field Camera Unit consists of three different optical channels, NUV, FUV, UVO, mounted on the FCU Optical Bench (Figure 65). Each channel has dedicated mirrors, filter wheels, detectors and mechanisms to allow imaging detection. There are three mechanical interfaces points to attach the FCU Optical Bench with the main OB developed by the Russian Team, each interface point is located on a circle with a diameter of 834 mm located on the bottom of the FCU OB.

In the center of the FCU Optical Bench is located the fore-optics that re-directs the optical beam from the telescope in the relative optical channel and the illumination to the FGS and slit spectrometers is guaranteed by three dedicated apertures on the FCU OB.

Protection of the detectors, optical channels and their elements against environmental harmful, affecting at telescope's operation, is provided by dedicated casing.

Overall dimension for the Field Camera Unit are 173 mm high with a diameter of the FCU OB of 1300 mm, the total mass is 31.2kg (20% contingency), excluding detectors, 48.2kg including them.

## 3.2   FCU coordinate system

The FCU coordinate system is defined as:

- X axis is along the telescope optical axis, oriented from the OB to toward the PMU
- Y axis is along the OB radius that intersect the rods used to support VUV Spectrometer
- Z axis is perpendicular to the other two axis, on clock wise looking the OB from positive X axis

## 3.3   FCU Optical Bench

The FCU optical bench (Figure 66) is the main structure that bears mechanical loads during launch phase and also supports all the components along the three optical paths.

The optical bench has to be designed to satisfy the requirements of the three channels, which operate over a large wavelength range from 115 nm to 700 nm, that is Far Ultraviolet to Visual. In the addition, the bench has to fit in the limited space between the primary mirror unit frame and the instrument compartment that accommodates the spectrometers.

The UVO and FUV channels match well in a OB smaller than 1300 mm diameter but the NUV channel optical design needs this larger dimension.

The main purpose is to provide support for the optical subsystems within the limited space available, while meeting very tight optical stability requirements. The bench has to provide easy access to the optical components throughout integration and has to provide an interface with adequate structural and thermal interfaces.

The bench has to meet a minimum frequency requirement and have enough strength to show positive margins of safety.

Very demanding manufacturing tolerances has to be placed on interfaces to ensure a proper fit with the optical components.

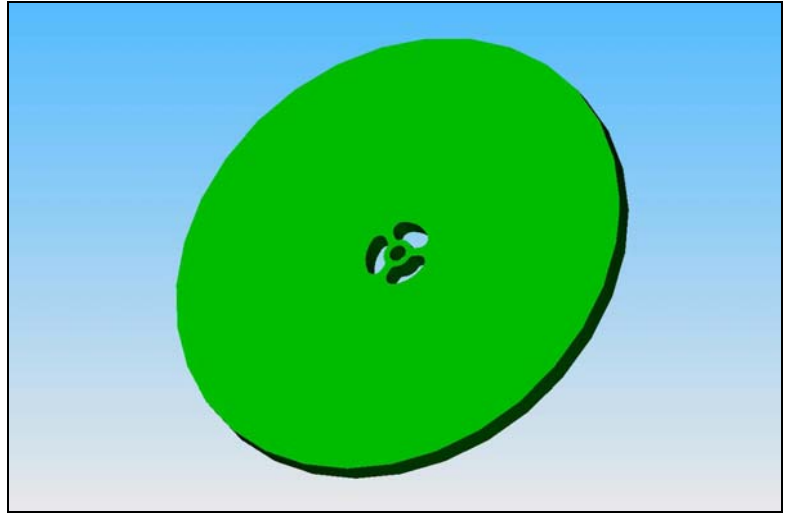

Figure 66. Preliminary design of FCU optical bench

### 3.3.1 Optical Bench Mechanical Design

The optical bench is design to support optics, mechanisms and detectors. The stiffness of the bench has to match a very high thermal stability. At this stage two different materials are under evaluation process for the construction of the bench: CeSic® and advanced composite honeycomb materials.

CeSic® and advanced composite honeycomb panels imply two different building process philosophies. Both materials have similar low coefficient of thermal expansion and mechanical properties, but they have differences concerning mass, costs, manufacturing, finishing, processing and deliverable time.

The composite face sheet/aluminum core honeycomb panels, are widely used in space projects and for the FCU could be used not only to build the bench but also to realize internal bulkheads, lateral and top instrumental cover. This material exhibits high strength, stiffness and thermal conductivity and possesses low moisture absorption and degassing. Dedicated inserts for mirrors and instruments fixing points have to be located inside the core.

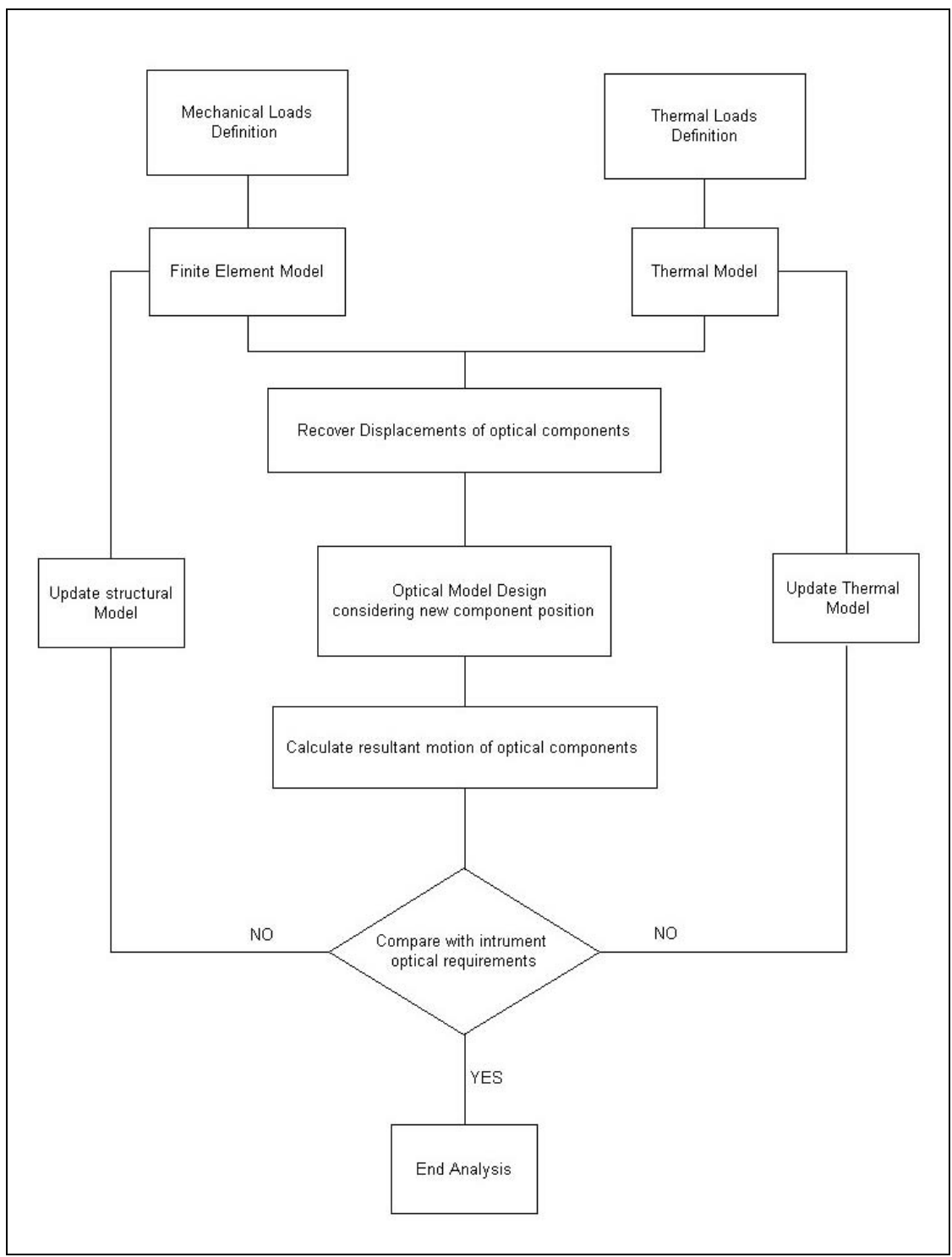

Figure 67. Conceptual flow of the structural and thermo/optical design activity

## 3.3.2 Structural and Thermo/Optical Design

In order to provide quantitative performances predictions of the optical systems based on thermal and mechanical loads effects, an iterative process between thermal, structural and optical design will be developed.

This process, presented in Figure 67, takes into account distortions produced by mechanical and thermal loads, the purpose is to perform it several times during the design of the optical bench, this process will create a mechanical structure with the tight stability requirements given by the optical layout.

### 3.3.3 OB mechanical interfaces

The FCU bench has three mechanicals interfaces with the S/C OB. Each point consists of a pad 30 x 30 mm$^2$ with a passing hole 8 mm diameter. These points are located at 120° one respect to the other near the rods that joints primary mirror structure to spectrometers holding points.

The bottom of the OB will provide the respective contacting pads. Attention has to be made during manufacturing process to minimize misalignments between the two systems.

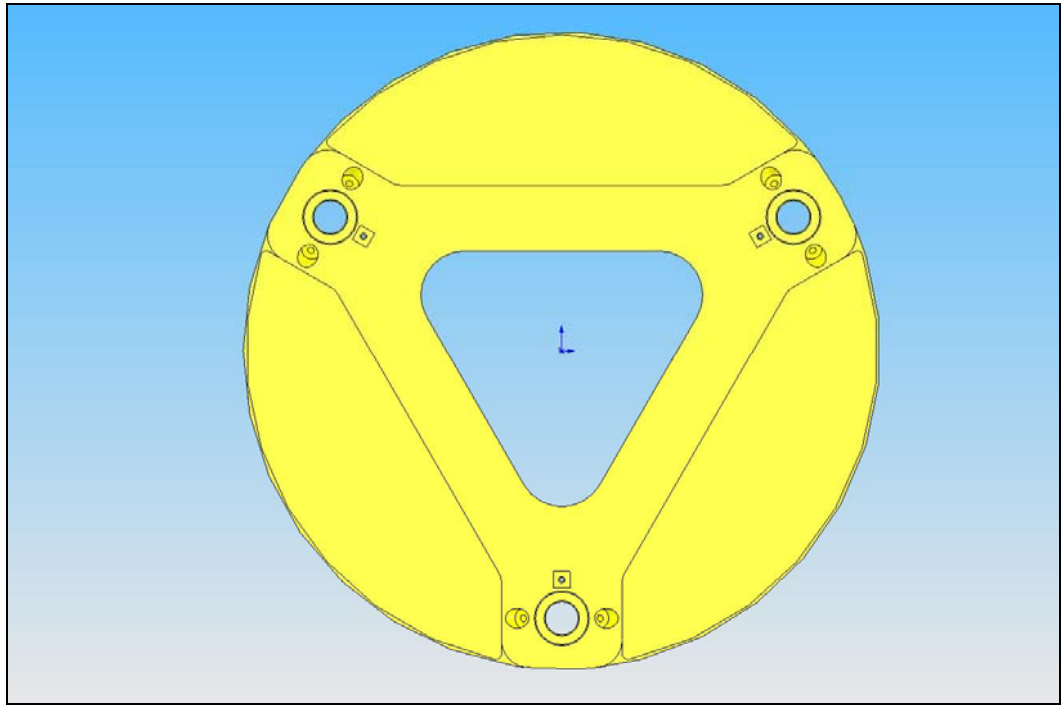

Figure 68. The S/C optical bench

The design of the Russian structure is developed to minimize mechanical loads given from the entire structure to the FCU Optical Bench.

On Figure 69 the fixing points from FCU OB to S/C OB and the relative rod cross section with the adjusting mechanism are shown.

The bench must provide three apertures to allow the beam to pass through and illuminate the FGS detectors and the slit of the spectrometers.

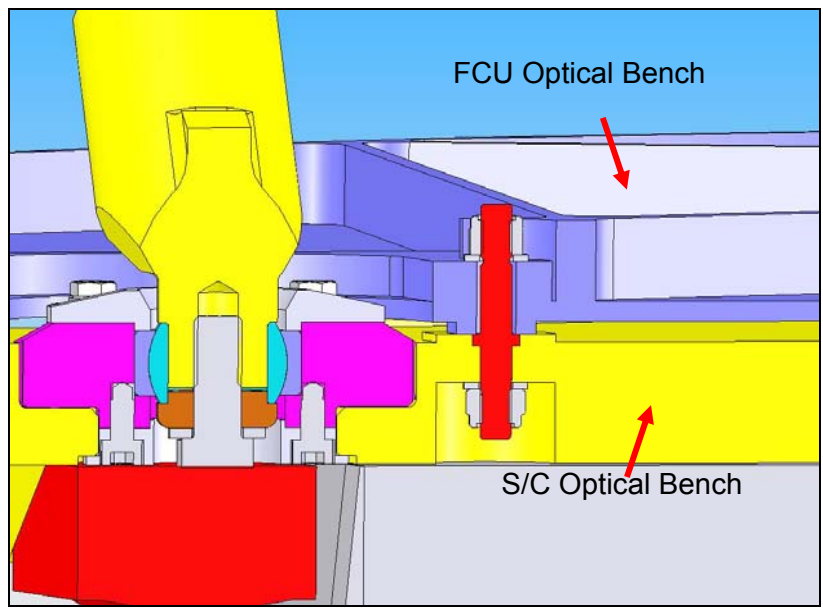

Figure 69. S/C optical bench: particular of the fixing points

In Figure 70 is shown the mechanical structure that supports the mechanism needed to re-direct the optical beam towards the various optical channels.

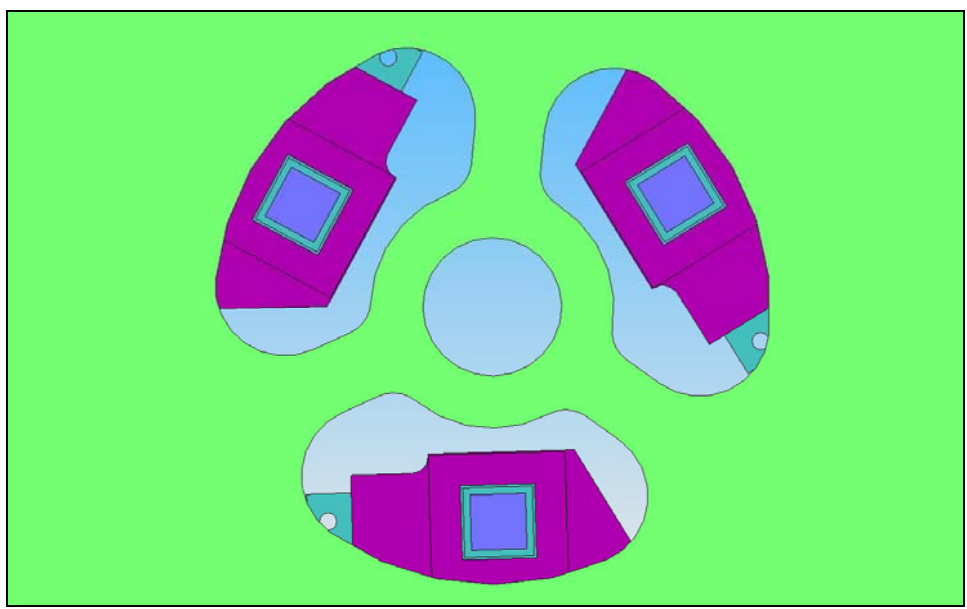

Figure 70. FCU optical bench: particular of the central part hosting the pick up mirror

## 3.4   Mechanisms

On the optical bench there will be different type of mechanisms to allow all the necessary operations during images acquisition.

In the design of these components great care will be dedicated to have devices as much as possible with similar characteristics to minimize costs, time delivery and software development.

The limited space available implies the development of custom devices but at the same time commercial components are considered as much as possible for time and costs reduction.

### 3.4.1 Rotating pick-up mirror

As shown in Figure 70, the centre of the bench has a hole and a frame structure that supports the pick-up mirror and, on the bottom, the motor used for the motion.

The main requirement for this mechanism consists in a high positioning accuracy and stability. To comply with the 3 mas requirement on the field distortion stability for the UVO channel an accuracy < 3 arc sec.

Two different designs are being analyzed to make this device: the first one uses in combination a stepper motor for rough positioning and piezoelectric actuator with encoder for fine positioning or in alternative a brushless motor with an encoder.

Both designs seem to match the mechanical position requirements.

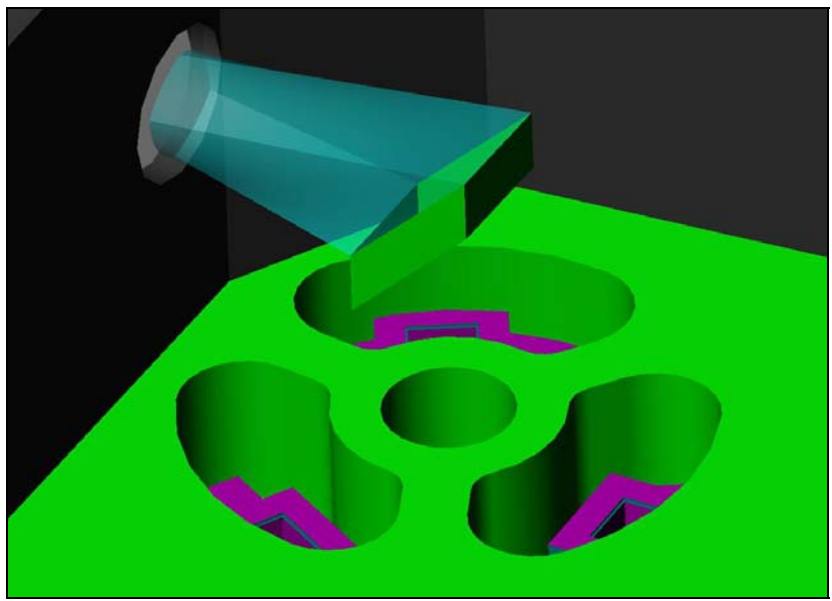

Figure 71. 3D lateral view of the rotating pick up mirror. No mechanism is shown.

Table 34 shows position tolerances and stability for central rotating mirror.

Table 34. Characteristics of the rotating mirror mechanism.

| Mechanism | Position tolerance | Stability tolerances | Motion |
|---|---|---|---|
| Rotating Mirror | < 3 arc sec | 0.1 arc sec | 3 positions |

### 3.4.2 Mosaic pick-up mirror

Figure 72 shows one of the possible solutions to position mosaic pick-up mirror in the center of the OB. In red and blue are highlighted the forbidden zones in the focal plane, these zones are occupied by FGS detectors and spectrometer slits.

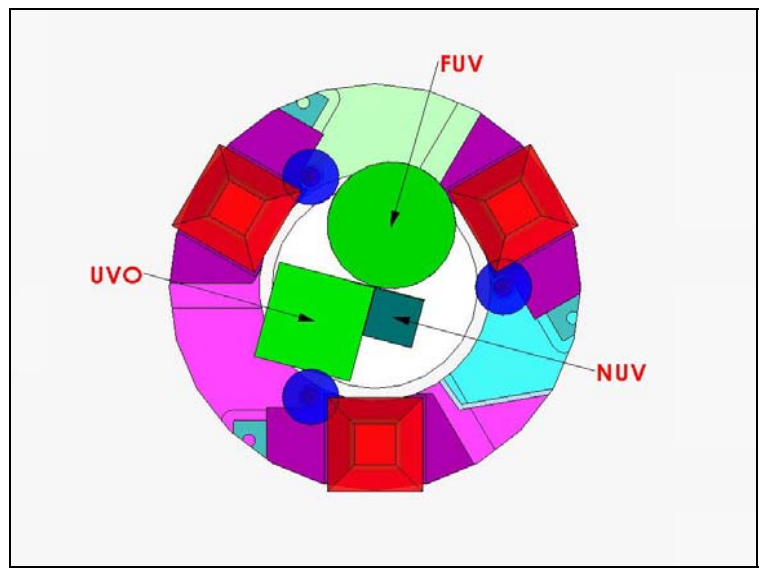

Figure 72. Preliminary design of mosaic pick up mirror layout.

## 3.5   Filters Wheels

Considering all three optical channels, three different designs have been performed for each filter wheel. All wheels are circular with a diameter of 150 mm.

The positions on the bench will be defined when the opto-mechanical design will be completed. The filters will be mounted tilted by TBD degrees to greatly reduce the effects of ghosts images. Figure 73 show a preliminary design for NUV filter wheel.

At the moment 2 wheels are dedicated for FUV channel, 2 for NUV channel and 3 for UVO channel. A space qualified commercial solution for UVO wheel channel is considered, this mechanism could have more than 3 wheels but dimensional solutions have to be found in order to install this device inside the UVO channel.

Table 35 shows the respectively the diameter, the number of wheels and filter diameter for each channel, accuracy position of the wheels.

Table 35. Filter wheels characteristics

| Channel | Wheel Diameter | No. Wheels | Optic Size |
|---------|----------------|------------|------------|
| FUV | 150 mm | 2 | 45 mm diam. |
| NUV | 150 mm | 2 | 5 mm diam. |
| UVO | 150 mm | 3 | 24x24 mm |

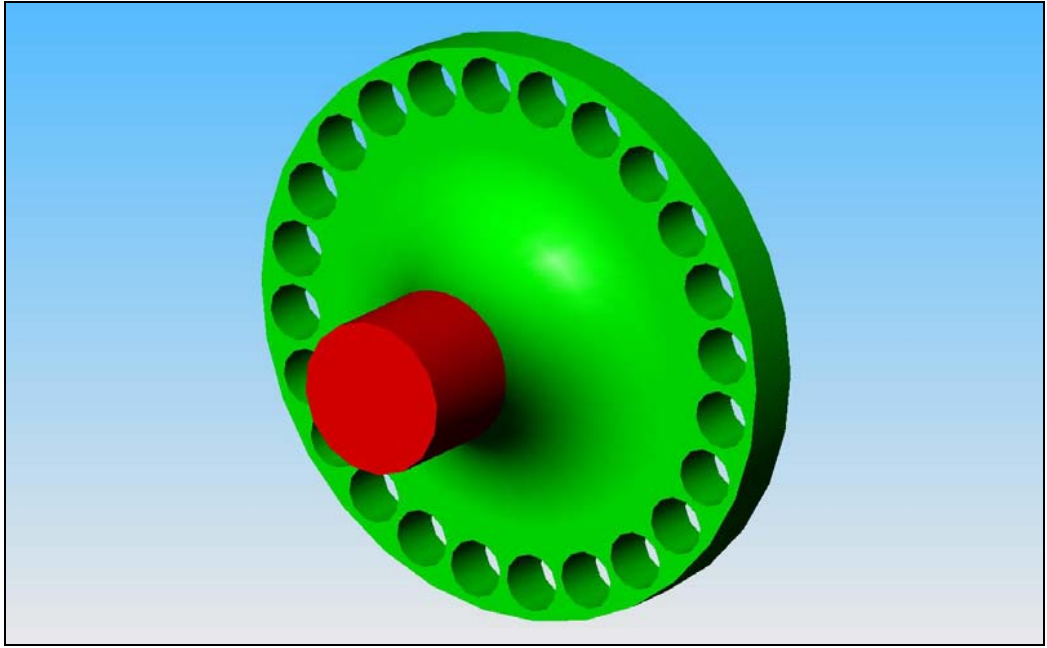

Figure 73. Preliminary design of one the NUV channel filters wheel.

Table 36 shows positioning and stability tolerances needed for the FCU filters wheels.

Table 36. Filter wheels mechanism requirements

| Channel | Position tolerances | Stability tolerances | Motion |
|---------|---------------------|----------------------|--------|
| FUV | 0.35 Degree | 0.15 Degree | Full rotation |
| NUV | 0.35 Degree | 0.15 Degree | Full rotation |
| UVO | 0.10 Degree | 0.05 Degree | Full rotation |

## 3.6   Shutters

### 3.6.1  NUV & FUV channel shutters

No specific shutter mechanisms are foreseen for the FUV and the NUV channel because photon counting detectors do not need shutter to set precisely the exposure time. However we are studying the possibility to block the optical beam in case of emergency, e.g. loss of spacecraft attitude. A possible solution could be to have a dark, i.e. full, position on the filter wheels. This will of course imply the loss of a slot on the corresponding wheel to be used for filters.

### 3.6.2  UVO channel shutter

There will be one CCD shutter mechanism to set the exposure times of the UVO camera. The shutter design is similar to the one used for the HST ACS camera (see Figure 74) with a blade whose size will fit the optical beam dimensions.

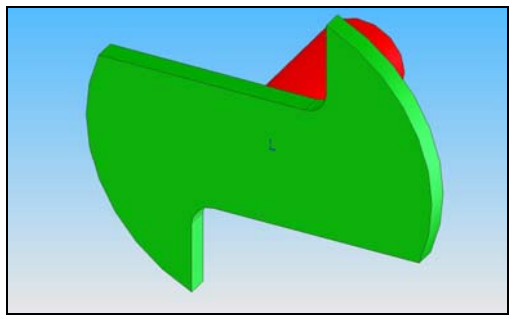

Figure 74. Preliminary design of the CCD shutter on the UVO channel.

A TBD (brushless, stepper) motor system will be used with a 2 degree of accuracy.

Table 37 lists positioning and stability tolerances need for the CCD shutter mechanism.

Table 37. CCD shutter mechanism requirements

| Mechanism | Position tolerance | Stability tolerances | Motion |
|---|---|---|---|
| CCD Shutter | 2 Degree | 0.15 Degree | 4 positions |

## 3.7  Focus & tip/tilt mechanism

A focus & tip/tilt mechanism for the FUV and UVO channels is being considered for alignment and focusing purposes. The degrees of freedom as well as the range of motion and the accuracy of the mechanism are TBD.

## 3.8  NUV Grating Wheel

As explained in section 2.4.3 to optimize both imaging and slitless spectroscopy mode in the NUV channel the disperser element (a grating) must be placed in the position of one of the two mirrors of the channel. This means that to keep both observing modes a mechanism must be introduced to switch between the mirror and the grating. The preliminary requirements of this mechanism are identical to that of the filters wheels of the same channel (see Table 35 and Table 36).

## 3.9  Baffling

Baffles structures have to be considered to reduce and possible eliminate scattered radiation during image acquisition. These baffles are developed inside each single channel and, in the centre of the OB, to minimize scattering in the FGS and Spectrometers.

Figure 75 shows a top view of part of UVO channel with relative baffles and in Figure 76 there is a lateral view where the baffles dedicated to the FGS are located under the primary rotating mirror.

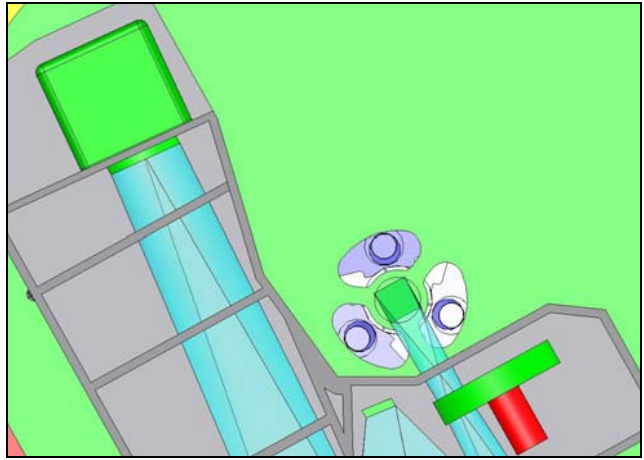

Figure 75. FCU baffling: example of internal baffling in the UVO channel

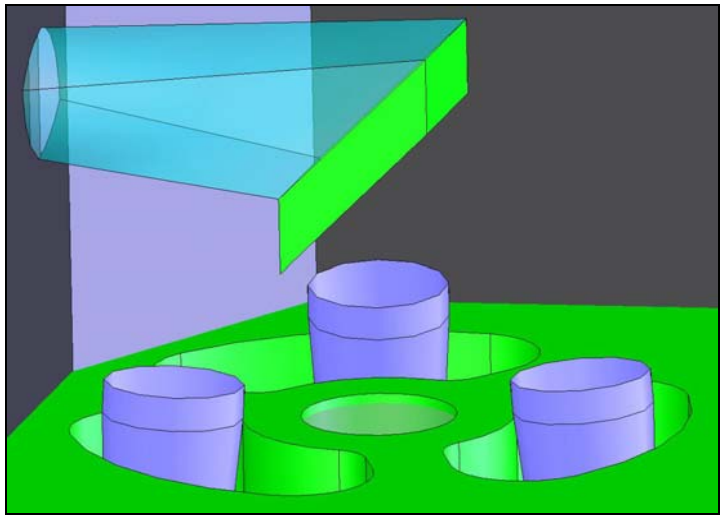

Figure 76. FGS baffles coming out the FCU optical bench

## 3.10 FEM analysis

After considering all the elements concerning the FCU camera, a Finite Element analysis will start. The analysis will take in consideration two different designs with two different building materials. Results of these simulations will be presented and discussed on September with the Russian team.

## 3.11 Mass Budget

| SUBSYSTEM | Weight (kg) |
|---|---|
| Optics & Mechanisms | |
| Optics | 1.5 |
| Optics Holders | 2.0 |
| Filters Wheels + Filters | 6.0 |
| Other Mechanisms (shutter etc) | 2.0 |
| **Subtotal** | 10.5 |
| Mechanical Structure | |
| Channels enclosures | 2.5 |
| Optical Bench | 10.0 |
| Calibration Unit enclosure | 3.0 |
| **Subtotal** | 15.5 |
| **Total Subsystems** | 26.0 |
| Contingency (20%) | 5.2 |
| **Total** | 31.2 |

# 4. DETECTORS SUBSYSTEM

## 4.1 FUV & NUV Detector Subsystem

### 4.1.1 FUV MCP performance specifications

The FUV detector will have the following characteristics:

- Range: 115-190 nm
- Active area size: 40 mm
- Pixel Size: 20 $\mu$m
- Local Dynamic Range (point-like source): > 10 counts/s
- Global Dynamic Range (diffuse source): > 200,000 counts/s
- Time resolution: $\leq$ 20 ms
- Dark noise: < 4·$10^{-5}$ counts/pixel/s at 20°C
- Large scale non-uniformity: < 10%
- Solar Blind: DQE<1.e-5 % @ $\lambda \geq$400nm

## 4.1.2 NUV MCP performance specifications

The NUV detector will have the following characteristics:

- Range:                                   150-280 nm
- Active area size:                        40 mm
- Pixel Size:                              20 $\mu$m
- Local Dynamic Range (point-like source): > 10 counts/s
- Global Dynamic Range (diffuse source):   > 200,000 counts/s
- Time resolution:               $\leq$ 20 ms
- Dark noise:                              < 4·10$^{-5}$ counts/pixel/s at 20°C
- Large scale non-uniformity              < 10%
- Solar Blind:                            DQE<0.1 % @ $\lambda \geq$ 400nm

## 4.1.3 Detector Read-Out System Trade-Off: MCP/WSA vs. MCP/PC-ICCD

### 4.1.3.1 Readout Systems Overview

Each detector that uses MCPs consists of three parts:

1) a photocathode – to convert photons in primary photoelectrons

2) a stack of MCPs – generally 2 or 3 to amplify the initial single electron into electron pulse of suitable amplitude

3) a readout device – a mechanism to detect the electron avalanche.

There are a large variety of readout systems for MCP's. Since none of them fully exploits all the characteristics of the MCP, a trade-off process has to be done considering the particular application for which the detector will be used.

A first classification can be done into two main groups:

1. optical readout
2. electronic readout

With optical readouts, the electrons cloud produced by the MCPs is converted into optical photons by a phosphor screen; an image sensor (CCD, CID or APS) is then used to detect the light emitted by the screen

Electronics readout, instead, are based on the direct detection of the electrons. These devices can be subdivided into:

2a. discrete anodes
2b. continuous position sensors

Discrete anode schemes include simple multi-anode pad arrays, CODACON and MAMA.

Continuous position sensors can operate by resistive, or conductive, charge division, or by signal timing methods. Charge division schemes include the resistive anode, the wedge & strip, Vernier anodes and the crossed wire encoders. Signal timing methods include the planar, multilayer and wire wound delay lines.

Next paragraphs will focus on the kinds of devices that at this moment seems available and providing reasonable performance: PC-ICCD and Wedge & Strip. Some notes about Delay Lines have been included but their availability is still under investigation, since the only suppliers which guarantee suitable performance are USA companies, subject to ITAR.

### 4.1.3.1.1    Optical readout (PC-ICCD)

These devices detect single photons by imaging the light spot produced by the intensifier for each input photon.

The phosphor screen is optically coupled to an image sensor, generally a fast scan (because the readout time of a frame set the dead time), frame transfer CCD (but also CMOS APS or CID can be used). Coupling is obtained with a lens or a fiber optic taper (also providing the demagnification ratio required to match the CCD format to the phosphor screen).

Each frame is searched for light spots originated by photon events; center of gravity of each spot is then computed at sub-pixel accuracy, thus the space resolution can be higher than the physical size of the CCD pixel.

### 4.1.3.1.2    Wedge & Strip

Wedge & Strip anode is a charge division readout, using charge measurement for position determination. The anode is subdivided in three (or more) segments, opportunely shaped. The center of mass of the charge cloud can be deduced from the ratio of charge on the anode segments.

### 4.1.3.1.3    Delay Line

Delay Line readouts determine the event position by measuring the relative time delay between signals propagated from the point of charge collection in either direction along the anode structure. The anode can either constitute the delay line itself, or may be connected to a tap on a remote delay line. Various geometries have been developed, two among the more popular being:

1) Wire wound delay line, formed by two helically wound wires constituting a transmission line.

2) Cross delay line (XDL), based on a crossed conductor layout, with period typically ~0.5mm. The electron cloud generated by the MCP intensifier is divided between upper and lower charge collectors (giving the two spatial coordinates). Event centroids are linearly proportional to signal arrival time difference at ends of delay lines connected at the edge of the collector anodes.

### 4.1.3.2 Detector characteristics with impact on the architecture choice

Characteristics affected by the choice of the architecture of the detectors are:

- Spatial resolution
- Time resolution
- Dynamic Range
    - LDR
    - GDR
- Geometric distortions
- Stability of geometry with flux

### 4.1.3.2.1    Spatial resolution

The MCP itself poses a limit to the spatial resolution, due to the finite size of the pores: resolution better than the pore pitch is obviously not possible. However, other mechanisms can worsen the performance before this limit, for example the proximity focus of the photocathode (if it is deposited on the entrance window) or the readout system.

The highest spatial resolution can be obtained with PC-ICCD: MCP pores can be clearly resolved, obtaining a resolution intrinsic to this readout system of few $\mu$m FWHM.

With W&S, position resolution is usually dominated by the electronic noise, which typically reaches a minimum at a shaping time of ~ 10 μs. With optimal signal-to-noise ratio the relative position resolution is generally limited to about 1:1000 (40 μm on 40 mm active area diameter), which does not meet the requirements.

With delay line (in particular, with XDL), pore limited resolution has been proved. With a 40 mm detector, 20 μm resolution could be feasible.

### 4.1.3.2.2    Time resolution

The intrinsic time resolution of MCPs is very high (~ 100 ps – 1 ns, even better with readouts optimized for that), so that the time resolution of the detector will depend essentially on the readout system and telemetry constraints.

For detector with optical readout, the time resolution is limited by the frame rate of the optical sensor: 30-1000 frame/s will translate, respectively, in 1-30 ms time resolution.

For detector with anode readout, the time resolution is usually in the μs range.

### 4.1.3.2.3    Dynamic range (LDR, GDR)

Microchannel plates are known to exhibit saturation effects due to the finite recharge time of MCP channels. Readout systems too have generally limitations on the maximum count rate.

As a result, the dynamic range of MCP-based detectors is dependent upon the scene to be imaged (i.e. on the size of the illuminated area). Two parameters are generally used:

– local dynamic range (LDR), quantified as the maximum count rate from a point like source at which the deviation from linearity of the detector response is less than 10%

– global dynamic range (GDR), quantified as the maximum count rate on the overall detector surface under uniform illumination (at 10% deviation from the linear response)

Typical values are:  LDR ~ 100 counts/s, GDR ~ several MHz.

#### 4.1.3.2.3.1    PC-ICCD

The bright limit on the dynamic range in this kind of photon-counting detectors is, in principle, governed by a number of factors:

1. frame rate of the sensor

2. size of an event on the CCD (which is related to the geometrical/electrical characteristics of the phosphor gap and on the demagnification ratio of the optical system between the intensifier and the sensor)

3. decay time of the phosphor

4. MCP saturation

Among these, generally the real limit is given by the first one (2 and 3 can be optimized, 4 can be dominant for the GDR). Generally, PC-ICCDs show high GDR but low LDR, being their weak point. The readout time of a CCD frame should avoid the confusion of multiple events at the same location, thus also the algorithm used for event recognition play a role.

With widely used algorithms and with optimized event size, the 10% loss of linearity on a point-like source occurs (as can be derived by Poissonian statistics) at $k=0.2*f$ ($k$ being the count rate and $f$ the frame rate). The requirement LDR > 10 counts/s translates in a request of frame rate > 50 frame/s.

The GDR limit (10% loss of linearity under diffuse illumination) can roughly be approximated by multiplying the LDR by $N/n$, where $n$ is the area in CCD pixels of a single event (typically n=9 pixels) and $N$ is the illuminated area (in CCD pixels). Assuming a 512x512 pixels CCD, the requirement GDR > 200,000 counts/s translates in a request of frame rate ≥ 35 frame/s.

### 4.1.3.2.3.2    WSA

With Wedge & Strip, the readout system itself introduces a limit to the maximum count rate, disregarding of the illumination being point-like or diffuse. In this case, the limit on the LDR is set by the saturation of the MCP, whereas the GDR is limited by the FEE (typically, count rate in the order of 100 KHz can be obtained).

However, W&S anodes are known for suffering from a tradeoff between count rate and spatial resolution, since high spatial resolution requires high SNR, which varies as the reciprocal of the square root of the electronic bandwidth: in order to improve the dynamic range, the time resolution should be increased, thus increasing the bandwidth and reducing the SNR, which results in a spatial resolution degradation.

### 4.1.3.2.3.3    Delay Lines

DL can achieve high spatial resolution and, due to the intrinsically fast nature of timing pulses, can operate at higher count rate than readout based on charge amplitude measurements. In fact, since there is no charge division encoding, and the anode propagation times are usually from 10ns to 100ns, high counting rates may be accommodated (~1Mhz). 20 $\mu$m on a 40 mm detector can be obtained even at high count rate.

### 4.1.3.2.4    Geometric distortion (and dependence on the count rate)

### 4.1.3.2.4.1    PC-ICCD

PC-ICCD benefits from a fixed (physically defined) pixel readout with inherent image stability. In terms of large scale geometric distortion and stability they are similar to CCD cameras.

However, photon event location is evaluated to sub-pixel accuracy by means of a centre of gravity algorithm and the choice to limit the sampling of photon event footprint to a few pixels and to employ a truncated weighting algorithm (to optimize the dynamic range) is known to introduce systematic shifts of the computed event coordinates with respect to the true ones. The overall effect of this mislocation of events results in a fixed pattern modulation in the accumulated image with the spatial frequency of the CCD pixels (produced by a distortion of the original image at sub-pixel scale). A correction can be applied to reduce this effect, but some residual nonlinearity on the pixel dimension could remain, due to the variability in the light spot which would require a spatially-variant correction.

### 4.1.3.2.4.2    WSA and DL

With WSA, geometric distortion on large scale (e.g. 1 mm on 25 mm size) is expected due to imperfection in W&S pattern. This distortion, however, is fixed and can be calibrated.

More troublesome is the fact that, as any devices based on continuous position sensors (for example Delay Lines), it is susceptible to variations in the correlation between the measured parameter, which is related to the spatial coordinate, and the reported digital value. A continuous readout computes the centroid of the charge cloud in analog circuitry, then the analog value is digitized, but the relationship between physical space and the reported pixel value is affected by outside influences including temperature, noise on the ground and power lines, and count rate.

### 4.1.3.3 Summary

In conclusion:

W&S provides high LDR, but (in the configuration currently available) the spatial resolution does not meet the requirements. Moreover, there are some concerns about stability at high count rate and geometric distortions;

PC-ICCD can guarantee the required spatial resolution and, with appropriate design, LDR and time resolution meeting the requirements can be obtained

Moreover, PC-ICCD needs a lower gain intensifier. Since the MCP degradation process is only dependent on the abstracted charge, the duration of the operational life will depends on:

1. the number of detected photons
2. the gain of the intensifier

Thus, readout systems working with lower gain, allow longer life of the detector.

For the reasons mentioned above, at the moment the PC-ICCD is considered as baseline for the FUV and NUV detectors.

Table 38. Summary of the performance of MCP-based detectors, for various readout systems (for the PC-ICCD, only the suppliers of the MCP intensifiers are listed; image sensors suppliers are listed in section 4.1.6)

| | MCP limit | PC-ICCD | W&S | DL |
|---|---|---|---|---|
| Spatial resolution | Pore pitch 12 μm (6 μm can have export license issues) | MCP limited (down to 2μm) | Limited by FEE noise 40 μm (on 40 mm) | 20 μm |
| Time resolution | ≤ ns | ~ ms | ~ μs | < μs |
| LDR | Depends on resistivity & gain High gain ~ 10 counts/p/s (~ 100 counts/s on PSF) Low gain ~ 300 counts/p/s (~ 3000 counts/s on PSF) | ~10 counts/s ~100 counts/s possible (and higher LDR on reduced FOV) | MCP limited (~100 counts/s) | MCP limited (~100 counts/s) |
| GDR | Depends on resistivity & gain ≥ 10 MHz | ~ 0.5-3 MHz | Readout limited ~ 100 KHz | ~ 1 MHz |
| Geometric distortion | | Non linearity on small scale (few pixels) due to centroiding artifacts | ~1mm on 25 mm FOV For the most part can be corrected by calibration, but some dependence on the count rate and outside influences | Some dependence on the count rate and outside influences |
| Gain (→ life time) | | Moderately low gain 5x10$^5$ | Moderately high gain 10$^7$ | Moderately high gain 10$^7$ |
| Suppliers | | Photek, UK Photonis, Fr Hamamatsu, Jp | Photek, UK | Sensor Sciences, USA |

## 4.1.4 PC-ICCD System

A scheme of the PC-ICCD is shown in Figure 77. A photocathode, tailored on the wavelength range of interest, converts incoming photons into primary photoelectrons. If the operative range include only wavelength above 1050 Å, it is possible to use a window of appropriate material and the intensifier can be sealed. In this case, the photocathode can be deposited on the inner face of the input window (semitransparent) or directly on the surface of the first MCP (opaque). If the detector is in open configuration, only the second option is available.

The high-gain Micro-Channel Plate (MCP) image intensifier virtually converts each primary photoelectron into a luminous spot on a phosphor screen that preserves the (x,y) location of the event. A fast scan CCD camera, coupled to the phosphor screen by means of a fiber optic taper, which also provides matching between the MCP and the CCD format, subsequently detects the footprint of each event. Event centroid coordinates can be determined with sub-pixel accuracy, if geometry and electrical parameters are configured as to obtain a proper sampling of photon events by the CCD matrix. A high speed digital processing electronics unit is thus needed in order to process in real time the high amount of data generated by the digital camera, without limiting the detector dynamic.

The NUV and FUV detectors will be identical, except for the photocathode, reducing the cost. A sealed configuration will be adopted for both intensifiers, since this greatly reduces the complexity of the system at expense of introducing, for the FUV detector, a cut-off just below the Lyman-$\alpha$. In fact to operate in open configuration, some other items should be added:

- a vacuum housing with a door to be opened on orbit
- an ion pump to maintain the operating level of vacuum, with the related high voltage supply
- an electronics controller for both the door and the pump

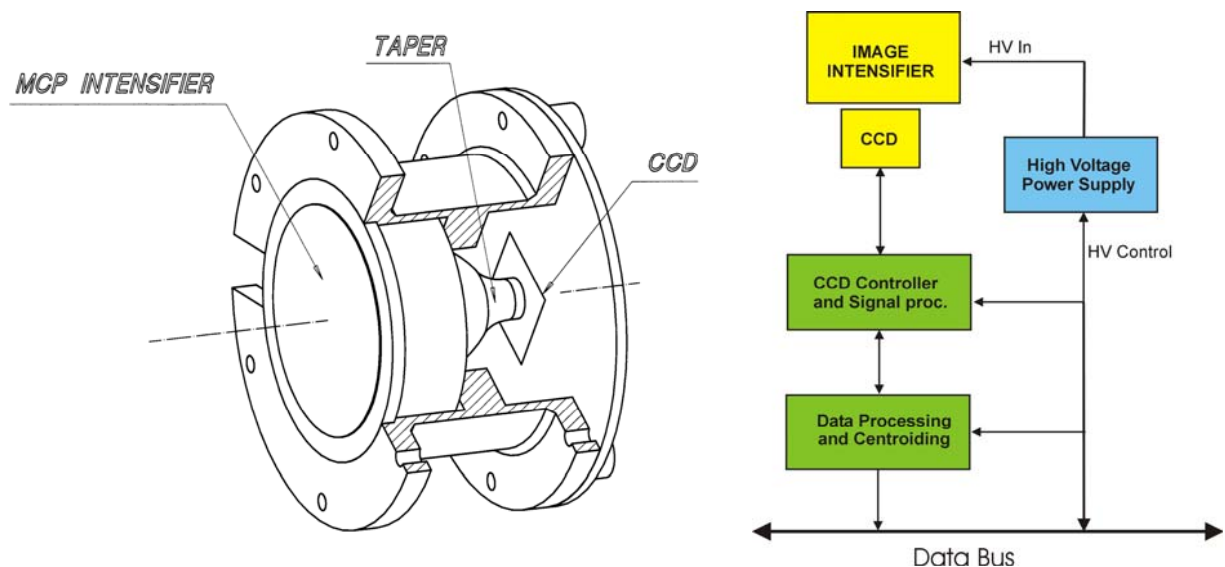

Figure 77. PC-ICCD head (left) and block diagram of the system (right).

## 4.1.5 MCP Intensifier

The drawing of a Microchannel Plate intensifier with active area diameter of 40 mm and space qualified housing is shown in Figure 78.

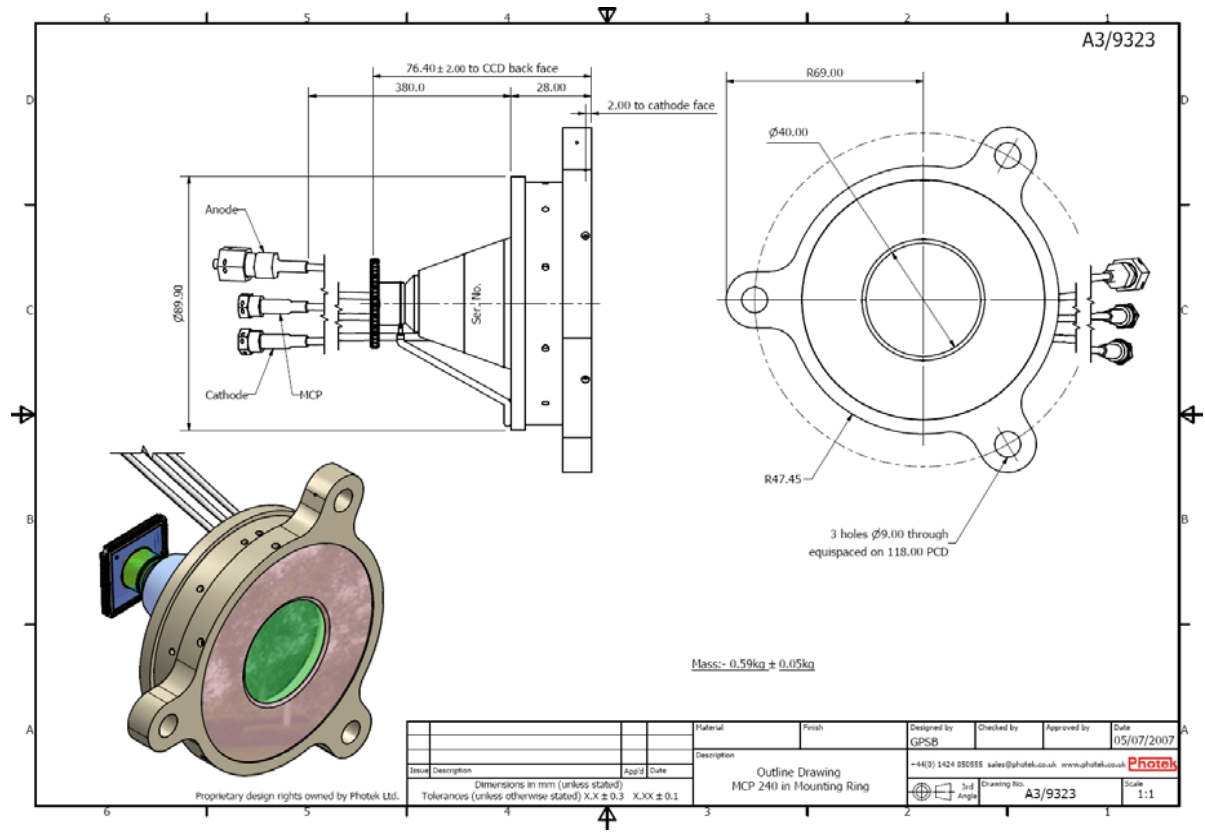

Figure 78. Space qualified MCP intensifier from Photek.

### 4.1.5.1 Window

Transmission curves of window materials are shown in Figure 79, cut-off wavelength are reported in Table 39.

A $MgF_2$ will be used for both the NUV and FUV intensifier since:

- The cut-off $\lambda$ is short enough to provide reasonable transmission in the FUV spectral domain, as specified in the TLR

- It also allows extension of the NUV detector range at shorter wavelength in order to guarantee overlap of the two channels

- It is a more practical material than LiF from the point of view of long-term stability; in fact LiF is not particularly stable: its transmission suffers exposition to damp air and it is susceptible to radiation damage (see e.g. [C.Coleman, Appl.Opt., vol.20, p.3693 (1981)])

Note that $MgF_2$ also can be susceptible to cosmic-ray induced phosphorescence, resulting in increased dark count: windows should be screened before launch to avoid this problem. An example of this effect is given by the high dark count rate of the NUV MAMA detector on HST-STIS (R.A. Kimble et al., ApJ, 492,L83L93, 1998): due to an error in the ground testing, a window with poor characteristics of phosphorescence has been selected: SAA particles were able to excite metastable states in impurities in the window and the subsequent phosphorescent de-excitation of these states produced UV photons that are detected by the MAMA photocathode, producing a background roughly an order of magnitude higher than the specifications.

Window thickness will be 5mm.

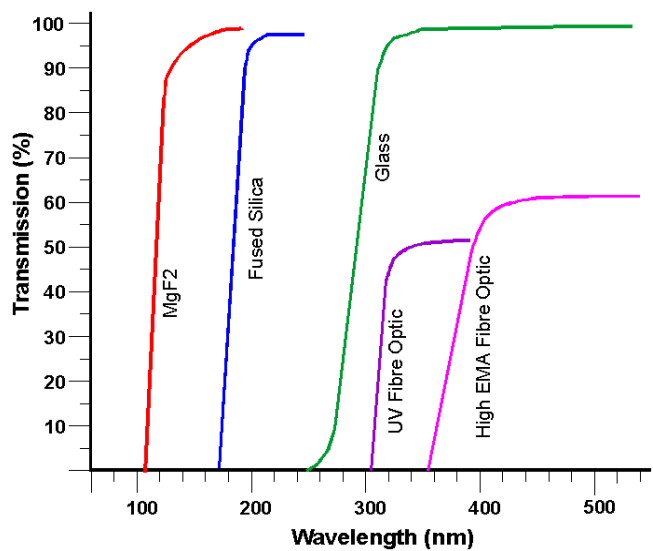

Figure 79. UV transmission of windows of various materials [Photek].

Table 39. UV-Cut-Off of window materials [Siegmund et al. - Appl.Opt. 26 (1987) 3607]

| Material | $SiO_2$ | $Al_2O_3$ | $BaF_2$ | $CaF_2$ | $MgF_2$ | LiF |
|---|---|---|---|---|---|---|
| UV Cut-Off | 1595 Å | 1435 Å | 1335 Å | 1230 Å | 1130 Å | 1050 Å |

### 4.1.5.2 Photocathodes

For both channels, a solar blind photocathode will be used, in order to limit the problem of red-leakage of the photometric filters.

#### 4.1.5.2.1    FUV Channel

Below 200 nm, the most used photocathodes are alkali halides. Table 40 lists the characteristics of the materials more suitable as photocathodes in the FUV.

Table 40. Photocathodes for the FUV range. $-\lambda_s$ = threshold wavelength; $\lambda_a$ = first peak wavelength; $\lambda_b$ = wavelength of the first minimum (transition from 1 to 2 electrons); $\lambda_c$ = wavelength of the second peak.

[1] Siegmund et al. - Appl.Opt. 26 (1987) 3607
[2] Siegmund et al.- ESA SP-356 (1992) Proc.Photon detectors for space instrumentation
[3] Siegmund et al-  - Appl.Opt. 27 (1988) 4323

| Photocathode | $E_g$ (eV) | Ea (eV) | $\lambda_s$ (Å) | $\lambda_a$ (Å) | DQE (a) | $\lambda_b$ (Å) | DQE (b) | $\lambda_c$ (Å) | DQE (c) | ref. |
|---|---|---|---|---|---|---|---|---|---|---|
| CsI | 6.2 | 0.2 | 1950 | 1100 | ~28% | 800 | <20% | 670 | ~35% | [1] |
| KBr | 7.4 | 0.8 | 1510 | 1050 | ~40% | 750 | <20% | 450 | ~55% | [1] |
| NaBr | 7.7 | | 1610 | 1000 | >35% | 750 | ~25% | 575 | >40% | [2] |
| RbBr | 7.4 | 0.5 | 1560 | 1100 | >40% | 800 | ~20% | 600 | ~50% | [2] |
| KCl | 8.4 | 0.4 | 1400 | 900 | ~34% | 670 | ~15% | 450 | ~40% | [3] |

Among these, the more stable and commercially available are CsI and KBr. Although KBr has higher efficiency, its long wavelength cut-off is around 160 nm, shorter than the requirement for the FUV channel. For this reason, a CsI photocathode is the baseline choice for the FUV detector.

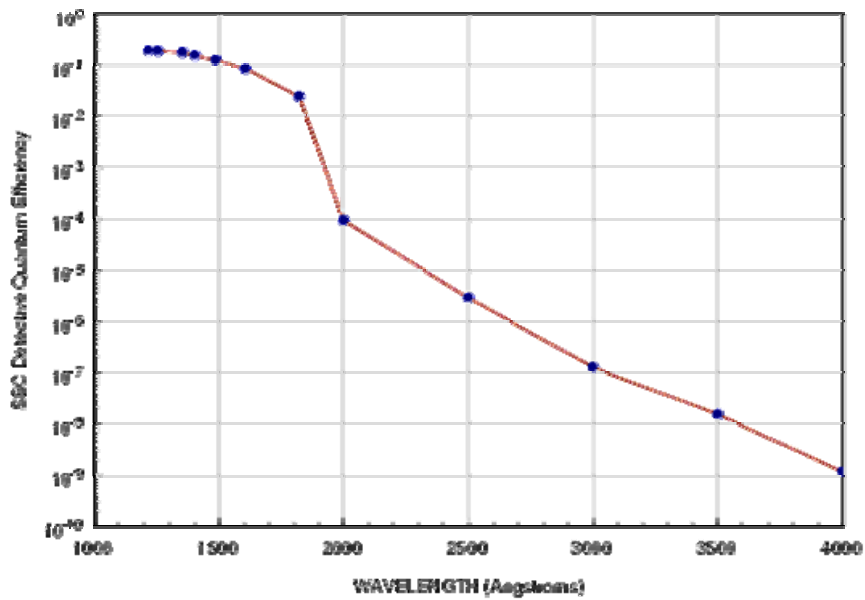

Figure 80. Detective Quantum Efficiency for the CsI coated ACS SBC MCP detector [ACS Instrument Handbook].

The photocathode will be deposited on top of the first MCP, in reflective configuration, because:

- the efficiency is considerably higher than in semitransparent configuration (with the photocathode deposited on the entrance window)
- the spatial resolution is not degraded by the proximity focus (see description for the NUV detector)

CsI coating can be provided by: Photek, Hamamatsu.

### 4.1.5.2.1.1   Repeller Grid

To improve the DQE a repeller grid should be used to repel electrons emitted away from the microchannel plate back into the channels (see

Figure 81). This provides a significantly increase in DQE at the price of a small enlargement of the detector PSF halo. For high resolution imaging, an observation mode with the repeller grid off could be implemented.

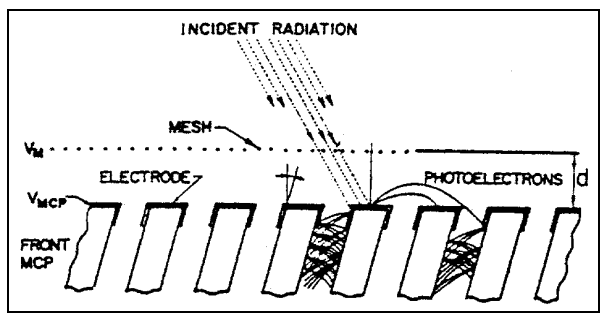

Figure 81. Repeller Grid.

#### 4.1.5.2.2    NUV Channel

The most common, and practically the only used, solar blind materials in the 200-300 nm range are $Cs_2Te$ and $Rb_2Te$, which exhibit very similar response, with the former having a longer long wavelength limit (300 nm vs. 250 nm).

In order to provide better overlap with the UVO channel, a $Cs_2Te$ photocathode will be used.

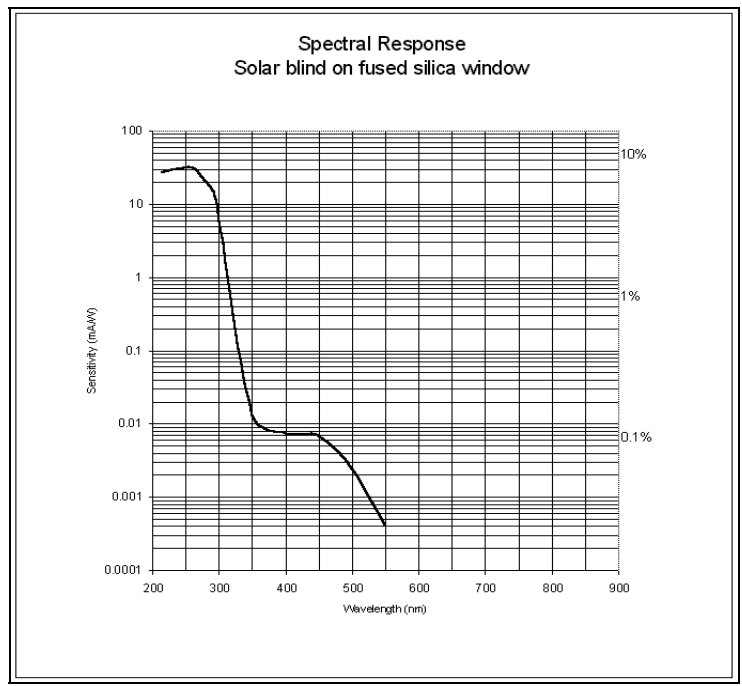

Figure 82. $Cs_2Te$ spectral response [Photek].

$Cs_2Te$ coating can be provided by: Photek, Photonis, Hamamatsu.

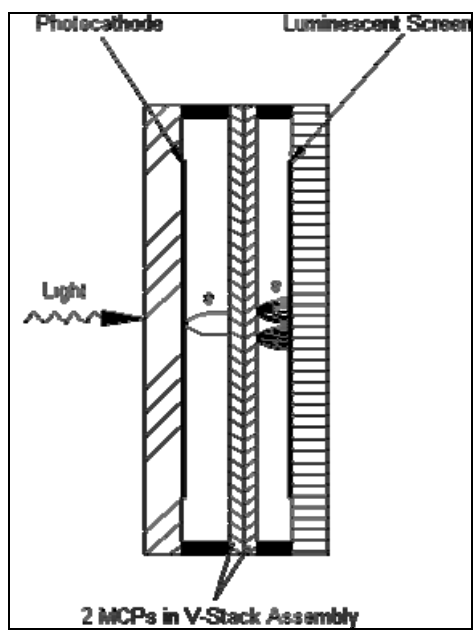

Figure 83. Intensifier in proximity focus configuration.

Due to chemical incompatibility between the MCP lead glass and the photocathode materials, in the NUV the photocathode has to be used in proximity focus configuration (deposited on the inner face of the window, separated by a narrow gap from the entrance face of the first MCP).The PSF will be dominated by the lateral dispersion of the primary photoelectrons traveling in the gap between the photocathode and the first MCP (Figure 83).

Some optimization of the gap parameters is possible (narrower gap and/or higher electric field along it can improve the PSF), but the electric field should not exceed 5 kV/mm to avoid the risk of high voltage breakdown between electrodes. Moreover, a too narrow gap is not suitable for space applications, because it does not have the mechanical stability to endure the vibration during the launch.

### 4.1.5.3 MCP stack

The MCP stack should provide a gain of about $\sim 5 \cdot 10^5$ e$^-$/photoelectron (including the phosphor screen efficiency, this should give $\sim 10^7$ optical photons/photoelectron)

The MCP stack will consist of 2 plates in a Chevron configuration with:

- circular format of 40 mm
- 2 MCP in Chevron configuration
- first MCP:
  - o  Aspect ratio              L/D: 50:1
  - o  Pore size:                10 μm pore on 12 μm pitch
- second MCP:
  - o  Aspect ratio              L/D: 80:1
  - o  Pore size:                12 μm pore on 15 μm pitch
- Pulse Height Distribution:     < 125% FWHM
- Bias angle:                    15° (TBC)
- PHD spatial variation:    <10% peak-to-peak
- Local count rate:              > 300 counts/s (@loss of linearity 10%, spot FWHM=40 μm)

The two different microchannel pitches are chosen to reduce fixed pattern noise in the resultant flat field. This pattern can be caused by Moiré beating between the hexagonal arrays of microchannels of the different plates in the electron amplification process [J.V. Vallerga, et al. Nucl. Instr. and Meth. A 310 (1991), 317]. In fact, due to the high spatial resolution requested for these detectors, close to the size of the channels pattern, the Moiré beat pattern could be an issue.

Low resistivity MCPs have to be considered. MCPs with lower resistance and/or operated at lower gain exhibit higher dynamic range, since the saturation is produced when the output current (proportional to the gain) is a fraction of few per cent of the strip current (inversely proportional to the MCP resistivity): in this case the strip current is unable to restore the neutrality of the wall of the channel after each impulse.

#### 4.1.5.3.1    Lifetime

The mechanism responsible for the loss of electron gain is believed to be a decrease in the secondary electron emission coefficient of the MCP glass, due to the removal of elements responsible for the conducting properties of the glass channels by electron scrubbing.

During the first part of the life of a MCP, the gain drops quickly with the extracted charge, then reaches a plateau. Thus MCP should be subject to scrub before the launch, in order to reduce the gain variation during operations.

The useful lifetime of any MCP detector can be extended by several orders of magnitude simply by progressively raising the plate bias voltages to compensate for the ongoing decrease in gain, but this procedure cannot be continued indefinitely. Typically the device could operate until the extracted charge is ~ 50 C/cm$^2$.

For space missions with several years of planned operations this can be an issue.

Since this degradation process is only dependent on the extracted charge, the duration of the operational life will depends on:

1)    number of detected photons

2)    gain of the intensifier

Since the PC-ICCD requires a low gain ($5 \cdot 10^5$ e$^-$/photoelectron), a higher number of photons can be detected during the detector life:  50 C/cm$^2$ corresponds to a total of $6 \cdot 10^{14}$ photons/cm$^2$. Even considering a total number of photon one order of magnitude less on a time span of 7 years, this corresponds to a continuous flux higher than the maximum tolerable by the detector.

### 4.1.5.4 Phosphor screen

Since the phosphor screen should be coupled to a fast scan camera, with integration times on the ms range, a fast phosphor should be selected (decay time < 1 $\mu$s). The decay time is a more critical parameter than the matching between the phosphor emission spectrum and the sensitivity response of the image sensor, since a low matching can be compensated by slightly raising the MCP stack gain. However, with phosphors with similar decay constant, preference should be accorded to the highest efficiency.

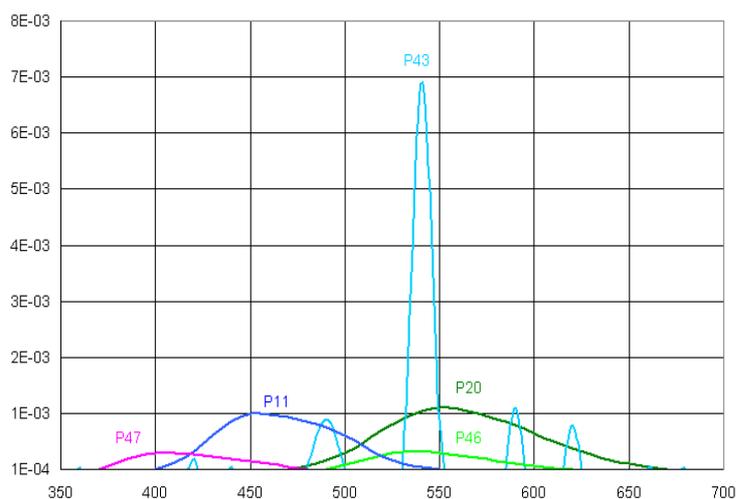
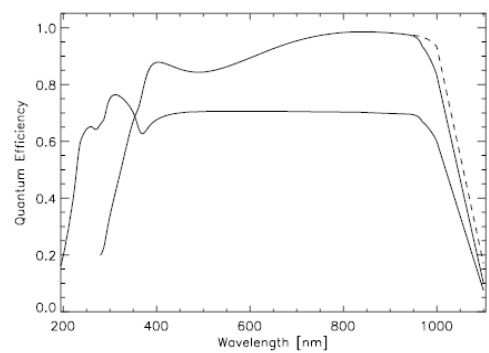
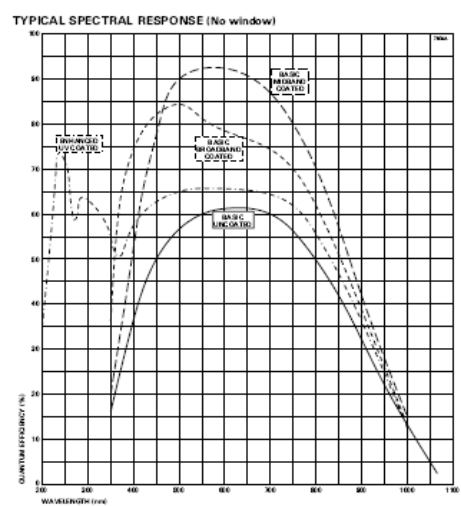

Figure 84. Left: Phosphor emission spectra. Right: efficiency of image sensors (top: pnCCD from pnsensor; bottom: CCD from e2v).

Table 41. Phosphor characteristics [Photek].

| Type ( Colour) | Screen efficiency % ( Optical Watts / Electrical Watt ) | Peak Wavelength (nm) | Photons / Electron @ 5 kV | Decay Characteristic |
|---|---|---|---|---|
| P11 (Blue) | 5.9 | 446 | 120 | Fast initial decay with long decay at low level. 50 ms to 1% |
| P20 (Green) Fast | 12[(1)] | 540 | 320 | Fast initial decay with long decay at low level. 1 ms to 1% |
| P24 (Blue) | 5.0 | 500 | 120 | 10 ms to 10% |
| P31 (Green) | - | 550 | - | 1 ms, good exponential |
| P43 (Green) | 8.7 | 548 | 240 | 1.2 ms/decade, true exponential |
| P46 (Green) | 1.8 | 530 | 55 | 300 ns |
| P47 (Blue) | 3 | 410 | 64 | 80 ns |
| FS | 4.2 | 513 668 768 | 96 | 12 $\mu$s to 10 % |
| GOS* | 0.7 | 420 | - | 50 ns to 10 % 110 ns to 1 % |

Under these considerations, P46 and P47 are both suitable (Figure 85 shows measurements of the decay of these two materials). P46 should be considered as baseline due to the higher efficiency.

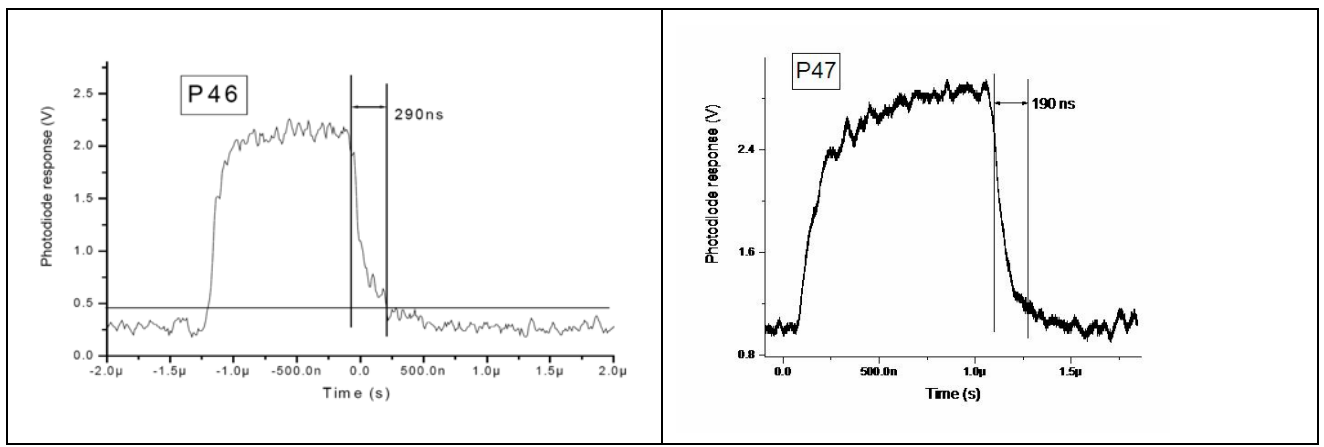

Figure 85. P46 and P47 Decay [Photek measurements].

## 4.1.6 CCD Readout Sensor

The image sensor to be used to read out the intensifier is a critical component, since its characteristics have a strong impact on the performance of the detector.

- Key characteristics that have to be evaluated in the sensor choice are:

- Frame rate: > 50 Hz (higher is better, because the frame rate limits the dynamic range of the detector)

- Active pixels: ≥ 256 x 256

- Fill factor: fill factor affects the subpixel centroiding interpolation; even if a lower fill factor could not preclude a moderate subpixel resolution (see e.g. M.Uslenghi et al., SPIE, Vol. 4854, pp. 583, 2003), 100% would be preferable

- Active area size: preferably in the range 10-40 mm; the different size of the image sensor and the intensifier requires a reducer fiber optic taper for optically coupling the two devices. Standard reducers are available (for example from Schott AG)  with a ratio up to 4:1, because the efficiency drop as the ratio increases and for higher ratios a lens-based optical system is more efficient (but less mechanically stable and more cumbersome)

- Frame Transfer time (FT CCD only): the time necessary to transfer the frame in the storage area should be as short as possible, being a dead time for the detector; in particular, the ratio between this time and the frame period should be no more than ∼ 10%

Moreover, due the critical nature of these devices, it would be preferable an ITAR-free sensor, to avoid exportation license issues (thus excluding the most part of sensors produced in US).

Secondary parameters are:

- Readout noise: the high level of the signal to be read out allows to overcome even a moderately high readout noise

- Full well capacity: it should guarantee an appropriate dynamic range (at least 8 bit)

- Dark current: due to the short integration time, dark current is not an issue

Table 42 lists fast image sensors with appropriate characteristics, available from suppliers with heritage in space applications. Most of them are space qualified or available in space qualified form, under request, in reasonable time. Table 43 shows the relevant characteristics of particular configurations of these sensors (using on-chip binning, if it is the case) fulfilling (or close to) the requirements.

Table 42 Fast image sensors. (1) Each column of this CCD should be readout independently. Asic, implementing the front-end electronics and multiplexing of columns are available (each Asic includes 132 channels, multiplexed on 2 output lines). (*) space qualified devices can be provided under request.

|  | device | technology | Active pixels | architecture | pixel size (μm) | Active area size (mm) | fill factor | Readout freq./output | Output ports | SQ |
|---|---|---|---|---|---|---|---|---|---|---|
| Fairchild (USA) | Hybrid 2013 | CCD/CMOS hybrid | 1280x1024 | splitted FT | 12 | 12 | 100% | 10-65 MHz | 4 |  |
| Fairchild (USA) | CCD456 | CCD | 512x512 | Interline | 17 | 9 | 19% | 25 MHz | 32 |  |
| Fairchild (USA) | CCD417 | CCD | 512x512 | splitted FT | 15 | 7.7 | 100% | 30 MHz | 4 | Yes |
| Dalsa (USA) | FT50M | CCD | 1024x1024 | FT | 5.6 | 6 | 100% | 60 MHz | 2 | * |
| Dalsa | FTT1010 | CCD | 1024x1024 | FT | 12 | 12 | 100% | 40 MHz | 2 |  |

| | device | technology | Active pixels | architecture | pixel size (µm) | Active area size (mm) | fill factor | Readout freq./output | Output ports | SQ |
|---|---|---|---|---|---|---|---|---|---|---|
| (USA) | M | | | | | | | | | |
| Cypress (ex Fill Factory) | STAR250 | CMOS APS | 512x512 | | 25 | 12.8 | 35% | 8 MHz | 1 | Yes |
| Cypress (USA) | LUPA-1300 | CMOS APS | 1280x1024 | | 14 | 14 | 50% | 40 MHz | 16 | |
| Sarnoff (USA) | CCD180-1M-SFT | CCD | 1024x1024 | splitted FT | 18 | 9 | 100% | 5 MHz | 32 | * |
| Sarnoff (USA) | CCD180-512-SFT | CCD | 512x512 | splitted FT | 18 | 18 | 100% | 10 MHz | 16 | * |
| E2V (UK) | CCD48-20 | CCD | 1024x1024 | FT | 13 | 13 | 100% | >10 MHz | 2 | Yes |
| E2V (UK) | CCD67 | CCD | 256x256 | FT | 26 | 6.7 | 100% | 5 MHz | 2 | * |
| PNSensor (DE) | pnCCD | pnCCD | 256x256 | splitted FT | 36-150 | 9-40 | 100% | up to 1 MHz | 512/8 (1) | Yes |

Table 43 Possible configurations using the sensors listed in the previous table.

| | sensor | binning | active pixels | centr.res. | pixel size (µm) | frame rate (Hz) | LDR (cts/s) | dead time | time res. |
|---|---|---|---|---|---|---|---|---|---|
| pnsensor | pnCCD | No | 256x256 | 8x | 36-150 | 1000 | 200 | 3,0% | 1 ms |
| sarnoff | CCD180-512-SFT | No | 512x512 | 4x | 18 | 500 | 100 | 3,6% | 2 ms |
| sarnoff | CCD180-512-SFT | 2x2 | 256x256 | 8x | 36 | 1000 | 200 | 7,2% | 1 ms |
| sarnoff | CCD180-1M-SFT | No | 1024x1024 | 2x | 18 | 150 | 30 | 7,6% | 6.7 ms |
| sarnoff | CCD180-1M-SFT | 2x2 | 512x512 | 4x | 36 | 300 | 60 | 15,2% | 3.3 ms |
| E2V | CCD48-20 | 4x4 | 256x256 | 8x | 52 | 80 | 16 | 16,0% | 12.5 ms |
| E2V | CCD67 | no | 256x256 | 8x | 26 | 150 | 30 | - | 6.7 ms |
| Dalsa | FT50M | No | 1024x1024 | 2x | 5.6 | 100 | 20 | 1,0% | 10 ms |
| Dalsa | FT50M | 2x2 | 512x512 | 4x | 11.2 | 200 | 40 | 2,0% | 5 ms |
| Dalsa | FT50M | 4x4 | 256x256 | 8x | 22.4 | 400 | 80 | 4,0% | 2.5 ms |
| Dalsa | FTT1010M | No | 1024x1024 | 2x | 12 | 60 | 12 | 6,0% | 16.7 ms |
| Dalsa | FTT1010M | 2x2 | 512x512 | 4x | 24 | 120 | 24 | 12,0% | 8.3 ms |
| Dalsa | FTT1010M | 4x4 | 256x256 | 8x | 48 | 240 | 48 | 24,0% | 4.2 ms |
| Fairchild | Hybrid 2013 | No | 1280x1024 | 2x | 12 | 200 | 40 | - | 5 ms |
| Fairchild | CCD456 | No | 512x512 | 4x | 17 | 1000 | 200 | 0,0% | 1 ms |
| Fairchild | CCD456 | 2x2 | 256x256 | 8x | 34 | 2000 | 400 | 0,0% | 0.5 ms |

| | sensor | binning | active pixels | centr.res. | pixel size (μm) | frame rate (Hz) | LDR (cts/s) | dead time | time res. |
|---|---|---|---|---|---|---|---|---|---|
| Fairchild | CCD417 | No | 512x512 | 4x | 15 | 500 | 100 | - | 2 ms |
| Fairchild | CCD417 | 2x2 | 256x256 | 8x | 30 | 1000 | 200 | - | 1 ms |
| Cypress | STAR250 | No | 512x512 | 4x | 25 | 30 | 6 | 0,0% | 33 ms |

Among these possibilities, one of the most interesting is the pnCCD by pnsensor (see Figure 86), based on the same technology used for XMM x-ray pnCCD (in fact pnsensor commercialize the technology developed by the Semiconductor Laboratory of the Max-Planck-Institutes/MPI-HLL). They can reach very high frame rate (due to the highly parallel readout architecture: each column is readout independently, no readout channel is present) with low dead time (due to their splitted Frame Transfer architecture). 1000 frames/s with 256x256 pixels are obtainable, allowing LDR = 200 counts/s. Space qualified ASIC are available for the front-end electronics (CAMEX).

Good performance could also be obtained with more classical CCD, like the Sarnoff CCD180-512-SFT (Figure 87). Also this device is a split Frame Transfer CCD, with 16 output ports. Without on-chip binning, it can reach 500 frames/s with 512x512 pixels, allowing LDR = 100 counts/s.

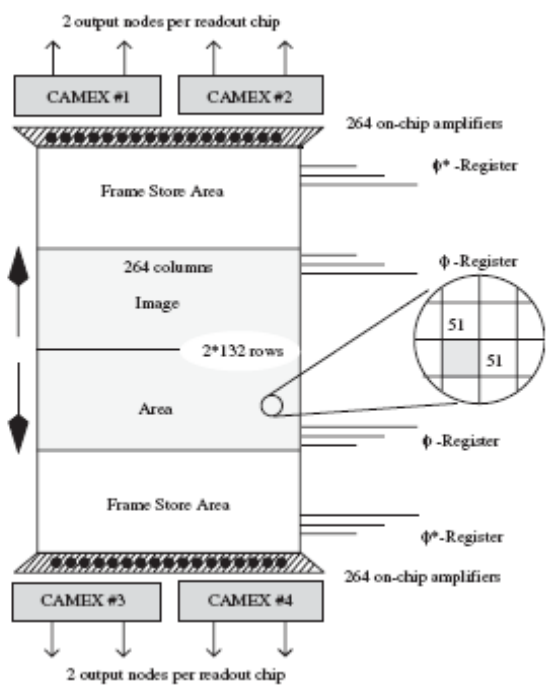 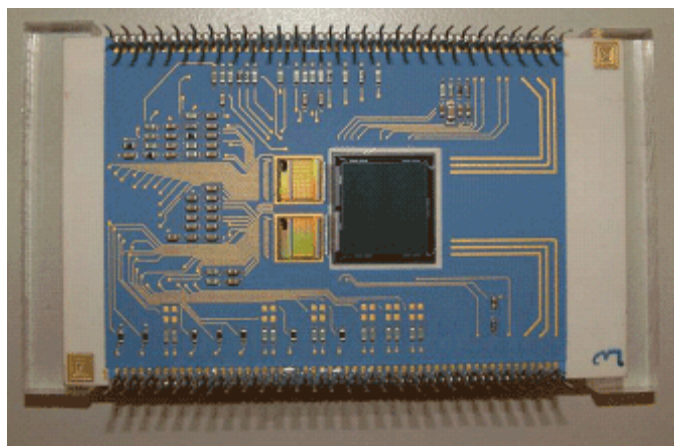

Figure 86. Left: schematic layout of the split-frame-transfer pnCCD for high-speed imaging [R.Hartmann et al., NIM A 568, p.118, 2006]. Two CAMEX chips are placed adjacently on each readout side of the detector. Right: picture of a board with non FT pnCCD and 2 CAMEX devices.

## 4.1.7 PC-ICCD Front-End Electronics

The detector electronics include two main blocks:

an Analog Front-End Electronics (AFEE) an high speed front-end CCD electronics

a Digital Front-End Electronics (DFEE) with a real time data processing unit, which will acquire the data from the CCD camera, search for the photon events and computes the coordinates of the detected photons with sub-pixel accuracy

Filters for the High Voltage Power Supply should also be located close to the detectors.

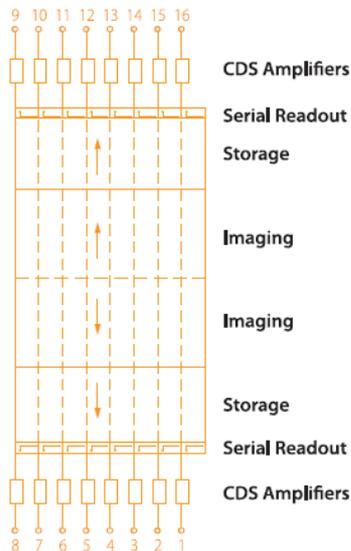 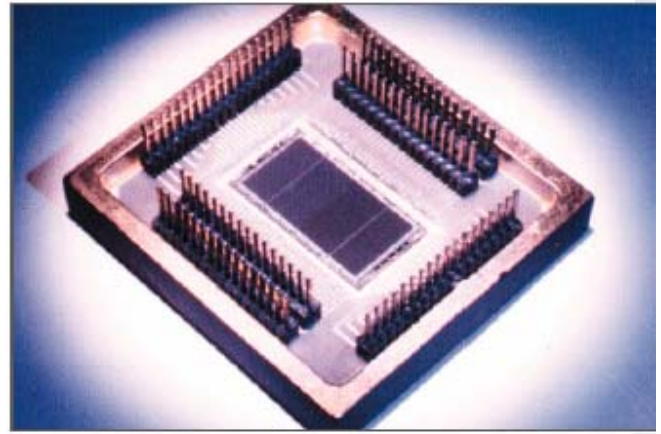

Figure 87. Sarnoff CCD180-512.

### 4.1.7.1 AFEE

The diagram for the CCD electronics is in Figure 88. Due to the high speed, the Correlated Double Sampling (CDS) will be not implemented. The clock drivers and the first stage differential amplifiers should be accommodated on the CCD board to reduce the noise. The analog to digital conversion is carried out by 8 bit ADC. The controller can be implemented in the same FPGA as the Digital Processing Unit.

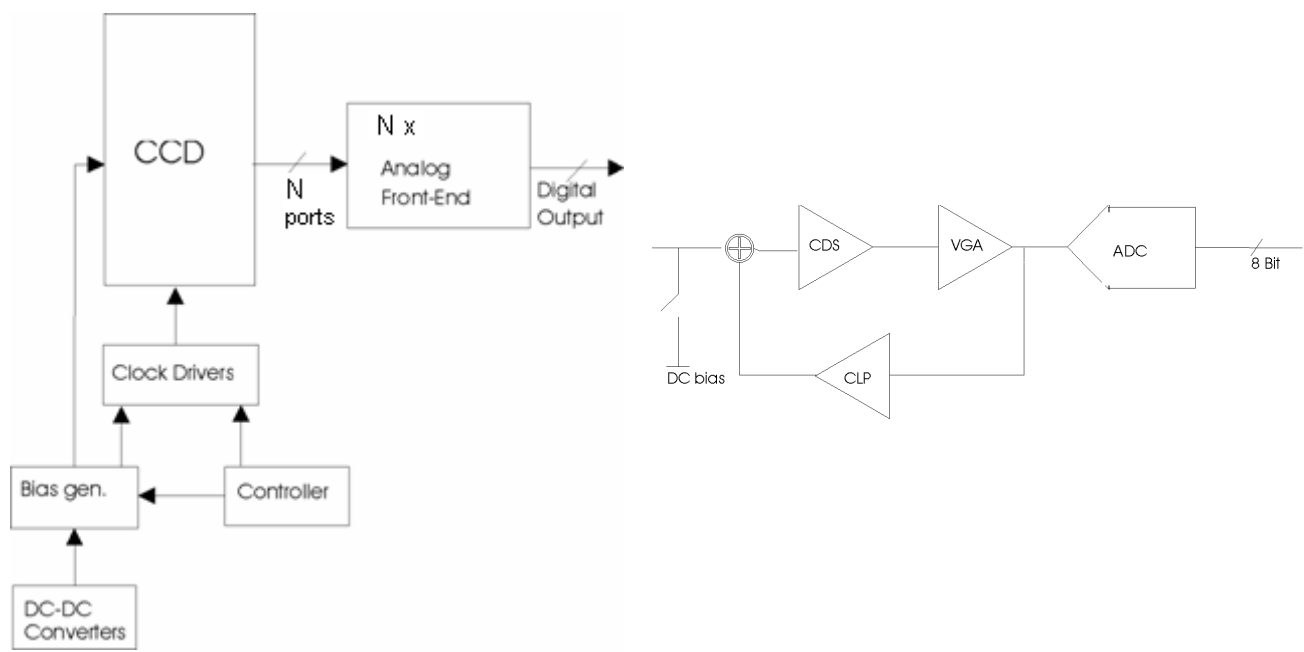

Figure 88. CCD Front-End Electronics. Left: block diagrams. Right: Analog Signal Processing: 1 chain

In the case of the pnCCD the number of analog signal processing chains is 512. However, ASIC devices (CAMEX) are available. Each chip implements 128 analog signal processing chains and multiplexing to 2 output lines, resulting in 8 analog lines to be digitalized.

### 4.1.7.2 DFEE

The DFEE will acquire serially the CCD 8-bit digitized output signals, along with three sync signals (frame, line and pixel sync). Through a proper system of delays, the system will generate 3×3 pixel windows that sweep dynamically the whole matrix at the pace of the pixel clock.  The processing system performs in parallel the following tasks on each window:

- check, according to appropriate discrimination and pile-up rejection criteria, for the presence of a charge distribution representing a photon event;

- compute the centroid coordinates, applying a centre of gravity algorithm to the charge distribution in the current window and correcting for systematic errors. The centroid coordinates of the charge distribution identified as a photon event are subsequently transferred to the ICU, along with the integrated amplitude of the event.

All these functions will be implemented in FPGA, adopting a pipeline architecture in order to process the data at the rate they are generated by the sensor, without introducing any limitation.

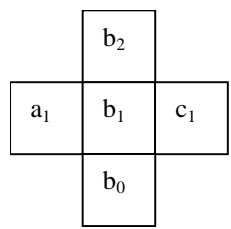

$$x_c = \frac{c_1 - a_1}{a_1 + b_1 + c_1}$$

$$y_c = \frac{b_2 - b_0}{b_0 + b_1 + b_2}$$

Figure 89. 3-point center of gravity algorithm.

## 4.1.8 Budgets

The mass and power budgets of NUV & FUV detectors are reported in Table 44 and in Table 45, respectively.

Table 44. FUV & NUV detectors mass budget

| (Kg) | NUV | | FUV | |
|---|---|---|---|---|
| | | max | | Max |
| Detector head | 0.60 | 0.80 | 0.60 | 0.80 |
| Front-end electronics | 2.00 | 2.25 | 2.00 | 2.25 |
| Harness | 0.50 | 0.75 | 0.50 | 0.75 |
| **Tot** | 3.10 | 3.80 | 3.10 | 3.80 |

Table 45. FUV & NUV detectors power budget

| (W) | NUV | | FUV | |
|---|---|---|---|---|
| | | max | | max |
| HVPS-MCP | 2.00 | 2.40 | 2.00 | 2.40 |
| CCD | 2.00 | 2.40 | 2.00 | 2.40 |
| CCD Electronics | 5.00 | 6.00 | 5.00 | 6.00 |
| Digital Front-end electronics | 5.00 | 6.00 | 5.00 | 6.00 |
| **Tot** | 14.00 | 16.80 | 14.00 | 16.80 |

## 4.2 UVO detector subsystem

## 4.2.1 CCD Detector Performance Specifications

The logical choice to comply with the requirements for the UVO detector given in Chapter II is a CCD detector.

The preliminary performance specifications for the UVO CCD derived from Chapter II are listed in Table 46.

Table 46. UVO channel CCD detector preliminary performance specifications

| Item | UVO CCD |
|---|---|
| Wavelength Range | 200 – 700 nm |
| Array Size | 4096 x 4096 pixels |

FCU phase A report – Optical, mechanical and electronics configurations

| Item | UVO CCD | |
|---|---|---|
| Detected Quantum Efficiency @ | | |
| 200 nm | $\geq 0.40$ | |
| 300 nm | $\geq 0.60$ | |
| 400 nm | $\geq 0.60$ | |
| 500 nm | $\geq 0.60$ | |
| 700 nm | $\geq 0.60$ | |
| Readout Noise | $\leq 3.0$ @ 200 kHz | |
| Dark rate, electrons/pixel/hour | $\leq 18$ | |
| Linearity over 50% of full well capacity. N=number of electrons | TBD | |
| Full Well Capacity, electrons | 350000 (NIMO) | 100000 (AIMO) |
| Dynamic Range, per readout | 65536 | 65536 |
| Pixel size | 15 x 15 microns | |
| Active Size | 61.4 X 61.4 mm | |
| Flat Field Non-uniformity Limit, peak-to-peak | $\pm 3.0\%$ | |
| Flat Field Stability Limit, peak-to-peak over one hour | $\pm 0.2\%$ | |
| Flat Field Stability Limit, peak-to-peak over one month | $\pm 0.1\%$ | |
| Spatial Resolution (FWHM) | 15 microns | |
| Charge Transfer Efficiency | $\geq 0.99999$ | |
| Residual Image, electrons | TBD | |
| Black & White Spots (pixels) | $< 800$ | |
| Maximum number of Column Defects | 5 | |
| Modulation Transfer Function (at 632.8 nm) | TBD | |
| Operational Temperature | $\leq -90C$ (NIMO) | $\leq -50C$ (AIMO) |
| Gain | 1,2,4,8 (NIMO) | 1,2,4 (AIMO) |
| Read Out Speed | 100 – 500 kHz | |

As it can be seen from the table there is still uncertainty on the kind of device, NIMO or AIMO that will be selected. This is mainly due to the method that will be chosen to cool the CCD and that is on the working temperature of the CCD. Higher working temperature implies AIMO CCD (lower dark current and smaller full well capacity), lower working temperature implies NIMO CCD.

CCD devices, space qualified, able to satisfy the requirements listed in Table 46 are commercially available. Figure 90 shows two examples of 4kx4k back illuminated CCD manufactured by e2v technologies (UK) and Fairchild Imaging (USA).

The response of the CCD detector shall be optimized in the UV wavelength range. Figure 91 shows a plot of the measured quantum efficiency of the two CCD chips (e2v technologies) of the WFC3 camera that will operate on board of HST next year. The high UV response is the result of specifically designed anti reflection coatings.

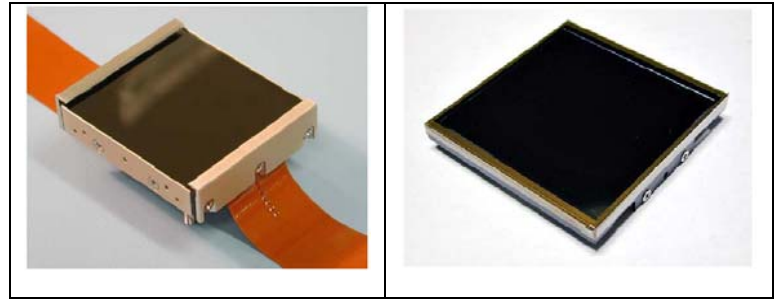

Figure 90. Left: e2v CCD203. Right: Fairchild CCD486

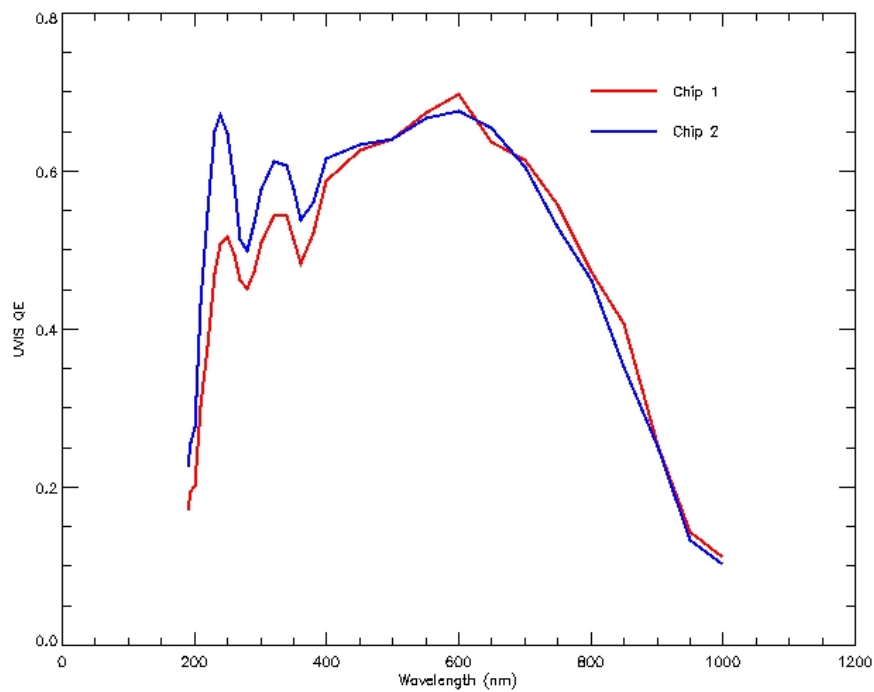

Figure 91. QE of HST/WFC3 CCD [www.stsci.edu/hst/wfc3/ins_performance/UVIS_sensitivity/ UVIS_QE.html].

## 4.2.2 CCD cooling system

Two different systems to cool down the CCD detector are under evaluation. An active cooling system based on TEC modules and a passive cooling system based on a radiator, external to the spacecraft. The adoption of the cooling system will impact the thermo-mechanical design of the detector head and then on the general power and mass budgets of the FCU instrument.

To perform the thermal analysis that will take to the final choice several boundary conditions have to be defined. The environmental temperature inside the SIC is 20±5 C (Moisheev et al. 2006). The CCD temperature can be in the range –60 C/ –50 C in case a AIMO CCD is used or in the range –100 C/ –90 C in case of a NIMO CCD. The total heat load on the CCD detector will be the result of the contribution of three sources of parasitic load: radiative heat transfer, conductive heat transfer, and power dissipation internal to the detector. Radiative heat transfer depends on the ambient temperature, the CCD temperature, the thermal properties of the used materials, and the sizes of the involved surfaces. Conductive heat loads originates from the cables containing the wires that connect the CCD to the FEE and from mounting supports for the detector. Finally the internal power dissipation originates from CCD preamps operations that is usually zero except during CCD readout. Figure 92 shows preliminary results for the heat transfer load and for internal power dissipation. The detailed analysis is in progress.

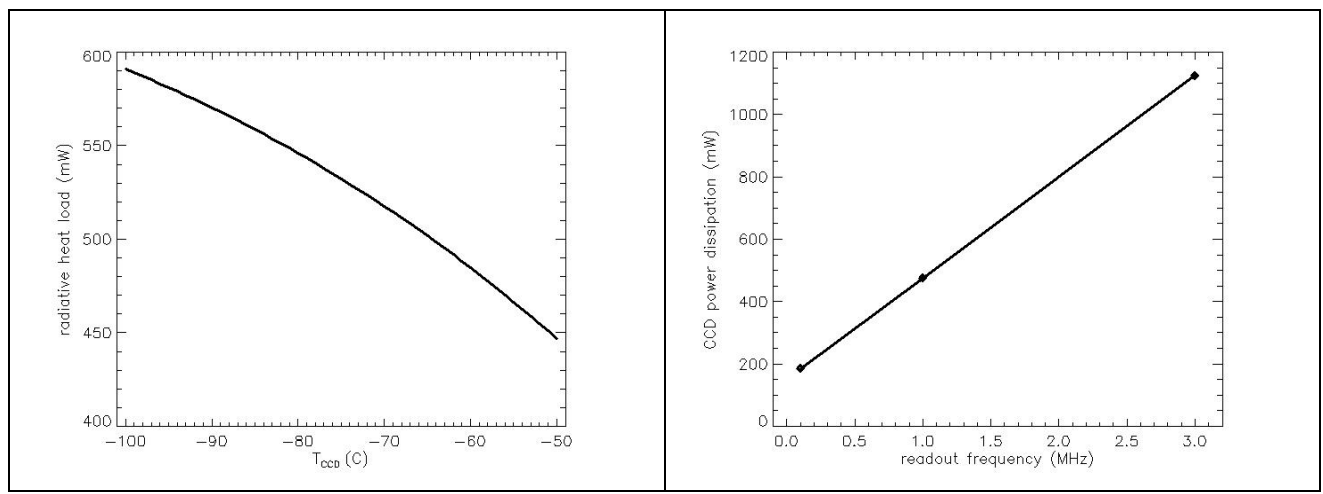

Figure 92. CCD parasitic loads. Left: radiative heat transfer. Right: internal power dissipation

### 4.2.2.1 Active cooling (TEC module)

Active cooling of the CCD detector can be achieved using a TEC module. To model and size (number of stages, dimensions of hot and cold faces) properly the TEC module, the temperature of the hot and cold side of the TEC and the heat load on the CCD detector must be defined. Fine regulation of the CCD temperature is obtained tuning properly the TEC current with the feedback of a temperature sensor.

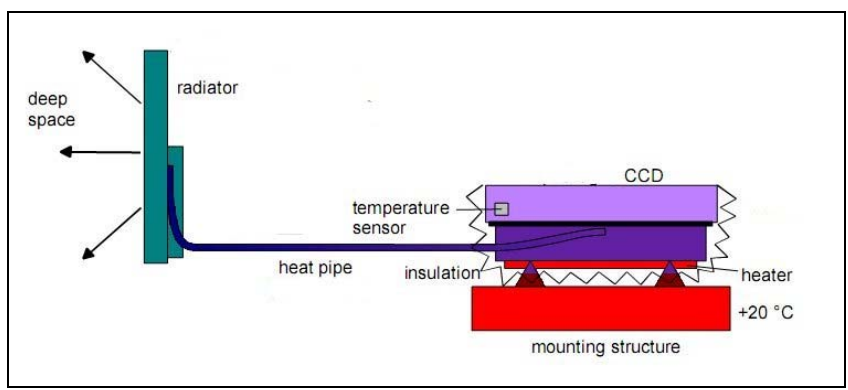

Figure 93. CCD passive cooling system (adapted from [RD3]).

### 4.2.2.2 Passive cooling (External radiator)

This method plans to use an external radiator kept at a temperature of –140 C (see Figure 93). An heat pipe connects the radiator to the CCD housing through an appropriate interface. The CCD temperature is then maintained at its working value using a heater and the feedback from a temperature sensor.

### 4.2.3  CCD Package Assembly

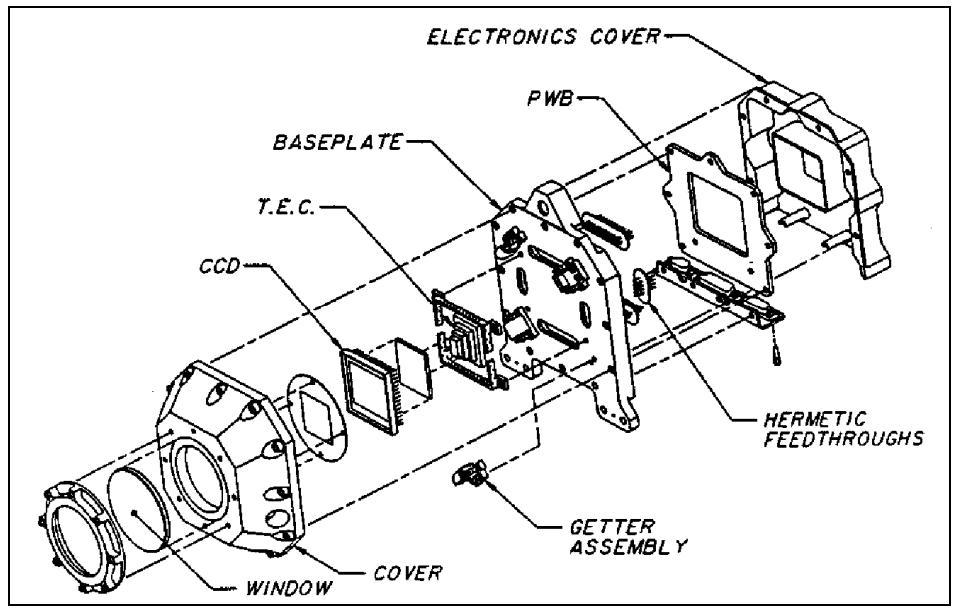

Figure 94. Exploded view of HST ACS/HRC CCD detector package. (Kimble et al.1994, SPIE, 2282, 169)

The design of the CCD package assembly or housing has not yet started. The main reason for this is, again, the choice of the detector cooling method. Figure 94 shows a possible design for the UVO CCD housing in case a TEC module will be used to cool down the CCD. The main components of the assembly will be:

A baseplate on which the detector and the TEC will mounted. The baseplate will have holes for connectors and will host getters to provide long term pumping against residual outgassing products.

A cover that will hold a flat MgF$_2$ window and baffles if necessary.

If the passive cooling method will be chosen, this design shall be modified removing all elements related to the TEC module and inserting a suitable support for the CCD and a system to exchange heat with the heat pipe connected to the external radiator.

In any case the material of cover and baseplate will be selected taking in consideration issues related to total mass and the shielding of the CCD.

### 4.2.4  CCD Detector Electronics

Figure 95 shows a block diagram of the UVO CCD electronics system. The main boards and their functions are the following:

Timing Board: Commands a CCD exposure to take place. Generates TTL level clocks that represent the CCD timing pattern. Generates biases digital data.

Clock Driver: translate the TTL level clocks to CCD clock voltages and send them to the CCD.

Bias Generator / Analog Signal Processing. Translate the biases digital data generating the bias voltages to operate the CCD. Digitize the signal from the CCD outputs to not less than 16-bit accuracy.

Low Noise Preamps. Receive the signal from each of the CCD outputs and send it to the analog signal processing board.

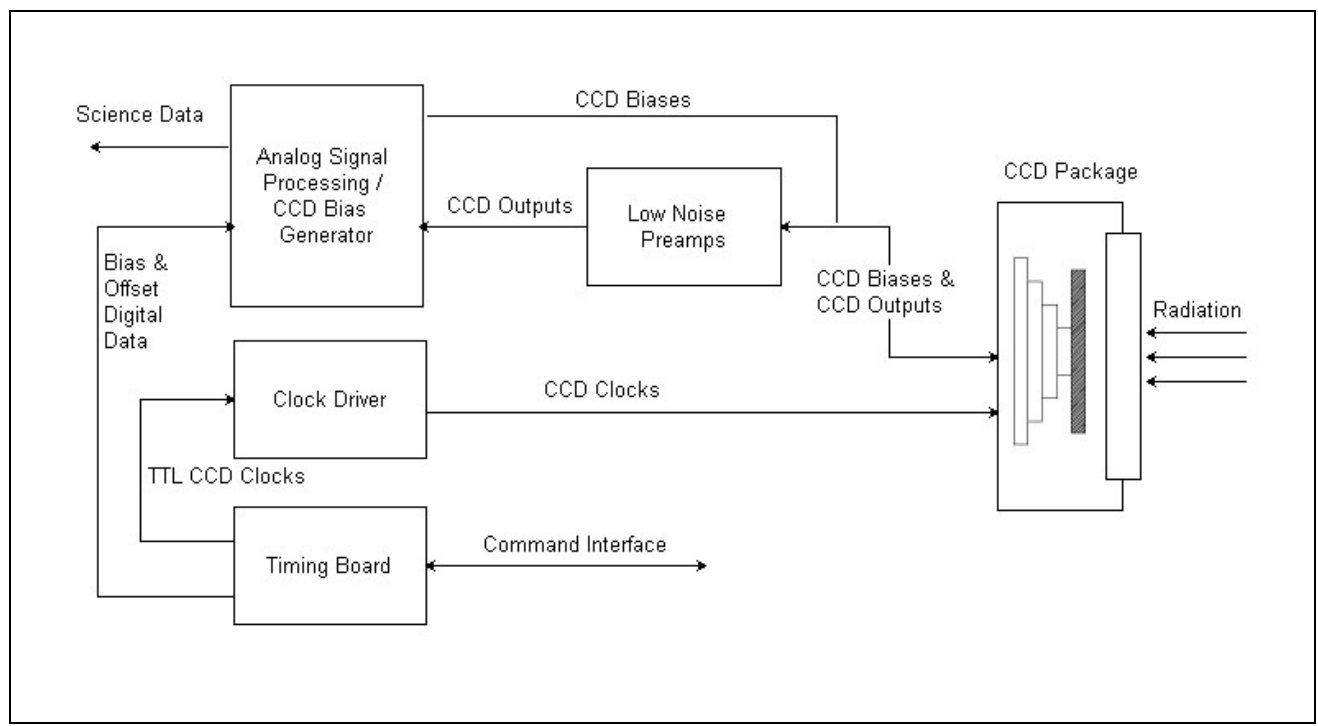

Figure 95. Preliminary schematics of FCU UVO CCD electronics

## 4.2.5 CCD Readouts

Selectable exposure times will be provided ranging from 0.5 sec (TBC) to 60 minutes. The rate to read out the full CCD array will be defined in the range 100 – 500 kHz. The CCD should have 4 output amplifiers. This implies a readout time varying according Table 47. The CCD will have also the capability for readout and storage subarrays with TBD requirements. Binning on chip (2x2) will be considered as an option to reduce the size of an image at expenses of spatial resolution. Different gains (1,2,4,8 e/ADU, TBC) will be provided to fully exploit the CCD well capacity. The number of gains will depend on the final choice (NIMO or AIMO) of the device.

Table 47. Full frame readout time of UVO CCD

|  | Readout Time (sec) | |
| --- | --- | --- |
|  | 100 kHz | 500kHz |
| Two amplifiers | 84 | 16 |
| Four amplifiers | 42 | 8 |

## 4.2.6 Budgets

The mass and power budgets of UVO detector are reported in Table 48 and Table 49, respectively. The TBC in the TEC row of the power budget table depends on the final choice of the CCD cooling system. If the passive cooling system will be selected no TEC module will be present.

Table 48. UVO CCD detector mass budget

| Item | Mass (kg) |
|---|---|
| Detector head | 7.5 |
| Front-end electronics | 0.3 |
| Harness | 0.2 |
| **Subtotal** | **8.0** |
| Contingency (20%) | 1.6 |
| **Total** | **9.6** |

Table 49. UVO CCD detector power budget

| Item | Power (W) |
|---|---|
| Front-end electronics | 10.0 |
| TEC (TBC) | 25.0 |
| **Subtotal** | **35.0** |
| Contingency (20%) | 7.0 |
| **Total** | **42.0** |

# 5. ELECTRONICS SUBSYSTEMS

## 5.1 Instrument Control Unit

The Instrument Control Unit (ICU) is the electronic unit that manages the whole FCU Instrument. Its preliminary architecture is shown in Figure 96. In particular the tasks that will have to perform can be listed as follows:

- Configuration of the three FCU channels
  This will include:
    - to switch on and off the relevant DC/DC converters
    - to manage the programmable bias and clock voltages of the CCD
    - to manage the programmable high voltage for the MCP
    - to manage the camera interfaces and the commands for their operations
- Acquisition and processing of the Science data
  See the detailed description of the scientific processing (Section 5.1.1).

- Acquisition and processing of the HK
  Analog HKs like voltages, currents and temperature of the various DC/DC converters, digital HK related to the status of the SW, other digital and analog HK related to the status of the three cameras.

- Monitoring and Autonomous functions
  The ICU shall monitor the health of the instrument and eventually intervene in case a quick response is needed.

- TM&TC and Op-Modes management

Including the enabling/disabling of the relevant TM and TC packets.

- Time management

    Including tagging of the acquired data and synchronization with the S/C on board time.

- Actuators management

    The Instrument shall manage in open or closed loop the following actuators:

    - 1, 2 or 3 calibration lamps
    - 1 motor for inserting the calibration mirror (TBC)
    - 1 Peltier cooler in case the CCD has to be actively cooled down
    - TBD heaters for the active thermal control of the Instrument (these could be managed directly in the PSU)
    - 1 high precision motor for the pick-up mirror (TBC)
    - 3 to 7 motors for the filter wheels
    - Focus & Tip/Tilt motor (TBC) up to 6 motors
    - 1 CCD shutter

    As a general rule the motors will not be used during the exposure of the camera except for the CCD shutter. Also motors and calibration lamps will not switched on simultaneously.

- Interface with the PSU

    Once the ICU is on, it will manage the PSU controlling the acquisition of the analogue HK physically located in the PSU and sending commands for the activation of the DC/DC converters and the programmable output voltages/currents.

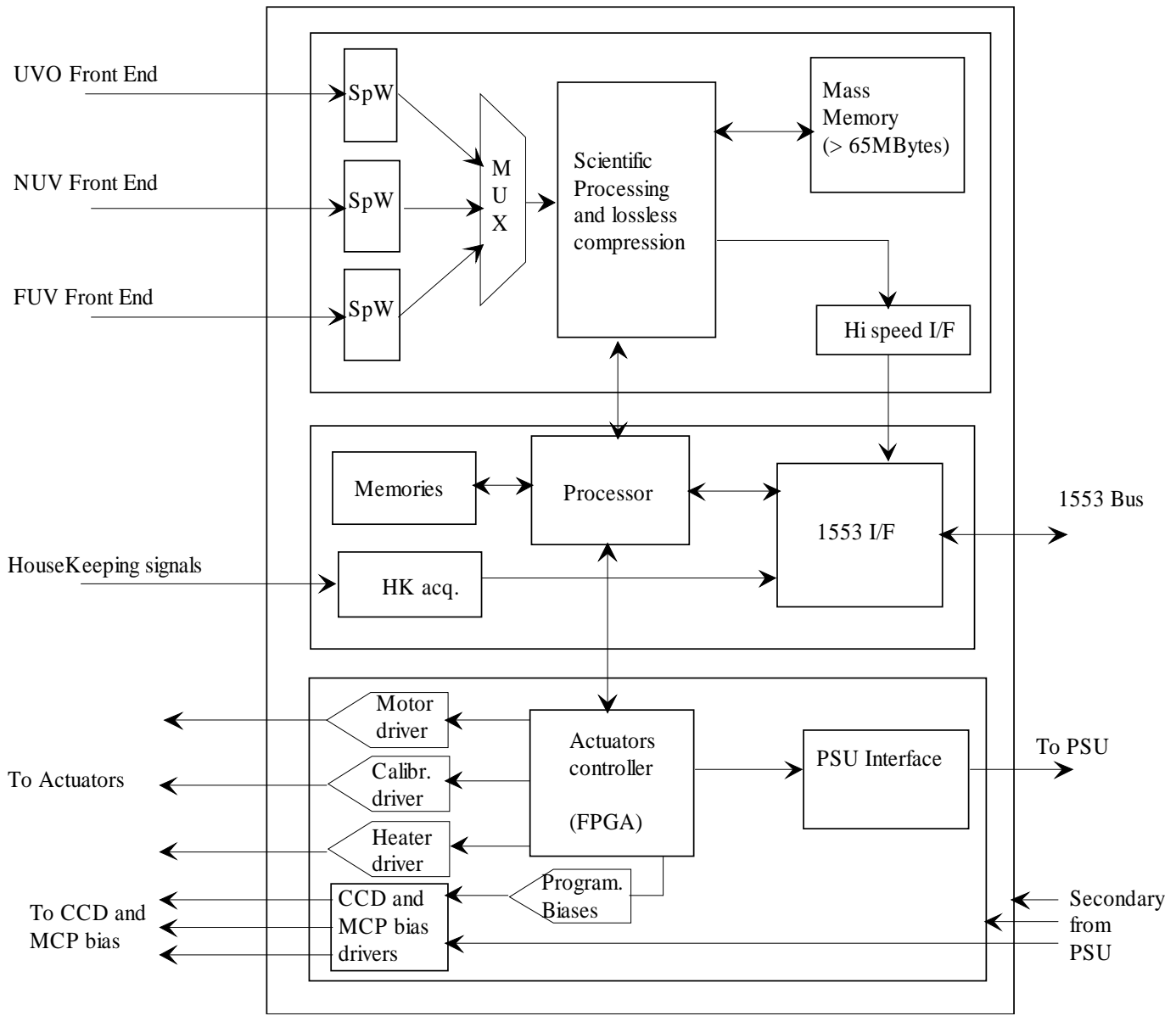

Figure 96. ICU Preliminary Architecture

## 5.1.1  Instrument scientific data acquisition and processing

The Instrument is composed by three channels, two of them are equipped with MCP detectors, the third one with a CCD detector.

The baseline configuration foresees the usage of one camera only per each observation.

The system shall be able to manage the detectors for what concerns all its programmable biases and parameters and to provide the necessary signals.

### 5.1.1.1 CCD

- 4k x 4k pixel image with a resolution of 16 bit per pixel.

- The image will be downloaded from the UVO Camera into the Electronic Unit (ICU) by means of a Spacewire link.

- The image is downloaded at the end of the CCD exposure, the downloading rate on the Spacewire link shall be limited only by the readout speed of the CCD.

- The image dimensions could be reduced according to a pre-programmed windowing configuration. The windowing can be performed at front end level.

- Once in the ICU the image shall be temporarily stored.

- The system shall be dimensioned to eventually perform a lossless compression on stored image.

    o A lossy compression option is not in the baseline.

- Once the processing on the acquired image is terminated, the image shall be splinted in data packets (usual format, with header, CRC, and a data header to allow the correct reconstruction of the image) and those packets queued for the transmission to the SDMU.

- The data packetization shall not be a limiting factor for the data throughput.

### 5.1.1.2 MCP (Accumulation mode)

- 2k x 2k pixel image with a resolution of 16 bit per pixel.

- The detected events (up to 500k event/s) are transmitted to the ICU through a Spacewire link.

- The resulting image dimensions could be reduced according to a pre-programmed windowing configuration. The events occurred out of the selected window will be filtered out and discarded.

- Each event will contain information about the arrived photon for about 48bit/event.

- The data are transmitted to the ICU during the MCP exposure.

- The incoming photon rate will be limited by reading rate of the MCP and by the number of operations to perform inside the ICU on each photon data. The goal is that the electronics is not going to limit the photon rate up to 500k event/s.

- The ICU shall reserve a 2k x 2k x 16bit memory area for the image accumulation and a 4k x 16bit memory area for the histogram accumulation.

- Once an event arrives to the ICU it shall update the amplitude histogram and accumulate (adding 1 to the event coordinate) the image.

    o It is likely that a correction factor for the coordinates shall be applied event by event before the imaging accumulation. This will be given in terms of Delta(x) and Delta(y) to the detected coordinates

    o The current attitude information will arrive through the S/C interface from 2 to 0.5 times per second (TBC), the ICU shall calculate the corresponding Delta(x,y) to apply to the scientific images.

- The system shall be dimensioned to possibly perform a lossless compression on accumulated image at the end of the acquisition time.

    o A lossy compression option is not in the baseline.

- Once the image acquisition and the relevant processing on the resulting image is terminated, the image shall be splinted in data packets (usual format, with header, CRC, and a data header to allow the correct reconstruction of the image) and those packets queued for the transmission to the SDMU.

- The data packetization shall not be a limiting factor for the data throughput.

### 5.1.1.3 MCP (Photon Count Mode)

- The reconstructed image will be 2k x 2k pixel with a resolution of 16 bit per pixel. The data going from the detector to the ICU will be related to the single photon events arrived at the detector, these data will include some time information that will allow the ICU to reconstruct the event time.

- The detected events are transmitted to the ICU through a Spacewire link.
- The resulting image dimensions could be reduced according to a pre-programmed windowing configuration. The events occurred out of the selected window will be filtered out and discarded.
- Each event will contain information about the arrived photon for about 48bit/event.
- The data are transmitted to the ICU during the MCP exposure.
- The incoming photon rate will be limited by reading rate of the MCP and by the number of operations to perform inside the ICU on each photon data. The goal is that the Electronic Unit is not going to limit the photon rate up to 500k event/s.
- The ICU shall reserve a 4k x 16bit memory area for the histogram accumulation.
- Once an event arrives to the ICU it shall update the amplitude histogram and store the complete event information for the subsequent generation of a Science TM packet.
  - o The FGS correction factor for the coordinates could (or not) be applied event by event, the correction data will be also added to the event information flow to allow the ground processing.
  - o The current attitude information will arrive through the S/C interface from 2 to 0.5 times per second, the ICU could calculate the corresponding Delta(x,y) or join the attitude information to the Science TM packets in a suitable way.
- A lossless compression on the event stream is considered not very efficient and unlikely to be requested, therefore, it is not considered in the baseline.
- During the image acquisition the events shall be fit in one or more Science TM packets (usual format, with header, CRC, and a data header to allow the correct reconstruction of the image) and made available for the transmission to the S/C Data Handling.
  - o In case the S/C link is not working during image acquisition the packets shall be temporarily stored in the Instrument on board memory dimensioned for the CCD case. In this case the maximum data rate and the maximum observation time will be somehow limited by the unavailability of the link. These limitations shall be evaluated.
  - o In case the S/C link is working during image acquisition but at a slower pace with respect to the incoming data rate the packets shall be again temporarily stored in the Instrument on board memory dimensioned for the CCD case. Also in this case the maximum data rate and the maximum observation time will be somehow limited by the low data rate of the link. These limitations shall be evaluated.

### 5.1.1.4 MCP (High resolution calibration mode)

- In this mode the MCP will produce high resolution data, higher than needed for the science purposes. This resolution will be needed for the centroid algorithm calibration.
- These higher resolution samples will be obtained by applying the uncorrected centroid algorithm and considering the centroid-increased spatial resolution for the resulting events on the whole detector 512k x 512k. The amplitude is not important for this calibration and can be coded with 1 bit.
- Once the detector is uniformly illuminated (flat field) the resulting events will show a distorted non-uniform image.
- In order to reduce the amount of memory needed for the Hi Res the images shall be windowed or "folded" keeping into account the periodicity of the CCD.
- The high resolution events shall be accumulated in a high resolution histogram that will precisely demonstrate the geometric distortions to be corrected.

- This histogram will be transmitted to ground the same way of another piece of Scientific data and once on ground will be used to generate the correction lookup table to store in the MCP Front End Electronics (TBC). This lookup table will be used for the on-board correction of the centroids at the level of the 2k x 2k image.
- The lookup table for the centroid correction will be composed of only 2 arrays of 256 cells (one for x correction and one for y correction) to be applied to any calculated centroid because of the periodicity of the CCD.

### 5.1.2 Science data processing

The front-ends shall communicate with the ICU through SpaceWire links.

In the baseline case, only one camera is active, while the other ones are idle.

In this case, the same resources are shared among the detectors, to occupy a single board of the ICU.

The microprocessor will configure the board for cameras commanding and relevant data processing, through UART interface.

The corresponding UART in the Science Data Processing block will be integrated in the FPGA.

### 5.1.3 Controls

Calibration lamps require a supply which starts as a current power supply and after warming up switch to a voltage mode power supply.

It is assumed that only one lamp is ON while the other ones are OFF and that the conditioning circuit can be the same.

In this case an area occupation of 10x18 cm$^2$ can be estimated for this block..

During calibration session all motors will be switched off.

For CCD shutter, filter wheels, focus/tip/tilt movements we assume to use stepper motors (TBC). For the pick up mirror mechanism different options are being evaluated based on stepper motors, brushless motors, piezo electric actuators or combinations of them.

The general driver block diagram for the all the motors is shown in Figure 97.

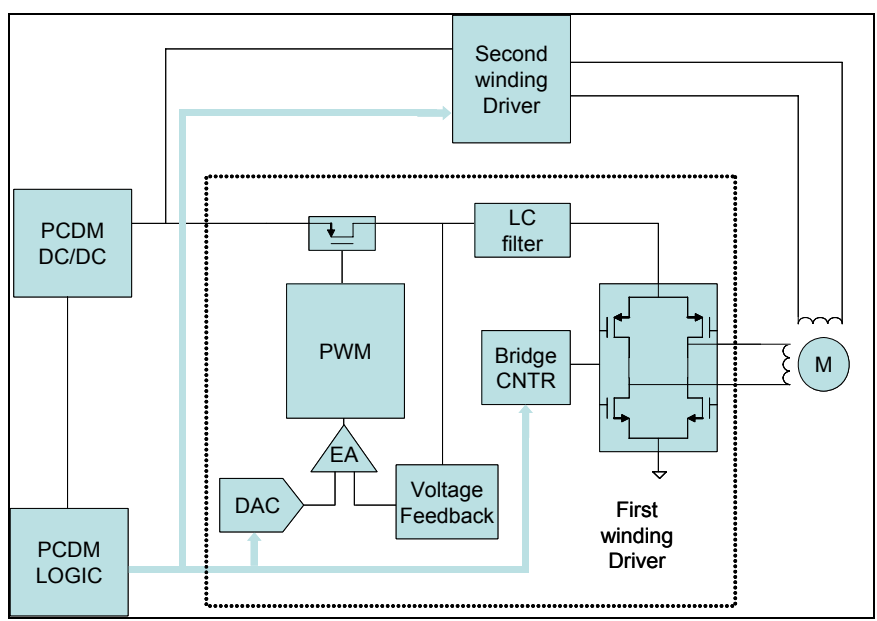

Figure 97. Motor driver block diagram

## 5.1.4 MIL1553B Bus Interface

### 5.1.4.1 Transfer data rate

Since the data transfer time cannot limit scientific observations, we will assume the typical exposure time as the upper data transfer time limit.

Taking into account a trade off between photon fluxes and cosmic rays effects, the estimated typical exposure time is ~ 6 minutes (maximum ~15 minutes) plus some overhead for data compression and image readout (the latter is about 33s and then negligible, if we assume a readout frequency of 500 kps).

Since the MCP data volume in time tagging mode could be very abundant (more than CCD), we must set a limit for such data amount. Beyond this limit we will use the MCP in accumulation or windowing mode. It is assumed that this limit will be at least the minimum for CCD data transfer; since the array is 4k x 4k @16 bits, the image is almost 270 Mbits. Assuming a transfer time of 6 minutes, it will give 750 kbps for CCD, that corresponds for MCP to 24 kevents per second in order not to fill the buffer, avoiding dead time between the exposures. For MCPs we assume this rate as the upper limit for time tag mode.

Although the physical data rate of MIL1553B is ~ 750 kbps, this number must be lowered by the protocol used to transfer data, to have handshaking and to exchange housekeeping and commands.

A basic scheme for transfer protocol is shown in Figure 99.

1. Since the Bus is managed in an asynchronous way, the SDMU shall poll the bus subscribers to check for data available for the transfer (according to Moisheev et al. 2006).

   - The BC interrogates the RT that in its Status Word include the Service Request flag
   - The BC repeats this with all the RTs (not with a broadcast) to check each of them

2. Once the SDMU has found one or more RT with data ready for the transfer, it shall understand the amount of data ready and then prepare its acquisition strategy from the RT.

   - The BC understand which RT has data ready for the transmission and interrogates them again to get the "Vector Word" that contains information about the amount of data ready for the transfer.
   - An intelligent machine shall organize the priority of download among the various RT that wait for a service not to exceed a maximum service time (in case of S/C intervention is needed)

3. After a certain time the acquisition of the available data (regardless their nature) starts as atomic blocks not larger than 2048 bytes.

   - The download is performed at the maximum speed thus the block is transferred in about 22.4ms for full length blocks.

4. The transfer of the blocks can (or shall) be interlaced with other activities on the bus. These activities are completely undefined.

   - The BC shall acknowledge the correct acquisition of the just transferred block and send to the RT the authorization to overwrite the correct block.
   - The BC shall poll all the RTs to update its acquisition strategy
   - The BC shall acquire data of other kinds (e.g. HK) from other RTs or even from the same.
   - The BC shall be able to manage commands toward any of the RTs.

In this hypothesis the data rate decreases down to 164 kbps, without any packetization.

FCU phase A report – Optical, mechanical and electronics configurations

Note that in agreement with the ESA standard packetization, the total data length (and then the total transfer time) changes only of few percents.

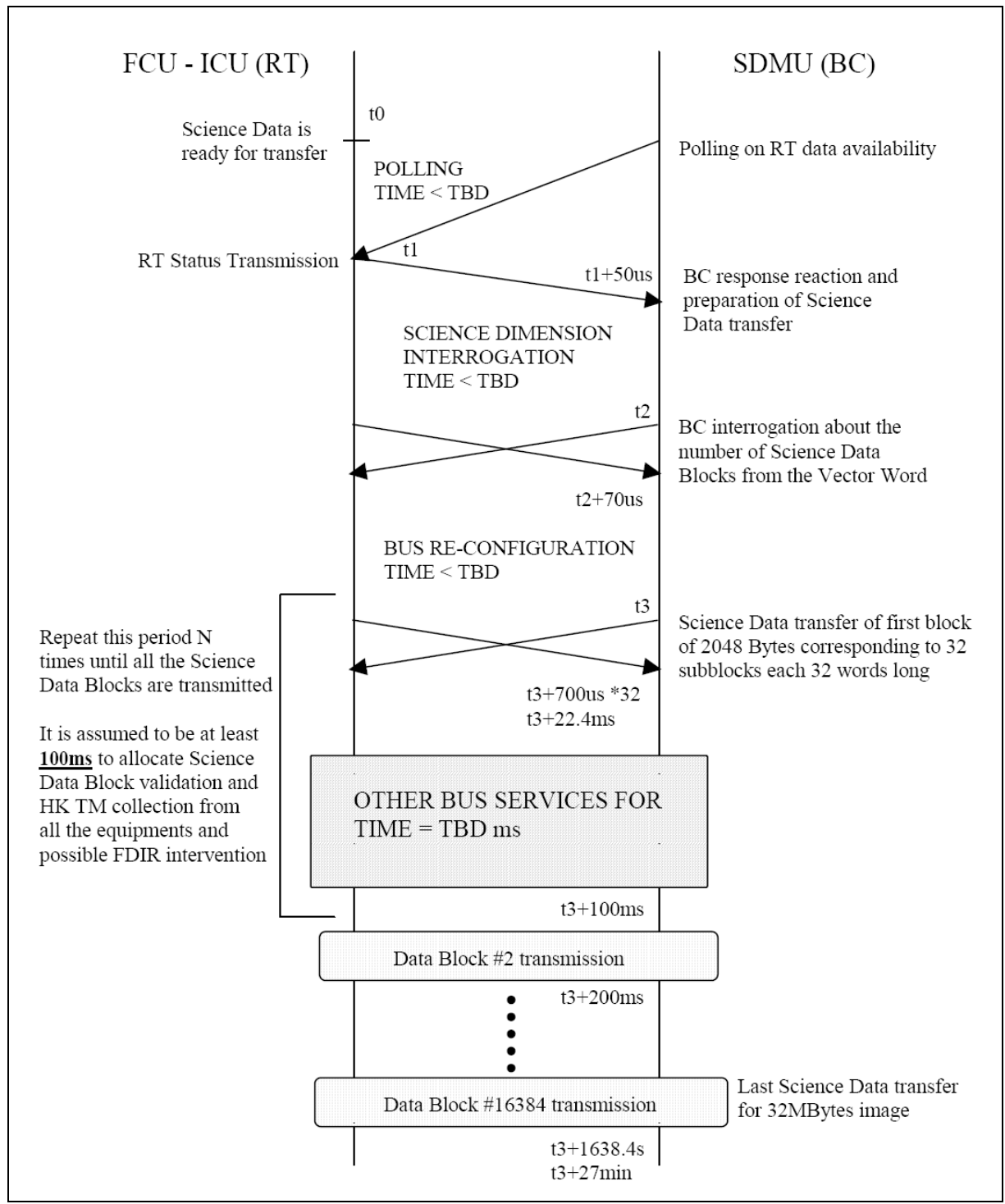

Figure 98. Basic scheme for transfer protocol

### *5.1.4.2 Compression*

Since MCP and CCD will produce a large quantity of data and both the transfer and telemetry data rate is low, a lossless compression has been considered.

Lossless compression algorithms exist or can be developed in order to obtain a compression factor of 2-4.

A suitable compression algorithm should have the following features:

- Only lossless compression method are acceptable
- The algorithm must execute in a time not much longer the CCD readout time (~ 33 s for a 4k x 4k CCD @ 500kpx/s of readout frequency).
- It must adapt to data statistical changes to maximize performances.
- The algorithm must minimize the process steps.
- Require minimum ground interaction during operation.
- Allow packetization for error containment.
- Offer adjustable data rate.

Among the lossless data compression algorithms, RICE and WHITE PAIR have been considered.

## 5.2   Interfaces description

### 5.2.1  Mechanical interface

The total number of boards in the ICU is 7 plus the motherboard. The boards will have the format of Extended Double Eurocard. The box external dimensions are 240 x 240 x 200 mm. The boards are subdivided as shown in Figure 99.

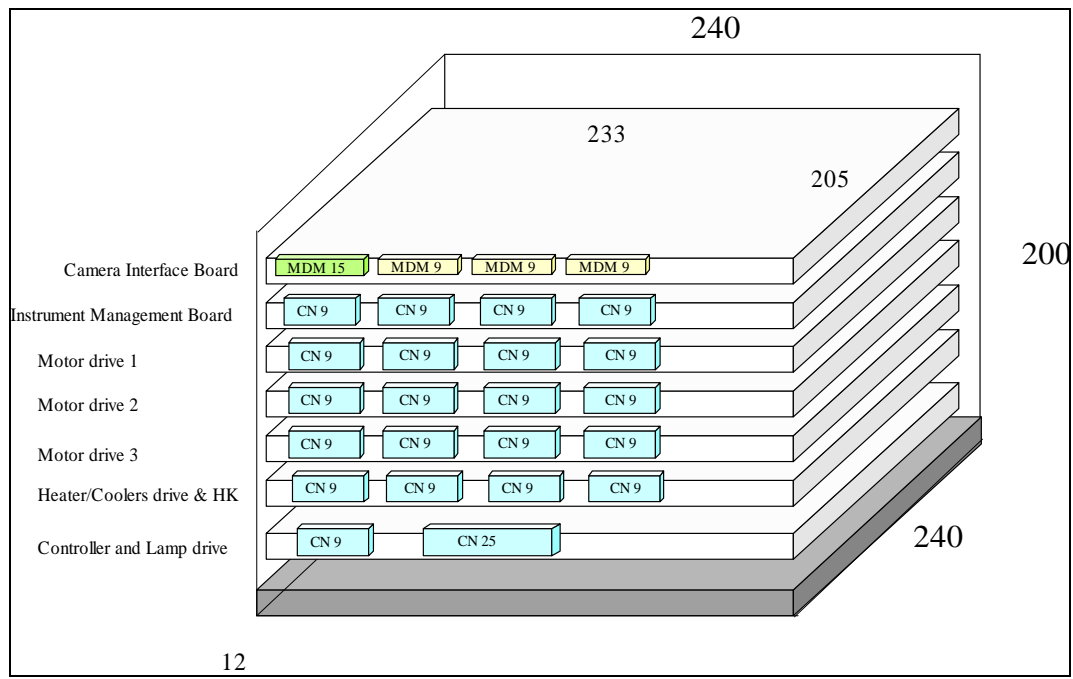

Figure 99. Boards schematic accommodation inside the ICU box

Table 50. ICU preliminary mass budget

| Item | Mass, g |
|---|---|
| Camera interface board | 600 |
| Microprocessor board | 600 |
| Motor drivers 1 board | 600 |
| Motor drivers 2 board | 600 |
| Motor drivers 3 board | 600 |
| Housekeeping and Heaters/Cooler driver board | 600 |
| Controller and lamp driver board | 600 |
| Motherboard | 600 |
| Mechanics | 1400 |
| **Total** | 6200 |
| Contingency 20% | 1200 |
| **Grand Total** | 7400 |

Table 50 list sthe preliminary mass budget of the ICU.

## 5.2.2 Power interfaces

The power consumption for the ICU can be different according to the operations executed.

Hereafter is reported the ICU consumption during the observation and when the cameras are not observing, but the actuators are working.

**NOTICE:** all the values referring to the motor drivers, heaters, coolers and calibration lamp controller are TBC due to the poor knowledge of the actual interfaces for these objects. The values reported here are estimated as "reasonable" values on the basis of the experience with similar equipments.

Nominal power supply budget during observation is shown in Table 51.

Table 51. Preliminary power budget during observation

| Item | Power, W |
|---|---|
| Camera interface board | 7 |
| Microprocessor board | 10 |
| Motor drivers 1 board | 0.5 |
| Motor drivers 2 board | 0.5 |
| Motor drivers 3 board | 0.5 |

| Item | Power, W |
|---|---|
| Housekeeping and Heaters/Cooler driver board | 10 |
| Controller and lamp driver board | 0.5 |
| **Total** | 29.0 |
| Contingency 20% | 5.8 |
| **Grand Total** | 34.8 |

Nominal power supply budget during calibration mode assuming the calibration unit as the highest power demanding actuator is shown in Table 52.

Table 52. Preliminary power budget during calibration

| Item | Power, W |
|---|---|
| Camera interface board | 7 |
| Microprocessor board | 10 |
| Motor drivers 1 board | 0.5 |
| Motor drivers 2 board | 0.5 |
| Motor drivers 3 board | 0.5 |
| Housekeeping and Heaters/Cooler driver board | 10 |
| Controller and lamp driver board | 12 |
| **Total** | 40.5 |
| Contingency 20% | 8.1 |
| **Grand Total** | 48.6 |

## 5.3   Power Supply Unit

### 5.3.1   Functional block diagram of the PSU

The schematic block diagram of the Power Supply Unit is shown in Figure 100..

The PSU is in charge of generating regulated supply voltages for all the subsystems of the FCU from the S/C Primary Bus voltages.

The Primary Supply goes into the Differential and Common Mode EMI filters, then the filtered voltage is distributed to the DC/DC Converters of the PSU.

All the DC/DC Converters are based on a standard Thales Alenia Space Italia - Milano design showing the following features:

- Galvanic isolation between primary and secondary voltages;
- EMI filtering;

- Inrush current limiting;
- Soft start;
- Latching overcurrent protection;
- Input undervoltage protection;
- Output overvoltage protection;
- Reset latching protection;
- Current mode control;
- On/Off capability
- PWM topology.

This Standard topology is used in a number of the Thales Alenia Space Italia – Milano projects such as Marsis on Mars Express, Agile, Sharad on MRO, Ibis on INTEGRAL, Epic on XMM, Cosmo SkyMed.

The PSU box is provided with 5 boards each hosting a DC/DC converter.

All the DC/DC converter outputs values, their voltage and their maximum power are TBC during the successive phases of the project.

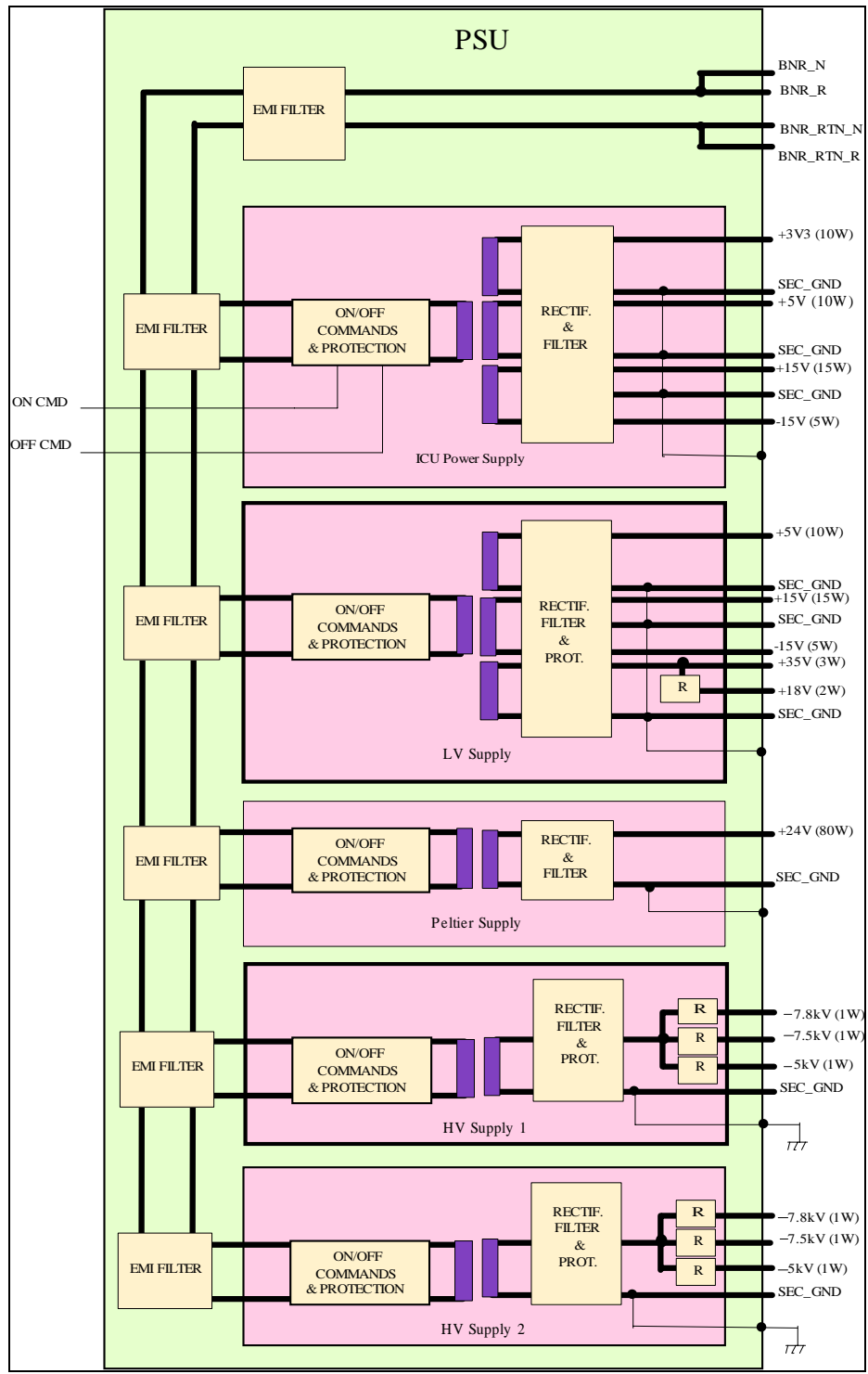

Figure 100. Power Supply Unit - Block Diagram

**ICU Power Supply** is in charge of generating four supply voltages (+3.3V max 10W,+5V max 10W, +15V max 15W and -15V max 5W) for the Control Logic of the PSU and the Heaters.

It can be switch on/off by two dedicated pulsed commands provided by the SDMU.

In case a protection is triggered the relevant DC/DC converter remains off until the SDMU releases the "Off" pulse command. This usually switches off the DC/DC converter but also resets the latching protection. The restart of the DC/DC converter can therefore be tried again simply sending again the "On" command from the SDMU.

When this DC/DC is off the ICU is off, so all the PCU is off.

**LV Power Supply** generates five voltages (+5V max 10W, +15V max 15W, -15V max 5W, +18V max 2W and 35V max 3W), which are required by the CCD Camera Supply and the Motors.

The ICU is in charge of switching it on/off by means of a bilevel command

As for the ICU DC/DC converter in case one of the protections is triggered an off/on cycle resets the latching protections.

**Peltier Supply** is in charge to generate a variable voltage from 0 to 24V max 80W for the Peltier cooler, it is digitally programmable by ICU.

The ICU is in charge of switching it on/off by means of a bilevel command

As for the ICU DC/DC converter in case one of the protections is triggered an off/on cycle resets the latching protections.

**HV Power Supply** two identical DC/DC converters are dedicated each to the bias of one MCP. Each one generates a -8kV bias voltage that is then post regulated by digitally programmable circuit and generates tree voltage at nominal value of -5kV, -7.5kV and -7.8kV with a max power available of 1W per output line.

The ICU is in charge of switching it on/off by means of a bilevel command

As for the ICU DC/DC converter in case one of the protections is triggered an off/on cycle resets the latching protections.

The efficiency of the DC/DC converters is typically around 75–80 %, depending on the kind of protections implemented and the actual power delivered with respect to the maximum value that is taken as the dimensioning case.

## 5.3.2  Mechanical interface

The total number of boards in the PSU is 5 plus the motherboard (TBC). The boards has the format of Single Eurocard (TBC). The box external dimensions are 100 x 230 x 200 mm.

Table 53. Preliminary Mass Budget for the PSU

| Item | Mass, g |
|------|---------|
| ICU Power Supply | 400 |
| LV Power Supply | 400 |
| Peltier Supply | 400 |
| HV Supply 1 | 400 |
| HV Supply 2 | 400 |
| Motherboard | 300 |
| Mechanics | 500 |
| **Total** | 2800 |
| Contingency 20% | 600 |
| **Grand Total** | 3400 |

## 5.4 Telemetry and telecommands

### 5.4.1.1 Telemetry packet structure

According with the ESA standard, the proposed structure of the telemetry packets is the one showed in Table 54.

Table 54. Telemetry structure

| SOURCE PACKET HEADER (48 bits) | | | | | | | PACKET DATA FIELD (VARIABLE) | | |
|---|---|---|---|---|---|---|---|---|---|
| PACKET ID | | | | PACKET SEQUENCE CONTROL | | PACKET LENGTH | DATA FIELD HEADER | SOURCE DATA | PACKET ERROR CONTROL |
| Version | Type | Data Field Header Flag | Application Process ID | Segmentation Flags | Source Sequence Count | | | | |
| 3 | 1 | 1 | 11 | 2 | 14 | | | | |
| 16 bits | | | | 16 bits | | 16 bits | 80 bits | up to 1015 x 16 bits | 16 bits |

### 5.4.1.2 Telecommand packet structure

According with the ESA standard, the structure of the telecommand packets will be the one shown in Table 55.

Table 55. Telecommand Structure

| PACKET HEADER (48 bits) | | | | | | | PACKET DATA FIELD (VARIABLE) | | |
|---|---|---|---|---|---|---|---|---|---|
| PACKET ID | | | | PACKET SEQUENCE CONTROL | | PACKET LENGTH | DATA FIELD HEADER | APPLICATION DATA | PACKET ERROR CONTROL |
| Version Number | Type | Data Field Header Flag | Application Process ID | Sequence Flags | Sequence Count | | | | |
| 3 | 1 | 1 | 11 | 2 | 14 | | | | |
| 16 bits | | | | 16 bits | | 16 bits | 32 bits | up to TBD x 16 bits | 16 bits |

## 5.5 Harness

Two possible options have been analyzed to connect the FCU instrument to the ICU subunit and then to SDMU (Figure 101).

The cables from FCU and ICU to the external panel are under responsibility of FCU team, including connectors on the panel. Lavochkin Association will provide the information on the length of the cables.

Wires are left open at the external panel and Lavochkin will provide connectors and perform soldering of the wires.

The first solution was selected.

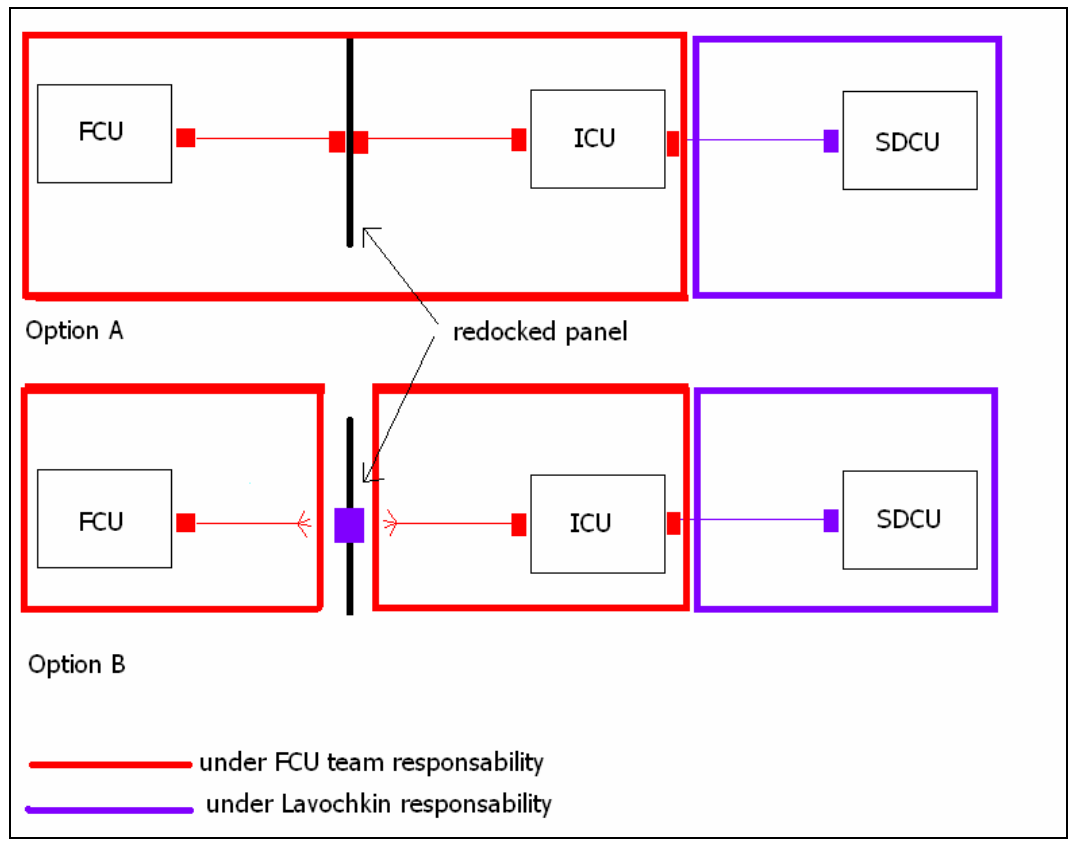

Figure 101. FCU – ICU – SDCU connections

# Chapter IV.

# FCU phase A report – Preliminary AIV and GSE plans

## 1. FLIGHT CONFIGURATION ITEMS AND AIV CONCEPT

The Flight Model of the FCU instrument is made of various Subsystems, which in turn consist of various Units, as sketched in Figure 102.

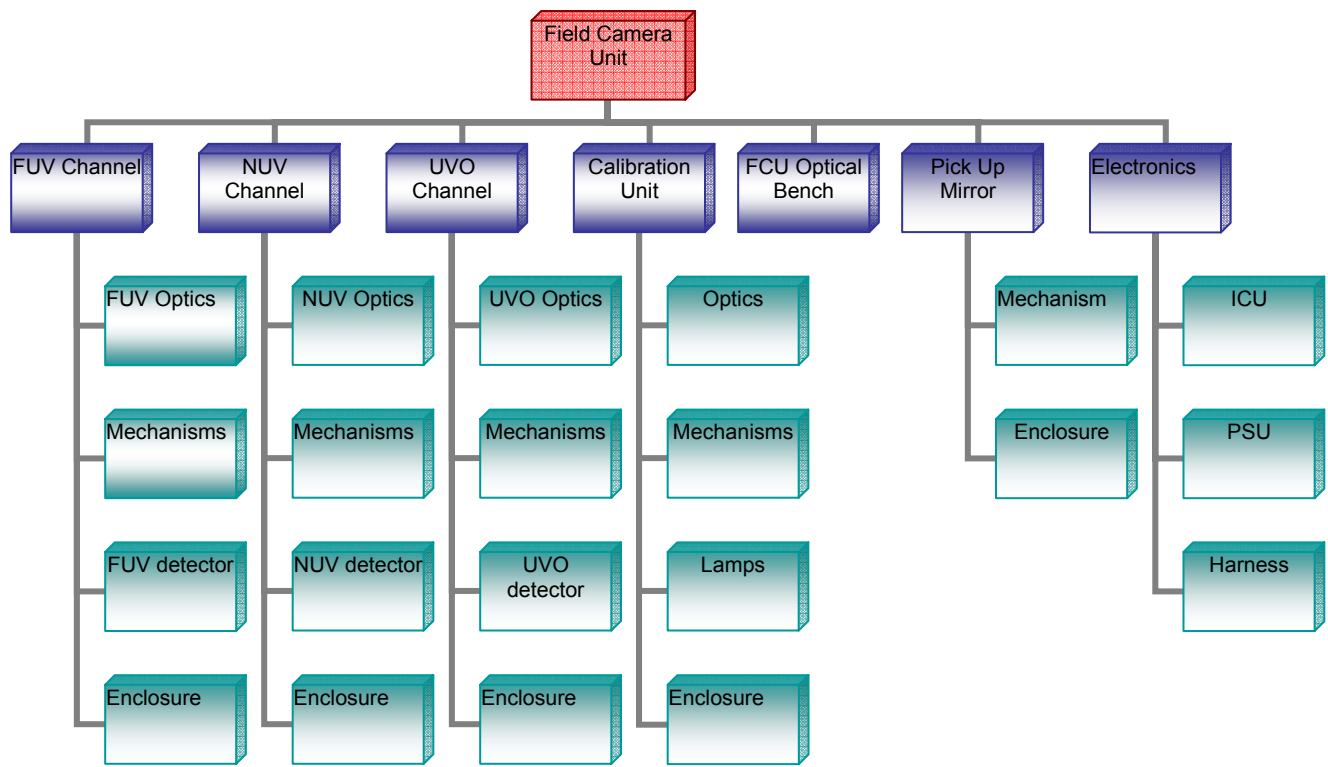

Figure 102 FCU Subsystems and Units

Another FCU system view, more suitable to the proposed FCU AIV flow, is provided by the diagram shown in Figure 103..

The vertical paths shown in this view correspond to independent AIV flows, which can be carried out in parallel.

In each flow, the AIV activities will be carried out at two levels:

- AIV on the single Unit (Level 1)
- AIV at Subsystem level (Level 2)

The AIV at instrument level (Level 3) will be carried out in parallel on two different paths:

- Optical Paths AIV: assembly, integration and test of the Optics and Mechanisms on the FCU Optical Bench
- Detectors&Electronics AIV: assembly, integration and test at bench level of the Detector and FEE elements of the three channels, the Electronics units (ICU, PSU), and the interconnecting harness.

Eventually, the Detectors shall be mounted on the Optical Bench and the FCU instrument shall undergo all the remaining tests foreseen by the instrument level AIV.

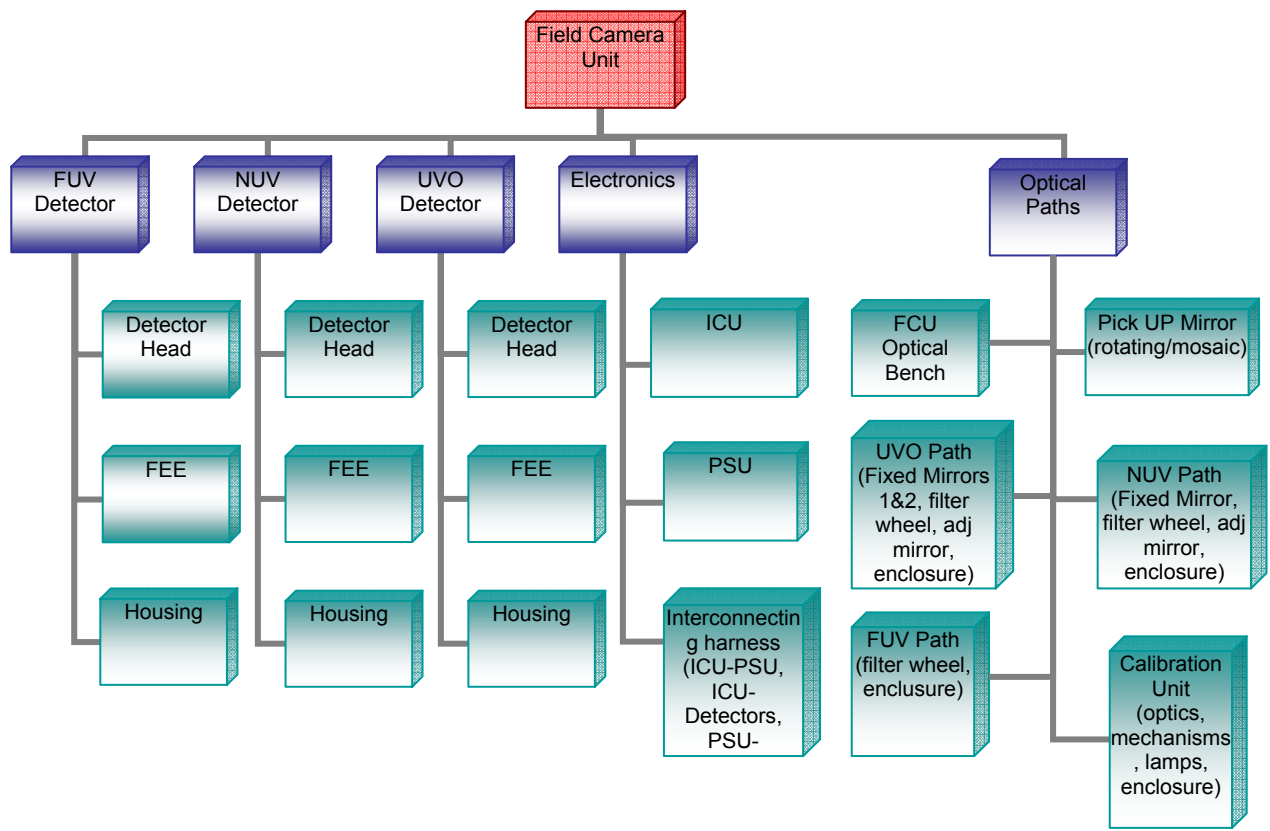

Figure 103 – AIV View of the FCU instrument

# 2. INSTRUMENT VERIFICATION PLAN

This section presents the overall approach to accomplish the instrument qualification and acceptance program. The Plan is divided into different phases.

## 2.1 Bread Boarding

In order to consolidate the design to be presented at the PDR, bread boarding activities will be carried out by the FCU Scientific Institutes during the B2 phase. The Bread Boarding items are not deliverables.

### 2.1.1 UVO Bread Boarding

#### *2.1.1.1 Objectives*
The development of various FEE designs.

#### *2.1.1.2 Configuration*
The UVO BB will consist of:
- One representative CCD
- Front-end electronics and Controller GSE configured to operate the CCD in TBD design modes
- A cryogenic GSE to support cold-running of the CCD

#### *2.1.1.3 Test Activities*
The proposed test plan includes functional and performance tests.

### 2.1.2 NUV & FUV Bread Boarding

#### *2.1.2.1 Objectives*
- The development of various FEE designs
- Study of the effects of the phosphor finite time decay:
  - o persistence of the events in subsequent frames
  - o gain sag due to incomplete signal collection during CCD integration period
  - o recognition of misallocated events occurring during the frame transfer
- Development of algorithms to be implemented in the real time digital signal processing unit of the DFEE and test of the effects on detector performance:
  - o Spatial resolution
  - o Flat field uniformity
  - o DQE, background rejection
- Study of the systematic errors introduced by the centroiding technique, their correction or minimization and stability
- Optimization of the photocathode/MCP gap parameters in order to improve the spatial resolution
- Study of the effect of the presence of bright sources in the Field of View

#### *2.1.2.2 Configuration*
- The NUV BB will consist of:
- One representative MCP intensifier
- One representative CCD
- One optical system for MCP to CCD coupling

- A CCD Front-end electronics boards
- A Programmable HVPS with gate unit
- A FPGA based digital programmable board implementing:
    - a controller configured to operate the CCD
    - DFEE implementing a real time digital signal processing unit with photon event recognition and centroiding

### 2.1.2.3 Test Activities

The proposed test plan includes functional and performance tests.

- CCD only:
    - Flat Field
    - Photon Transfer Curve (PTC)
    - Modulation Transfer Function (MTF)
    - Extended Pixel Edge Response (EPER)
- Intensifier only:
    - Pulse Height Distribution (PHD)
    - Dynamic range (local, global) of the MCP stack
- Detector system:
    - Pulse Height Distribution of the detector
    - Dynamic range (local, global) of the detector
    - Spatial resolution (@254 nm, @550 nm)
    - Flat Field (and its stability) @ optical light

## 2.1.3  Optical Paths Bread Boarding

### 2.1.3.1 Objectives

During phase A, two different philosophies in the AIV and design for the OB have been developed. In order to evaluate the best design, bread boarding activities have to be carried out to verify all the requirements needed.

### 2.1.3.2 Configuration and Test Activities

A simplified OB has to be developed to consider thermo-mechanical loads and interfaces, and in particular bread boarding activities will be performed on the central part off the OB.

The overall test concept is sketched in Figure 104.

The purpose is to develop a smaller OB with the central pick-up mirror that it will be used for environmental tests to evaluate the very tight requirements needed for this mechanism. These tests will consider a rotating mirror with mechanical structure, movement  and positioning system device. Furthermore we will test also mosaic mirrors layout. At the same time this OB will be used to test interfaces with the FGS concerning possible stray light problems.

Bread boarding activities will be done with filter wheels, possible piezo electric actuators, single optical mounting and dedicated optical path to verify all the requirements.

Dedicated studies have to be performed for housing elements in order to verify thermo-mechanical, moisture and degassing properties, stray light characteristics and interfaces regions.

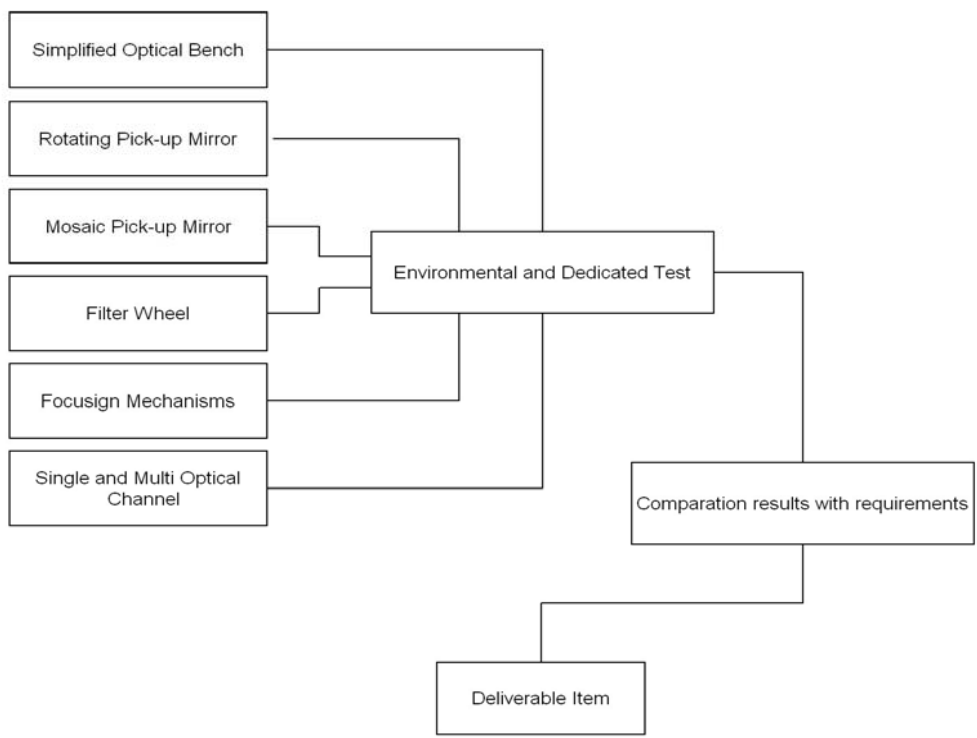

Figure 104 - Optical Paths Bread Boarding  Test Concept

## 2.2    Structure Thermal Model (STM)

### 2.2.1  AIV Objectives

This model will comply with the requirements stated in Moisheev et al. (2006) section 17 for the following models:

- the Full-size mass model
- the Thermal equivalent model.

The required characteristics are summarized in Table 56 below.

The STM compliance with the requirements shall mainly be verified by analysis or by measurements.

Table 56 - Requirements on the STM

| Model type | Flight Similarity | Accuracy |
|---|---|---|
| Full-size mass | Dimensions | |
| | Configuration | |
| | Center of Gravity location | 5-10% |
| | Eigen frequencies | 5-10% |
| | Primary momentum of inertia | 5-10% |
| | Mass | 5-10% |
| | Mounting units | |
| | Connectors | |
| | Coating | |
| | Load-bearing design | |
| | Alignment units design | |
| | Lifting units | |
| | Stiffness | |
| Thermal equivalent | Size | |
| | Case material | |
| | Coating | |
| | Heat dissipation | Full equivalent |
| | Thermal capacity | 5-10% |
| | Electric connection | |
| | Mounting units | |

## 2.2.2  Model Configuration

The STM will consist of:

- Optical Paths Assembly:
    - FCU optical bench
    - dummy masses to simulate the mirror and mechanisms
    - heaters to simulate the thermal behavior
- Detectors&Electronics:
    - dummy masses to simulate the three Detector assemblies, and the Electronics (ICU, PSU, harness)
    - heaters to simulate the thermal behavior

As far as the STM deliverables are concerned, at the time of writing two hypothesis are envisaged.

1) The STM shall be deliverable at design level (at the level of Phase B), in order to allow the completion of the preliminary design of Instrumental Compartment in the current year. This shall allow the WSO Spacecraft Contractor to produce the Instrumental Compartment structure and systems on the work documentation, to provide the Telescope model for vibration and static testing, and to perform the testing.

2) The STM shall be design and produced by the FCU Contractor.

## 2.2.3  AIV activities

No AIV activity is required in case 1).

In case 2), a limited set of tests shall be carried out separately on each of the two assemblies detailed in Section 2.2.2.

## 2.3 Engineering Qualification Model (EQM)

### 2.3.1 AIV Objectives

The EQM model will be fully equivalent to the Flight Model, as required in Moisheev et al. (2006).

Prior to delivery, the EQM will be subjected to a full environmental and functional qualification program, based on the AIV concept presented in Section 1.

### 2.3.2 Configuration

Being fully equivalent to the FM, the EQM will consist of all the FCU components presented in Section 1.

### 2.3.3 AIV Activities

All qualification testing and software verification will be carried out on the complete FCU.

A suitable calibration campaign is planned for the EQM. This will allow also to set up, test and validate all the specific procedures, tools and facilities to be exploited for the calibration of the Flight Model.

The EQM test matrix shown in Table 57 and Table 58 summarize the proposed test plan.

Table 57 - FCU EQM Test Matrix at Unit Level (Level 1)

| Legenda:<br>A = Accept. levels<br>Q = Qualif. levels | Vibration | Alignment | Thermal Balance | Thermal Vacuum. | Mass Properties | UV Calibration | Optical Calibration | Conducted EMC | Radiated EMC | ESD | Functional Test | Performance Test | Delivery to WSO |
|---|---|---|---|---|---|---|---|---|---|---|---|---|---|
| Optical Bench – Mechanical | | | | | X | | | | | | | | |
| Pick Up Mirror (rotating / mosaic) | Q | X | | Q | X | | X | | | | X | X | |
| NUV Fixed Mirror | | | | | X | | X | | | | X | X | |
| UVO Fixed Mirror 1 | | | | | X | | X | | | | X | X | |
| UVO Fixed Mirror 2 | | | | | X | | X | | | | X | X | |
| UVO Filter Wheel | Q | | | Q | X | | X | | | | X | X | |
| NUV Filter Wheel | Q | | | Q | X | | X | | | | X | X | |
| FUV Filter Wheel | Q | | | Q | X | | X | | | | X | X | |
| Calibration Unit | Q | | | Q | X | | X | | | | X | X | |
| UVO adj mirror | | | | | | | | | | | | | |
| NUV adj mirror | | | | | | | | | | | | | |
| UVO Detector Head | | | | | | | | | | | | | |

| Legenda:<br>A = Accept. levels<br>Q = Qualif. levels | Vibration | Alignment | Thermal Balance | Thermal Vacuum. | Mass Properties | UV Calibration | Optical Calibration | Conducted EMC | Radiated EMC | ESD | Functional Test | Performance Test | Delivery to WSO |
|---|---|---|---|---|---|---|---|---|---|---|---|---|---|
| UVO FEE | | | | | | | | | | | | | |
| NUV Detector Head | Q? | | | | | X | | | | | X | X | |
| NUV FEE | | | | | | | | | | | X | | |
| FUV Detector Head | Q? | | | | | X | | | | | X | X | |
| FUV FEE | | | | | | | | | | | X | | |
| ICU | Q | | | Q | X | | | | | X | X | | |
| PSU | Q | | | Q | X | | | | | X | X | | |
| Interconnecting Harness | | | | | | | | | | | | | |

Table 58 - FCU EQM Test Matrix at Subsystem Level (Level 2) and at FCU Level (Level 3)

| Legenda:<br>A = Accept. levels<br>Q = Qualif. levels | Vibration | Alignment | Thermal Balance | Thermal Vacuum. | Mass Properties | UV Calibration | Optical Calibration | Conducted EMC | Radiated EMC | ESD | Functional Test | Performance Test | Delivery to WSO |
|---|---|---|---|---|---|---|---|---|---|---|---|---|---|
| Optical Paths | Q | X | | Q | X | | | | | | X | X | |
| UVO Detector | Q | | | Q | X | X | X | | | | X | X | |
| NUV Detector | Q? | | | Q | X | X | | | | | X | X | |
| FUV Detector | Q? | | | Q | X | X | | | | | X | X | |
| Electronics&Detectors | | | | Q | | X | X | X | X | | X | X | |
| | | | | | | | | | | | | | |
| FCU Assembly | | | | | | X | X | | | | X | X | X |
| | | | | | | | | | | | | | |
| FCU System EGSE | | | | | | | | | | | X | X | ? |
| FCU System Science Console | | | | | | | | | | | X | X | X |

| Legenda:<br>A = Accept. levels<br>Q = Qualif. levels | Vibration | Alignment | Thermal Balance | Thermal Vacuum. | Mass Properties | UV Calibration | Optical Calibration | Conducted EMC | Radiated EMC | ESD | Functional Test | Performance Test | Delivery to WSO |
|---|---|---|---|---|---|---|---|---|---|---|---|---|---|
| FCU MGSE | | | | | | | | | | | X | X | ? |
| FCU MOGSE | | | | | | | | | | | X | X | |
| FCU Cryo GSE | | | | | | | | | | | X | X | ? |
| | | | | | | | | | | | | | |
| | | | | | | | | | | | | | |

## 2.3.4 Integration Sequence and required GSE items

### 2.3.4.1 FUV Detector

The MCP sensor shall be pre-evacuated. The Level 2 verification activities shall consist of:

- Dimensional tests
- I/F Tests
- Functional Tests:
    - Ambient temperature and pressure
- Environmental tests: EMC, Thermal-Vacuum, Vibrations
- Performance Tests
- UV Calibration using TE+OGSE

The integration sequence together with the required GSE items is shown inFigure 105.

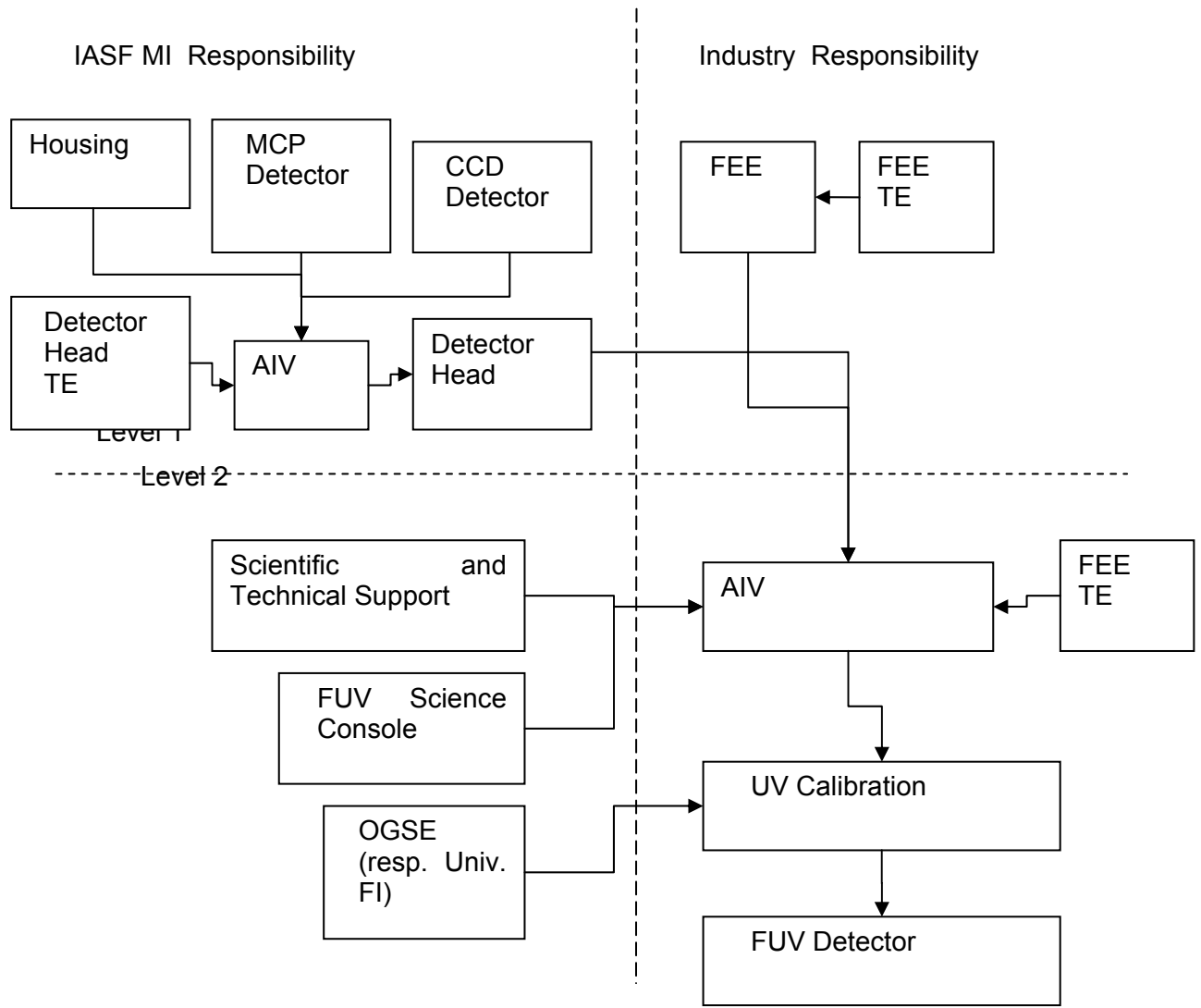

IASF MI  Responsibility                    Industry  Responsibility

Figure 105 - FUV Detector AIV Flow (Leve1 & Level 2)

### 2.3.4.2 NUV Detector

The MCP sensor shall be pre-evacuated. The Level 2 verification activities shall consist of:

- Dimensional tests
- I/F Tests
- Functional Tests:
  - Ambient temperature and pressure
- Environmental tests: EMC, Thermal-Vacuum, Vibrations
- Performance Tests
- UV Calibration using TE+OGSE

The integration sequence together with the required GSE items is shown in Figure 106.

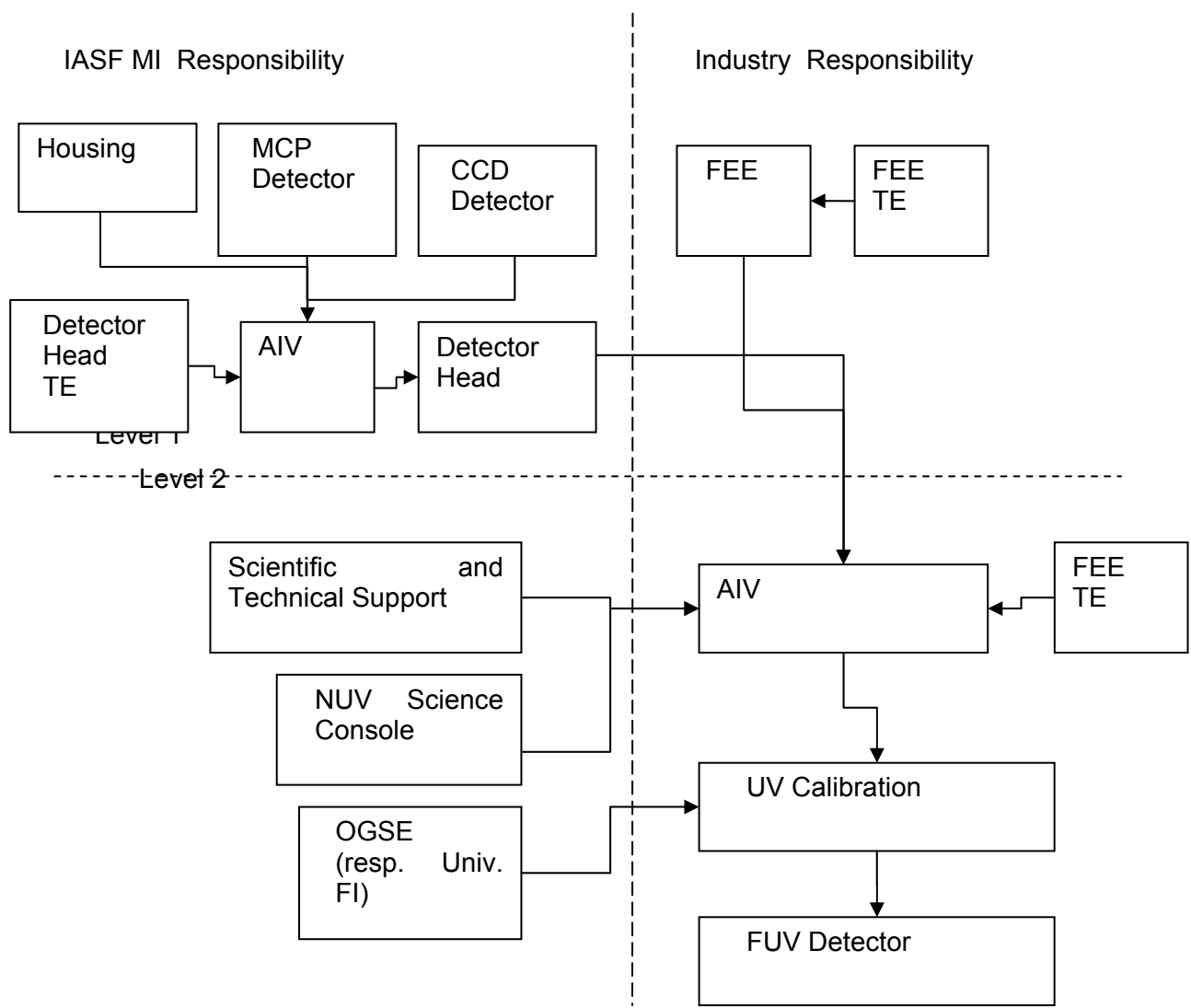

Figure 106 - NUV Detector AIV Flow (Leve1 & Level 2)

### 2.3.4.3 UVO Detector

The Level 2 verification activities shall consist of:

- Dimensional tests
- I/F Tests
- Functional Tests:
    - o With evacuated housing, @ -100 °C (CCD no MPP)
    - o Or
    - o Ambient temperature and pressure, included housing (CCD MPP)
- Environmental tests: EMC, Thermal-Vacuum, Vibration
- Performance Tests
- UV-Visual Calibration using TE+OGSE
    - o evacuated housing, @ [-100, -50] °C

The integration sequence together with the required GSE items is shown Figure 107.

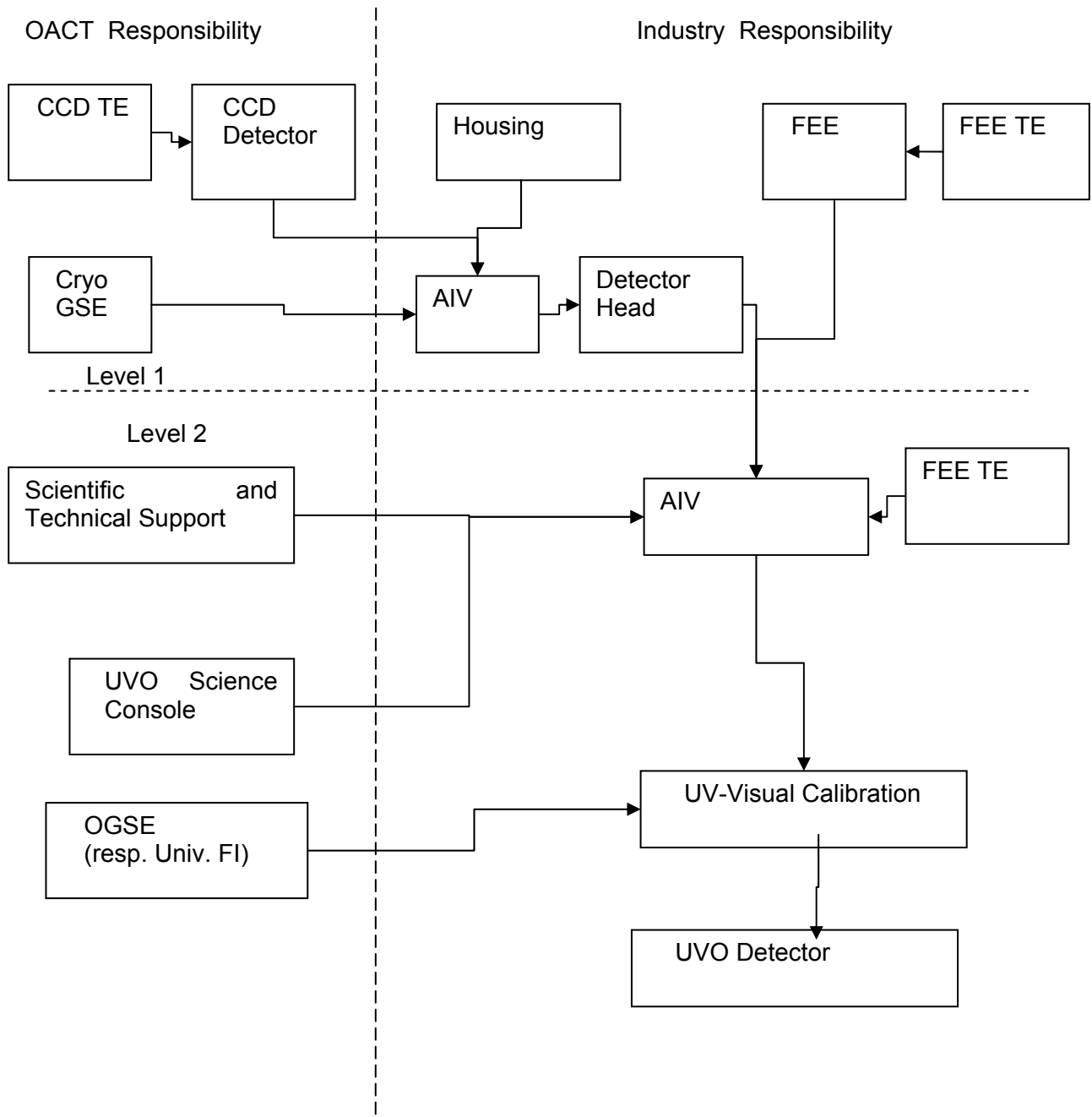

Figure 107 - UVO Detector AIV Flow (Leve1 & Level 2)

### 2.3.4.4 Electronics

The Level 2 verification activities shall consist of:

- Dimensional tests
- I/F Tests
- Functional Tests:
  - o Ambient temperature and pressure
- Environmental tests: EMC, Thermal-Vacuum, Vibrations
- Performance Tests

The integration sequence together with the required GSE items is shown in Figure 108,

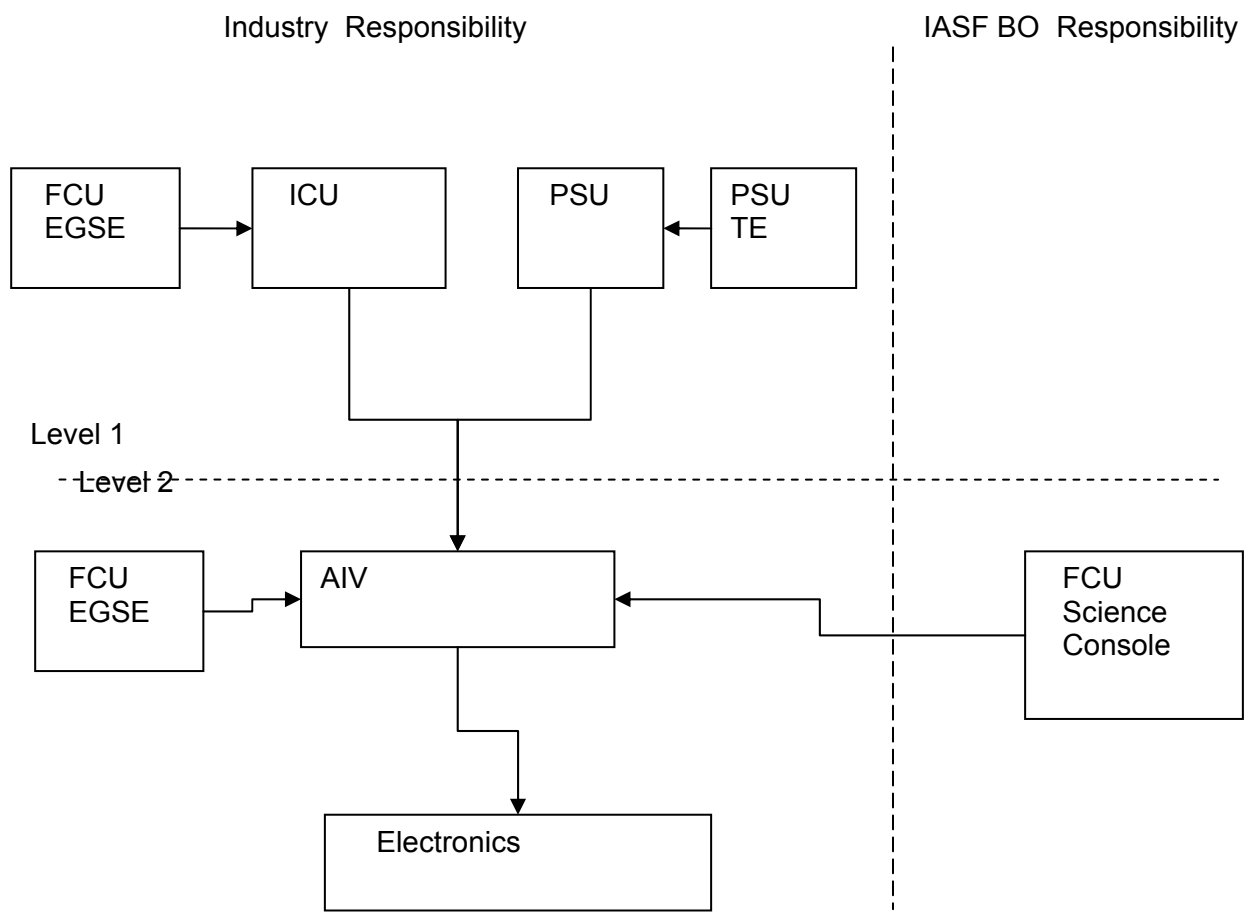

Figure 108 - Electronics AIV Flow (Leve1 & Level 2)

It is noted that, in case the PSU is not available, the integration of the ICU at Level 3 could proceed anyway as the FCU EGSE will be able to provide simulation of the missing PSU.

### 2.3.4.5 Optical Paths

The two proposed AIV flows corresponds to the two different philosophies which are being developed in the phase A study:

1) to mount the various elements directly on the FCU Optical Bench and test the FCU as a whole

2) to foresee one enclosure for each of the FUV path, NUV path, UVO path. In this case, each path is assembled and tested in standalone mode, and then mounted on the Optical Bench.

**Flow 1**

The Level 2 verification activities shall consist of:

- I/F tests
- Environmental tests: EMC, Thermal-Vacuum, Vibration
- Performance Tests
- Functional tests
- Optical alignment (OGSE)

This AIV will be responsibility of the industry with a contribution of the UNIFI for the optical alignment (FCU OGSE) and IASF BO for the FCU Science Console. Scientific and technical support will be given by the institutions having responsibility of each FCU path.

### Flow 2

FUV, NUV and UVO path will be assembled and tested in a stand alone mode. The activities consist of:

- I/F tests
- Performance Tests
- Functional tests
- Optical alignment (OGSE)

AIV activities assembling the paths on the OB will follow. The level 2 verification activities shall consist of:

- I/F tests
- Environmental tests: EMC, Thermal-Vacuum, Vibration
- Performance Tests
- Functional tests
- Optical alignment (OGSE)

All the above AIV activities  will be responsibility of the industry with a contribution of the UNIFI for the optical alignment (FCU OGSE) and IASF BO for the FCU Science Console. Scientific and technical support will be given by the institutions having responsibility of each FCU path.

### *2.3.4.6 FCU*

The Level 3 verification activities are identified in Table 58. The related integration sequence together with the required GSE items is shown in Figure 109.

Industry  Responsibility                                Institutes  Responsibility

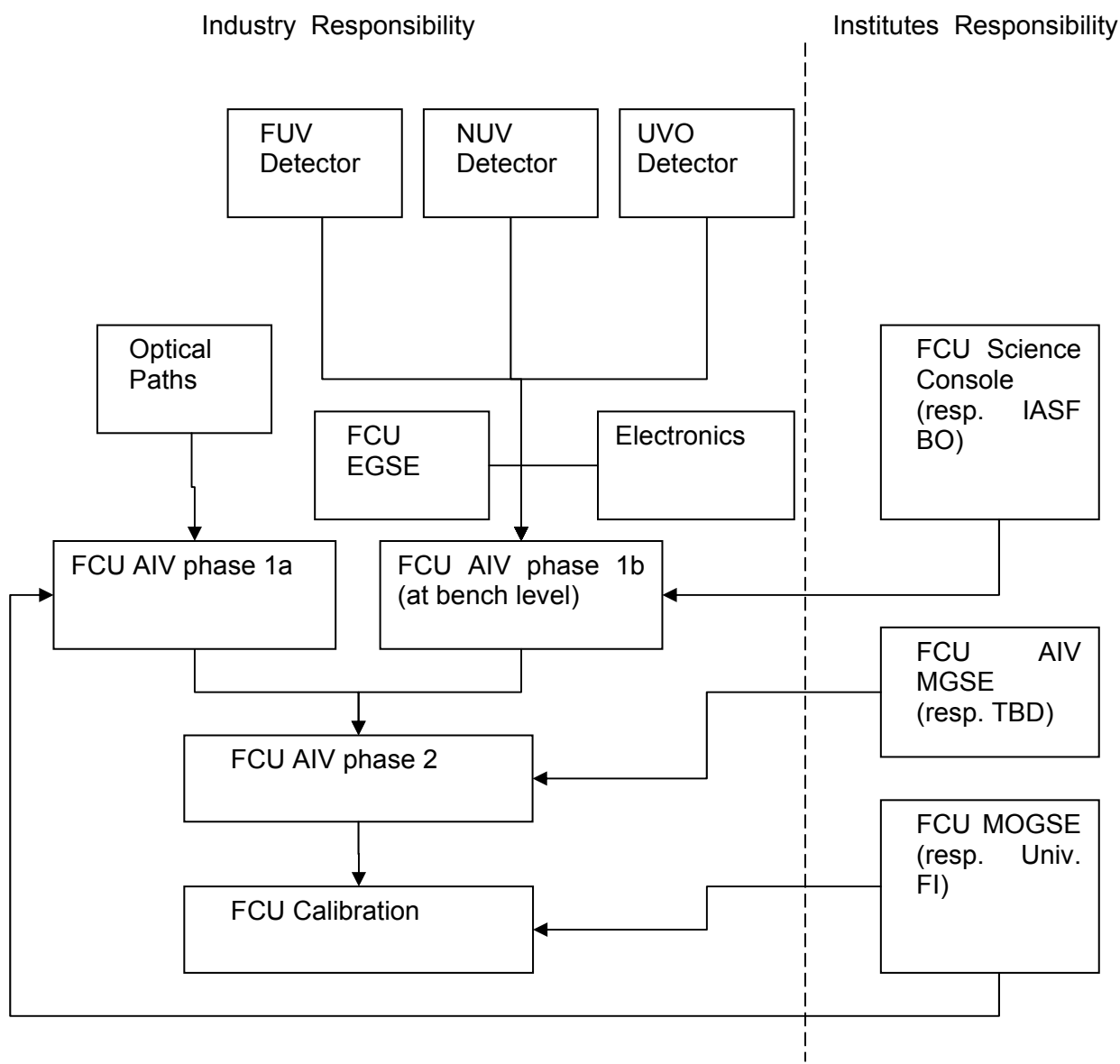

Figure 109 - FCU AIV Flow (Level 3)

## 2.4    Flight Model (FM)

### 2.4.1  AIV Objectives

As the FM is built in accordance with a design that has been qualified by the EQM, it will be subjected to the environmental and functional program required for flight acceptance certification. The AIV activities shall be based on the concept presented in Section 2.

### 2.4.2  Configuration

The FM will consist of all the FCU components presented in Section 2.

## 2.4.3 AIV Activities

Prior to delivery, the FM will be subjected to a full environmental and functional acceptance program, based on the AIV concept presented in Section 2.

All qualification testing and software verification will be carried out on the complete FCU.

A suitable calibration campaign is required for the FM.

The FM test matrix shown in Table 59 and Table 60 summarize the proposed test plan.

Table 59 - FCU FM Test Matrix at Unit Level (Level 1)

| Legenda: A = Accept. levels Q = Qualif. levels | Vibration | Alignment | Thermal Balance | Thermal Vacuum. | Mass Properties | UV Calibration | Optical Calibration | Conducted EMC | Radiated EMC | ESD | Functional Test | Performance Test | Delivery to WSO |
|---|---|---|---|---|---|---|---|---|---|---|---|---|---|
| Optical Bench – Mechanical | | | | | X | | | | | | | | |
| Pick Up Mirror (rotating / mosaic) | A | X | | A | X | | X | | | | X | X | |
| NUV Fixed Mirror | | | | | X | | X | | | | X | X | |
| UVO Fixed Mirror 1 | | | | | X | | X | | | | X | X | |
| UVO Fixed Mirror 2 | | | | | X | | X | | | | X | X | |
| UVO Filter Wheel | A | | | A | X | | X | | | | X | X | |
| NUV Filter Wheel | A | | | A | X | | X | | | | X | X | |
| FUV Filter Wheel | A | | | A | X | | X | | | | X | X | |
| Calibration Unit | A | | | A | X | | X | | | | X | X | |
| UVO adj mirror | | | | | | | | | | | | | |
| NUV adj mirror | | | | | | | | | | | | | |
| UVO Camera | | | | | | | | | | | | | |
| UVO FEE | | | | | | | | | | | | | |
| NUV Detector head | | | | | | X | | | | | X | X | |
| NUV FEE | | | | | | | | | | | X | | |
| FUV Detector head | | | | | | X | | | | | X | X | |
| FUV FEE | | | | | | | | | | | X | | |
| ICU | A | | | A | X | | | | | X | X | | |

| Legenda:<br><br>A = Accept. levels<br><br>Q = Qualif. levels | Vibration | Alignment | Thermal Balance | Thermal Vacuum. | Mass Properties | UV Calibration | Optical Calibration | Conducted EMC | Radiated EMC | ESD | Functional Test | Performance Test | Delivery to WSO |
|---|---|---|---|---|---|---|---|---|---|---|---|---|---|
| PSU | A | | | A | X | | | | | X | X | | |
| Interconnecting Harness | | | | | | | | | | | | | |

Table 60 -  FCU FM Test Matrix at Subsystem Level (Level 2) and at FCU Level (Level 3)

| Legenda:<br><br>A = Accept. levels<br><br>Q = Qualif. levels | Vibration | Alignment | Thermal Balance | Thermal Vacuum. | Mass Properties | UV Calibration | Optical Calibration | Conducted EMC | Radiated EMC | ESD | Functional Test | Performance Test | Delivery to WSO |
|---|---|---|---|---|---|---|---|---|---|---|---|---|---|
| Optical Paths | A | X | | A | X | | | | | | X | X | |
| UVO Detector | A | | | A | X | X | X | | | | X | X | |
| NUV Detector | A | | | A | X | X | | | | | X | X | |
| FUV Detector | A | | | A | X | X | | | | | X | X | |
| Electronics&Detectors | | | | A | | X | X | X | X | | X | X | |
| | | | | | | | | | | | | | |
| FCU Assembly | | | | | | X | X | | | | X | X | X |
| | | | | | | | | | | | | | |
| FCU System EGSE | | | | | | | | | | | X | X | ? |
| FCU System Science Console | | | | | | | | | | | X | X | X |
| FCU MGSE | | | | | | | | | | | X | X | ? |
| FCU MOGSE | | | | | | | | | | | X | X | |
| FCU Cryo GSE | | | | | | | | | | | X | X | ? |
| | | | | | | | | | | | | | |
| | | | | | | | | | | | | | |

## 2.4.4  Integration Sequence and GSE Requirements

As for the FCU EQM (see Section 2.3.4).

## 2.5    Flight Spares (FS)

### 2.5.1  AIV Objectives

As the FS is built in accordance with a design that has been qualified by the EQM, it will be subjected to the environmental and functional program required for flight acceptance certification.

### 2.5.2  Configuration

The FS will provide TBD spare parts of the FCU components presented in section 1.

### 2.5.3  AIV Activities

Prior to delivery, the FS will be subjected to a full environmental and functional acceptance program.

A suitable calibration  campaign is required for TBD FS items.

The FM test matrix applies herein for the corresponding FS parts.

### 2.5.4  Integration Sequence and GSE Requirements

The FM test sequence and GSE requirements applies herein for the corresponding FS parts.

# 3.    FACILITIES

## 3.1    AIV Facilities

The mechanical, electrical, functional, performance tests activities identified in section 2 will be carried out under specific environmental conditions.

The Level 1 integration and test environment will be based on clean rooms of class TBD featuring the following environmental conditions:

- Temperature: TBD°C ± TBD°C,
- Pressure STP Ambient Condition
- Relative Humidity TBD% ± TBD%

The Level 2 integration and test environment will be based on clean rooms of class TBD featuring the following environmental conditions:

- Temperature: TBD°C ± TBD°C,
- Pressure STP Ambient Condition
- Relative Humidity TBD% ± TBD%

The Level 3 integration and test environment will be based on clean rooms of class 100,000 featuring the following environmental conditions:

- Temperature: 22°C ± 3°C,
- Pressure STP Ambient Condition
- Relative Humidity 55% ± 10%

Additional facilities will be required for the environmental tests (vibration, thermal vacuum, EMC).

## 3.2 Calibration Facilities

Optical tests and calibration will be performed at the DXR-2 synchrotron beam facility at the INFN Laboratori Nazionali at Frascati where a VIS-UV radiation beam cover the 115-650 nm spectral range. The facility can arrange large optical systems, such as FCU, in a 10000-class clean room with the availability of vacuum facility and cooling systems. The radiation feature are unique because of the high intensity radiation is emitted on a continuous spectrum and time response can be investigated by means of pulsed radiation.

# 4.   GROUND SUPPORT EQUIPMENT

This section identifies the mechanical, electrical and optical tools required to support the AIV and Calibration activities presented in section 2.

## 4.1   EGSE

### 4.1.1  Concept and Configuration

The various AIV flows presented in section 2 identify a number of EGSE items:

#### 4.1.1.1 Detectors TE

Level 1 activities require the Test Equipments (TE) to support the development and test activities to be carried out on the Detector Units, namely:

- Detector Head TE: one for each FUV, NUV, and UVO
- FEE TE: one for each FUV, NUV, and UVO

The present baseline concept for the Detector Head TE is to have a self supporting unit incorporating:

- The Detector Front-End electronics functions
- The Detector Controller Electronics functions
- All the other functions needed in order to gather, process and display image data from the detector, namely:
  - o   Real time data acquisition, filing/archiving and quick look
  - o   Commanding

It will be self supporting in the sense that it will perform the required functions working as standalone equipment directly interfaced to the  CCD socket.

The present baseline concept for the FEE TE is to have a self supporting unit incorporating:

- The Detector Controller Electronics functions
- All the other functions needed in order to command and control the Detector Subsystem under test (NUV Detector / FUV Detector / UVO Detector) and to gather the image and HK Data from the FEE, namely:
  - o   Real time data acquisition, filing/archiving

- o Near real time Quick Look of the HK data
- o Commanding

In addition, the FEE TE will buffer and forward in real time to the Science Console all the data acquired from the FEE. The buffer function will include the addition of a suitable header to the acquired data. The resulting pseudo TM packets of data will be sent to the Science Console through a TCP/IP connection established on the LAN.

This approach allows to limit the FEE TE to the commanding and HK data processing.

All the remaining functionalities required to support the Level 2 activities will be provided by the Science Console, namely:

- Near real time data acquisition and archiving of the pseudo TM packets received from the FEE TE
- Near real time processing and quick look of the archived engineering and scientific data
- Off-line processing and quick look of the archived engineering and scientific data.

### 4.1.1.2 Electronics TE

Level 1 activities on the PSU will require standard laboratory test equipment.

Level 1 & 2 AIV activities on the ICU will be provided by the FCU EGSE described in Section 4.1.1.3.

### 4.1.1.3 FCU EGSE

As mentioned in section 4.1.1.2, the FCU EGSE will be exploited starting with the development and AIV activities of the ICU.

Hence, in conjunction with the Science Console, it will support all the AIV and Calibration activities at FCU Instrument level (Level 3).

As sketched in Figure 110, the FCU EGSE will consist mainly of:

- Power SCOE
- Satellite Interface Simulator (SIS)
- Central Checkout Equipment (CCOE)

The Power SCOE will provide to the FCU the missing Spacecraft elements related to the electrical power supply.

- The SIS will simulate the On Board Data Handling Bus. In near real time, it will:
- forward to the FCU the commands received from the CCOE
- forward to the Science Console the echo of the commands received from the CCOE
- forward to both the CCOE and to the Science Console the TM packets received from the FCU.

The CCOE will be the operator console which will:

- sequence the test operations by sending the Telecommand packets to the instrument
- archive all the Telemetry packets
- monitor the instrument HK data

All the remaining functionalities required to support the Level 3 activities will be provided by the Science Console, namely:

- Near real time data acquisition and archiving of the pseudo TM packets received from the SIS
- Near real time processing and quick look of the archived engineering and scientific data
- On-line processing and quick look of the archived engineering and scientific data.

In some case (e.g. Calibration), additional computers shall be exploited to perform on-site some specific off-line scientific analysis.

Time to time, the Science Console shall transfer via Internet the data to a remote central archive from where the test data will be accessible to the rest of the FCU collaboration for further analysis.

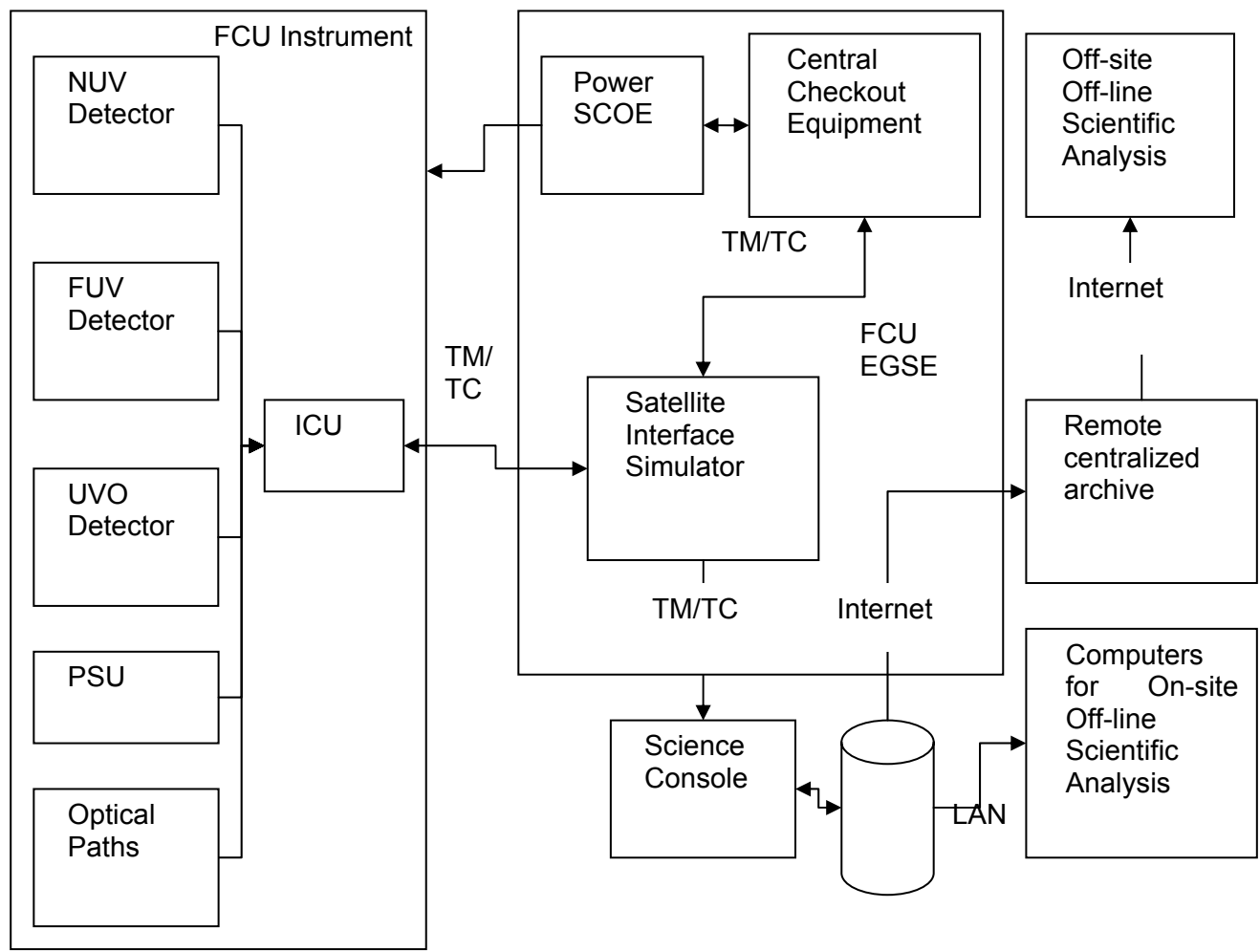

Figure 110 - FCU EGSE Design Concept

## 4.1.2 Utilization and deployment Plan

TBW.

## 4.2 MGSE

### 4.2.1 Concept and Configuration

The MGSE will include:

- the reusable instrument and EGSE transport containers required for the FCU AIV, included the delivery to the WSO satellite Contractor premises

- and the mechanical items required in order to handle the instrument during the level 1, 2 and 3 AIV, included the lifting device to be delivered to the WSO satellite Contractor premises.

TBD additional MGSE items will be required to support the Calibration activities.

## 4.2.2 Utilization and deployment Plan

TBW

# 4.3 FCU OGSE

## 4.3.1 Concept and Configuration

OGSE will provide for the level 2 verification the instrumentation in a controlled environment (clean room) to perform

- VIS and UV calibration of the detector subsystem of FUV, NUV and UVO paths
- Optical alignment of each of the FUV, NUV, UVO paths in air (following flow 2 of the Optical Path)
- Optical alignment of the FUV, NUV, UVO paths mounted on the optical bench in air
- Optical alignment of the FUV, NUV, UVO paths on the optical bench in vacuum
- Verification of the optical alignment of the FUV, NUV, UVO paths mounted on the optical bench in air after vibration tests
- Testing the optical performance (PSF, throughput, polarization analysis, spectral dispersion, reproducibility of the image positioning) of each of the FUV, NUV, UVO paths in vacuum (following flow 2 of the Optical Path)
- Testing the optical performance (PSF, throughput, polarization analysis, spectral dispersion, reproducibility of the image positioning) of each of the FUV, NUV, UVO paths in vacuum mounted on the optical bench
- Thermal variation of the optical alignment and performance in air and in vacuum

OGSE configuration will consist of:

- A mechanical support for the rough and fine positioning of the FUV, NUV and UVO paths (following flow 2 of the Optical Path)
- A mechanical support for the rough and fine positioning of the Optical bench
- A test chamber containing the Optical bench to perform optical testing in vacuum
- A laser beam and optical references for the optical alignment
- A polychromatic source emitting on a spectrum from 115 nm to 800 nm
- Calibrated detectors to assess the optical performances

## 4.3.2 Utilization and deployment Plan

TBW.

# Chapter V.

# Summary and Conclusions

We have illustrated the science case that move the Italian community to call for a national participation to WSO-UV.

As we have shown, the scientific plans for WSO-UV are very ambitious, and span all of the astronomical research branches. WSO-UV will be operating in the second decade of this century, and it will be a fundamental tool for the development of astronomical knowledge, fully integrated with the many other space and ground-based observatories operating in the same temporal interval.

The performances of the WSO-UV HIRDES and LSS spectrographs will provide the required science goals outlines in Chapter I.

The scientific objectives requiring imaging photometry, slitless low resolution field spectroscopy, as well as polarimetric field measurements have driven the design of the Field Camera Unit (FCU). The main purpose of the camera is direct observation of specific targets, and survey observations. The FCU will include three different channels to provide imaging, polarimetric, and low resolution slit-less spectroscopy facilities from far-UV to visual bands:

- A far ultraviolet high sensitivity and large field of view channel with the possibility of low resolution slitless spectroscopy (**FUV** channel)

- A near ultraviolet high angular resolution imager, with the possibility of low resolution slitless spectroscopy and polarimetric facilities (**NUV** channel)

- A near ultraviolet-visual diffraction limit imager, with the possibility of low resolution slitless spectroscopy and polarimetric facilities (**UVO** channel)

Being one of the focal plane instrument of a general-purpose orbital telescope as WSO-UV is, the FCU will provide high resolution and high sensitivity imaging available to the international community in a wavelength range otherwise uncovered during the years of operation.

The **science** for FCU requires:

- the possibility to obtain diffraction limited images for λ>230 nm (goal λ>200 nm).

- S/N=10 per pixel in one hour exposure time down to:
    - V=26.2 (21.5 – 11.1) for a O3 V (A0 V - G0 V) star in the band F150W
    - V=24.6 (21.9 – 19.5) for a O3 V (A0 V - G0 V) star in the band F250W
    - V=26.3 (26.3 – 26.3) for a O3 V (A0 V - G0 V) star in the band F555W

- Spectral resolution R=100 at λ=150 nm, R≥100 at λ=250 nm, and R=250 at λ=500 nm.

- Polarimetric filters

- High time resolution imaging
- The possibility to obtain wide field (~ 5x5 arcmin$^2$) images from Far-UV to visual wavelengths.
- Partial overlap between spectral range covered by FUV and NUV and NUV and UVO, for relative calibration purposes.

The throughput of the UVO channel is much better than that of WFPC2 which, by the way, is the oldest of the HST cameras, and is comparable to that of ACS/HRC and WFC3/UVIS being better in the UV range.ACS/WFC, being optimized in the visual range, has a much better throughput than the UVO channel.

Due to its large field of view the UVO channel of the FCU has a discovery efficiency, defined as system throughput times the area of the FOV as projected onto the sky, equal or greater than that of ACS/WFC. The performance of the FUV channel is even better when compared to HST because, in this case no HST camera working in this wavelength range has a large field of view.

The optical, mechanical and electronic configurations of the Field Camera Unit can be summarized as follows:

**Optics**

Two alternative layouts have been analyzed. The main difference is the fore optics system that is an on-axis rotating mirror for the first layout and an off-axis fixed mirrors mosaic in the second layout. From the point of view of the optical performances both layouts satisfy the requirements with the exception of the off-axis design of the NUV channel that does not comply with the requirements on the throughput and on the optical quality of the images. A slitless spectroscopic mode has been explicitly analyzed for the NUV channel.

**Mechanics & Mechanisms**

Several mechanical layouts based on the different optical designs have been developed. Different options for the FCU optical bench structure and material have been proposed and are under evaluation. A preliminary list of mechanisms has been identified. In particular detailed analysis on the design of the central rotating pick up mirror has been performed. Different solutions that could comply with the mechanism requirements have been analyzed.

**Detectors**

Two different kinds of detectors have been selected to be used in the FCU instrument. The FUV & NUV channels will use MCP based photon counting detectors with a CCD as readout system. The difference between the two detectors will be in the photocathode that will be specialized for the channel wavelength range. The UVO channel will use a UV-optimized back illuminated CCD.

**Electronics**

Preliminary architecture has been designed for the ICU and PSU subsystems. The FCU instrument operational modes have been identified and a preliminary scientific data acquisition and processing flow has been developed. Preliminary definition of telemetry and telecommands packets has been developed. The characteristics of the MIL1553B spacecraft bus interface have been thoroughly analyzed and compared with the transfer data rate requirements for the FCU instrument. The results of this analysis have been sent to the Russian team responsible for the bus interface for discussion and common analysis.

The following trade off process have been finalized:

- MCP vs. CCD for the NUV channel

Summary and Conclusions

The performances of two detectors in the NUV channel wavelength have been compared leading to the selection of the MCP detector.

- CCD as MCP readout system

The performances of different MCP readout systems have been analyzed and compared to the top level requirements. In particular the characteristics of CCD, Wedge & Strip Arrays and Delay Lines were investigated leading to the selection of CCD based readout system.

- Connections among FCU, ICU and SDMU

Several issues have to go through trade off processes that will be finalized in the following phases. Here is a list of the main ones:

- Opto-mechanical layout
    o Fore optics design

    To proceed with the opto-mechanical design the fore optics layout has to be frozen.

    o NUV & UVO Focusing Tip/Tilt Mechanisms

    The tolerance analysis of the optical design will allow to confirm the presence and the requirements of these mechanisms.

    o NUV Grating wheel

    The finalization of the NUV optical design will tell if this mechanism is necessary.

- CCD cooling system

A choice between active and passive CCD cooling system must be done before proceeding with the thermo-mechanical design of the CCD detector head.

- Optical bench material & structure

The analysis of the different options for the material and the structure of the FCU optical bench will proceed in parallel with the development of the FCU thermo-mechanical design.

- MIL1553B interface

Definitive adoption at spacecraft level of MIL1553B interface will have a great impact on the FCU ICU architecture (needs for data compression algorithm implementation, greater memory buffers, interfaces boards).

AIV and GSE plans have been presented in a preliminary version and will be finalized in the next B1 Phase.

The instrument verification plan has been developed assuming a Model Philosophy which is under evaluation at the time of writing.

The current status of the instrument design led to identify two different AIV flows.

The proposed AIV plan have been used in order to identify the preliminary concept and configuration of the GSE items to be procured to support the FCU instrument qualification and acceptance programs.

# ACKNOWLEDGEMENTS

We thank for the useful discussions and for providing contributions to the present documents J. Anderson, L. Bianchi, A. Bonati, G. Bonanno, G. Bono, N. Brosch, G. Di Cocco, P. Giommi, A.I. Gómez de Castro, E. Held, M. Huang, N. Kappelmann, J. Larruquert, J.L. Linsky, M. Miccolis, Y. Momany, M. Orio, G. Raimondo, D. Recupero, M. Sachkov, N. Schneider, E. Skripunov, A. Shugarov, B. Shustov, M. Stiavelli, G. Strazzulla, E. Tommasi, G.P. Tozzi, and the whole WSO-UV Russian team.

# REFERENCES

Anderson et al. 2006, A&A, 454, 1029

Anderson and King 2000, 112, 1360

Anderson and King 2003, PASP, 115, 113

Anderson and King 2006, ACS ISR 2006-01

Ajhar & Tonry 1994 ApJ 429, 557

Antonucci, R.R.J., 1993, ARA&A 31, 473

Araujo-Betancor S. et al. 2005, ApJ 622, 589

Ardila, D.R.; Basri, G.; Walter, F.M.; Valenti, J.A.; Johns-Krull, C.M. 2002 ApJ 567 1013

Ayres, T.R, Brown, A., and  Harper, G.M., 2003b, ApJ 598, 610

Ayres, T.R, Brown, A., Harper, G.M., Osten, et al. 2003a, ApJ 583, 963

Ayres, T.R., 1997, JGR (Planets) 102, 1641

Ayres, T.R., 1999, ApJ 525, 240

Bacciotti, F., Mundt, R., Ray, T. P., Eislöffel, J., Solf, J.  & Camezind, M. 2000, ApJ, 537, L49

Bacciotti, F., Ray, T. P., Mundt, R., Eislöffel, J. & Solf, J.,  2002, ApJ, 576, 222

Bacciotti, Francesca; Eislöffel, Jochen; Ray, Thomas P., 1999 A&A 350 917

Bailyn 1995, ARA&A 33, 133

Bally, John; Sutherland, Ralph S.; Devine, David; Johnstone, Doug, 1998, AJ 116, 293

Barbero et al. 1990 ApJ 351, 98

Bedin et al. 2000, A&A, 363, 159

Bedin et al. 2001, ApJ, 560, L79

Bedin et al. 2003, AJ, 127, 247

Bedin et al. 2004, ApJ, 605, L125

Behr et al. 1999, ApJ 517, L135

Bellazzini et al. 2002, AJ 123, 1509

Belle K. et al. 2003, ApJ 587, 373

Bentz, M., et al., 2006, ApJ 651, 775

Benz & Hills 1987, ApJ 323, 614

Best, P.N., et al. 2005, MNRAS, 362, 25

Blakeslee et al. 2001 MNRAS 320, 193

Bond, H.E., et al.: Wide Field Camera 3 Instrument Mini-Handbook, Version 3.0, Baltimore: STScI (2006)

Brown T.M. 2004 Astroph Space Science 291, 215

Brown T.M. et al 1998 ApJ 504, 113

Brown T.M. et al 2000 ApJ 532, 308

Burstein et al. 1988 ApJ 328,440

Butler, C.J., 1994, Sol. Phys. 152, 35

Buzzoni, 1989, ApJS 71, 817

Campana, S., et  al. 2006, Nature, 442, 1008

Cantiello et al. 2003 AJ 125, 2783

Cantiello et al. 2005 ApJ 634, 239

Cantiello et al. 2007 ApJ accepted

Capetti, A., Axon, D.J, Macchetto, F.D., Marconi, A., Winge, C., 1999, ApJ 516, 187

Catelan et al. 2001, AJ 122, 3171

Catelan et al. 2006, ApJ 651, L133

Charbonneau, D., Brown, T., Noyes, R., & Gilliland, R. 2002, ApJ, 568, 377

Chernoff and Shapiro 1987, ApJ, 322, 113

Clayton G. C., Gordon K. D., Salama F., Allamandola L. J., Martin P. G., Snow T. P., Whittet D. C. B., Witt A. N., Wolff M. J., ApJ 592, 947 (2003)

Coffey, D., Bacciotti, F., Ray, T. P., Eislöffel, J., Woitas, J., 2007 ApJ in press

Coffey, D., Bacciotti, F., Woitas, J., Ray, T. P., & Eislöffel, J., 2004, ApJ, 604, 758

Cool A. et al. 1998, ApJ 439, 695

Cote, P. et al 2004 ApJS 153, 223

Cuntz M., Saar S.H., Musielak Z.E., 2000, ApJ 533, L151

D'Antona et al. 2002, A&A 395, 69

D'Antona et al. 2005, ApJ 631, 868

Davies & Hansen 1998, MNRAS 301, 15

Davies, Piotto & de Angeli 2004, MNRAS 349, 129

De Marco et al. 2005, ApJ 632, 894

de Martino D. et al. 1999, A&A 350, 517

de Mello,D, et al.,A.J., 2006,132:"Near-Ultraviolet Sources in the Great Observatories Origins Deep Survey Fields

Dere, K. P., Bartoe, J.-D. F., and Brueckner, G. E., 1989, Sol. Phys. 123, 41

Devine, David; Grady, C. A.; Kimble, R. A.; Woodgate, B.; Bruhweiler, F. C.; Boggess, A.; Linsky, J. L.; Clampin, M. 2000, ApJ 542 L115

Di Matteo T., Springel V., Hernquist L., 2005, Nature, 433, 604

Dieball A. et al. 2005, ApJ 634 L105

Dorman & Rood 1993, ApJ 409, 387

Dorman et al. 1993, ApJ 419, 596

Doschek, G. A., et al., 1975, ApJ  196, L83

Dougados, C.; Cabrit, S.; Lavalley, C.; Ménard, F. 2000 A&A 357 L61

Ehrenfreund P., Rasmussen S., Cleaves J., Chen L., Astrobiology 6, 490 (2006)

Eriksson M.S. et al. 2004, A&A 422, 987

Federman S. R., Cardelli J. A., van Dishoeck E. F., Lambert D. L., Black J. H., ApJ 445, 325 (1995)

Federman S. R., Lambert D. L., Sheffer Y., Cardelli J. A., Andersson B.-G., van Dishoeck E. F., Zsargó J., ApJ 591, 986 (2003)

Feldman, U., Curdt, W., Landi, E., and Wilhelm, K., 2000, ApJ 544, 508

Ferlet R., André M., Hébrard G., Lecavelier des Etangs A., Lemoine M., Pineau des Forêts G., Roueff E., Rachford B. L.,  et al.  ApJ 538, L69 (2000)

Ferrarese, L. et al. 2006 ApJS 164, 334

Ferrarese, L., & Merritt, D., 2000, ApJL 539, 9

Ferraro 2006, astro-ph/0601217

Ferraro et al 2001, ApJ, 561,337

Ferraro et al. 1998, ApJ 500, 311

Ferraro et al. 1999, ApJ 522, 983

Ferraro et al. 2000, ApJ 537, 312

Ferraro et al. 2003, ApJ 588, 464

Ferraro et al. 2006, ApJ 647, L53

Fruchter, A.S. 2003, priv. comm.

Fusi Pecci et al. 1992, AJ 104, 1831

Gaensicke B.T. et al. 2005, ApJ 622 589

Gaensicke B.T. et al. 2006, MNRAS 365, 969

Gaensicke B.T. et al., 1998, ApJ, 333, 127

Gallagher, J.S., & Code, A.D., 1974, ApJ, 179, 303

Gallagher, J.S., & Starrfield, S., 1976, MNRAS, 176, 53

Gallagher, J.S., & Starrfield, S., 1978, ARAA, 16, 171

Gebhardt K., et al., 2000, ApJL 539, L13

Giavalisco, M, et al.,Ap.J., 2004,600L: "The Great Observatories Origins Deep Survey: Initial Results from Optical and Near-Infrared Imaging"

Gomez de Castro, Ana I.; Lamzin, S. A. 1999 MNRAS 304 L41

Gómez de Castro, Ana I.; Lecavelier, Alain; D'Avillez, Miguel; Linsky, Jeffrey L.; Cernicharo, José, 2006, Ap&SS 303, 33

Gómez de Castro, Ana I.; Verdugo, Eva 2001 ApJ, 548 976

Gonzales Delgado, M., et al., 1998, ApJ 505, 174

Greggio & Renzini 1990, ApJ 364, 35

Grindlay et al 2001, Science, 292, 2290

Grundahl et al. 1999, ApJ 524, 242

Guinan, E. F.; Ribas, I.; Engle, S. G., 2007, in "UV Astronomy: Stars from Birth to Death", Proceedings of the IAU-GA/JD4, A.I. Gómez de Castro & M. A. Barstow (eds.), Editorial Complutense, in press

Hartigan, P., Edwards, S., Pierson, R., 2004 ApJ 609, 261

Hartigan, P., Morse, J. 2007 ApJ 660, 426

Hartigan, P., Morse, J. A., Tumlinson, J., Raymond, J. & Heathcote, S., 1999, ApJ 512, 901

Hartigan, P., Raymond, J. & Hartmann, L., 1987, ApJ 316, 323

Hartley L. E. et al. 2005, ApJ 623, 425

Haswell C. et al. 1997, ApJ 476, 847

Hawley S.L., and Johns-Krull, C.M., 2003, ApJ 588, L112

Hawley S.L., et al., 2003, ApJ 597, 535

Holland, S.T. 2006, GCN GRB Observation Report N. 5255

Horne K. et al. 1994, ApJ 426, 294

Hut & Verbunt 1983, Nature 301, 587

Hutchings, J.B., et al., 1995, AJ, 110 2394

Jensen et al. 2003 ApJ 583, 712

Jordan, A. et al 2007 astro-ph 2496

Judge, P.G., Saah, S.H., Carlsson, M., and Ayres, T.R., 2004,ApJ, 609, 392

King et al. 1998, ApJ, 492, L37

Knauth D. C., Andersson B.-G., McCandliss S. R., Warren Moos H., Nature 429, 736 (2004)

Knigge C. & Drew J. 1997, ApJ 486 445

Knigge C. et al. 2002, ApJ 579, 752

Königl, A. & Pudritz, R., 2000, in Protostars and Planets IV, V. Mannings, A. P. Boss, & S. S. Russell (Tuscon: Univ. Arizona Press), 759

Kormendy, J., & Richstone, D., 1995, ARA|&A 33, 581

Lammer, H., Selsis, F., Ribas, I., et al 2003, ApJ, 598, L121

Larsson S. 1995, ASP Conf. Ser. 85, p.311

Lee et al. 1994, ApJ 423, 248

Leonard & Livio 1995, ApJ 447, L121

Li J. & Wickramasinghe D.T. 1998, MNRAS 300, 718

Linsky, J. L., Wood, B. E., Judge P., et al. 1995, ApJ 442, 381

Linsky, J.L., Wood, B.E., Brown, A., & Osten, R.A. 1998, ApJ, 492, 767

Liu et al. 2002 ApJ 564, 216

Lombardi et al. 1995, ApJ 445, L117

Long K. et al. 2004, ApJ 602, 948

Mackey & Nielsen 2007, astro-ph 0704.3360

Malloci, Giuliano; Joblin, Christine; Mulas, Giacomo, Chemical Physics 332, 353 (2007)

Mapelli et al. 2007, arXiv:0706.3311

Maran, S. P., et al., 1994, ApJ 421, 800

Marconi, A., & Hunt, L.K., 2003, ApJL, 589, 21

Mathys 1991, A&A 245, 467; Shetrone & Sandquist 2000, AJ 120, 1913

Mengel et al. 2002, A&A, 383, 137

Milone et al. 2006, A&A, 456, 517

Milone et al. 2007, ApJ, submitted

Moehler et al. 2004, A&A 415, 313

Moisheev A.A. et al., 2006, Summary of the "Spectrum UV/WSO-UV" space complex and its components: Initial data on the design of the telescope T-170M main scientific instrumentation and the "Spectrum UV/WSO-UV" scientific instrumentation complex equipment – Lavochkin Association (112110-T170M-4-06).

Moni Bidin et al. 2006, A&A, 451, 499

Momany et al. 2007, A&A, 468, 973

Mulas G., Malloci G., Joblin C., Toublanc D., A&A 446, 537 (2006)

Narayan, R. & Yi, I., 1995, ApJ 452, 710

Natta, A.; Testi, L.; Muzerolle, J.; Randich, S.; Comerón, F.; Persi, P. 2004, A&A 424 603

Neill, J.D., & Shara, M.M., 2004, ApJ, 127, 816

Orosz & Wade R. 2003, ApJ 593, 1032

Pace et al. 2006, A&A, 452, 493

Pagano I., Linsky, J. L., Carkner, C., et al. 2000, ApJ 432, 497

Pagano, I., Rodonò, M., Linsky, J. L., Neff, J. E., Walter, F. M., Kovári, Zs., & Matthews, L. D. 2001, A&A 365, 128

Pagano, I., Linsky, J. L., Valenti, J., Duncan, D.K, 2004, A&A 415, 331

Pagano I., Ayres, T.A., Lanzafame, A.C., Linsky J.L., Montesinos, B., and Rodonò, M., 2006, ApSS 303, 17-31

Pallé Bagó, E., Butler, C. J., 2001, IAU Symp. 203, 602

Parker, E. 1988, ApJ, 330, 474

Peng E.W., et al 2006 ApJ 639, 838

Peter, H., 2001, A&A 374, 1108

Peter, H.; Judge, P. G, 1999, ApJ 522, 1148

Peterson, B. M. & Bentz, M. C., 2006, NewAR 50, 796

Pillitteri, I., Micela, G., Reale, F., Sciortino, S. , A&A, 2005, 430, 155

Piotto et al. 1999, AJ, 118, 1727

Piotto and Zoccali, 1999, A&A, 345, 485

Piotto et al. 2004, ApJ, 604, 409

Piotto et al. 2005, ApJ, 621, 777

Piotto et al. 2007, ApJ, 661, L53

Pizzolato, N., Maggio, A., Micela, G., Sciortino, S. and Ventura, P., 2003, A&A 397, 147

Podio, L., Bacciotti, F.; Nisini, B.; Eislöffel, J.; Massi, F.; Giannini, T.; Ray, T. P. 2006 A&A 456, 189

Rachford B. L., Snow T. P., Tumlinson J., Shull J. M., Blair W. P., Ferlet R., Friedman S. D., Gry C., Jenkins E. B., Morton D. C., et al. ApJ 577, 221 (2002)

Rachford B. L., Snow T. P., Tumlinson J., Shull J. M., Roueff E., Andre M., Desert J.-M., Ferlet R., Vidal-Madjar A., York D. G., ApJ 555, 839 (2001)

Raimondo et al. 2005 AJ 130, 2625

Raimondo et al. 2007 in preparation

Ray et al. 1996, ApJ, 468, L103

RD2 – Top Level User Requirments for WSO-UV, WSO-ITA-SYS-URD-0001

Rebull L. M., Stauffer J. R., Ramirez S. V., et al. 2006. AJ, 131, 2934

Recio-Blanco et al. 2006, A&A, 452, 876

Redfield, S., Ayres, T.R., Linsky, J.L., et al., 2003, ApJ 585, 993

Reipurth 1989, Nature 340, 42

Renzini A., 2006 ARAA 44, 141

Ribas, I.; Guinan, E.F.; Güdel, M.; Audard, M., 2005, ApJ 622, 680

Richter P., Sembach K. R., Wakker B. P., Savage B. D., ApJ 562, L181 (2001)

Robinson, R.D., Linsky, J.L., Woodgate, B., and Timothy G., 2001, ApJ 554, 368

Robinson, R.D., Wheatley, J.; Welsh, B. Y.; et al, 2005 ApJ 633, 447

Rodonò, M., Messina, S., Lanza, A. F., Cutispoto, G., Teriaca, L., 2000, A&A 358, 624

Rood & Crocker 1989, IAU Colloq. 111, 103

Rood 1973, ApJ 184, 815

Rutledge, R. E., Basri, G., Martín, E. L., and Bildsten, L. 2000, ApJ 538, L141

Sandage 1953, AJ 58, 61

Sarna & de Greve 1996, QJRAS 37, 11

Schenker K.A. et al. 2002 MNRAS 337 1105

Schmidtke, P.C., et al., 2004, ApJ, 127, 469

Shara et al.  1997,ApJ,489,59

Shara M.M. et al. 1986, ApJ 311 163

Shara M.M. et al. 2007 Nature 446, 159

Shkolnik E., Walker G.A.H., Bohlender D.A., 2001, AAS 33, 1303

Shkolnik, E.; Walker, G. A. H.; Bohlender, D. A.; Gu, P.-G.; Kürster, M., 2005, ApJ 622, 1075

Shore S. 2002 in Proc. "Classical Nova Explosions" p. 175

Shu, F. H., Najita, J. R., Shang, H. & Li, Z.-Y., 2000, in Protostars and Planets IV, V. Mannings, A. P. Boss, & S. S. Russell (Tuscon: Univ. Arizona Press), 789

Shull J. M., Beckwidth W., ARA&A 20, 163 (1982)

Shull J. M., Tumlinson J., Jenkins E. B., Moos H. W., Rachford B. L., Savage B. D., Sembach K. R., Snow T. P., Sonneborn G., York D. G.,  et al.  ApJ 538, L73 (2000)

Silk, J., & Rees, M.J., 1998, A&A 331, 1

Sion E. 1999, PASP 111, 532

Sion E. et al. 2004, ApJ 614, L61

Smette,  A., et al. 2001, ApJ, 556, 70

Snow T. P., Mc Call B. J., ARA&A 44, 367 (2006)

Snow T. P., Rachford B. L., Tumlinson J., Shull J. M., Welty D. E., Blair W. P., Ferlet R., Friedman S. D., Gry C., Jenkins E. B. et al. ApJ 538, L65 (2000)

Snow T. P., Smith W. H., ApJ 250, 163 (1981)

Soderblom, D., Jones, B.F. and Fisher, D., 2001, ApJ 563, 334

Spaans M., Neufeld D., Lepp S., Melnick G. J., Stauffer J., ApJ 503, 780 (1998)

Stelzer B., Flaccomio E., Briggs K., et al., 2007, A&A, 468, 463

Strigari et al. 2007, ApJ, 657, L1

Sweigart et al. 2002, "Omega Cen: A Unique Window into Astrophysics", eds. van Leeuwen, Hughes & Piotto, ASP -OCUW, p. 261

Szkody P., et al. 2007, ApJ 658, 1188

Tappe A., Rho J., Reach W. T., ApJ 653, 257 (2006)

Tielens A. G. G. M., "The Physics and Chemistry of the Interstellar Medium", Cambridge University Press (2005)

Tonry et al. 2001 ApJ 546, 681,

Townsley D.M & Bildstein L. 2003, ApJ 565, L227

Tumlinson J., Shull J. M., Rachford B. L., Browning M. K., Snow T. P., Fullerton A. W., Jenkins E. B., Savage B. D., Crowther P. A., et al. (2002)

Urban & Sion E. 2006, ApJ 642, 1029

van den Heuvel, E., et al., 1992, A&A, 262, 97

van Teeseling, A., et al. 1999, A&A, 351, L31

Vanzella, E. et al. A\&A, 2006,454:"The great observatories origins deep survey. VLT/FORS2 spectroscopy in the GOODS-South Field: Part II"

Verbunt & Meylan 1988, A&A 203, 297

Vidal-Madjar, A.; Désert, J.-M.; Lecavelier des Etangs, A.; Hébrard, G.; Ballester, G. E.; Ehrenreich, D.; Ferlet, R.; McConnell, J. C.; Mayor, M.; Parkinson, C. D., 2004, ApJ 604, 69

Vidal-Madjar, A.; Lecavelier des Etangs, A.; Désert, J.-M.; Ballester, G. E.; Ferlet, R.; Hébrard, G.; Mayor, M., 2003, Nature 422, 143

Villanova et al. 2007, ApJ, 663, 296

Vijh U. P., Witt A. N., Gordon K. D., ApJ 633, 262 (2005)

Vijh U. P., Witt A. N., York D. G., Dwarkadas V. V., Woodgate B. E., Palunas P., ApJ 653, 1336 (2006)

Viti S., Williams D., Casu S., Cecchi-Pestellini C., in preparation (2007)

Warner B. & van Zyl L. 1998, IAUS 185, 321

Warner B. et al. 2003, MNRAS 344 1193

Welsh, B. Y.; Wheatley, J.; Browne, S. E.; et al, 2006, A&A 458, 921

Whelan, Emma T.; Ray, Thomas P.; Bacciotti, Francesca; Natta, Antonella; Testi, Leonardo; Randich, Sofia 2005 Natur. 435 652

Whitney et al. 1998, ApJ 495, 284

Williams D. A., Brown W. A., Price S. D., Rawlings J. M. C., and Viti S., Astronomy & Geophysics, 48, 1.25 (2007)

Witt A. N., Gordon K. D., Vijh U. P., Sell P. H., Smith T. L., Xie R.-H., ApJ 636, 303 (2006)

Woitas, J., Ray, T. P., Bacciotti, F., Davis, C. J. \& Eisl\"offel, J., 2002, ApJ, 580, 336

Woitas, J.; Bacciotti, F.; Ray, T. P.; Marconi, A.; Coffey, D.; Eislöffel, J. 2005 A&A 432, 149

Wood, B. E., Linsky, J. L., and Ayres, T. R., 1997, ApJ 478, 745

Wood, B.E., Müller, H.-R., Zank, G.P., and Linsky, J.L., 2002, ApJ 574, 412

Worthey G., 1994 ApJS 95, 107

Woudt P. & Warner B. 2004, MNRAS 348, 599

Yuan, F. Quataert, E., Narayan, R., 2003, ApJ 598, 301

Yungelson, L., et al., ApJ, 481, 127.

# APPENDIX A: COMPOSITION OF THE WSO-UV ITALIAN SCIENCE TEAM

| Name | Affiliation | Research | Role in the FCU Team |
|------|-------------|----------|----------------------|
| Francesca Bacciotti | INAF-OA Arcetri, Firenze | Accretion and outflows in young stars | Science |
| Luigi Bedin | STScI-ESA | Globular clusters, photometric and astrometric techniques | Science |
| Enzo Brocato | INAF-OA Teramo | Stellar Evolution, Population synthesis, Surface Brightness Fluctuations. | Science, Filters, Gratings & Polarizers WG |
| Lucio Buson | INAF-OA Padova | UV Emission of Galaxies, Stellar Populations, ISM | Science |
| Carla Cacciari | INAF – OA Bologna | Stars and stellar populations - Instruments & technology | Science, Calibration WG |
| Angelo Cassatella | INAF-IFSI Roma | Hot Stars and Mass-Loss phenomena, Interactic binaries, symbiotic stars, GRB | Science, Calibration WG |
| Alessandro Capetti | INAF-OA Torino | Active Galaxies | Science, Filters, Gratings & Polarizers WG |
| Riccardo Claudi | INAF-OA Padova | Extrasolar Planets, Astrobiology, Stellar variability, Astronomical instrumentation | Science, Calibration WG |
| Domitilla De Martino | INAF-OA Capodimonte, Napoli | Cataclismic Variables (CV) | Science |
| Francesco Ferraro | Dip. di Astronomia – Univ. di Bologna | Stars and stellar populations | Science, Filters, Gratings & Polarizers WG |
| Simone Marchi | Dip. di Astronomia - Università di Padova | Solar System | Science, Filters, Gratings & Polarizers WG |
| Giacomo Mulas | INAF – OA Cagliari | Interstellar medium | Science |
| Mario Nonino | INAF-OA Trieste | Cosmology | |
| Emanuele Pace | Dip. di Astronomia e Scienza dello Spazio – Università di Firenze | Optics & Detectors. Small bodies of the Solar System. Astrobiology | Science, NUV channel resp. |
| Isabella Pagano | INAF-OA Catania | Cool Stars and Magnetic Activity, Young stars in clusters. | Principal Investigator |
| Elena Pian | INAF-OA Trieste | GRB, afterglow physics | Science |
| Giampaolo Piotto | Dip. di Astronomia - Università di Padova | Globular Clusters. Photometric and Astrometric techniques. | Instrument Scientist |
| Salvatore Scuderi | INAF-OA Catania | Detectors, Hot Stars and Mass Loss-Phenomena, SNe | System Engineer, UVO channel resp., CCD resp. |
| Steve Shore | Università di Pisa | Novae, Interacting binaries, Astronomical Instrumentation | Science |

| Name | Affiliation | Research | Role in the FCU Team |
|------|-------------|----------|----------------------|
| Massimo Turatto | INAF-OA Padova | Supernovae, nucleosynthesis, interstellar medium, planetary nebulare, Astronomical Instrumentation | Science |
| Michela Uslenghi | INAF-IASF Milano | Detectors, Cataclysmic Variables, LMXRB, AGN | Photon counting detectors, FUV channel resp. |

# GLOSSARY

| ACS | Advanced Camera for Surveys |
|-----|------------------------------|
| AFEE | Analog Front End Electronics |
| AGB | Asymptotic-Giant-Branch |
| AGN | Active Galactic Nuclei |
| AIMO | Advanced Inverted Mode |
| APS | Active Pixel Sensor |
| ASI | Agenzia Spaziale Italiana |
| BB | Broad Band |
| BC | Bus Controller |
| BH | Black Hole |
| BSS | Blue Stragglers Stars |
| CCD | Charge Coupled Device |
| CDS | Correlated Double Sampling |
| CID | Charge Injection Device |
| CMB | Cosmic Microwave Background |
| CMD | Colour Magnitude Diagram |
| CME | Coronal Mass Ejection |
| CNSA | China National Space Administration |
| COS | Cosmic Origin Spectrograph |
| CSM | Circum Stellar Material |
| CV | Cataclysmic Variables |
| DC | Direct Current |
| DFEE | Digital Front End Electronics |
| DHU | Data Handling Unit |
| DL | Delay Lines |
| DLR | Deutschen Zentrum fur Luft- und Raumfahrt |
| DNO | Dwarf Novae Oscillations |
| DQE | Detected Quantum Efficiency |
| EGSE | Electrical Ground Support Equipment |
| EHB | Extreme Horizontal Branch |
| ELT | Extremely Large Telescope |
| EQM | Engineering Qualified Model |
| ESA | European Space Agency |
| ETC | Exposure Time Calculator |
| FCU | Field Camera Unit |
| FEE | Front End Electronics |
| FGS | Fine Guidance System |
| FM | Flight Model |

| FP | Focal Plane |
|---|---|
| FPGA | Field Programmable Gate Array |
| FS | Flight Spares |
| FUV | Far UltraViolet (Channel) |
| FWHM | Full Width Half Maximum |
| GAIA | Global Astrometric Interferometer for Astrophysics |
| GALEX | Galaxy Evolution Explorer |
| GDR | Global Dynamic Range |
| GGCs | Galactic Globular Clusters |
| GHRS | Goddard High Resolution Spectrograph |
| GRB | Gamma Ray Burst |
| HAR | High Angular Resolution |
| HB | Horizontal Branch |
| HIRDES | High Resolution Double Echelle Spectrograph |
| HK | Housekeeping |
| HST | Hubble Space Telescope |
| IB | Interacting Binaries |
| ICM | Inter Clusters medium |
| ICU | Instrument Control Unit |
| IGM | Inter Galactic Medium |
| IMBH | Intermediate Mass Black Holes |
| IMF | Initial Mass Function |
| INASAN | Institute of Astronomy Russian Academy of Science |
| INTA | Instituto Nacional de Tecnica Aerospacial |
| ISM | Interstellar Medium |
| ITAR | International Traffic in Arms Regulations |
| IUE | International Ultraviolet Explorer |
| JPL | Joint Propulsion Laboratory |
| JWST | James Webb Space Telescope |
| L2 | Lagrangian Point 2 |
| LDR | Local Dynamic Range |
| LEO | Low Earth Orbit |
| LG | Local Group |
| LLAGN | Low Luminosity AGN |
| LMXRB | Low Mass X-Ray Binaries |
| LSS | Long Slit Spectrograph |
| MAMA | Multi Anode Microchannel Array |
| MC | Magellanic Clouds |
| MCP | Micro Channel Plate |

| MGSE | Mechanical Ground Support Equipment |
|------|-------------------------------------|
| MOC | Mission Operation Center |
| MoU | Memorandum of Understanding |
| MSP | MilliSecond Pulsar |
| MS-TO | Main Sequence Turn Off |
| NASA | National Aeronautics and Space Administration |
| NB | Narrow Band |
| NIMO | Non Inverted Mode |
| NUV | Near Ultraviolet Camera |
| OB | Optical Bench |
| OC | Open Clusters |
| ORFEUS | Orbiting and Retrievable Far and Extreme Ultraviolet Spectrometer |
| PAH | Polycyclic Aromatic Hydrocarbon |
| PI | Polarimetry Image |
| PMS | Pre Main Sequence |
| PMU | Primary Mirror Unit |
| PSF | Point Spread Function |
| PSU | Power Supply Unit |
| QE | Quantum Efficiency |
| QPO | Quasi Periodic Objects |
| QSO | Quasi Stellar Object |
| RIAF | Radiative Inefficient Accretion Flows |
| RON | Read Out Noise |
| RT | Remote Terminal |
| S/C | Spacecraft |
| SagDIG | Sagittarius Dwarf Irregular Galaxy |
| SBF | Surface Brightness Fluctuation |
| SDMU | Science Data Management Unit |
| SED | Spectral Energy Distributon |
| SFR | Star Formation Rate |
| SIC | Science Instrumentation Compartment |
| SIS | Slitless imaging spectroscopy |
| SMBH | Super Massive Black Hole |
| SMU | Secondary Mirror Unit |
| SN | Supernovae |
| SOC | Science Operation Center |
| SpW | Space Wire |
| ST | Science Team |
| STIS | Space Telescope Imaging Spectrograph |

| STM | Structural Thermal Model |
|-----|--------------------------|
| SW | Software |
| TBC | To Be Confirmed |
| TBD | To Be Defined |
| TC | Telecommands |
| TEC | Thermo Electric Cooler |
| TLR | Top Level Requirement |
| TM | Telemetry |
| UART | Universal Asynchronous Receiver-Transmitter |
| UN | United Nations |
| UVES | Ultraviolet Echelle Spectrograph |
| UVO | Ultraviolet Optical Camera |
| UVOT | Ultraviolet and Optical Telescope |
| VUVES | Vacuum Ultraviolet Echelle Spectrograph |
| WFC3 | Wide Field Camera 3 |
| WFPC2 | Wide Field Planetary Camera 2 |
| WIC | WSO-UV Implementation Committee |
| WSA | Wedge & Strip Array |
| WSO | World Space Observatory |
| WSO-UV | World Space Observatory/UltraViolet |
| XDL | Crossed Delay Lines |